\documentclass[epj]{svjour}
\usepackage{graphics}
\usepackage{comment}
\usepackage[normalem]{ulem}
\usepackage{amsmath}
\usepackage{amssymb}
\usepackage{txfonts} 
\usepackage{epsfig} 
\usepackage{enumerate}
\usepackage{xcolor}
\usepackage{ulem}

\def\nuc#1#2{\relax\ifmmode{}^{#1}{\protect\text{#2}}\else${}^{#1}$#2\fi}
\newcommand{\bi}{\begin{itemize}}
\newcommand{\ei}{\end{itemize}}
\newcommand{\be}{\begin{equation}}
\newcommand{\ee}{\end{equation}}

\newcommand{\Jp}{J} 
\newcommand{\Mp}{M} 

\newcommand{\bR}{\vec{R}}
\newcommand{\br}{\vec{r}}
\newcommand{\vecR}{{\vec R}}
\newcommand{\vecr}{{\vec r}}
\newcommand{\vecrp}{\vec{r}\mkern2mu\vphantom{r}'}

\newcommand{\vecRp}{\vec{R}\mkern2mu\vphantom{R}'}

\newcommand{\psixprior}{\varphi^\mathrm{prior}_x} 
\newcommand{\psixpost}{\varphi^\mathrm{post}_x} 
\newcommand{\psixaus}{\varphi^\mathrm{3b}_x} 
\newcommand{\psixno}{\varphi^\mathrm{HM}_x} 
\newcommand{\PsiTB}{\Psi^\mathrm{3b(+)}}

\newcommand{\images}{figs}

\usepackage{xcolor}

\begin{document}
\title{The art of modeling nuclear reactions with weakly bound nuclei: status and perspectives}
\author{Antonio M. Moro  \and Jes\'us Casal  \and Mario G\'omez-Ramos
}                     
%
%
\institute{Departamento de F\'{\i}sica At\'omica, Molecular y Nuclear, Facultad de F\'{\i}sica, Universidad de Sevilla, Apartado 1065, E-41080 Sevilla, Spain}
\date{Received: date / Revised version: date}
%
\abstract{
We give an overview of the theoretical description of nuclear reactions involving weakly-bound nuclei. Some of the more widespread reaction formalisms employed in the analysis of these reactions are briefly introduced, including various recent  developments. We put special emphasis on the continuum-discretized coupled-channel (CDCC) method and its extensions to incorporate core and target excitations as well as its application to three-body projectiles. The role of the continuum for one-nucleon transfer reactions is also discussed. The problem of the evaluation of inclusive breakup cross sections is  addressed within the Ichimura-Austern-Vincent (IAV) model. 
Other methods, such as those based on a semiclasical description of the scattering process, are also briefly introduced and some of their applications are discussed and a brief discussion on topics of current interest, such as nucleon-nucleon correlations, uncertainty evaluation and non-locality is presented.   
\PACS{
      {24.10.-i }{Nuclear reaction models and methods}   \and
      {24.50.+g}{Direct reactions}
       \and
       {25.45.De}{ Elastic and inelastic scattering}
       \and
       {25.70.De}{Coulomb excitation}
       \and
         {25.70.Hi} {Transfer reactions}
     } 
} 

\maketitle
\setcounter{tocdepth}{4} 
\tableofcontents
\section{Introduction}
\label{intro}
Nuclear reactions are key tools to extract information about the structure of atomic nuclei, and the dynamical phenomena arising from nucleus-nucleus forces. When one of the colliding partners is weakly bound, new phenomena and mechanisms arise, requiring in some cases specific reaction frameworks tailored to the peculiarities of these systems and their interactions.  

Weakly-bound nuclei appear in the proximity of the neutron and proton driplines, a region where new exotic structures and phenomena are found. Promiment examples are {\it halo nuclei}, weakly bound nuclei composed of a compact core and one or two loosely bound nucleons with an unusually large matter radius, or {\it Borromean systems}, three-body systems with no bound binary subsystems, such as $^{9}\text{Be}~(\alpha+\alpha+n)$ or $^6\text{He}~(\alpha+n+n)$.

Although the field has experienced a great impulse in the last decades, the physics of nuclear scattering with weakly-bound projectiles is not new and many of the problems that are being addressed were already recognized much earlier as, for example, in the context of low-energy deuteron scattering.  The deuteron, while being a stable nucleus, displays many halo-like features, such as weak binding ($E_b=2.22$~MeV), large spatial extension (the proton-neutron separation is about 3.8~fm) and no bound excited states.

\begin{figure}
\centering
\resizebox{0.6\columnwidth}{!}{\includegraphics{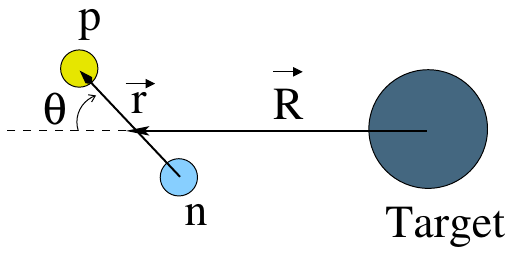}}
\caption{Relevant coordinates for a deuteron+target scattering problem.}
\label{fig:dA_coord}  
\end{figure}

To illustrate the role of the weak binding on the scattering observables, let us consider the scattering of low-energy deuterons (a few MeV) by a heavy target nucleus, like \nuc{208}{Pb}. Initially, when the deuteron is far apart from the target nucleus, it is in an internal state given by the ground-state of the proton-neutron Hamiltonian. As the deuteron approaches the target, it will feel its Coulomb repulsion. If the deuteron were a point-like particle, the effect of this interaction would be to distort the {\it trajectory} of the deuteron, without altering its internal structure. However, this is not the case. The Coulomb interaction acts on the proton, whose distance from the center-of-mass of the deuteron is $\sim$2~fm. Consequently, in addition to the monopole Coulomb potential ($\propto 1/R$) the deuteron will feel higher-order Coulomb multipoles arising from the expansion, valid for $r\ll R$, given by \cite{Cle62b} 
\be
V_C(\br,\bR)=\frac{Z_t e^2}{|\bR + \br/2|} = \frac{Z_t e^2}{R} - \frac{Z_t e^2}{2 R^2} r \cos(\theta) + \ldots
\label{eq:Vp}
\ee
where $\bR$ is the coordinate from the target to the deuteron c.m.\,  $\br$ the proton-neutron relative coordinate and $\theta$ is the angle between them (see Fig.~\ref{fig:dA_coord}). The main deviation from the point Coulomb interaction is caused by the dipole term which depends on the orientation of the proton-neutron relative coordinate with respect to the deuteron-target coordinate. Specifically, when the proton is closer to the target nucleus ($\cos(\theta)<0$), the dipole potential will add a positive contribution, whereas when the proton is farther from the c.m.\ of the deuteron the contribution will be negative. Classically, we may think of this problem as an electric dipole moving in a slowly varying electric field, in which the tidal force exerted on the dipole will favour the configuration of Fig.~\ref{fig:dA_coord}. Quantum-mechanically, the perturbation induced by the dipole force will couple the deuteron ground state (positive parity) with negative-parity states. Since the ground-state is the only bound state, these states necessarily appear in the continuum. Therefore, the polarization effect will modify the state of the system, producing a new one in which the ground state is mixed with continuum states of negative parity. This has a twofold effect on the outcome of the scattering process. First, it will produce deviations of the elastic scattering with respect to the Rutherford formula. Second, the coupling with the positive-energy states will give rise to some dissociation probability, that is, the breakup of the deuteron into a proton and a neutron. 
In the next sections, we will discuss several methods to incorporate these two effects in the reaction formalisms.

The discussed example, albeit describing a specific problem, exhibits some general features commonly found in the scattering of weakly bound nuclei. In particular, the coupling to the breakup channels will play a role, to a larger or smaller extent,  in essentially all reaction observables. Therefore, reaction models employed to describe these  observables will have to incorporate this effect. 

We enumerate some fingerprints of the weak binding on reaction observables:
\begin{itemize} 
\item {\it Large interaction cross sections in nuclear collisions at  high energies.} 
Historically, the first evidence of the unusual properties of halo nuclei came from the pioneering experiments performed by Tanihata and co-workers at Berkeley using very energetic (800 MeV/nucleon)  secondary beams of radioactive species \cite{Tan85a,Tan85b}. At these high energies,  interaction cross sections are approximately proportional to the size of the colliding nuclei. It was found that some exotic isotopes of light nuclei ($^{6}$He, $^{11}$Li, $^{14}$Be) presented much higher interaction cross sections than their neighbour isotopes, which was interpreted as an abnormally large radius. 

\item {\it Narrow momentum distributions of residues following fast nucleon removal}. Momentum distributions of the residual nucleus following the removal of one or more nucleons of a energetic projectile colliding  with a target nucleus are closely related to the momentum distribution of the removed nucleon(s) in the original projectile. Kobayashi {\it et al.} \cite{Kob88} found that the momentum distributions of  $^{9}$Li following the fragmentation process $^{11}\text{Li}+{^{12}}\text{C} \rightarrow {^{9}}\text{Li} + \text{X}$ were abnormally narrow, which, according to the Heisenberg's uncertainty principle, suggested a long tail in the density distribution of the $^{11}$Li nucleus. This result was later found in other weakly bound nuclei. 

\item {\it Abnormal elastic scattering cross sections}. Elastic scattering is affected by the coupling to non-elastic processes. In particular, when coupling to breakup channels is important, elastic scattering cross sections are depleted with respect to the case of tightly bound  nuclei. Some other key signatures are the departure of the elastic cross section from the Rutherford cross section at sub-Coulomb energies and the disappearance of the Fresnel peak at near-barrier energies in reactions induced by halo nuclei on heavy targets \cite{San08,Pie12,Cub12,Pes17}. 

\item {\it Enhanced near-threshold breakup cross section in Coulomb dissociation experiments of neutron-halo nuclei}. When a neutron-halo nucleus, composed of a charged {\it core} and one or two weakly-bound neutrons ($^{11}$Be, $^{6}$He, $^{11}$Li,\ldots) collides with a high-$Z$ target nucleus, the projectile structure is heavily distorted due to the tidal force originated from the uneven action of the Coulomb interaction on the charged core and the neutrons. This produces a stretching which may eventually break up the loosely bound projectile. This gives rise to a large population of the continuum states close to the breakup threshold. 

\item {\it Complete fusion suppression}

Experiments with weakly bound light stable nuclei (such as $^{6,7}$Li and $^{9}$Be) have shown a systematic suppression of  complete fusion (CF)  cross sections   (defined as capture of the complete charge of the projectile) of  $\sim$20-30\% compared to the case of tightly bound nuclei \cite{Das99,Tri02,Das02,Das04,Muk06,Rat09,LFC15}.  The effect has been attributed to the presence of strong competing channels, such as the breakup of the weakly bound projectile prior to reaching the fusion barrier, with the subsequent reduction of capture probability. This interpretation is supported by the presence of large $\alpha$ yields as well as target-like residues which are consistent with the capture of one of the fragment constituents  of the projectile, a process which is usually termed as {\it incomplete fusion} (ICF). 

\end{itemize}

A proper and quantitative understanding of these and other phenomena requires the use of an appropriate reaction theory. 
When dealing with weakly-bound systems, one expects a certain decoupling between the degrees of freedom describing the relative motion between the weakly-bound nucleon (or cluster) from that of the internal excitations of the clusters themselves. Following this argument, one may be tempted to adopt an extreme model in which only the degree of freedom for the relative motion of the weakly-bound nucleon(s) or clusters are considered, while the others are simply ignored or {\it frozen}. This approach, which reminds of the separation between active nucleons and core nucleons in the shell-model, has in fact become very useful to understand the main features of the structure of halo, and other weakly-bound, nuclei and their dynamics. In some cases, such as the deuteron system or one-nucleon halos, such as $^{19}$C or $^{11}$Be, a two-body model may provide a reasonable starting point for the description of these nuclei. However, in some other systems, such as the Borromean nuclei $^{11}$Li, $^{6}$He or $^{9}$Be, it  will be in general mandatory to resort to (at least) a three-body model for a meaningful description of their structure and reactions.

One of the most successful models for describing reactions involving weakly-bound nuclei, which incorporates in a natural way the few-body structure of these systems as well as its breakup, is the  Continuum Discretized Coupled-Channels (CDCC) \cite{Aus87}. The model and its recent extensions will be discussed in Sec.~\ref{sec:cdcc}. Other models, tailored to specific processes, will be discussed along this review.

It is the purpose of this paper to review some of the theories developed and applied for describing nuclear reactions with weakly bound nuclei. As any review, the present one will be necessarily incomplete and possibly biased by the expertise and personal taste of the authors but with the hope that the reader will obtain, at least, a flavour of how reaction theory has helped the interpretation of experiments with these nuclear species.

\section{The impact of weak-binding on the elastic cross section: optical model approach}
\label{sec:om}
Elastic scattering provides a valuable source of information on the structure and dynamics of nuclei. As explained in the Introduction, it was long realised that the elastic scattering of deuterons, admittedly the simplest example of weakly bound nucleus, does not follow the expected behavior for tightly bound nuclei.

As in the case of well-bound nuclei, the natural framework to study elastic scattering is the Optical Model (OM), which consists in solving a two-body Schr\"odinger equation with an effective nucleus-nucleus potential, referred to as the {\it optical model potential} (OMP) \cite{Sat83}. The form of this OMP can be formally derived from the microscopic nucleus-nucleus interaction. In his seminal work, Feshbach \cite{Fesh58} showed that the OMP can be formally written as a sum of two terms, a {\it bare potential}, which is the expectation value of the nucleus-nucleus potential in the ground state of the projectile+target system, and an additional term, the {\it polarization potential}, which accounts for the effect of non-elastic channels (inelastic scattering, transfer, breakup, fusion, \ldots) on elastic scattering: Explicitly:
\begin{equation}
\mathcal{V} = V_{00} + \mathbf{V}_0 \frac{1}{E^{(+)} - \mathbf{H}} \mathbf{V}_0^{\dagger}.
\label{eq:u_fesh}
\end{equation}
where the first and second terms correspond, respectively, to the bare and polarization potentials. Here,  $\mathbf{H}$ stands for the full, microscopic Hamiltonian, $\mathbf{V}_0$ is an operator with  non-vanishing matrix elements between the ground-state and other states of the system, $\mathbf{V}_0=(V_{01}, V_{02}, \ldots)$. This effective effective potential gives rise to the Schr\"odinger equation for the relative-motion wavefunction:

\begin{equation}
\left[ T_{\bR} + \mathcal{V} - E \right] \chi_0(\bR) = 0
\label{eq:sch_upol}
\end{equation}
where $T_{\bR}$ is the kinetic energy operator   and $\chi_0$ the wave-function for the relative motion between projectile and target in their ground states.

The effect of weak binding of any of the colliding partners affects both the bare and polarization potentials. The ground-state of the projectile will exhibit an extended tail (as compared to the case of tightly-bound nuclei) and this will influence the form of $V_{00}$. In addition, the weak-binding will enhance certain couplings described by the $\mathbf{V}_0$ operator, very prominently the breakup channels, and this will also modify the polarization term. In general, the evaluation of the Feshbach operator associated with the polarization potential is very involved and very often it is replaced by some phenomenological form. A few cases exist however in which a explicit evaluation is feasible, at least for specific nonelastic processes. One such example is the so-called adiabatic Coulomb polarization potential, briefly introduced in the next subsection.

\subsection{Evaluation of the polarization potential in a simple case: the adiabatic polarization potential}
\label{sec:vad}
In the introductory section, we described the case of deuteron scattering by a heavy-target nucleus. It was argued that the tidal force caused by the action of the Coulomb field on the proton produced a modification of the Coulomb-target interaction with respect to the point-Coulomb case [c.f.~Eq.~(\ref{eq:Vp})]. 
Quantum-mechanically, the effect of this modified Coulomb interaction will be to induce coupling to deuteron breakup states. This is in fact one of the contributions included in the Feshbach polarization potential given by the second term of 
Eq.~(\ref{eq:sch_upol}). This can be analytically evaluated in the so-called adiabatic limit, which assumes that the excitation energies are high enough so the characteristic time for a transition to a state is small compared to the characteristic time for the collision \cite{Cle62b}. Applying
second-order perturbation theory, one gets the following expression for this {\it adiabatic polarization potential} \cite{Cle62b,Ald75} :
\begin{equation}
  V_\mathrm{pol}(R) = -\sum_{n \neq 0} \frac{|\langle n|V_\mathrm{dip}|0 \rangle|^2}{E_n - E_0} = - \frac{1}{2} \alpha \frac{(Z_T e)^2}{R^4},
  \label{eq:vad}
\end{equation}
where $V_\mathrm{dip}$ is the second term of Eq.~(\ref{eq:Vp}) and $\alpha$ is the dipole polarizability parameter \cite{Cle62b}, which measures the electric response of the nucleus (the deuteron in this case) to an external electric field. It is therefore a structure property of any nucleus.

It is to be noted that the adiabiatic polarization potential given by Eq.~(\ref{eq:vad}) is purely real and, as such, does not describe the effect of the deuteron breakup that would remove flux from the elastic channel. 
 This point will be revisited in Sec.~\ref{sec:DPP}, where another version of the polarization potential, which accounts for deuteron breakup, will be presented.

The polarization potential (\ref{eq:vad}) adds an attractive contribution to the Coulomb point-particle potential that, when used in the Schr\"odinder equation, will produce a small, but measurable deviation of the elastic cross section with respect to the well-known Rutherford formula. An accurate measurement of sub-Coulomb elastic scattering data can therefore be used to extract the dipole polarizability of the deuteron (and other polarizable systems). This idea was used by Rodning {\it et al.} \cite{Rod82} to infer the deuteron polarizability parameter from the analysis of sub-Coulomb deuteron elastic scattering on lead.

\begin{figure}
\centering
\resizebox{0.75\columnwidth}{!}{\includegraphics{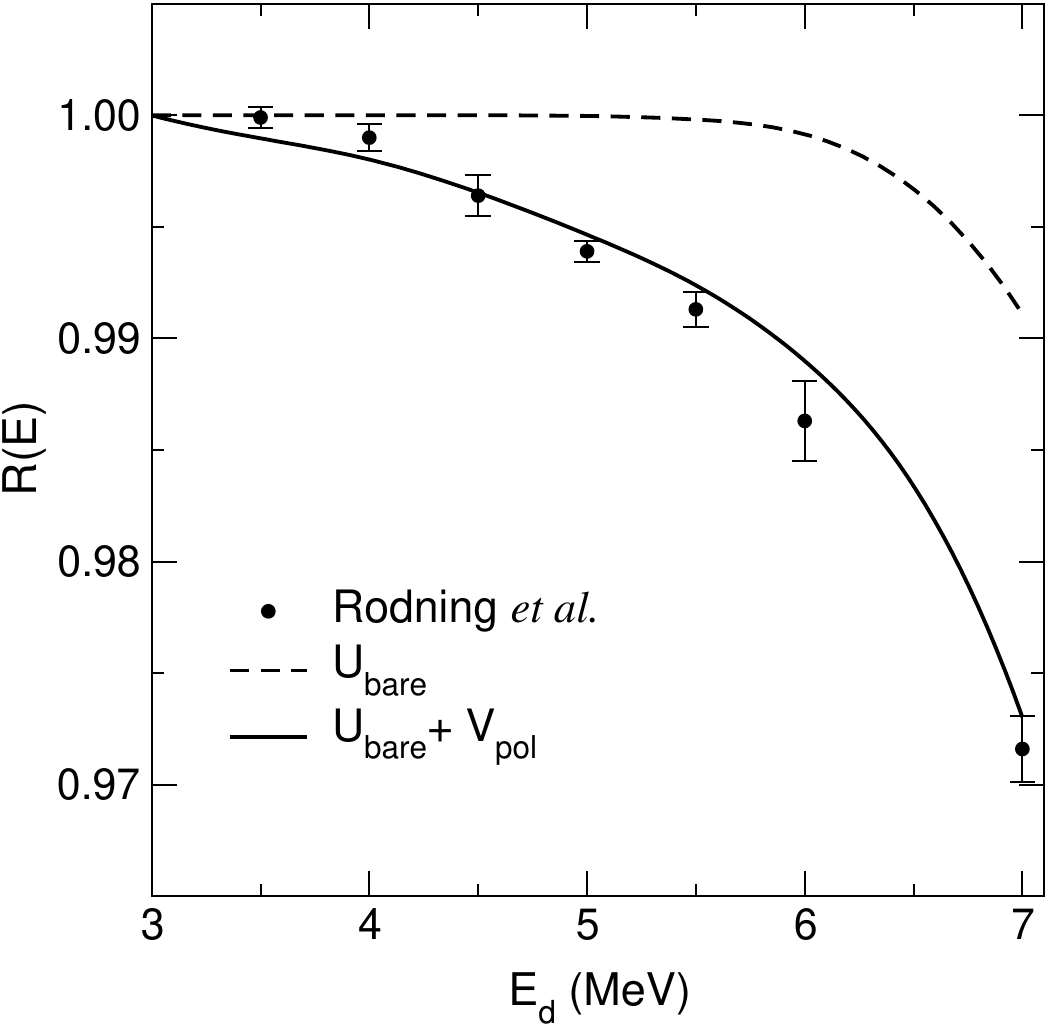}}
\caption{Experimental \cite{Rod82} values for the quantity $R(E)$, defined in Eq.~(\ref{eq:RE}) for deuteron scattering on $^{208}$Pb at sub-Coulomb energies. The dashed line is a single-channel calculation including only the nuclear potential. The solid line is the single-channel calculation include also the effect of the Coulomb dipole polarizability by means of the adiabatic polarization potential of Eq.~(\ref{eq:vad}). See text for details.}
\label{fig:RE}  
\end{figure}

To avoid the determination of absolute cross sections, in their analysis the authors of Ref.~\cite{Rod82} introduced the adimensional quantity
\begin{equation}
  R(E) = \frac{\sigma(E = 3 \text{ MeV}, \theta_1 = 60^\circ)}{\sigma(E = 3 \text{ MeV}, \theta_2 = 150^\circ)} \frac{\sigma(E, \theta_2 = 150^\circ)}{\sigma(E, \theta_1 = 60^\circ)}.
  \label{eq:RE}
\end{equation}
which is identically unity for pure Rutherford scattering. The deuteron polarizability will give rise to values of $R(E)$ which are less than unity and the deviation increases with increasing scattering energy. In Fig.~\ref{fig:RE} we show the measured  values of $R(E)$ from \cite{Rod82} along with two calculations. One in which only a nuclear potential is included, which plays the role of a bare part of Eq.~(\ref{eq:u_fesh}) and a second calculation including the effect of the deuteron poloarizability in the adiabatic approximation, Eq.~(\ref{eq:vad}), with $\alpha=0.70 \pm 0.05$~fm$^3$, the value extracted in  \cite{Rod82} by comparison with the data.

This constitutes a neat and beautiful example in which an important structure quantity can be inferred from reaction observables.   


\subsection{The phenomenological optical model}
Most practical applications of the OM rely on approximate forms of the Feshbach potential, consisting typically of a central potential (with possibly spin-orbit and tensor terms) whose radial parts are parametrized in terms of simple analytical forms, such as the popular Woods-Saxon potential. Because the Feshbach potential is complex, so is the phenomenological optical potential. In principle, the potential should be also non-local and angular-momentum and energy dependent. While the energy dependence is a common feature of phenomenological potentials, nonlocalilty and angular-momentum dependence is more rarely taken into account (see Sec.~\ref{sec:nonlocal}).



\begin{figure}
\centering
\begin{minipage}[t]{.48\columnwidth}
\includegraphics[width=0.95\columnwidth]{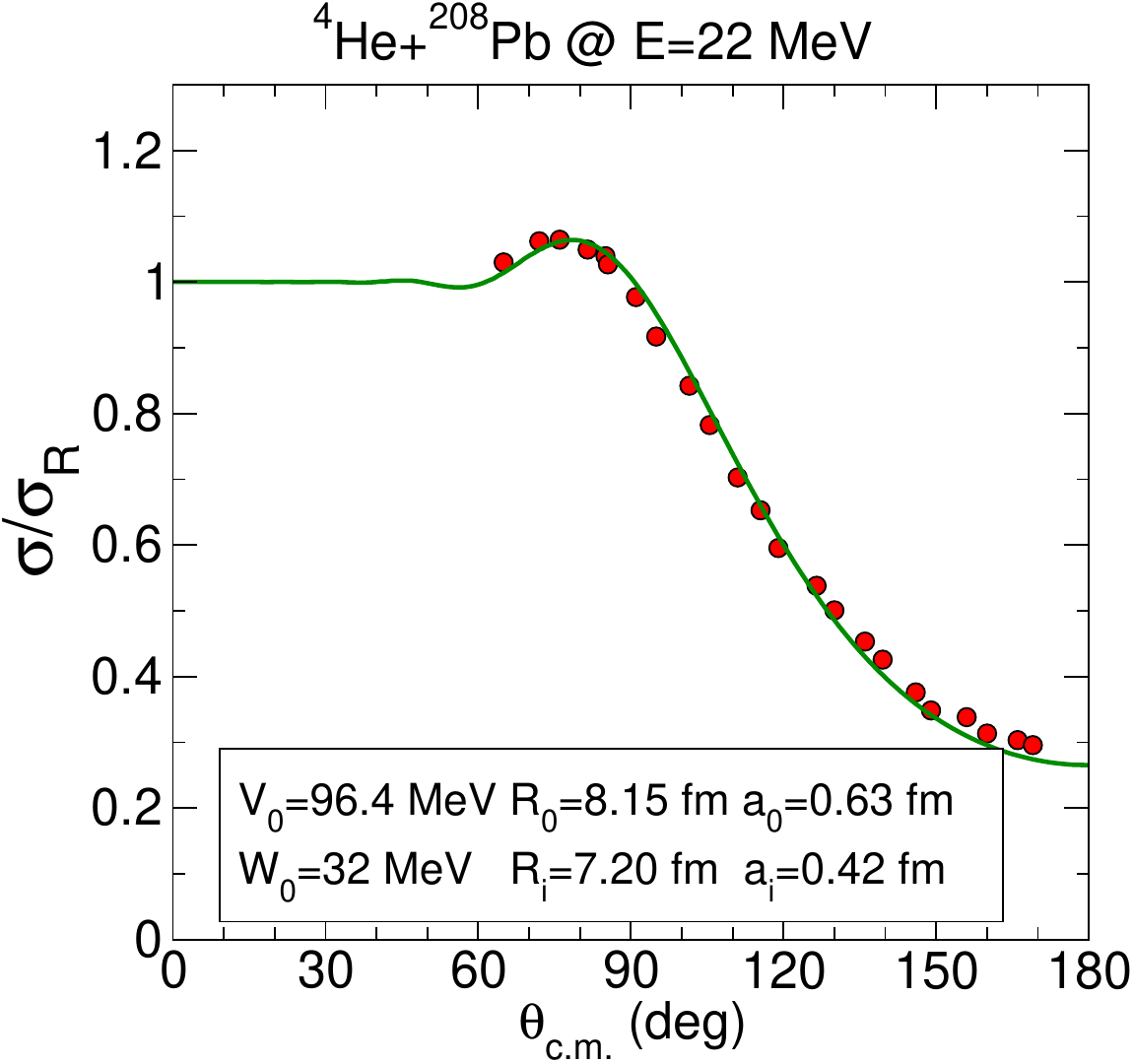} 
\end{minipage}
\begin{minipage}[t]{.48\columnwidth}
\includegraphics[width=0.95\columnwidth]{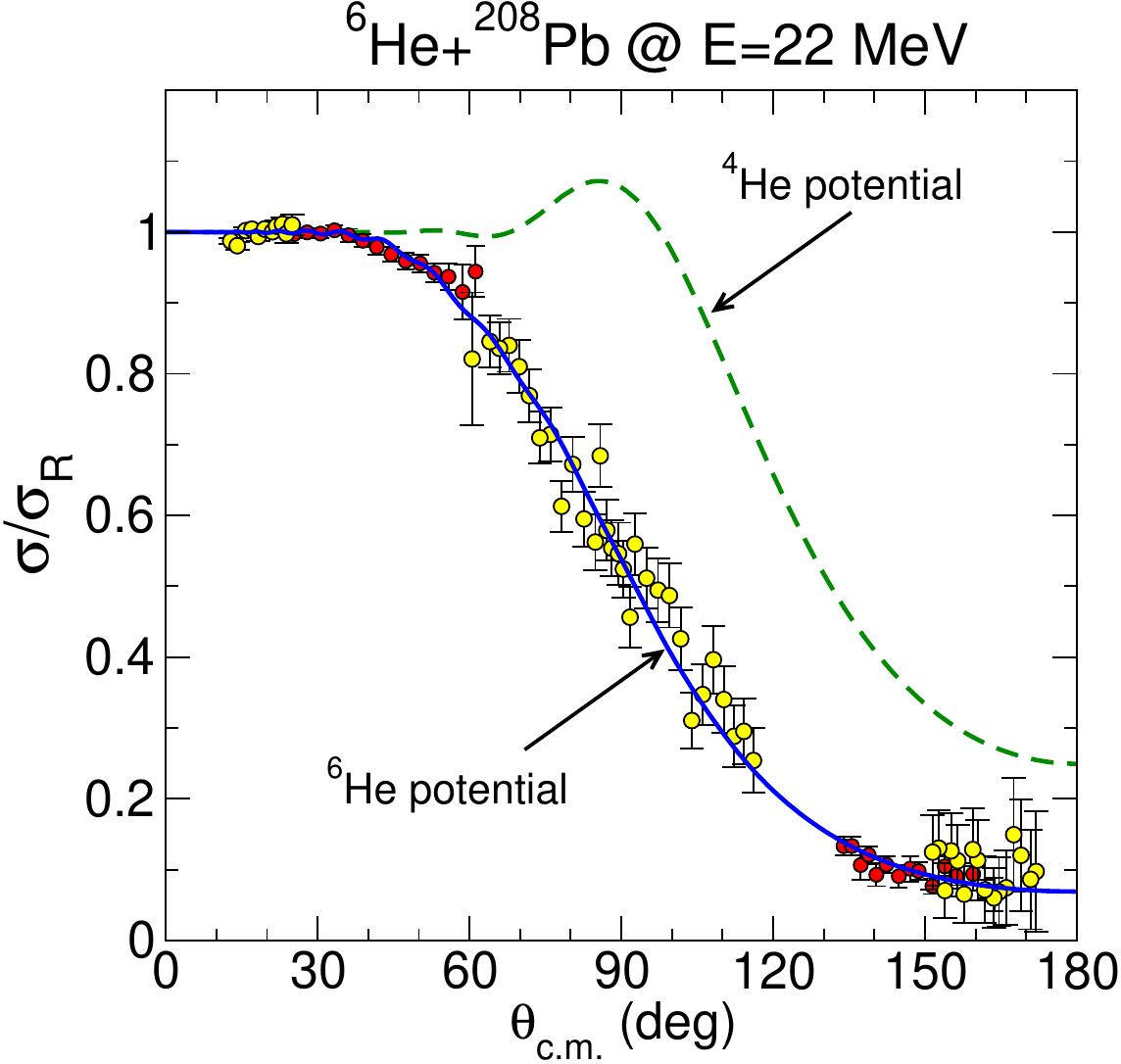}
\end{minipage}
\caption{ \label{fig:he46pb_om} Experimental data for $^{4,6}$He+$^{208}$Pb compared with OM calculations, with Woods-Saxon forms with parameters given in Table \ref{Table:he46pb_omp}. Experimental data from \cite{Bar74,Aco11}}
\end{figure}

In the presence of strong absorption, elastic scattering angular distributions between composite nuclei display some common features regardless of the interacting nuclei. At near-barrier energies, elastic distributions present two distinct regions. At small angles, where the Coulomb interaction dominates, the cross section remains close to the Rutherford formula prediction, but oscillates around it. 
The amplitudes of the oscillations increase as the scattering angle increases but remain roughly 25\% within the Rutherford prediction. This is the so-called ``illuminated'' region. After this maximum, the cross section drops rapidly, becoming much smaller than the Rutherford cross sections at large angles (the ``shadow'' region).  An example of this behaviour can be seen in the elastic scattering data of $\alpha$ particles on lead, shown in the left panel of Fig.~\ref{fig:he46pb_om}. 

These features have been commonly interpreted by invoking classical models. The oscillatory behaviour in the illuminated region can be qualitatively interpreted within a Fresnel picture as an interference between distant pure Coulomb {\it trajectories} with closer trajectories affected by the nuclear attractive potential, while the shadow region can be understood as a consequence of the existence of a maximum angle in the classical deflection function \cite{Sat83}. A complete understanding of the observed pattern requires also the introduction of refractive effects which arise in nuclear scattering as a result of absorption~\cite{Fri89}. 

All these effects are naturally accommodated within the OM. 
For example, the solid line in the left panel of Fig.~\ref{fig:he46pb_om} represents an OM fit of the data using a standard (complex) Woods-Saxon potential. Although the OM is not unique, one may find potentials with radii close to the sum of the geometrical radii of the interacting nuclei and diffuseness parameters close to those of the nuclear densities.

In the case of reactions involving weakly bound nuclei, the observed elastic scattering angular distributions display notable deviations with respect to the aforementioned behaviour. The effect becomes more evident in the case of halo nuclei, such as $^6$He or $^{11}$Li. The oscillations in the illuminated region are damped (or even absent). The drop of the cross section with respect to the Rutherford prediction starts at smaller angles and hence the cross section exhibits a smoother angular dependence.  An example is given by the $^{6}$He+$^{208}$Pb data taken at $E_\mathrm{lab}=22$~MeV shown in the right panel of Fig.~\ref{fig:he46pb_om}.

This behaviour can also be described within the OM but with potential parameters very different from those used in standard parametrizations extracted from well-bound nuclei. In particular, one needs very diffuse potentials with long-range absorptive tails.  The effect of this long-range absorption is twofold: (i) it damps or suppresses the elastic Coulomb amplitude at small angles (i.e., at very large impact parameters), and (ii) it reduces the nuclear amplitude. 
The result is that the forward cross-section not only becomes smaller overall, but also loses its Coulomb-nuclear oscillations.

In order to reproduce the elastic scattering data of weakly-bound nuclei within the OM framework using standard Woods-Saxon potentials, a very long-range imaginary part is needed. Table \ref{Table:he46pb_omp} lists the OMP parameters for  $^4$He+$^{208}$Pb and  $^6$He+$^{208}$Pb at the same incident energy (22 MeV). The radii of the real and imaginary parts are kept the same for both systems. To reproduce the data, the $^{6}$He OM requires a significantly larger imaginary diffuseness parameter. The result of the OM calculation for $^{6}$He is compared with the data for this reaction in the right panel of Fig.~\ref{fig:he46pb_om}.


\begin{table}
\begin{center}
\begin{tabular}{ ccccccc }
\hline
\hline
System   &   $V_0$    & $r_0$    & $a_0$  &   $W_v$  & $r_i$  & $a_i$    \\  
	   &   [MeV]  & [fm] &  [fm]  & [MeV] & [fm]  & [fm]     \\  
\hline
$^{4}$He+$^{208}$Pb &   96.44 & 1.085  & 0.625  &  32   & 0.958 & 0.42  \\
$^{6}$He+$^{208}$Pb &   124.8 & 1.085  & 0.564  &  6.8 & 0.958 & 1.91    \\
\hline  
\end{tabular}
\caption{\label{Table:he46pb_omp} Woods-Saxon parameters for  $^{4,6}$He+$^{208}$Pb optical models. Reduced radii ($r_x$) are converted into absolute (physical) radii as 
$R_x=r_x (A^{1/3}_p + A^{1/3}_t)$.}
\end{center}
\end{table}

This property is a clear indication of the presence of reaction channels dominated by long-range couplings. Natural candidates are the breakup of the projectile and neutron transfer (either one or two) from the projectile to the target nucleus. Both processes are of a peripheral nature, which explains the long-range imaginary potential. To identify the actual channels producing this effect one must go beyond the single-channel OM scheme and incorporate those channels into the reaction framework. In the case of the breakup of the projectile, this can be efficiently accomplished with the Continuum-Discretized Coupled-Channels (CDCC) method, which is described in the next sections.  

\section{Inclusion of breakup: the CDCC method}
\label{sec:cdcc}

\subsection{Brief resum\`e of the CDCC method}

\begin{figure}
\centering
\begin{minipage}[t]{.47\columnwidth}
\includegraphics[width=0.85\columnwidth]{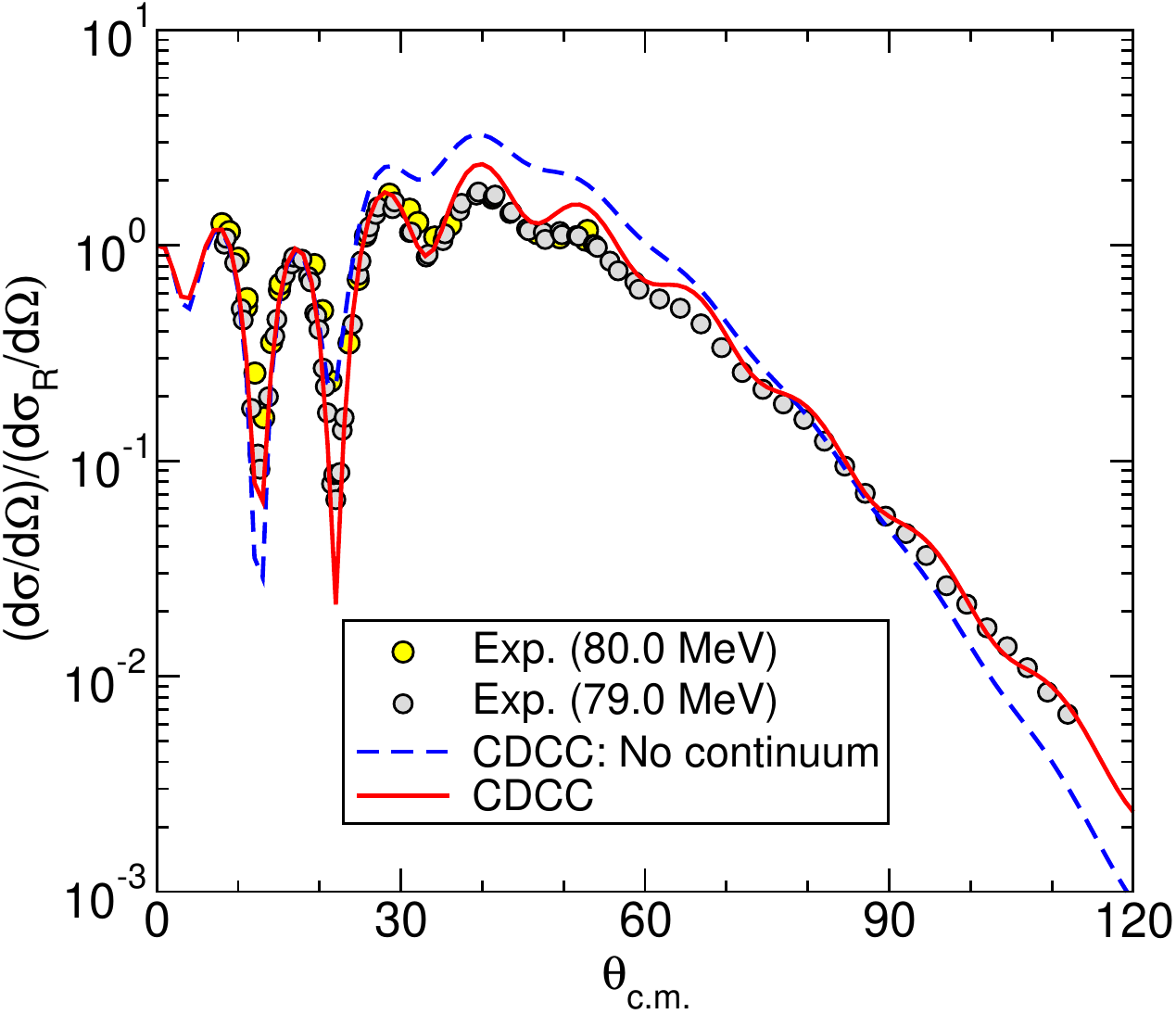} 
\end{minipage}
\begin{minipage}[t]{.47\columnwidth}
\includegraphics[width=0.85\columnwidth]{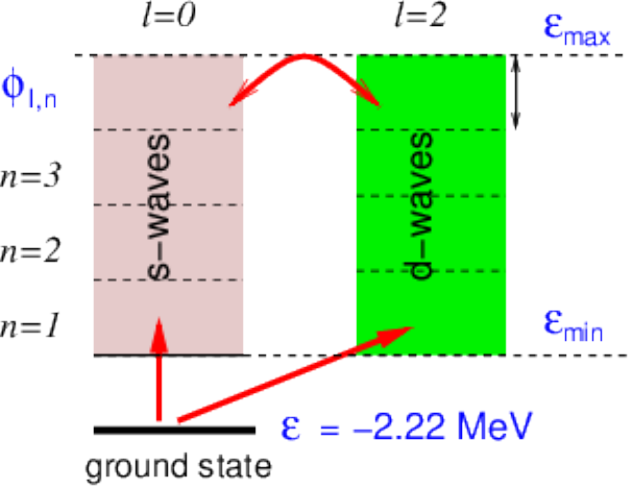}
\end{minipage}
\caption{Left: Application of the CDCC method to $d$+$^{58}$Ni elastic scattering at $E_d=80$~MeV. The solid line is the full CDCC calculation. The dashed line is the calculation omitting the breakup channels. The calculations were performed ignoring the internal spins of the proton and neutron and so $\ell=j$. Right: Illustration of continuum discretization for the same problem. \label{fig:dni_cdcc}} 
\end{figure}

The CDCC method was originally introduced by G.\ Rawitscher \cite{Raw74} and later refined by the Pittsburgh-Kyushu collaboration \cite{Yah86,Aus87} to describe the effect of the breakup channels on the elastic scattering of deuterons. Denoting the reaction by $a+A$, with $a=b+x$ (referred hereafter as the {\it core} and {\it valence} particles, respectively), the method assumes the effective three-body Hamiltonian
\begin{equation}
\label{eq:Heff}
H= H_\text{proj} + T_{\bR}   + U_{bA}(\br_{bA}) + U_{xA}(\br_{xA}) ,
\end{equation}
with $H_\mathrm{proj}=T_{\br}+V_{bx}$ the projectile internal Hamiltonian,   $T_{\br}$ and $T_{\bR}$ are kinetic energy operators, $V_{bx}$ the inter-cluster interaction and $U_{bA}$ and $U_{xA}$ are the core-target and valence-target optical potentials (complex in general) describing the elastic scattering of the corresponding $b+A$ and $x+A$ sub-systems, at the same energy per nucleon of  the incident  projectile. 
In the  CDCC  method  the  three-body wave function of the system is expanded in terms of the eigenstates of the Hamiltonian $H_\mathrm{proj}$ including both bound and unbound states. Since the latter form a continuum, a procedure of discretization is applied, consisting in representing this continuum by a finite and discrete set of square-integrable functions. In actual calculations, this continuum must be truncated in excitation energy and limited to a 
finite number of partial waves associated with the relative co-ordinate $\vec r$. Normalizable states representing the continuum should be obtained for each set of orbital and total angular momenta. Two main methods are used for this purpose:
\bi
\item {\it The pseudo-state method}, in which the $b+x$ Hamiltonian is diagonalized in a basis of square-integrable functions, such as Gaussians \cite{Kaw86a} or transformed harmonic oscillator functions \cite{Mor09b}. Negative eigenvalues correspond to the bound states of the systems, whereas positive eigenvalues are regarded as a finite representation of the continuum. 
\item {\it The binning method}, in which  normalizable states are obtained by constructing a wave packet ({\it bin}) by linear superposition of the actual continuum states over a certain energy interval \cite{Aus87}.
\ei


We describe the latter method in some more detail.  Assuming for simplicity a spinless core, these discretized functions are denoted as 
%
\begin{align}
\label{eq:PsiCDCC}
 \phi_{n} (\br)  = \frac{u_{\ell_n  j_n}(r)}{r} [Y_{\ell}(\hat{r}) \otimes \chi_s  ]_{j_n m_{j_n}} ,
\end{align}
where $n \equiv \{[k_n,k_{n+1}] \ell_n, s, j_n, m_{j_n} \}$ specifies the $n$-th bin, with $[k_n,k_{n+1}]$ the wavenumber  interval of the bin, $\ell$ the valence-core orbital angular momentum, $s$ the valence spin, and $\vec{j}=\vec{\ell}+ \vec{s}$ the total angular momentum. 
The symbol $\otimes$ denotes angular momentum coupling. The radial part of the bin is obtained as a linear combination (i.e., a wave packet) of  scattering states  as 
\be
u_{\ell_n  j_n} (r) = \sqrt {\frac{2 }{\pi N_n}} ~~
            \int _ {k_n} ^ {k_{n+1}} w_n(k) u_{k,\ell  j} (r) dk ,
\ee
where $u_{k,\ell  j}(r)$ are the scattering states for a continuum energy $\varepsilon=\hbar^2 k^2/2\mu$ and angular momentum quantum numbers $\ell,s,j$,  $w_n(k)$ is a weight function (for non-resonant continuum $w_n(k)$ is usually taken as $e^{i \delta_{\ell}}$, where $\delta_{\ell}$ are the phase shifts \cite{Joa75} of the scattering states within the bin) and $N_n$ is a normalization constant.  The effect of this averaging is to damp the oscillations at large distances, making the bin wavefunction normalizable. 

Assuming a single bound state for simplicity, the CDCC wavefunction can be written  
\be
\Psi^{\mathrm{CDCC}}(\bR,\br) =\chi_{0}^{(+)}(\bR) \phi_{0}(\br) + \sum_{n=1}^{N} \chi_{n}^{(+)}(\bR) \phi_{n}(\br) ,
\label{PsiCDCC}
\ee 
where the index $n=0$ denotes the ground state of the $b+x$ system.  

This model wave function  must verify the Schr\"odinger equation: $[H-E] \Psi^{\mathrm{CDCC}}(\bR,\br)=0$. This gives rise to a set of coupled differential equations for the unknowns $\chi_{n}^{(+)}(\bR)$
\begin{equation}
\label{eq:cc}
\left[E-\varepsilon_{n}-T_{\bR} -  U_{n n}(\bR) \right] \chi^{(+)}_{n}(\bR)  = 
\sum_{m \neq n} U_{nm}(\bR) \chi^{(+)}_{m}(\bR) ,
\end{equation}
where $\varepsilon_{n}= \langle \phi_n | H_\text{proj} | \phi_n \rangle$ and $U_{nm}(\bR)$ are the coupling potentials given by 
\be
\label{eq:Vij_3bCDCC}
U_{nm}(\vecR) = \int d\vec r \; \phi^*_{n}(\br)  \left [U_{bA}+U_{xA} \right ]  \phi_{m}(\br) \, .
\ee

The inclusion of the breakup channels in expansion (\ref{PsiCDCC})  will affect the elastic channel wavefunction through the coupling potentials $U_{0n}$ ($n>0$). The entrance flux, given by the norm of the plane wave associated with the entrance channel, will be distributed among the elastic and breakup channels and this will naturally reduce the elastic cross section with respect to the situation in which those breakup channels are omitted. This effect turns out to be essential for a correct description of the elastic cross sections, as illustrated in Fig.~\ref{fig:dni_cdcc} for the d+$^{58}$Ni reaction at 80~MeV. The figure also depicts the discretization scheme employed in these calculations, which comprises $\ell=0,2$ continuum states.

\begin{figure}
\begin{center}
\includegraphics[width=0.6\columnwidth]{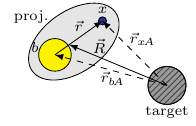} 
\caption{\label{fig:cvt} Relevant coordinates for the description of the scattering of a two-body composite nucleus by a target.}
\end{center}
\end{figure}

In actual calculations, such as those performed by the popular coupled-channels code FRESCO \cite{fresco}, the total three-body wavefunction is expanded in states with good total angular momentum ($J_T$), i.e.,
\begin{equation}
\Psi(\vecR,\vecr)=\sum_{\beta_i,J_T, M_T}  C_{\beta_i,J_T,M_T} \Psi_{\beta_i,J_T, M_T}(\vec{R},\vec{r})   
\end{equation}
with
\begin{equation}
\Psi_{\beta_i,J_T, M_T}(\vec{R},\vec{r})=\sum_{\beta}
\frac{\chi_{\beta,\beta_i}^{J_T}(R)}{R} 
\left[Y_L({\hat{R}})\otimes\phi_{n,\Jp}(\vec{r})\right]_{J_T,M_T},  
\label{f3b}
\end{equation}
where, for the internal states, $\phi_{n,\Jp,\Mp}$, we have included an additional subscript $\Jp$ to write explicitly the projectile spin and hence $n$ is now an index to enumerate states with the same $\Jp$.  Also, in the former equation  $\vec{R}$ is the relative coordinate between the projectile
center of mass and the target (assumed by now to be structureless), see
Fig.~\ref{fig:cvt}. 
The index $\beta$ gathers the quantum numbers compatible with a given total angular momentum $J_T$, $\beta\equiv\{L, \Jp, n\}$,
where $L$ (projectile-target orbital angular momentum) and $\Jp$ both couple
to the total spin of the three-body system $J_T$. The spin of the
target is ignored for simplicity of notation.  Each of these sets is called a  {\it channel}. In particular, $\beta_i$ denotes the channels compatible with the initial state of the system (typically, the ground state of the projectile and target nuclei).

The radial coefficients,  $\chi_{\beta,\beta_i}^{J_T}(R)$, from which the
scattering observables are extracted, are calculated
by inserting Eq.~(\ref{f3b}) in the Schr\"odinger equation, giving rise to
a system of coupled differential equations:

\begin{align}
\left(-\frac{\hbar^2}{2 \mu}{d^2 \over dR^2}+ \frac{\hbar^2 L(L+1)}{ 2 \mu R^2} +
 \varepsilon_n -E \right)  \chi^{J_T}_{\beta,\beta_i}(R)  &   \nonumber \\
 +   \sum_{\beta'} U_{\beta,\beta'}^{J_T}(R) \chi^{J_T}_{\beta',\beta_i}(R)  &=  0 
\label{eq:cc_radial}
\end{align}
where $\varepsilon_n$ is the nominal energy of function $\phi_n$ and with the coupling potentials: 
\begin{equation}
U_{\beta,\beta^\prime}^{J_T}(R)=
\langle \beta; J_T|V_{bA}(\vec{R},\vec{r})+V_{xA}(\vec{R},\vec{r})|\beta^\prime;
J_T\rangle , 
\label{cpot}
\end{equation}
where 
\begin{equation}
\langle \hat{R}, \vec{r} |\beta; J_T\rangle =
\left[Y_L({\hat{R}})\otimes\phi_{n,\Jp}(\vec{r})\right]_{J_T}. 
\label{coupledbasis}
\end{equation}

The system of equations (\ref{eq:cc_radial}) is to be solved  numerically  following the methods described elsewhere (see, e.g.~\cite{Hag22} and references therein) and subject to the asymptotic boundary conditions:
\begin{align}
\label{eq:chi_asym}
\chi^{J_T}_{\beta:\beta_i}(K_\beta,R) & \rightarrow   
   \frac{i}{2} \left[H^{(-)}_L(K_\beta R) \delta_{\beta,\beta_i} -  S^{J_T}_{\beta,\beta_i} H^{(+)}_{L}(K_\beta R) \right]  \\
   & \rightarrow 
   \left[  F_L(K_\beta R) \delta_{\beta,\beta_i} + T^{J_T}_{\beta,\beta_i} H^{(+)}_{L}(K_\beta R) \right]   
\end{align}
where $K_\beta=\sqrt{2 \mu (E-\epsilon_n)}$,  $S^{J_T}_{\beta,\beta_i}$ are the S-matrix elements,  $T^{J_T}_{\beta,\beta_i}$ are the T-matrix elements,  $H^{\pm)}_L(KR)$ are the ingoing ($-$) and outgoing ($+$) Coulomb functions and $F_L(KR)$ is the regular Coulomb function \cite{Sat83}.  Because of the relation between the Coulomb functions, the two equations above are fully equivalent. In fact, the coefficients $S^{J_T}_{\beta,\beta_i}$  and $T^{J_T}_{\beta,\beta_i}$ are related by $S^{J_T}_{\beta,\beta_i}= \delta_{\beta,\beta_i}+ 2 i T^{J_T}_{\beta,\beta_i}$. Scattering amplitudes and differential cross sections are expressed in terms of these coefficients (see, e.g., \cite{Glen04,Sat83}).

\begin{figure}
\begin{center}
\includegraphics[width=0.75\columnwidth]{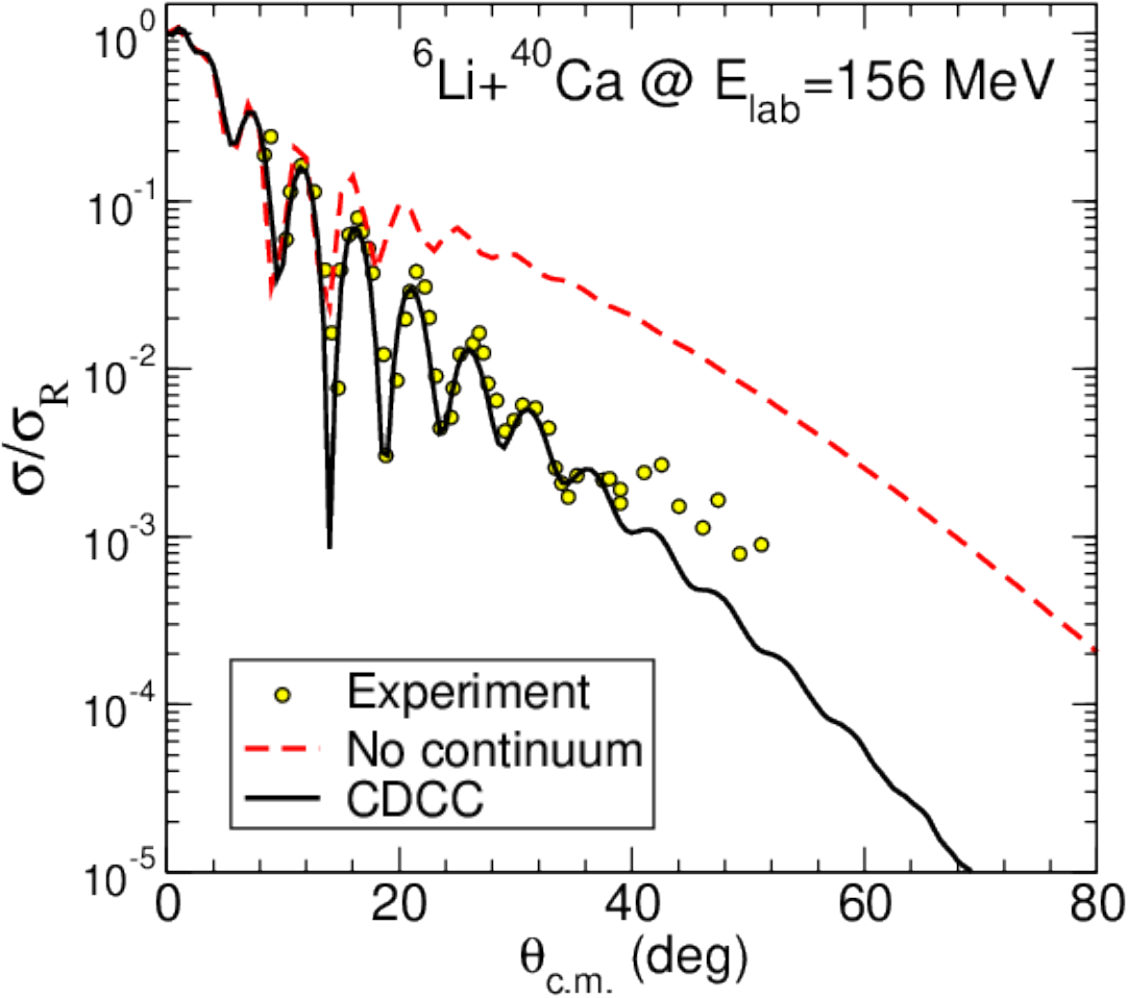} 
\caption{\label{fig:li6_cdcc}Application of the CDCC method to $^6$Li+$^{40}$Ca elastic scattering at 156 MeV. The  solid and dashed lines are the CDCC calculations with and  without inclusion of the $^{6}$Li ($\alpha$+d) continuum. Experimental data are from Ref.~\cite{Maj78}. Adapted from \cite{Mor09b}}.
\end{center}
\end{figure}

The standard CDCC method is based on a strict three-body reaction model ($b+x+A$), and has proven rather successful in describing elastic and breakup cross sections of deuterons and other weakly bound two-body nuclei, such as $^{6,7}$Li and  $^{11}$Be (see Fig.~\ref{fig:li6_cdcc}). However, it has limitations. The assumption of inert bodies is not always justified, since excitations of the projectile constituents ($b$ and $x$) and of the target ($A$) may take place along with projectile dissociation. Furthermore, the two-body picture may be inadequate for some nuclei as, for example, in the case of the  Borromean systems (e.g.~$^{6}$He,  $^{11}$Li).  Some extensions of the CDCC method to deal with these situations are outlined below.

\subsection{The role of closed channels}
In principle, expansion (\ref{PsiCDCC}) may include internal states with excitation energies below ($\varepsilon_i < E$) or above
($\varepsilon_i > E$) the total energy, which are referred to as open and closed channels, respectively. Closed channels violate energy conservation and, as such, cannot contribute asymptotically to the scattering wavefunction. However, due to the couplings with other channels, they can affect the open channels and hence the cross sections. 

\begin{figure}[!ht]
\begin{center}
\includegraphics*[width=0.86\columnwidth]{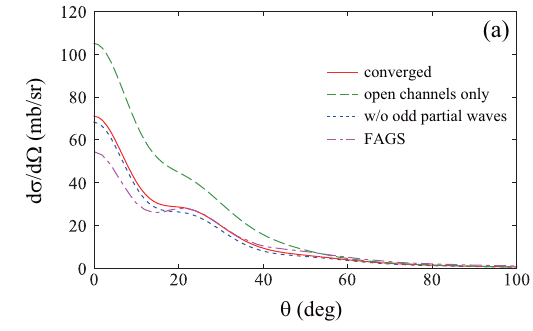}
\includegraphics*[width=0.86\columnwidth]{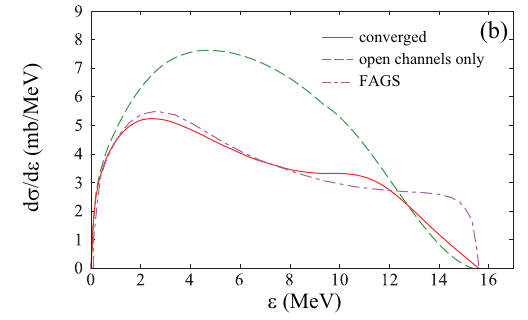}
\caption{
(a) Angular distribution and (b) breakup energy distribution of the elastic breakup cross section for $^{12}$C($d,pn$)$^{12}$C at 12~MeV. The solid, dashed, and dash-dotted lines in each panel show the converged CDCC result, the result of the CDCC method calculated with only the open channels, and the result of the Faddeev-AGS (FAGS) theory taken from Ref.~\cite{Upa12}, respectively. The dotted line in (a) is the same as the solid line but omitting the odd partial waves between $p$ and $n$. Taken from~\cite{Oga16}, with authorization from APS.
}
\label{fig3-8}
\end{center}
\end{figure}

The solution of the CDCC equations (\ref{eq:cc_radial}) in the presence of closed channels is totally analogous to the case with open channels, except for the modification of the boundary condition of the radial functions associated with closed channels, which now reads
\begin{align}
\label{eq:chi_asym_closed}
\chi^{J_T}_{\beta:\beta_i}(K_\beta,R) & \rightarrow   
     C_{\beta,\beta_i} W_{-\eta_i,L+1/2}(-2 i K_\beta R) ,
\end{align}   
where $W_{-\eta_i,L+1/2}$ is a Whittaker function with $\eta$ the  Sommerfeld parameter, defined as $\eta=Z_p Z_t e^2 /\hbar v$. In general, the effect of closed channels is small at sufficiently high energies. However, at low energies, they can have a strong influence on the scattering observables. An example is shown in Fig.~\ref{fig3-8}, corresponding to the reaction $d+^{12}\mathrm{C} \rightarrow p+n + ^{12}\mathrm{C}$ at 12~MeV \cite{Oga16}. CDCC calculations with and without the inclusion of closed channels are compared with Faddeev calculations performed in the momentum-space formulation of Alt, Grassberger and Sandhas (FAGS) \cite{AGS}, which provide an essentially exact solution of the same Hamiltonian and therefore also account for the effect of closed channels. The upper and lower panels are for the angular and excitation energy distributions. The importance of closed channels is clearly seen. Only when closed channels are included, the CDCC calculation approaches the Faddeev solution.

\subsection{Trivially equivalent polarization potential}
\label{sec:telp}

From a coupled-channel calculation (in particular, CDCC) one may extract an effective polarization potential, also called {\it trivial equivalent local polarization potential} (TELP). For simplicity, we assume that there is only one entrance channel ($\beta_i=\beta_0$) so that the label $\beta_i$ in (\ref{eq:cc_radial_el}) can be omitted. Then the radial equation for the elastic channel becomes:
\begin{align}
\left(-\frac{\hbar^2}{2 \mu}{d^2 \over dR^2}+ \frac{\hbar^2 L_0(L_0+1)}{2 \mu R^2} + U_{\beta_0,\beta_0}^{J_T}(R) + \varepsilon_0 -E   
 \right)  \chi^{J_T}_{\beta_0}(R)  &   \nonumber \\
 = -  \sum_{\beta'} U_{\beta_0,\beta'}^{J_T}(R) \chi^{J_T}_{\beta'}(R)  &=  0 
\label{eq:cc_radial_el}
\end{align}

The right-hand side of this equation is then used to define the angular-momentum dependent polarization potential:
%
\begin{equation}
U_\text{pol}(R)=\frac{\sum_{J_T}w_{J_T}(R)U_{J_T}^\text{TELP}(R)}{\sum_{J_T} w_{J_T}(R)}, 
\label{eq:Ueff}
\end{equation}
where $U_{J_T}^{TELP}(R)$ is the  ``trivial equivalent potential'' for the total angular momentum $J_T$ defined by
%
\begin{equation}
U_{J_T}^{TELP}(R)=\frac{1}{\chi^{J_T}_{\beta_0}(R)}\sum_{\beta^{\prime}\neq \beta_0} U_{\beta_0,\beta^{\prime}}^{J_T}(R)\chi^{J_T}_{\beta^{\prime}}(R),  
\label{eq:TELP}
\end{equation}
and $w_{J_T,\beta_i}(R)$ are weight factors chosen as
\begin{equation}
w_{J_T}(R)=(2J_T+1)(1-\mid S^{J_T}_{\beta_0,\beta_0} \mid ^{2})\mid \chi^{J_T}_{\beta_0}(R)\mid ^{2},  \label{Eq15}
\end{equation}
where $S^{J_T}_{\beta_0,\beta_0}$ are the elastic S-matrix elements for each $J_T$ value.
A single channel calculation (elastic
channel) using the sum of ``$U_{\beta_0,\beta_0}^{J_T}(R)$ + $U_\text{pol}(R)$'' should
approximately reproduce the elastic scattering cross sections.

In Fig.~\ref{fig:be11p_xcdcc}, we show for illustration the polarization potential derived from the CDCC calculation for the reaction d+$^{58}$Ni reaction at 80 MeV (c.f.~Fig.~\ref{fig:dni_cdcc}). For comparison, we show the  ground state diagonal monopole coupling potential ($U_{00}$) In this case, the effect the coupling to the breakup channels produces is mostly  repulsive and absorptive.  
\begin{figure}
\begin{center}
\includegraphics[width=0.85\columnwidth]{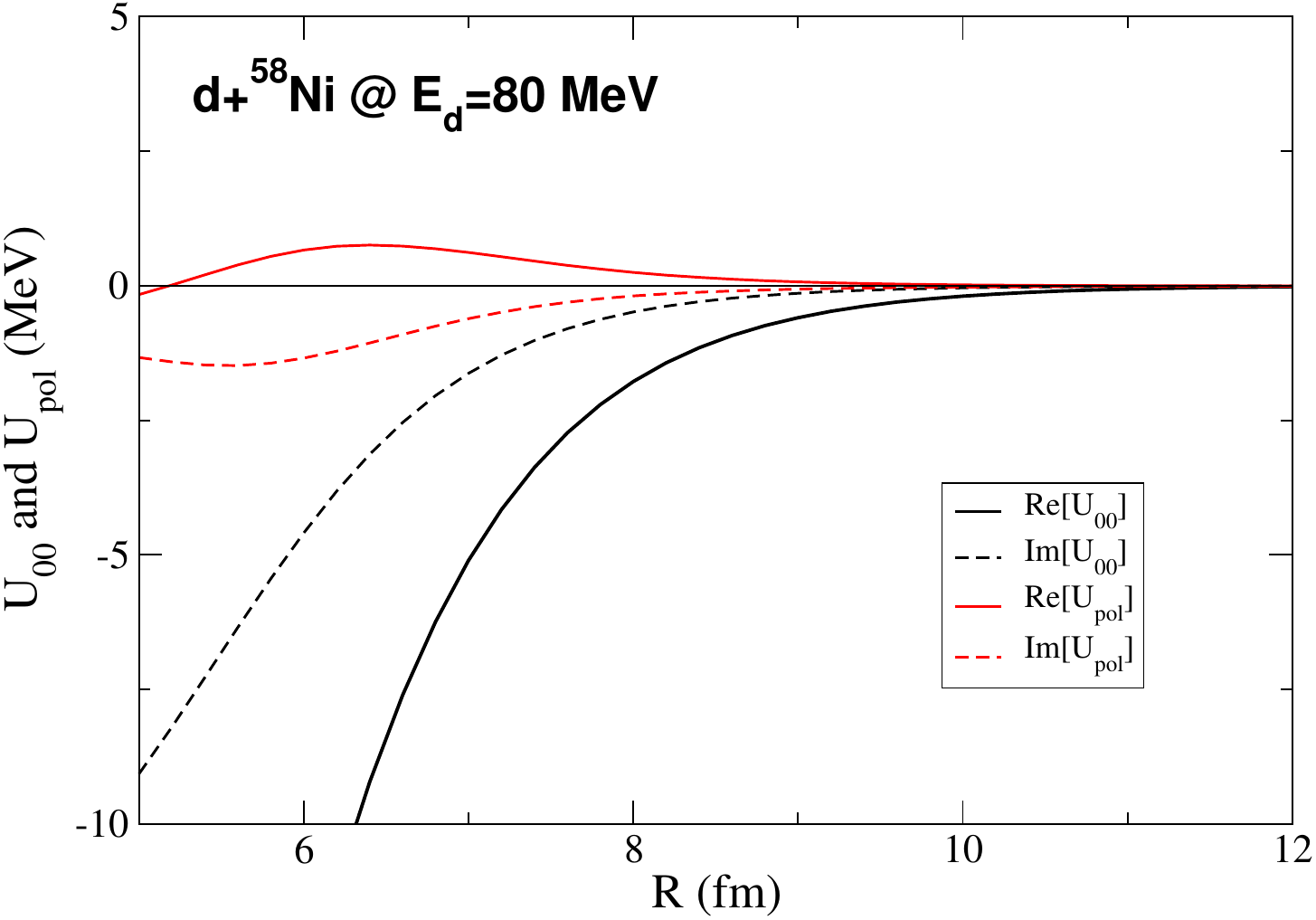}
\caption{\label{fig:dni_e80_telp} 
Average polarization potential $U_\text{pol}$ (solid lines), compared with the ground state diagonal monopole potential $U_{00}$ (dashed) for the d+$^{58}$Ni reaction at 80 MeV.
}
\end{center}
\end{figure}

\subsection{Inclusion of core excitations\label{sec:corex}}
Excitations of the projectile constituents ($b$ and $x$ in our case) may take place along with projectile breakup. This mechanism is neglected in the standard formulation of the CDCC method. For example, for the scattering of halo nuclei, the {\it core} fragment $b$ is assumed to be inert and, as such,  the projectile states are described by pure single-particle or pure cluster states. Except for very specific cases (e.g.\ when $b$ is an alpha particle), the structure of a composite system of the form ($a=b+x$) is known to contain in general admixtures of core-excited components. Additionally, the interaction of the core with the target will produce excitations and deexcitations of the former during the collision, thereby modifying the reaction observables to some extent. These two effects (structural and dynamical) have been investigated within extended versions of the DWBA and CDCC methods \cite{Cre11,Mor12,Sum06,Die14}. For example, considering only possible excitations of $b$, the effective three-body Hamiltonian is generalized as follows:
\begin{equation}
\label{eq:Heff_corex}
H= H_{\rm proj}(\br,\xi_b)   + T_{\bR} + U_{bA}(\br_{bA},\xi_b) + U_{xA}(\br_{xA}) .
\end{equation}
The potential $U_{bA}(\br_{bA},\xi_b)$ is now meant to describe both elastic and inelastic scattering  of the $b+A$ system (for example, it could be represented by a deformed potential as those used in the context of inelastic scattering with collective models). 
Note that the core degrees of freedom ($\xi_b$) appear in the projectile Hamiltonian (structure effect) as well as in the core-target interaction (dynamic effect). 

In the weak-coupling limit, the projectile Hamiltonian can be written more explicitly as
\be
H_{\rm proj}= T_{\br} + V_{bx}(\br,\xi_b) + h_{\rm core}(\xi_b) , 
\label{Hproj}
\ee
where $h_{\rm core}(\xi_b)$ is the internal Hamiltonian of the core. The eigenstates of this Hamiltonian are of the form 
\be
\phi_{n,\Jp,\Mp}(\xi_b, \br ) 
\equiv \sum_{\alpha}  
\left[   \varphi^{n}_\alpha(\br) \otimes \Phi_{I}(\xi_b) \right]_{\Jp \Mp} ,
\label{wfrot}
\ee
where $n$ is an index labeling the states with angular momentum $\Jp$ (with projection $\Mp$),  
$\alpha \equiv \{\ell,s, j,I \}$, with $I$ the core intrinsic spin, $\vec{j}=\vec{\ell} + \vec{s}$ and $\vec{J}=\vec{j}+\vec{I}$. 
The functions $\Phi_{I}(\xi_b)$ and $\varphi_\alpha(\br)$ describe, respectively, the core states and the valence--core relative motion.  This expansion is in principle also valid for continuum states, in which case a procedure of continuum discretization must be used. 

Once the projectile states in Eq.~(\ref{wfrot}) have been calculated, the three-body wave function is expanded in a basis of such  states, as in the standard CDCC method (c.f.\ Eq.~(\ref{f3b})),
\begin{equation}
\Psi_{\beta_i,J_T, M_T}(\vec{R},\vec{r},\xi)=\sum_{\beta}
\chi_{\beta,\beta_i}^{J_T}(R)\left[Y_L({\hat{R}})\otimes\phi^{(N)}_{n,\Jp}(\vec{r},\xi)\right]_{J_T,M_T},  
\label{f3b_2}
\end{equation}
where the superscript $N$ is the number of internal states retained in the unavoidable basis truncation.

The first calculations using this extended CDCC method (referred to as XCDCC) were  performed by Summers {\it et al.}~\cite{Sum06,Sum07} for  $^{11}$Be and $^{17}$C on $^{9}$Be and  $^{11}$Be+$p$, finding a very little core excitation effect in all these cases. However, later calculations for the $^{11}$Be+$p$ reaction based on a alternative implementation of the XCDCC method using a pseudo-state representation of the projectile states \cite{Die14} suggested much larger effects.  The discrepancy was found to be due to an inconsistency in the numerical implementation of the XCDCC formalism presented in Ref.~\cite{Sum06}, as clarified in \cite{Sum14}. For heavier targets, such as $^{64}$Zn or $^{208}$Pb, the calculations of \cite{Die14} suggest that the core excitation mechanism plays a minor role in the reaction dynamics, although its effect on the structure of the projectile is still important.  

\begin{figure}
\begin{center}
\includegraphics[width=0.7\columnwidth]{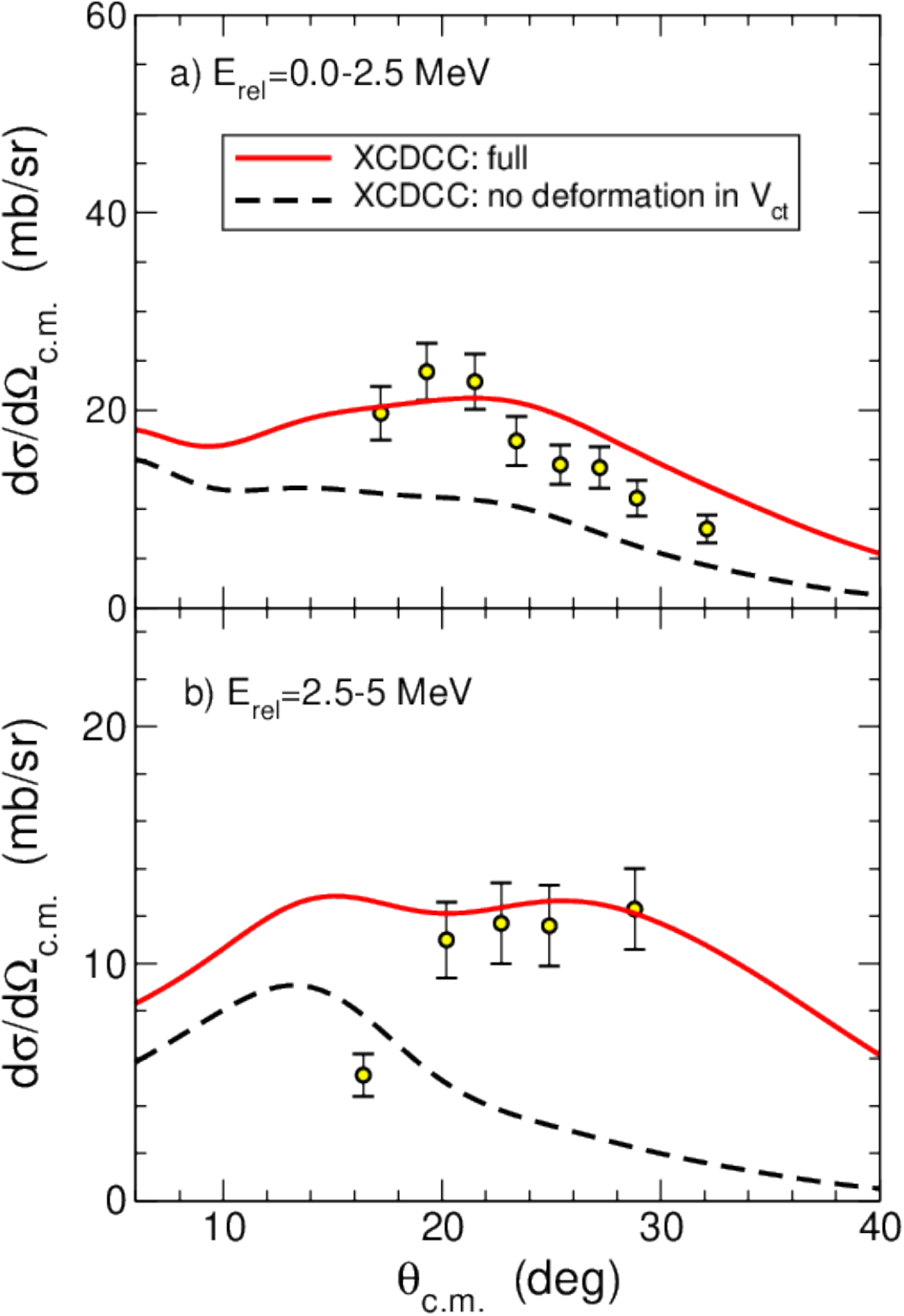}
\caption{\label{fig:be11p_xcdcc} 
Breakup angular distributions for the relative-energy cuts indicated in the labels. Dashed line are the calculations in which the $^{10}$Be+p potential includes only a central part and hence ignores possible excitations  of $^{10}$Be. Solid lines are the full XCDCC calculations, including $^{10}$Be excitations within a collective model. Adapted from Ref.~\cite{Die14,Die17}.
}
\end{center}
\end{figure}

 As an example of these XCDCC calculations we show in Fig.~\ref{fig:be11p_xcdcc} the breakup angular distributions for the reaction $^{11}$Be+$p$ at 63.7~MeV/nucleon. Details of the structure model and potentials are given  in Ref.~\cite{Die14}. Continuum states with angular momentum/parity $\Jp=1/2^\pm$, $3/2^\pm$ and $5/2^+$ were included using a pseudostate representation in terms of transformed harmonic oscillator (THO) functions \cite{Lay12}. Each panel corresponds to a relative-energy internval, as indicated by the labels. The energy interval considered in panel (a) contains the narrow $5/2^+$ resonance whereas that of the lower panel contains the $3/2^+$ low-lying resonance.
 The solid line is the full XCDCC calculation, including the $^{10}$Be deformation in the structure of the projectile as well as in the projectile-target dynamics. The dashed line is the XCDCC calculation omitting the effect of the core-target  excitation mechanism. It is clearly seen that the inclusion of this mechanism significantly increases the break-up cross sections, particularly in the region of the $3/2^+$ resonance, due to the dominant $^{10}$Be(2$^+$)$\otimes$$2s_{1/2}$ configuration of this resonance \cite{Lay12,Cre11,Mor12}.

\begin{figure*}
\begin{center}
\begin{minipage}[c]{.32\textwidth}
{\par \resizebox*{0.85\textwidth}{!}
{\includegraphics{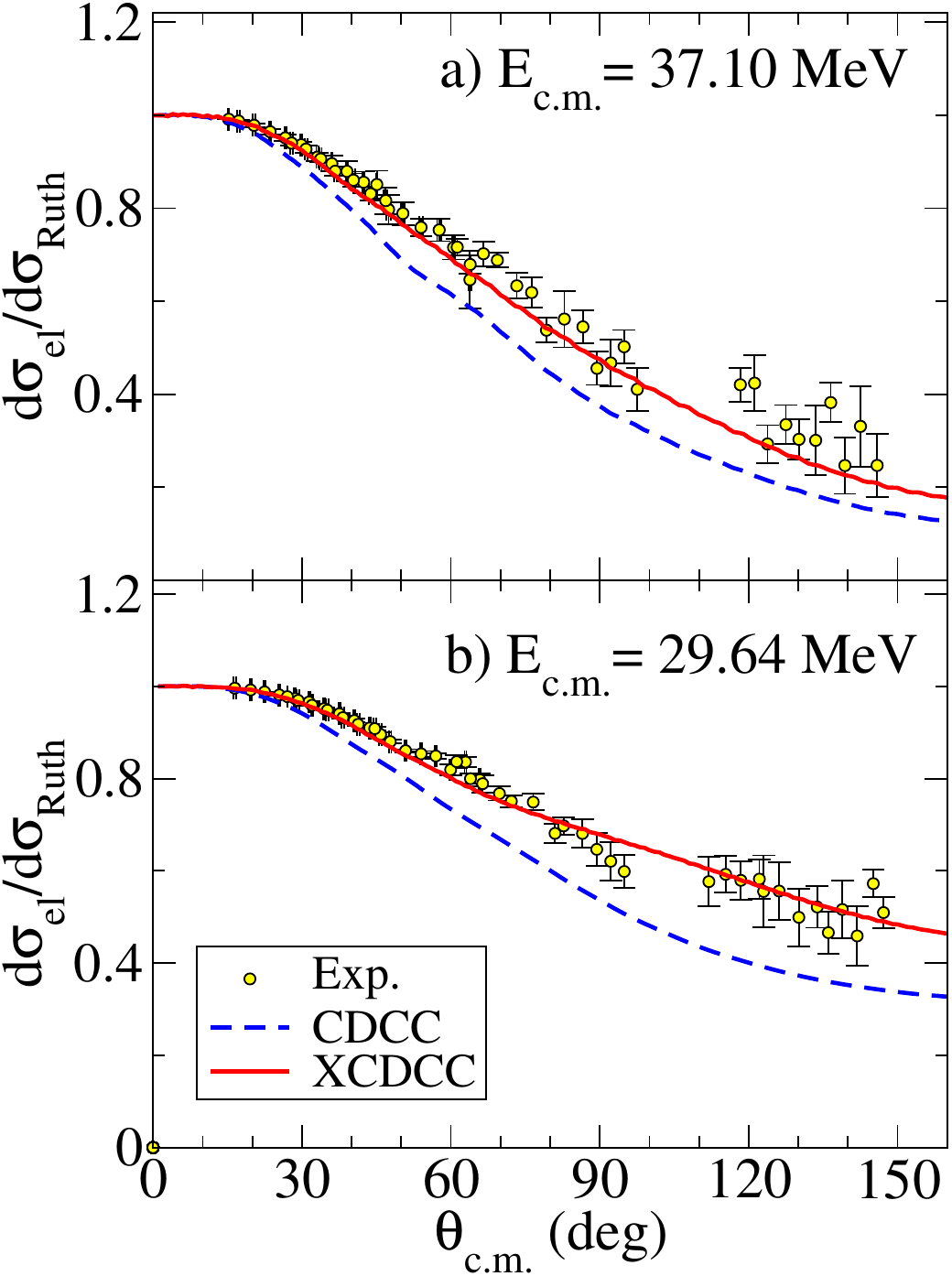}} \par}
\end{minipage}
\begin{minipage}[c]{.32\textwidth}
{\par \resizebox*{0.85\textwidth}{!}
{\includegraphics{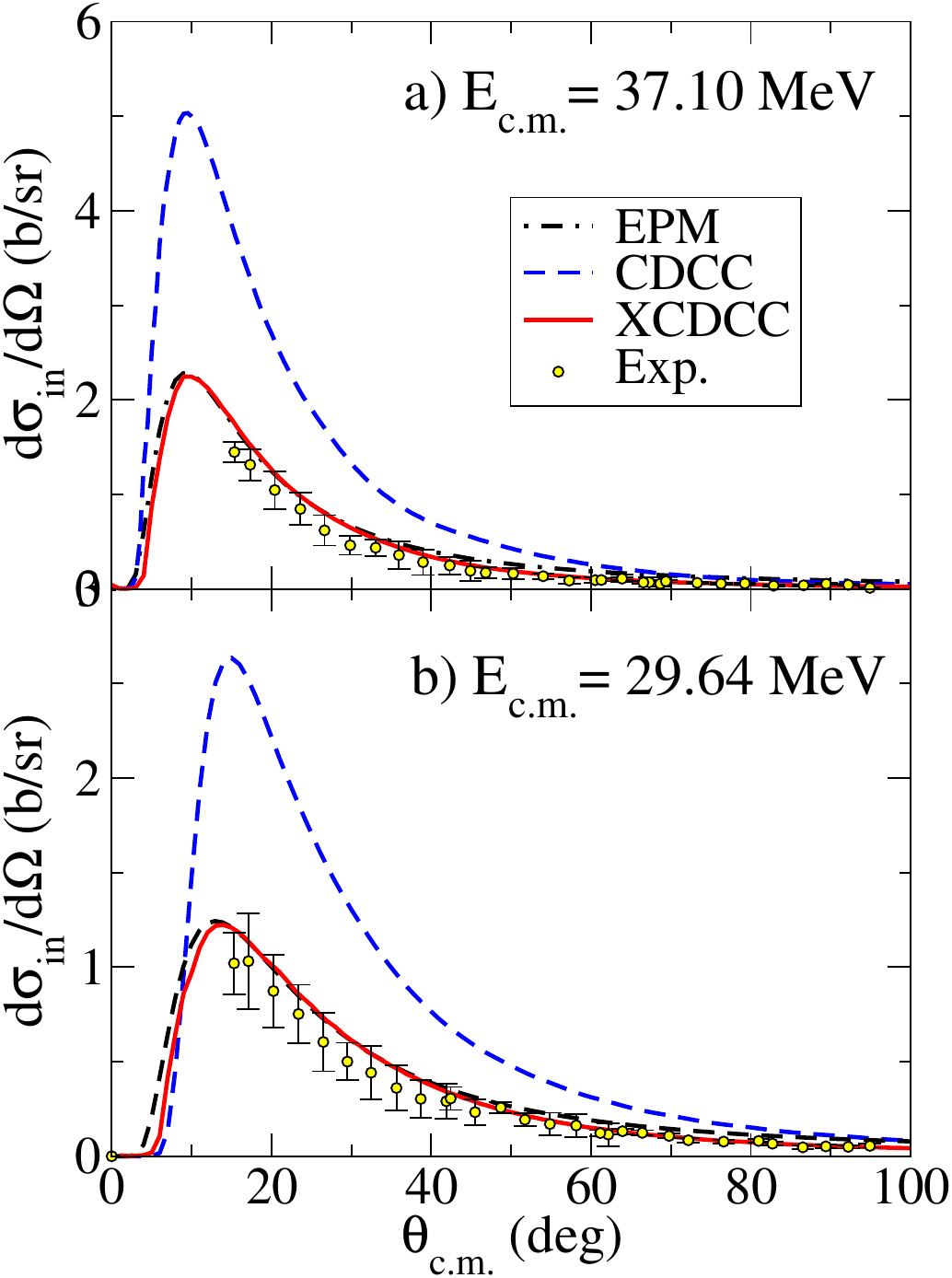}} \par}
\end{minipage}
\begin{minipage}[c]{.32\textwidth}
{\par \resizebox*{0.85\textwidth}{!}
{\includegraphics{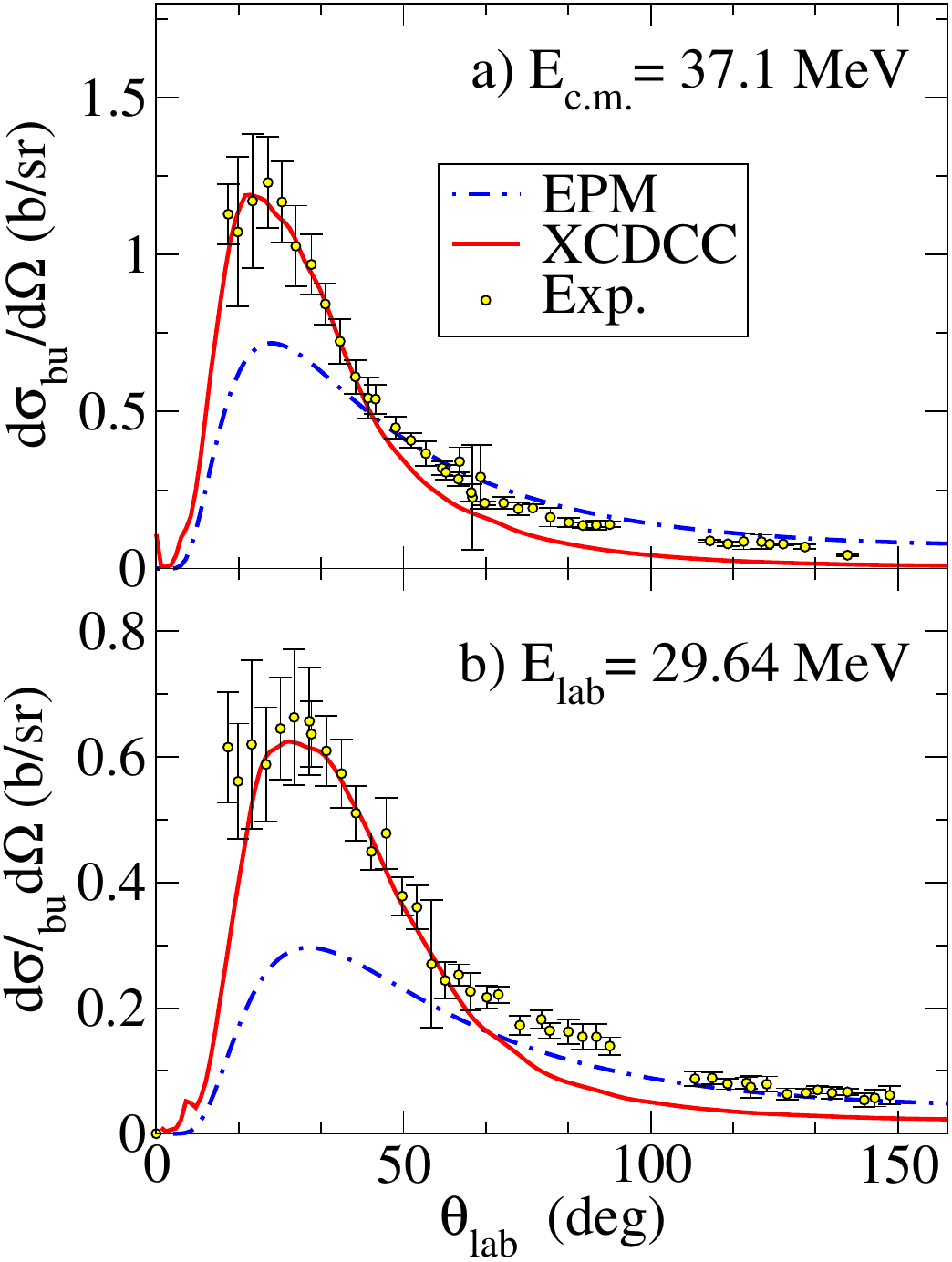}} \par}
\end{minipage}
\caption{\label{fig:be11au} Elastic (left), inelastic (middle) and breakup (right) cross sections for the reaction $^{11}$Be+$^{197}$Au at near barrier energies ($V_b \sim 40$~MeV). Experimental data are compared with CDCC, XCDCC and first-order semiclassical pure $E1$ calculations (labeled EPM). Adapted from Ref.~\cite{Pes17}.}
\end{center}
\end{figure*}

The importance of the deformation on the structure of the projectile is clearly evidenced in the elastic and inelastic scattering of $^{11}$Be on $^{197}$Au at near-barrier energies \cite{Pes17}, as shown in Fig.~\ref{fig:be11au}. XCDCC calculations based on the particle-plus-rotor model of Ref.~\cite{Tar03} (solid lines) are able to reproduce simultaneously the elastic, inelastic and breakup angular distributions. In contrast, standard CDCC calculations using single-particle wave functions fail to describe the elastic and inelastic data, despite describing well the breakup. This is due to the overestimation of the $B(E1)$ connecting the ground state with the bound excited state, as discussed in \cite{Pes17}.  

\begin{figure}
\begin{center}
\includegraphics[width=0.75\columnwidth]{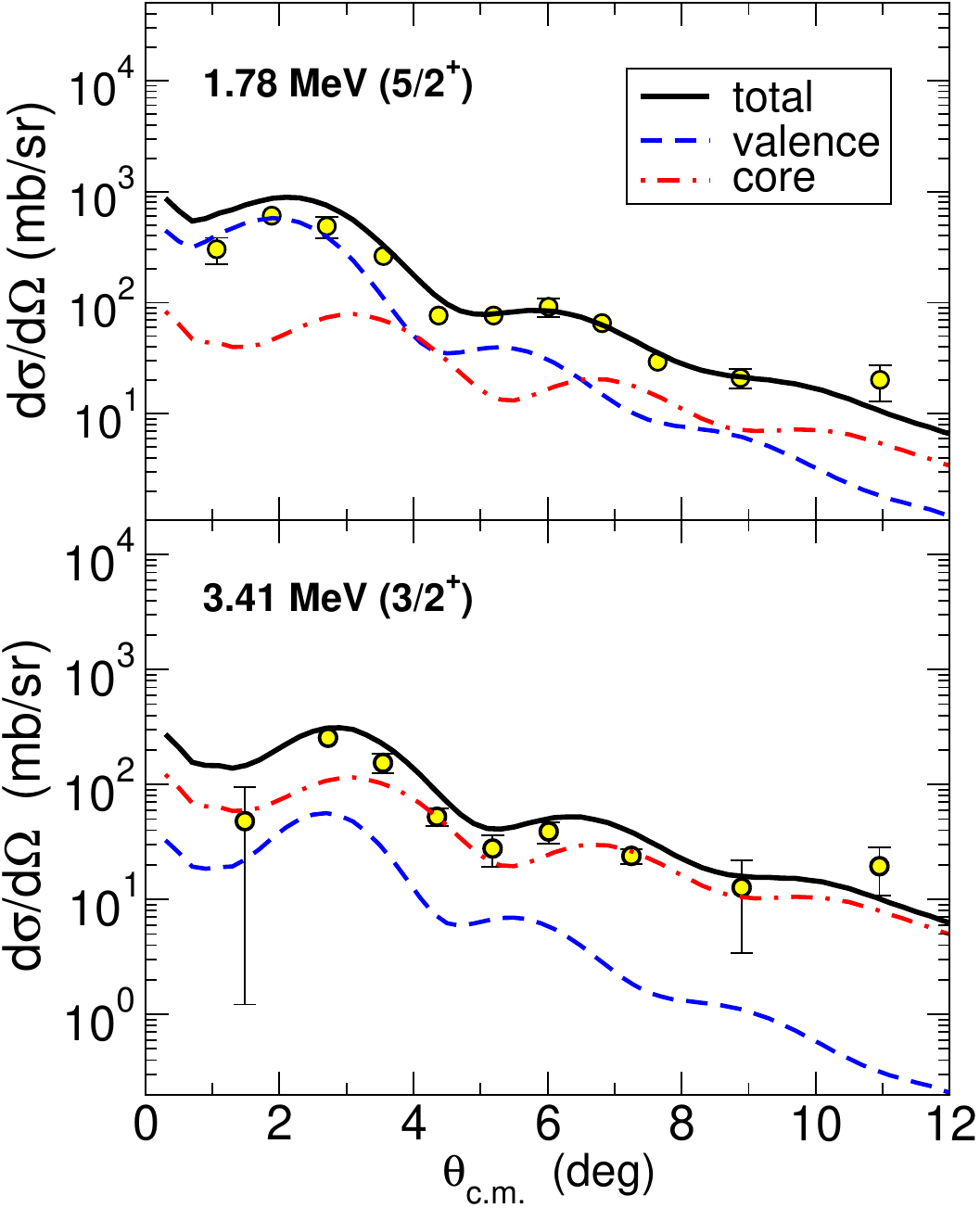} 
\caption{\label{fig:be11c_xdwba} Valence (dashed line) and core (dot-
dashed line) contributions to the breakup of the 1.78 and
3.41~MeV resonances populated in the $^{11}$Be+$^{12}$C  reaction at
70~MeV/nucleon, using a particle-core description of the $^{11}$Be
nucleus. The solid line is the coherent sum of both contribution. An overall normalization factor is included in both plots, to match the magnitude of the data. Adapted from Ref.~\cite{Mor12b}, with authorization from APS.}
\end{center}
\end{figure}

A simpler DWBA, no-recoil  version of the formalism (XDWBA) has also been proposed in refs.~\cite{Cre11,Mor12}. An application of this formalism to the $^{11}$Be+$^{12}$C reaction at 69~MeV/u showed that the core excitation mechanism may interfere with the single-particle excitation mechanism, producing a conspicuous effect on the interference pattern of the resonant breakup angular distributions, as illustrated in Fig.~\ref{fig:be11c_xdwba} for the population of the low-lying $5/2^+$ and $3/2^+$ resonances of $^{11}$Be \cite{Mor12b}.

\subsection{Inclusion of target excitations}
In addition to the excitations of the projectile constituents, excitations of the target nucleus may also take place and compete with the projectile breakup mechanism. Note that, within CDCC, projectile breakup is treated as a inelastic excitation of the projectile to its continuum states and, thus, inclusion of target excitation amounts to including, simultaneously, projectile plus target excitations so their relative importance, and mutual influence, can be assessed. These target excitations can be treated with the collective models mentioned in the previous section. It is worth noting that, within this three-body reaction model, target excitation arises from the non-central part of the valence-target and core-target interactions.  To incorporate this effect, the  effective Hamiltonian, Eq.~(\ref{eq:Heff}), must be now generalized as \cite{Yah86}:  
\begin{equation}
\label{eq:Heff_tarx}
H= H_\text{proj}(\br)+ T_{\bR} + U_{bA}(\br_{bA},\xi_t) + U_{xA}(\br_{xA},\xi_t) + H_A(\xi_t)
\end{equation}
where $H_A(\xi_t)$ is the target internal Hamiltonian, and the $b-A$ and $x-A$ interactions depend now, in addition to the corresponding relative coordinate, on the target degrees of freedom (denoted as $\xi_t$).  
 Ideally, these $U_{xA}$ and $U_{bA}$ potentials should reproduce simultaneously  the elastic and inelastic scattering for the $x+A$ and $b+A$ reactions, respectively. Once a specific model has been chosen for the target, its eigenstates  ($\Phi^i_{J_t}(\xi_t)$) can be computed. They are characterized by their angular momentum $J_t$ and the index $i$.

 In the presence of these target excitations, the CDCC wavefunction is now expanded on the generalized channel basis \cite{Mor12b,Die14,fresco} containing the target eigenstates 
\begin{align}
 \Psi^\text{CDCC}_{J_T,M_T}(\vec{R},\vec{r},\xi)&=  \sum_{\beta} \chi_\beta^{J_T} (R)
 \left\lbrace \left[ Y_L(\hat{R}) \otimes \phi^n_{J_p}(\vec{r}) \right]_J \otimes \Phi^i_{J_t}(\xi_t) \right\rbrace_{J_T,M_T},
 \label{eq:CDCC_tarx}
\end{align}
where $\beta$ denotes all the quantum numbers necessary to define the channel $\beta=\lbrace L,n,J_p,J,J_t,i \rbrace$. 
The explicit inclusion of target excitation was first done by the Kyushu group in the 1980s \cite{Yah86}, which considered the case of deuteron scattering. The motivation was to compare the roles of target excitation and deuteron breakup in the elastic and inelastic scattering of deuterons. They applied the formalism to the $d$+$^{58}$Ni reaction at $E_d=22$ and 80~MeV, including the ground state and the first excited state of $^{58}$Ni ($2^+$) and finding that, in this case, the deuteron breakup process is more important than the target-excitation. 

\begin{figure}
{\par\centering \resizebox*{0.75\columnwidth}{!}{\includegraphics{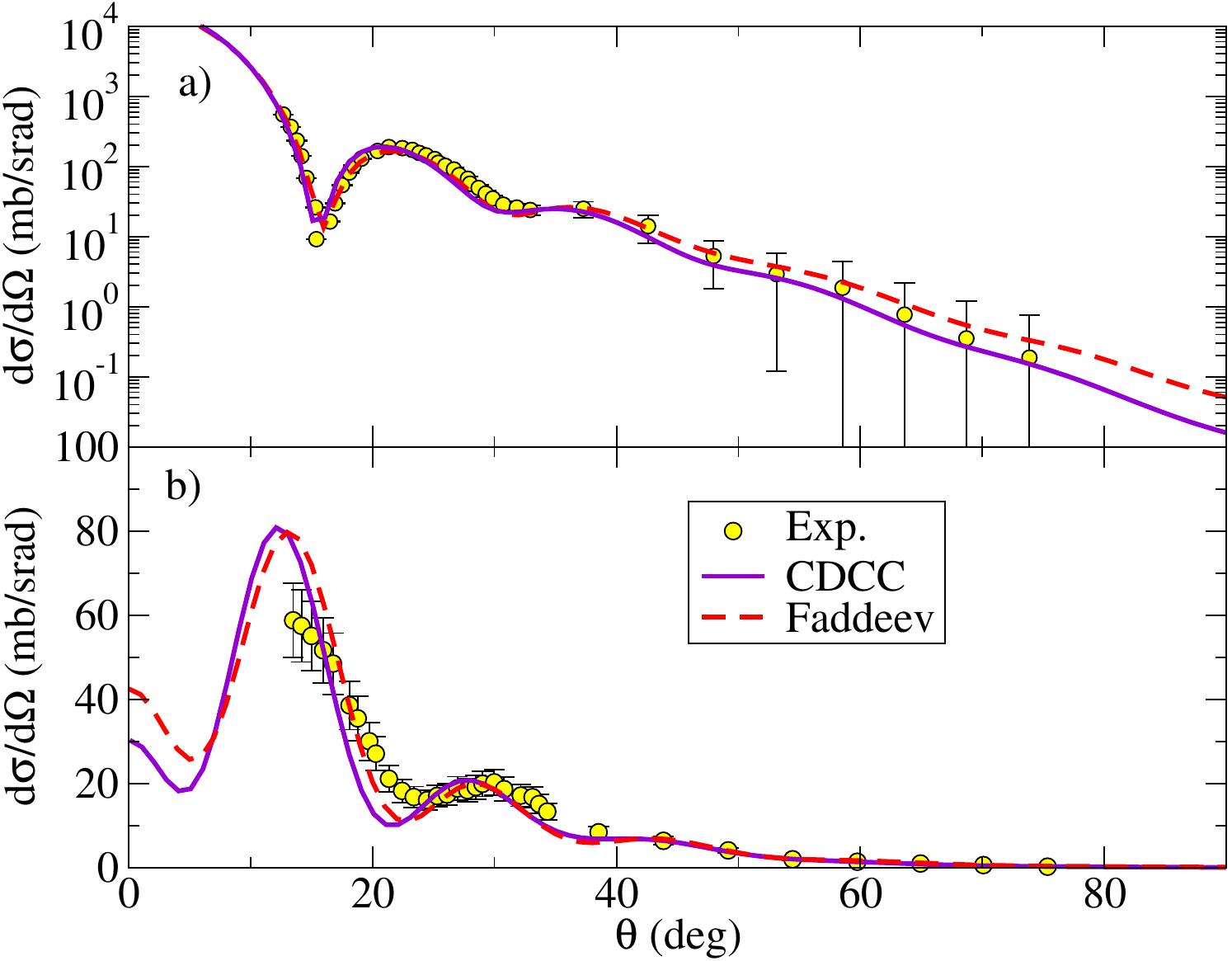}}\par}
 \caption{\label{fig:24Mg}Elastic (upper) and  $^{24}$Mg$(2^+)$ excitation (lower) differential angular cross sections for $d$+$^{24}$Mg at $E_d=70$ ${\rm MeV}$. CDCC calculations are compared with Faddeev calculations from Ref.~\cite{Del16} and with the data of Ref.~\cite{Kis76}. The calculations  employ CH89 \cite{CH89} parameterization for the $p$+$^{24}$Mg and $n$+$^{24}$Mg potentials and treat the target excitation with a collective model, assuming a deformation of $\beta=0.5$  for the $^{24}$Mg nucleus.  The plot is adapted from Ref.~\cite{Gom17a}.}
\end{figure}

Recently, the problem has also been addressed by some authors \cite{Chau11,Gom17a}, also in the context of elastic and inelastic deuteron scattering. 
A recent application of the formalism is shown in Fig.~\ref{fig:24Mg}, which corresponds to the reaction $^{24}{\rm Mg}(d,d)^{24}{\rm Mg}^*$ at  $E_d=70$~MeV, including the ground and first excited states of $^{24}{\rm Mg}$, in addition to the deuteron breakup. The data are from Ref.~\cite{Kis76}. The target excitation was treated within the collective model, using a quadrupole deformation parameter of $\beta_2=0.5$. Also included are Faddeev calculations performed by A.~Deltuva \cite{Del16}. Both calculations reproduce equally well the elastic differential cross section. The calculated  inelastic angular distributions are slightly out of phase with the data, but they agree well with each other, pointing to some inadequacy of the structure or potential inputs. 

Note that the traditional procedure to analyze these reactions employs DWBA or coupled channel methods, based on a deformed deuteron-target potential. However, it is not guaranteed that the deformation parameters extracted by this procedure are consistent with those derived from nucleon-nucleus inelastic scattering. The three-body approaches outlined in this section (CDCC and Faddeev) have the advantage of treating both types of reactions (deuteron-nucleus and nucleon-nucleus) from the same underlying (deformed) nucleon-nucleus potentials. In fact,  it was shown in \cite{Del16} that the extracted deformation parameter obtained with the three-body approach is more consistent with that derived from nucleon-nucleus inelastic scattering.

\subsection{Four-body CDCC}\label{sec:4bcdcc}

The extension of the CDCC formalism to include three-body projectiles is typically referred to as four-body CDCC and was developed in Refs.~\cite{Mat04a,Mat04b,manoli08,manoli09} to describe reactions induced by the Borromean two-neutron halo nuclei $^6$He and $^{11}$Li. Within the few-body scheme, the method amounts to incorporate the three-body structure in the corresponding coupling potentials of the CDCC equations, thus solving the same coupled-channel problem but typically with a considerably larger model space. 
\begin{figure}
    \begin{center}
        \includegraphics[width=0.5\linewidth]{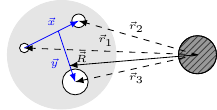}
    \end{center}
    \caption{\label{fig:jacobi4b} Schematic representation of the four-body CDCC coordinate system, where the projectile is described in Jacobi coordinates $\{\vec{x},\vec{y}\}$. Adapted from Ref.~\cite{casal15}.}
\end{figure}

In simple terms, the generalization implies the inclusion of three-body projectile wave functions in the channel states of Eq.~(\ref{f3b}). These can be written in Jacobi coordinates $\left\{\vec{x},\vec{y}\right\}$ (see Fig.~\ref{fig:jacobi4b}), in such a way that the coupled basis~in Eq.~(\ref{coupledbasis}) becomes
$$
\langle \hat{R}, \vec{x}, \vec{y} |\beta; J_T\rangle =
\left[Y_L({\hat{R}})\otimes\Phi_{n,J}(\vec{x},\vec{y})\right]_{J_T},
\label{coupledbasis_4b}
$$
and the coupling potentials, now involving the interaction between each of the three bodies and the target, reads schematically as
\begin{equation}
U_{\beta,\beta^\prime}^{J_T}(R)=
\langle \beta; J_T|V_{1t}+V_{2t}+V_{3t}|\beta^\prime;
J_T\rangle_{\vec{x},\vec{y}}.
\label{cpot_4b}
\end{equation}
As discussed, for instance, in Ref.~\cite{manoli08,casal15}, the fragment-target potentials can be expressed each in a different $q$-Jacobi system, so that its radial dependence involves only $\vec{R}$ and $\vec{y}_q$. Then, a multipole expansion of these potentials is introduced to obtain the corresponding form factors. The effect of different multipole couplings can therefore be studied. 

The construction of the three-body projectile states can be performed using various techniques. In Refs.~\cite{Mat06,Matsumoto19}, for instance, a Gaussian expansion method in Jacobi coordinates is employed. Here we outline another typical choice, based on the standard hyperspherical harmonics formalism~\cite{Zhu93}. By introducing the hyperspherical coordinates $\rho=\sqrt{x^2+y^2}$ and $\alpha=\arctan\left(x/y\right)$, one can write
\begin{equation}
    \Phi_{n,J,M}(\rho,\alpha,\hat{x},\hat{y})=\frac{1}{\rho^{5/2}}\sum_{\eta}u_{n\eta}(\rho)\mathcal{Y}_{\eta J M}(\alpha,\hat{x},\hat{y}),
\end{equation}
with
\begin{equation}
    \mathcal{Y}_{\eta J M}(\alpha,\hat{x},\hat{y})=\varphi_{K}^{l_x,l_y}(\alpha)\left\{\left(\left[Y_{l_x}(\hat{x})\otimes Y_{l_y}(\hat{y})\right]_l\otimes \kappa_{S_x}\right)_j \otimes \chi_I\right\}_{J M}.
\end{equation}
Here, $\eta=\left\{K,l_x,l_y,l,S_x,j,I\right\}$ are the possible sets of quantum numbers associated with the orbital angular momenta and spins coupled to a total $J$, where $K$ is the so-called hypermomentum, and $\varphi_K^{l_x,l_y}(\alpha)$ are obtained in terms of Jacobi polynomials. The three-body hyperradial functions $u_{n\eta}(\rho)$ can be obtained within different discretization methods, including the binning procedure~\cite{manoli09,Cub12} and the pseudostate method with different bases~\cite{manoli08,Desc15,casal15}. This is done for a fixed value of $K_{max}$, up to convergence. Note that the above wave-function expansion and its coupling order is not unique, and other choices exist in the literature.

\begin{figure}[tb]
\begin{center}
 {\centering \resizebox*{0.75\columnwidth}{!}{\includegraphics{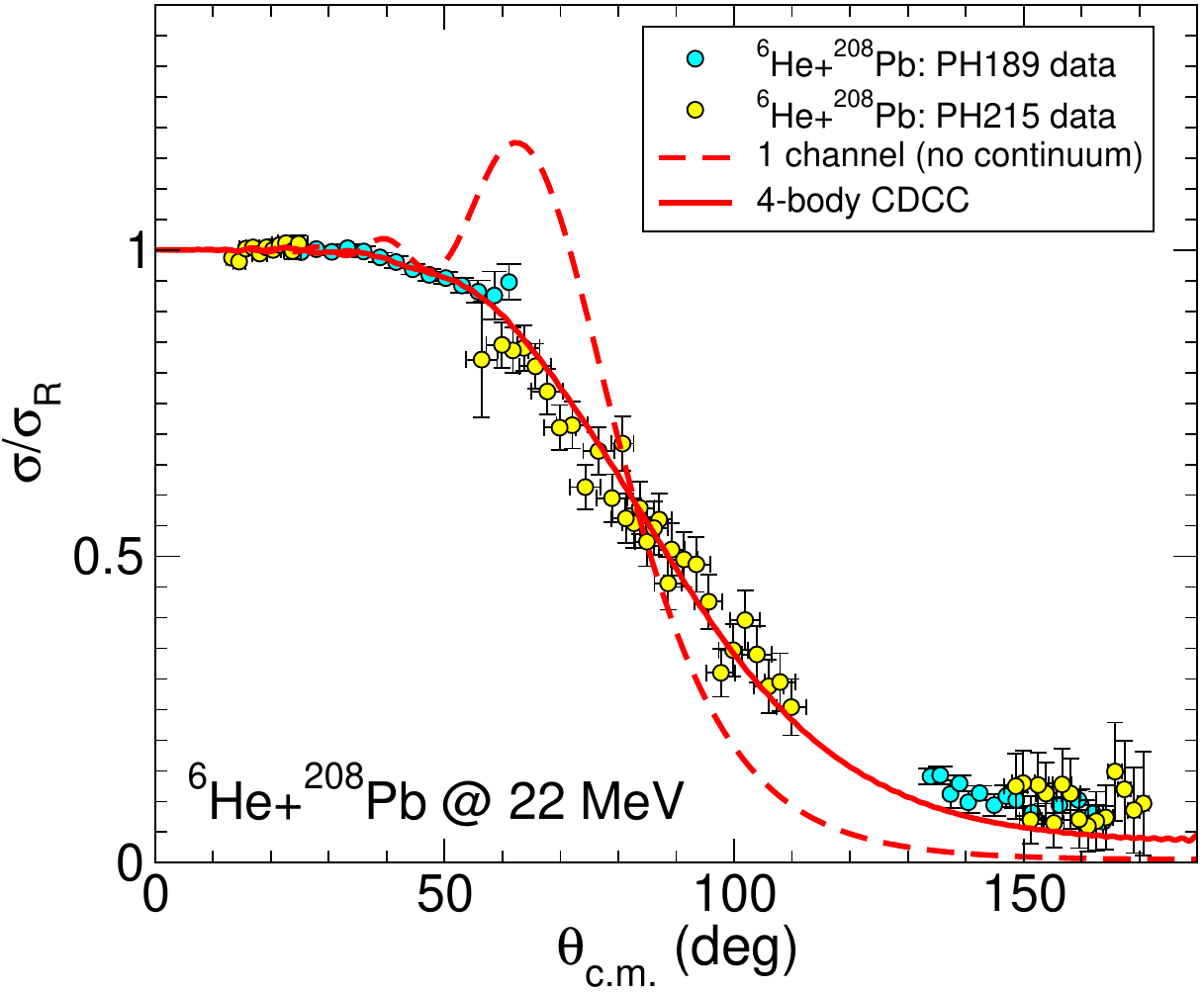}} \par}
\caption{\label{fig:he6pb_el_4bcdcc} Elastic differential cross section, relative to Rutherford, for the  $^{6}$He+$^{208}$Pb reaction at 22 MeV. The data are from Refs.~\cite{San08,Aco11}. The lines are four-body CDCC calculations, using a binning discretization procedure, from Ref.~\cite{Rod09}. The dashed line is the CDCC calculation omitting the breakup channels.}
\end{center}
\end{figure}

Figure~\ref{fig:he6pb_el_4bcdcc} shows an application of the four-body CDCC method to the $^{6}\text{He}+{^{208}}\text{Pb}$ reaction at 22 MeV. The combined data from two experiments \cite{San08,Aco11}  are compared with four-body CDCC calculations (solid line) from Ref.~\cite{Rod09} assuming a $^{4}\text{He}+n+n$ model for the $^{6}$He states and  using the binning discretization method for the continuum. The dashed line is the CDCC calculation omitting the coupling to the unbound states of the projectile. The comparison of the two curves evidences the importance of the breakup channels on the elastic cross section. 

\begin{figure}
    \begin{center}
        \includegraphics[width=0.8\linewidth]{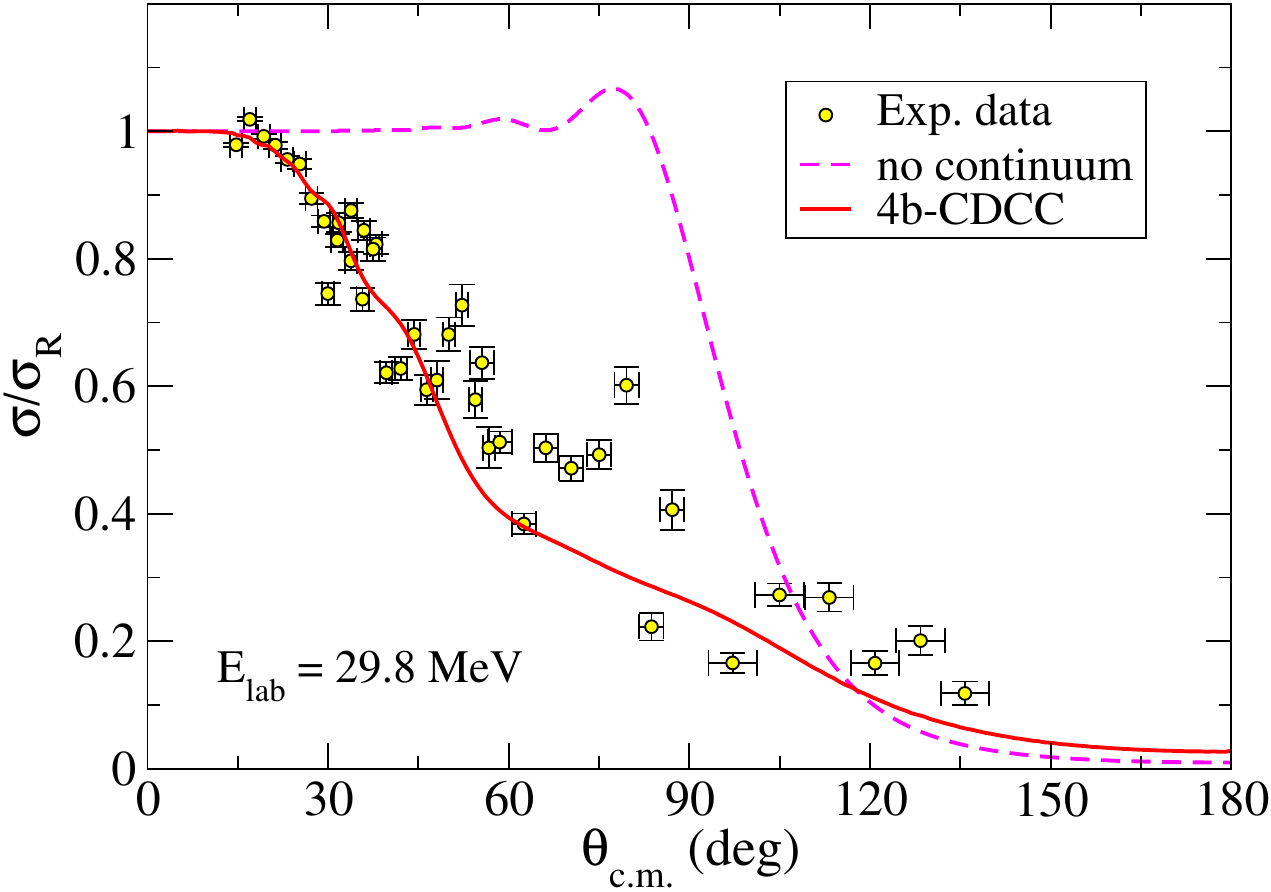}
    \end{center}
    \caption{\label{fig:4b-11Li} Experimental $^{11}\text{Li}+{^{208}}\text{Pb}$ elastic cross section at $E_\text{lab}=29.8$ MeV, relative to Rutherford, compared to four-body CDCC calculations (solid) and the result without the coupling to continuum states (dashed). Adapted from Ref.~\cite{Cub12}.}
\end{figure}

As discussed in the introductory section (see also \cite{Rod08}) a significant part of the suppression of the elastic scattering cross section stems from the Coulomb dipole polarizability of the projectile which, in the CDCC calculation, is included in the coupling potentials connecting the projectile ground state with the  $J=1^-$ continuum states. Thus, the effect is expected to be enhanced for more weakly bound halo nuclei, which exhibit a larger $E1$ strength at low excitation energies. This is confirmed in Fig.~\ref{fig:4b-11Li}, which shows the experimental and calculated elastic differential cross section for the $^{11}\text{Li}+{^{208}}\text{Pb}$ reaction at a near-barrier energy~\cite{Cub12}. 
 In that work, it is shown that the coupling to low-energy continuum states is crucial to explain the rapid decrease of the elastic differential cross section, and the calculations indicate the presence of a low-lying dipole resonance in $^{11}$Li for a proper description of the experimental data. 

\begin{figure}
    \begin{center}
        \includegraphics[width=0.9\linewidth]{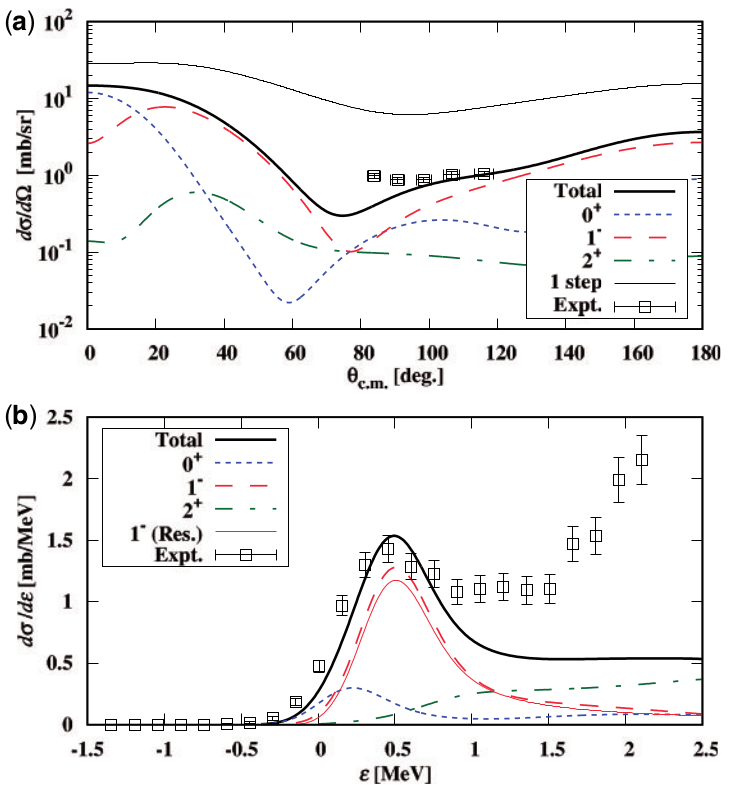}
    \end{center}
    \caption{\label{fig:pp-11Li} (a) Angular distribution and (b) breakup-energy distribution for $^{11}\text{Li}(p,p')$. The calculation labeled `1 step' represents the result without continuum states, and the solid black line is the total calculation including $0^+,1^-,2^+$ states. Taken from Ref.~\cite{Matsumoto19}. Data are from Ref.~\cite{tanaka17}.}
\end{figure}

Figure~\ref{fig:pp-11Li} presents results for the $^{11}\text{Li}(p,p')$ reaction, focusing on the breakup cross section extracted from four-body CDCC~\cite{Matsumoto19}. These calculations employ pseudostates to describe the $^{11}$Li states, and the so-called complex scaling method is adopted to build the corresponding energy distributons. Again, a dipole resonance is found to be crucial for a proper understanding of the data. Other calculations for the same reaction exist in the literature~\cite{Descouvemont20}. Note that the complex-scaling method has been applied in four-body CDCC calculations for other projectiles, such as $^9\text{C}({^7}\text{Be}+p+p)$~\cite{Singh21}, thus providing a valuable tool in this kind of studies.

\begin{figure}
    \begin{center}
        \includegraphics[width=0.8\linewidth]{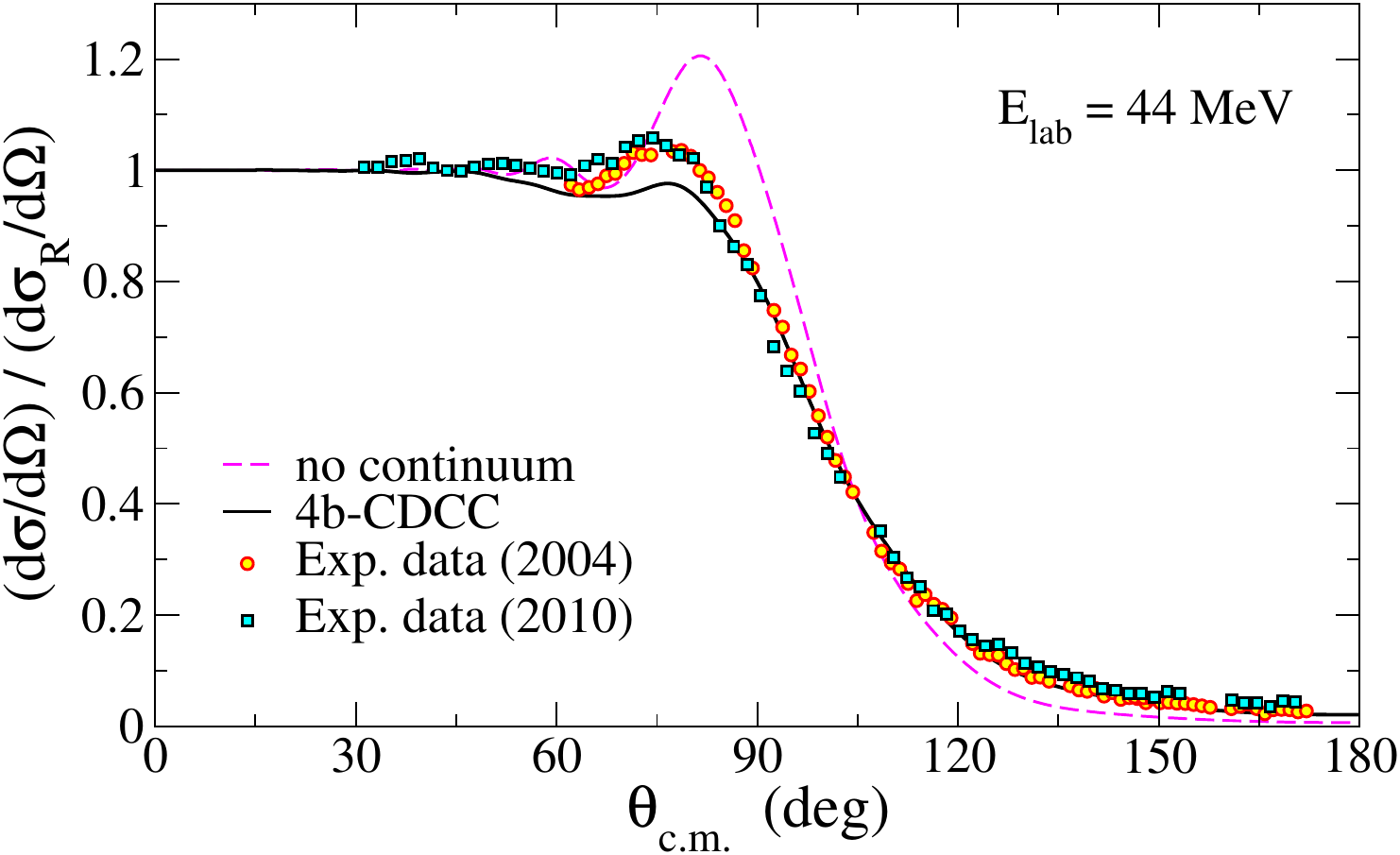}
    \end{center}
    \caption{\label{fig:4b-9be} Experimental $^{9}\text{Be}+{^{208}}\text{Pb}$ elastic cross section at $E_\mathrm{lab}=44$~MeV, relative to Rutherford, compared to four-body CDCC calculations (solid black) and the result without the coupling to continuum states (dashed magenta). Adapted from Ref.~\cite{casal15}. Data are from Refs.~\cite{Wooll2004,Yu2010}. }
\end{figure}

Four-body CDCC calculations typically involve very large channel sets, for which the pseudostate method is computationally less demanding than the binning method. However, it was shown in e.g., Ref.~\cite{mrg11proc} that the results for weakly bound projectiles on heavy targets may exhibit a slow convergence when using pseudostates, due to the very strong continuum couplings involving low-energy states (see Sec.~\ref{sec:PSvsbin} below). Still, pseudostates are the preferred choice in many cases, especially for three-body projectiles comprising several charged particles, for which the calculation of the true three-body continuum states (and the corresponding bins) is not feasible. In Refs.~\cite{Desc15,casal15}, four-body CDCC calculations were performed for the scattering of $^9\text{Be} (\alpha+\alpha+n)$, showing reasonable agreement with the available experimental data. Figure~\ref{fig:4b-9be} shows some $^{9}\text{Be}+{^{208}}\text{Pb}$ results from Ref.~\cite{casal15}, where  the relevance of dipole couplings for a proper description of the observations was highlighted.

\begin{figure}[tb]
\begin{center}
 {\centering \resizebox*{0.9\columnwidth}{!}{\includegraphics{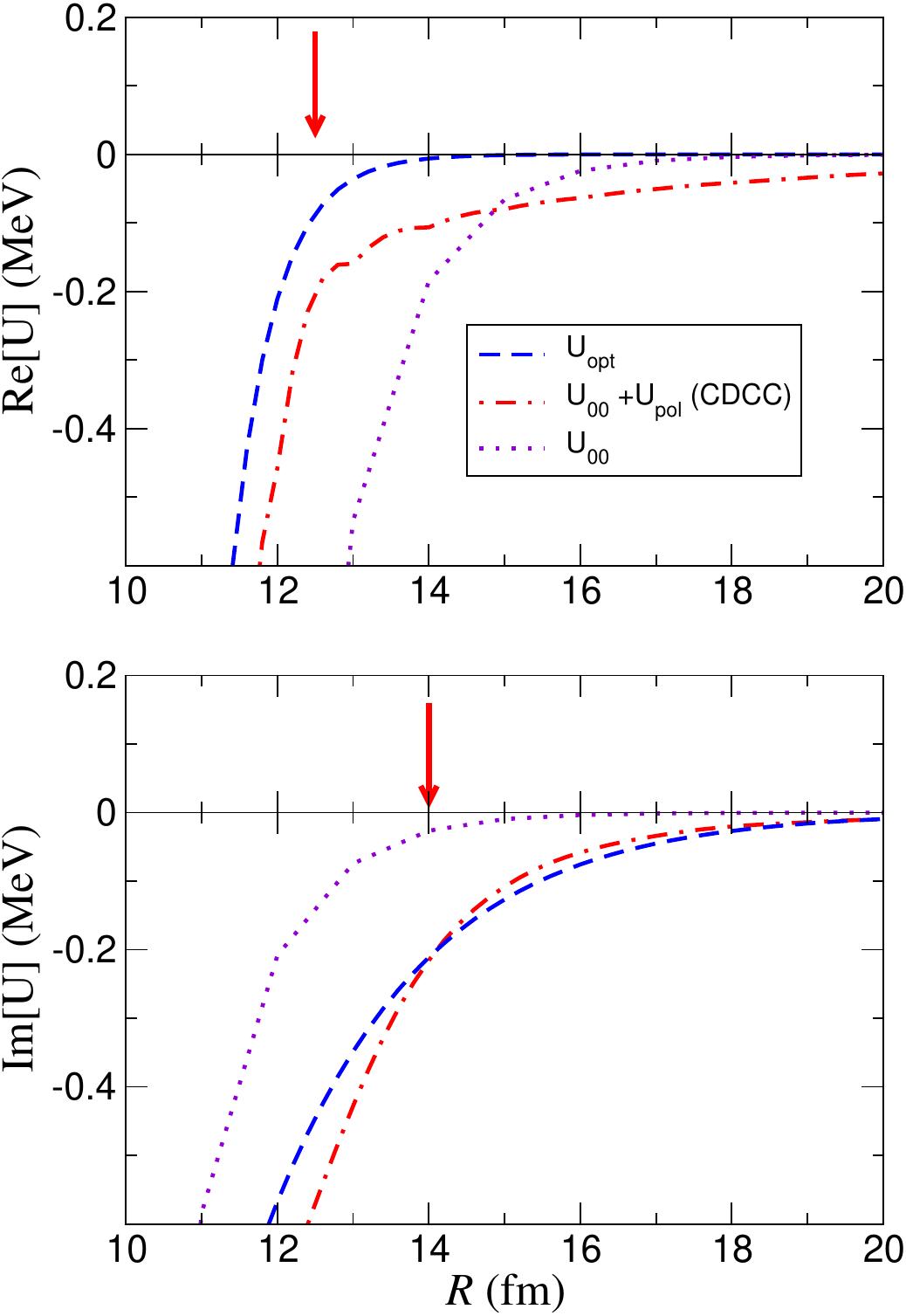}} \par}
\caption{\label{veff} Bare $U_{00}$ (dotted line) and effective potential,  $U_\mathrm{eff}=U_\mathrm{00}+U_\mathrm{pol}$ (dot-dashed line),  for $^{6}$He+$^{208}$Pb at 22 MeV, extracted from 
the CDCC. The dashed line is the phenomenological optical model potential listed in Table \ref{Table:he46pb_omp}. The top and bottom panels correspond to the real and imaginary parts, respectively. The vertical arrows denote the radius of maximum sensitivity, as estimated in \cite{San08}.}
\end{center}
\end{figure}

In Sec.~\ref{sec:om} it was shown that, in order to reproduce the elastic cross section of halo nuclei in terms of the conventional optical model framework, one requires an unusually large diffuseness parameter for the imaginary part of the potential. This can be linked to the CDCC predictions for these systems making use of the local polarization potential introduced in Sec.~\ref{sec:telp} which is meant to encode the net effect of the full coupling scheme on elastic scattering. 
As an illustration, in Fig.~\ref{veff} we compare the phenomenological potential of Table~\ref{Table:he46pb_omp} with the effective potential ($U_\mathrm{00}+U_\mathrm{pol}$) extracted from the four-body CDCC  calculations of \cite{Rod09} for $^{6}\text{He}+{^{208}}\text{Pb}$ at $E_\mathrm{lab}=22$ MeV. The top and bottom panels correspond to the real and imaginary parts, respectively. The arrows indicate the radii of sensitivity of the real and imaginary parts of the optical potential, according to the OM analysis performed in \cite{San08}.  It is seen that both the phenomenological and CDCC potentials display similar behavior, and, in particular, the effective potential derived from CDCC has the long-range imaginary contribution obtained in the phenomenological OM analysis.

\subsection{Microscopic CDCC}

\begin{figure}
\begin{minipage}[c]{.8\columnwidth}
\begin{center}\includegraphics[width=0.4\textwidth]{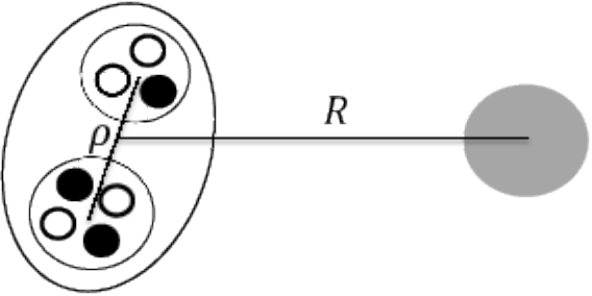} \end{center}
\end{minipage}
\begin{minipage}[c]{.9\columnwidth}
\begin{center}\includegraphics[width=0.85\textwidth]{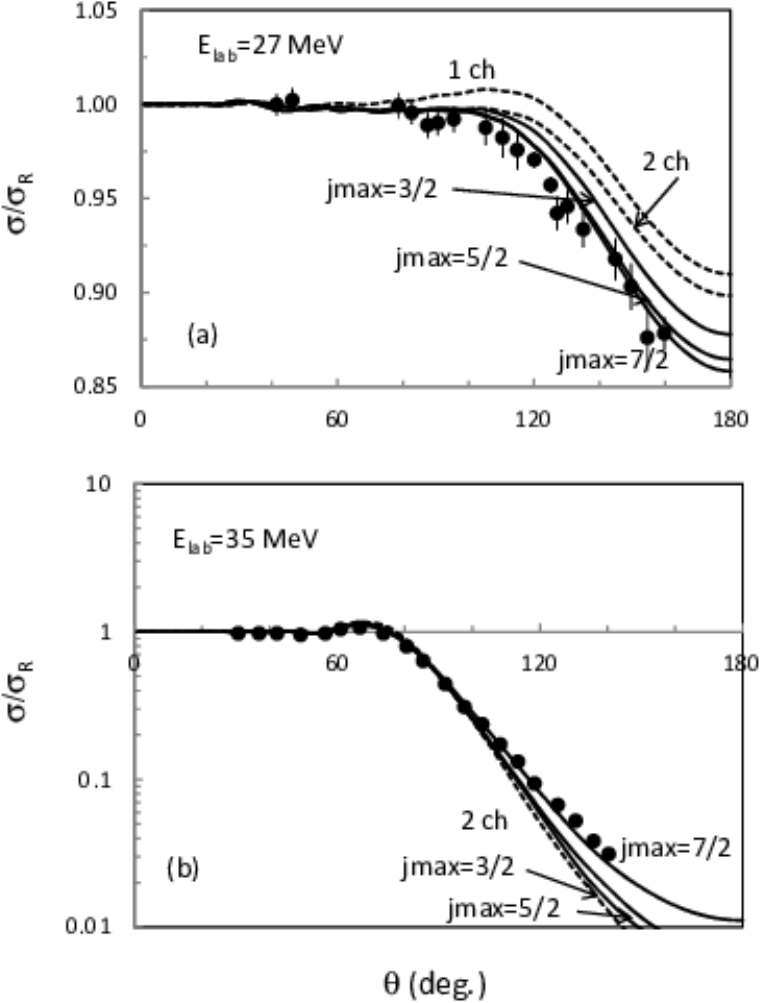} \end{center}
\end{minipage}
\caption{\label{fig:li7pb_mcdcc}Top:  Schematic picture of the projectile-target system, with a microscopic cluster structure of the projectile. Bottom: $^{7}$Li+$^{208}$Pb elastic cross section \cite{Mar96}, relative to Rutherford, compared with microscopic CDCC calculations. Dotted lines represent the calculations without breakup channels  and the solid lines are the full calculations with increasing  $\alpha+t$ maximum angular momentum of the projectile. Taken from Ref.~\cite{Des13}, with authorization from APS.}
\end{figure}

The standard CDCC method assumes a cluster (two-body or three-body) description of the projectile nucleus. This simplification has of course limitations and drawbacks. For example: (i) it requires cluster-target optical potentials, which are not always well determined; (ii) the extension to more than three bodies is very challenging and currently not available; (iii)  excitations of the fragments are ignored altogether or, at most, approximately included with some collective model. To overcome these problems, a microscopic version of the CDCC method (MCDCC) has been proposed by Descouvemont and co-workers \cite{Des13,Des18}. The method uses a many-body description of the projectile states, based on a cluster approximation, known as resonating group method (RGM). In the RGM, an eigenstate of the projectile Hamiltonian is written
as an antisymmetric product of cluster wave functions. For example, for a $^{7}$Li projectile, described as $\alpha+t$, the RGM wave function is expressed as:
%
\begin{align}
\Phi_{i,J,M}(\xi_p)& ={\mathcal A}\bigl[ [\phi_{\alpha} \otimes \phi_t]^{1/2} \otimes 
Y_{\ell}(\Omega_{\rho})\bigr]^{J ,M} g^{\ell J}_i(\rho),
\label{eq:li7rgm}
\end{align}
where $\phi_{\alpha}$ and $\phi_{t}$ are  shell model wave functions of the $\alpha$ and $t$ clusters, $\ell$ their relative orbital angular momentum and $J$ the total spin.  In Eq.~(\ref{eq:li7rgm}), $\vec{\rho}$ is the relative coordinate (see Fig.~\ref{fig:li7pb_mcdcc}), and ${\mathcal A}$ is the 7-body antisymmetrization operator which takes into account the Pauli principle among the 7 nucleons of the projectile.  The function $g^{\ell J}_i(\rho)$ is determined from a Schr\"odinger equation associated with the projectile Hamiltonian.  Continuum states are included using a pseudo-state basis.

In MCDCC, the projectile-target interaction is given by the sum of nucleon-nucleus interactions (instead of cluster-target interactions), for which reliable parametrizations are available. Thus, the full projectile+target Hamiltonian is written as:
\be
H=H_0+T_{\bR}+\sum_{j=1}^A U_{tj}(\vecr_j-\vecR),
\ee
where $U_{tj}(\vecr_j-\vecR)$ is the interaction of the $j$-th projectile constituent with the target.

Figure \ref{fig:li7pb_mcdcc} shows an application of the method to the reaction $^{7}$Li+$^{208}$Pb at near-barrier energies. Experimental data are compared with a one-channel calculation (only $^{7}$ Li g.s.), two-channel calculation (ground plus first excited state) and several CDCC calculations including continuum states up to a certain projectile angular momentum. It is seen that a good description of the data is achieved when a sufficiently large number of continuum states are included.

\begin{figure}
\centering
   \includegraphics[width=0.85\linewidth]{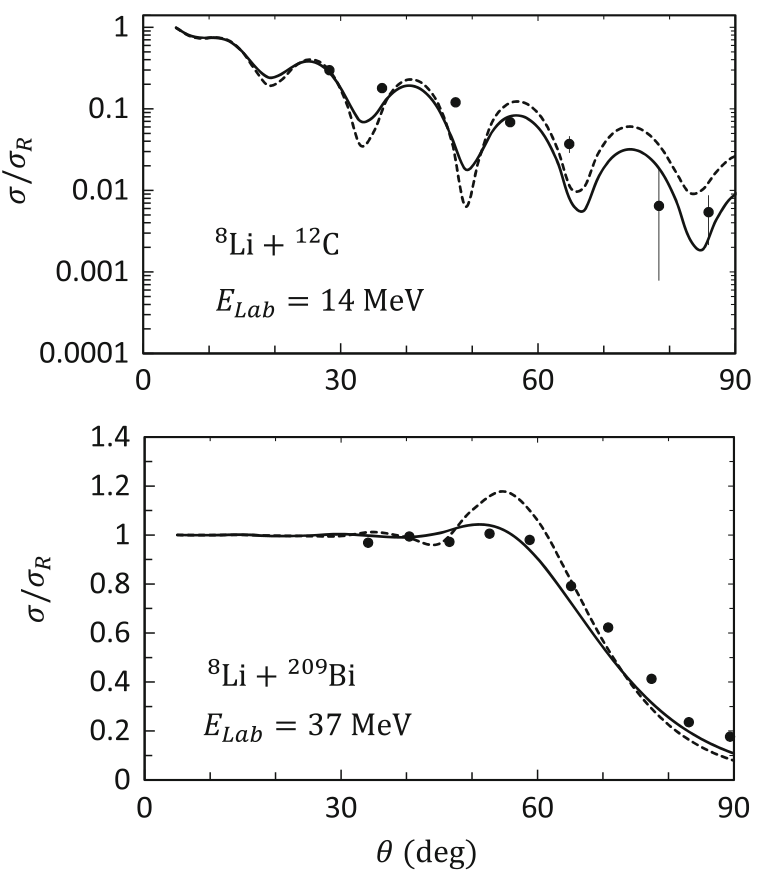}
    \caption{\label{fig:li8_mcdcc} $^{8}$Li+$^{12}$C (top panel) and  $^{8}$Li+$^{209}$Bi (bottom panel) elastic cross section, relative to Rutherford, computed within the microscopic CDCC approach for three-body projectiles. Dotted lines represent the calculations without breakup channels and the solid lines are the full calculations. Experimental data from Refs.~\cite{Coo18,Barioni09}. Figure from Ref.~\cite{Descouvemont2018}, with authorization from Springer.}
\end{figure}

Other applications of the microscopic CDCC method include its extension for three-body projectiles including a single nucleon as one of the three particles, such as $^8$Li or $^8$B as $(^4\text{He}+t+n)$ and $(^4\text{He}+{^3}\text{He}+p)$, respectively~\cite{Descouvemont2018}. Figure~\ref{fig:li8_mcdcc} shows results for the elastic scattering of $^7$Li on $^{12}$C and $^{209}$Bi targets at low energies. The dashed line corresponds to the calculation including only the projectile ground state, while the solid line is the full microscopic CDCC calculations that exhibit a better agreement with the data. 

\subsection{Interpretation of the CDCC wavefunction and cross sections}\label{sec:interp}
For a meaningful comparison with experimental data, it is important to properly understand which information and observables are (and which are not!)  provided by the CDCC wavefunction. Of course, elastic scattering is the most direct outcome of the method and the one that can be compared with data more easily. By construction, the elastic scattering produced by the CDCC calculation  incorporates the effect of the coupling to the inelastic and breakup channels included in the CDCC modelspace. According to our former notation, these breakup channels correspond to the three-body final states of the form $b+x+A$ in which the target and the projectile subsystems $b$ and $x$ escape the interaction region and remain in their ground state. Morover, in addition to these channels explicitly included in the modelspace, the CDCC includes effectively the effects of other channels associated with possible excitations of the target and/or of the fragments via the optical potentials. 

Regarding breakup observables, in principle, the CDCC wavefunction provides only the elastic breakup ones. The most immediate observables are the breakup angular distributions for specific bin intervals, i.e., 
\begin{equation}
\frac{d\sigma_n}{d\Omega_\text{c.m.}} = |f_{0,n}(\theta)|^{2}.
\end{equation}
When the binning method is employed, 
an approximate double differential cross section (with respect to the c.m. angle and relative energy) can be obtained by dividing each angular distribution by the bin width, as follows:
\begin{equation}
\frac{d^2\sigma}{d\Omega_\text{c.m.}  d \epsilon}  \simeq \frac{1}{\Delta_{n}}\frac{d\sigma_n}{d\Omega_\text{c.m.}}.
\label{eq:2body_bu}
\end{equation}

Regarding the nonelastic breakup contributions, which are effectively included in the fragment-target imaginary potentials, the CDCC method by itself does not provide detailed breakup cross sections involving excitations. However, one can define and compute a total absorption cross section that accounts for their overall contribution. For that, one first derives a reaction cross section in terms of the CDCC S-matrices as%
\footnote{See Ref.~\cite{Des17} for a different convention.}:
\begin{align}
\sigma_\text{reac} & = \frac{\pi}{K^2_i}\sum_{J_T \pi} \sum_{L_i} \frac{2J_T+1}{(2J^i_p+1)(2 J^i_t+1)}
 \left (1 - |S^{J_T}_{\beta_i,\beta_i}|^2 \right ),
\end{align}
and an integrated elastic breakup cross section
\begin{align}
\sigma_\text{bu} &= \frac{\pi}{K^2_i} \frac{K}{K_i} \sum_{J_T \pi} \sum_{L L_i} \frac{2J_T+1}{(2J^i_p+1)(2 J^i_t+1)}
 |S^{J_T}_{\beta,\beta_i}|^2 .
\end{align}

Then, one may define an {\it absorption} cross section as $\sigma_\text{abs} = \sigma_\text{reac} -\sigma_\text{bu}$. This absorption cross section will account for all nonelastic processes associated with the imaginary part of the fragment-target potentials, such as target excitation (i.e., inelastic scattering), complete fusion (i.e., capture of the whole projectile) or incomplete fusion (i.e., capture of one of the projectile fragments). Very often, comparison with experimental data requires more detailed cross sections, such as the angular or energy distributions of the fragment that survives after the capture (absorption) of the other fragment. As noted above, these observables are not directly provided by the CDCC method, at least in its standard formulation. However, as will be shown in Sec.~\ref{sec:3b-obs}, the CDCC wavefunction can be used as an input for a suitable formalism capable of providing such observables. 

\begin{figure}
\centering
\includegraphics[width=0.75\columnwidth]{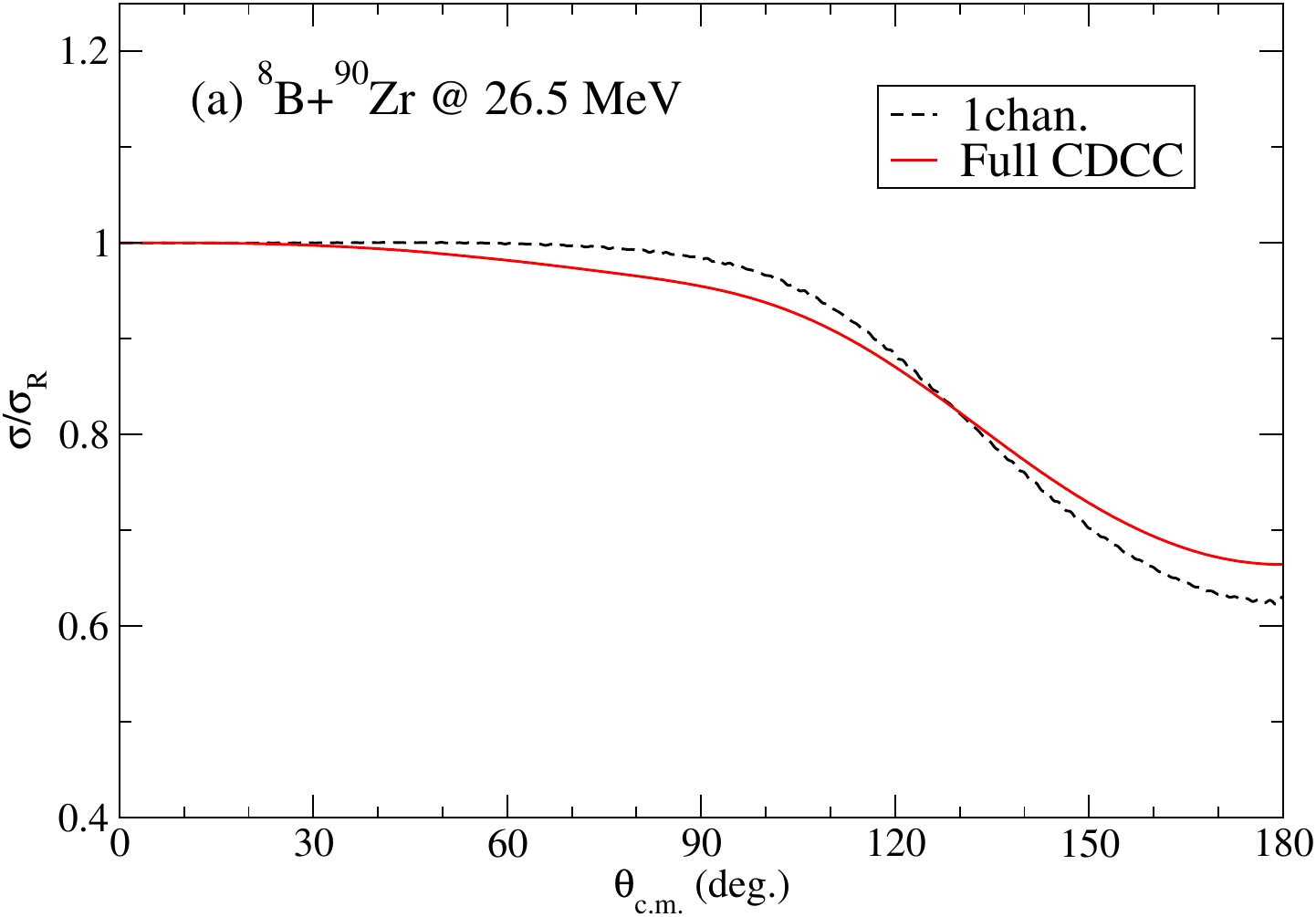} 
\includegraphics[width=0.75\columnwidth]{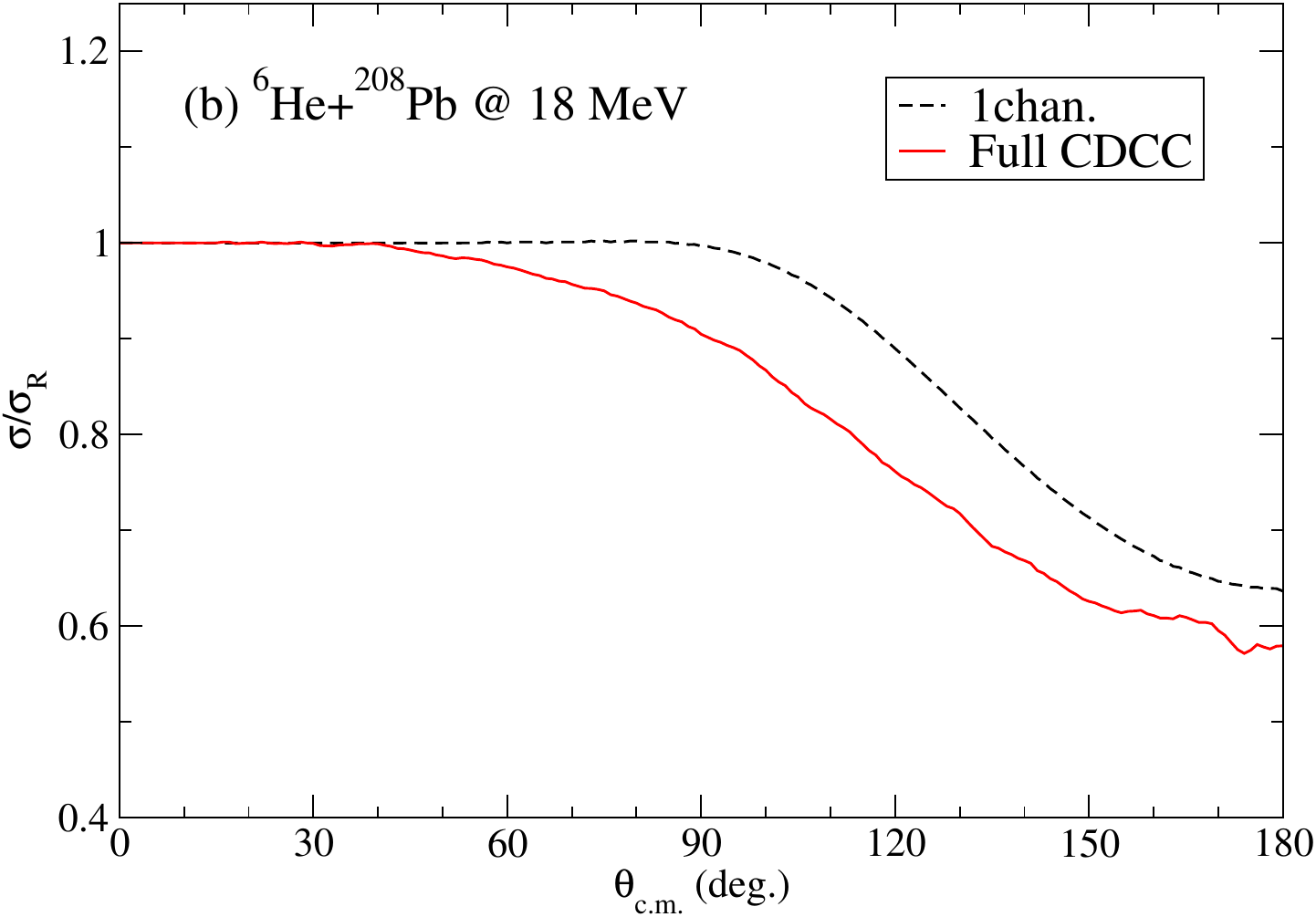}
\caption{Comparison of one-channel (i.e., no continuum) calculations and full CDCC calculations for the elastic scattering of (a) $^{8}$B+$^{90}$Zr 
and (b)  $^{6}$He+$^{208}$Pb at near-barrier energies. See text for details. }
\label{fig:b8_vs_he6} 
\end{figure}

The absorption cross section plays an important role in understanding the CDCC results. To quantify the importance of the continuum on the elastic cross sections it is common to compare full CDCC results with no-continuum calculations, in which only the ground-state to ground-state coupling potential is retained. An example is shown in Fig.~\ref{fig:b8_vs_he6}, which depicts the results for the $^{6}$He+$^{208}$Pb reaction at 18 MeV and the $^{8}$B+$^{90}$Zr reaction at 26.5 MeV. Dashed and solid lines denote the no-continuum and full CDCC calculations, respectively (in the $^{6}$He case, the improved dineutron model of Ref.~\cite{Mor07} was used).  Table \ref{tab:b8_vs_he6} lists the reaction, absorption, and (elastic) breakup cross sections from the CDCC calculations. Clearly, inclusion of the continuum has a much larger effect on the elastic cross section for $^{6}$He. This is in apparent contradiction with the fact that the elastic breakup cross section is larger in the $^{8}$B case, as shown in the Table. The apparent inconsistency can be explained by looking at the absorption and reaction cross sections. In the $^{8}$B reaction, the inclusion of the continuum couplings produces only a small increase of the absorption cross section, and the reaction cross section is essentially increased by the elastic breakup cross section. By contrast, in the $^{6}$He reaction, the absorption cross section increases by almost a factor of 3 when adding the continuum couplings and this results also in an equivalent increase of the reaction cross section, hence the enhanced effect on the elastic cross section. One may anticipate that the increased absorption cross section for $^{6}$He is associated with processes in which the valence neutrons experience some kind of nonelastic interaction with the target, such as transfer to bound states. These processes will be discussed in a subsequent section.

\begin{table}[ht]
    \centering
    \begin{tabular}{ccccc}
        \hline       
        Reaction &      & $\sigma_\mathrm{reac}$  & $\sigma_\mathrm{abs}$  & $\sigma_\mathrm{bu}$  \\
                &      & (mb) & (mb) & (mb) \\
                        \hline \hline
        $^{8}$B+$^{90}$Zr & Full  & 362 & 136 & 226 \\ 
                           & no cont. & 136   & 124  & -- \\ \hline
       $^{6}$He+$^{208}$Pb & Full  & 477  & 359  & 117 \\ 
                           & no cont. & 134   & 134 & -- \\ \hline
    \end{tabular}
    \caption{Comparison of reaction, absorption and elastic breakup cross sections for  $^{6}$He+$^{208}$Pb at 18 MeV and  $^{8}$B+$^{90}$Zr  at 26.5 MeV.}
    \label{tab:b8_vs_he6}
\end{table}

The different effect of the breakup channels on the elastic channel for neutron versus proton halo nuclei has been addressed in the literature by several authors \cite{Bon04,Kum12,Yan16}.

\subsubsection{Smoothing procedure for two-body observables}
Owing to the discretization procedure inherent to the CDCC method, the breakup cross sections are given only for the discrete energies of the bins or pseudostates. In reality, breakup cross sections should be continuous functions of the relative energy between the fragments. Several procedures have been proposed to convert the discrete breakup energy distributions into continuous distributions for arbitrary values of the inter-fragment relative energy. We discuss here two such methods, one based on the S-matrices and the other on the scattering amplitudes.

The S-matrices calculated in CDCC correspond to
discrete values of the energy. 
In the binning method, an approximation to this continuous $S-$matrix can be obtained  dividing
the discrete $S-$matrix  by the square root of the bin width, i.e.,
\begin{equation}
S^{J_T}_{\beta,\beta_i}(k)  \approx \frac{1}{\sqrt{\Delta E_n}} \hat{S}^{J_T}_{\beta,\beta_i} 
\label{eq:Scont_bin}
\end{equation}
with $\beta=\{L,J,n \}$, $\beta_i=\{L_i,J_0, n_0 \}$ and  $\Delta E_n$ is the width of the $n$-th bin.  

In the PS method, one could apply a similar procedure by assigning
a width to each pseudostate. In fact,  
this approach was used, for example, in Ref.~\cite{Mor02} to calculate the differential
breakup cross section from the cross section to individual
pseudostates, assuming that
the width of the $i$th pseudostate is approximately given by
$\Delta E_n=(\varepsilon_{n+1}-\varepsilon_{n-1})/2$. A more accurate procedure, 
previously proposed in Ref.~\cite{Mat03} is to obtain continuous $S-$matrix elements $S_{\beta,\beta_i}(k)$, depending
on the continuous variable $k$, as well as on the initial and final
angular momenta,  by an appropriate superposition of  the discrete
$S-$matrix elements   
$\hat{S}_{\beta,\beta_i}$ resulting from 
the solution of the coupled channel equations. Ignoring for simplicity intrinsic spins of the projectile constituents as \cite{Mat03,Tos01} one gets
\begin{equation}
\label{scont}
S^{J_T}_{\beta,\beta_i}(k) 
\approx
\sum_{n=1}^{N} 
\langle \phi^{(-)}_{k,\ell} | \phi^{(N)}_{n,\ell} \rangle 
\hat{S}^{J_T}_{\beta,\beta_i} ,
\end{equation}
%
where  $\phi^{(-)}_{k,\ell}(r)$ and $\phi^{(N)}_{n,\ell}(r)$ are the radial parts of the \emph{exact} and  pseudostate wavefunctions, respectively.  
The sum runs over the set pseudostates included in the  
coupled--channels calculation.

From the S-matrices, the double differential cross section for a $b-x$ breakup state with orbital angular momentum $\ell$ and wavenumber $k$ is given by:
\begin{equation}
\frac{d^2 \sigma_{\ell}(k)}{dk \, d\Omega_\text{c.m.}} = \sum_{m=0}^{\ell} \frac{\pi}{K_0^2} \left| \sum_{J,L} (2J+1) \langle \ell m L - m | J 0 \rangle 
|Y_{L-m}(\Omega_{\alpha}) S^{J_T}_{\beta,\beta_i}(k) \right |^2,
\label{eq:placeholder}
\end{equation}

\begin{figure}[t]
{\par\centering \resizebox*{0.85\columnwidth}{!}
{\includegraphics{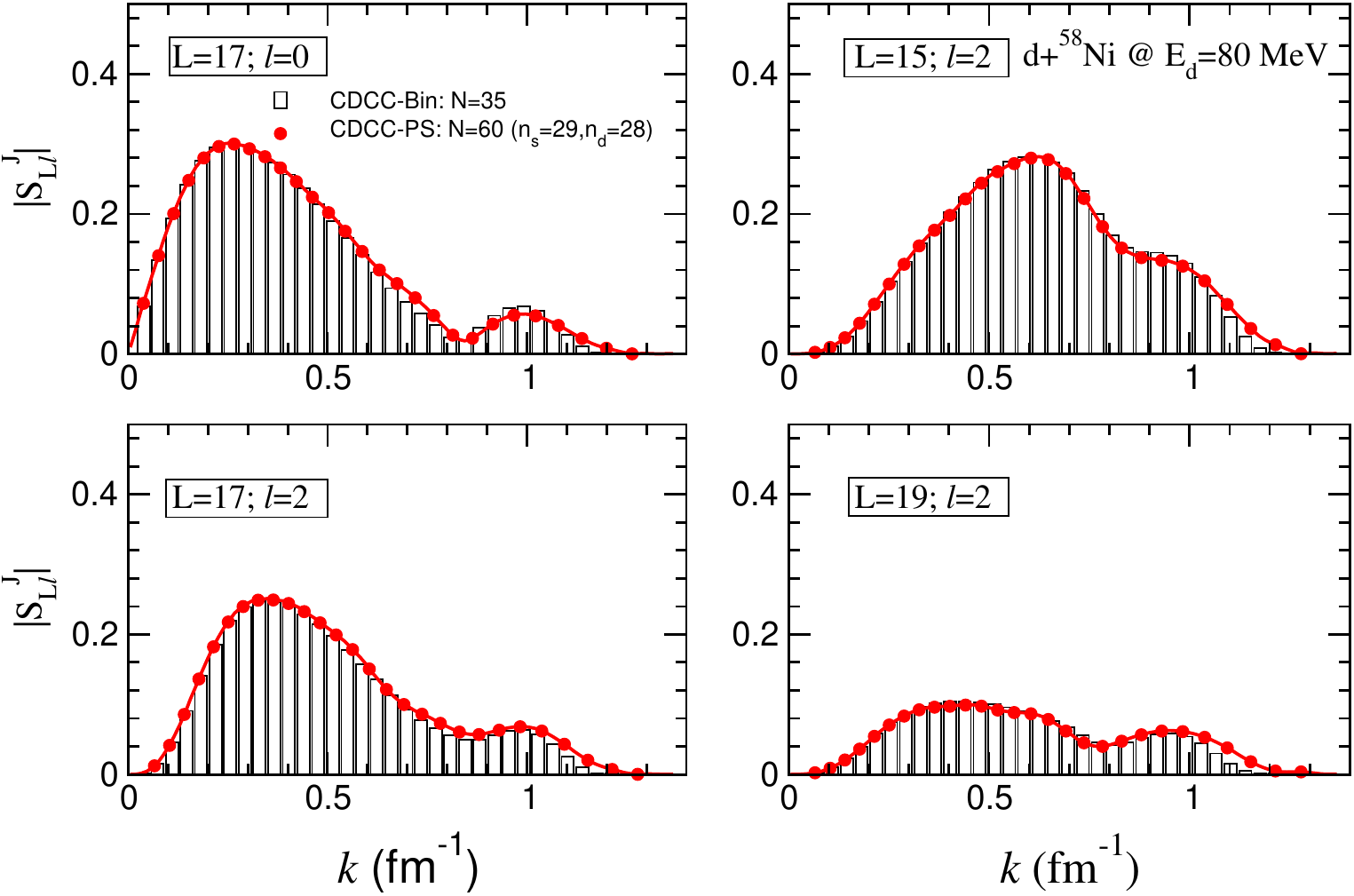}}\par}
\caption{\label{sbu_sd_j17} (Color online) Modulus of the breakup $S$ matrix elements for the 
total angular momentum $J=17$ for the reaction $d$+\nuc{58}{Ni} at 80 MeV
as a function of the p--n relative momentum in the final state. The histogram is the 
CDCC calculation with binning method. The filled and open circles represent the CDCC  calculations using the analytical THO pseudostate basis of Ref.~\cite{Mor09b}. The solid lines 
are obtained folding the discrete $S$-matrices with the {\em true} continuum wavefunctions. Adapted from Ref.~\cite{Mor09b}. 
}
\end{figure}

As an application of this method, in Fig.~\ref{sbu_sd_j17} we plot the modulus of the 
breakup $S-$matrix elements for the reaction d+$^{58}$Ni$\to$p+n+$^{58}$Ni for a total angular momentum $J=17$. Intrinsic spins of the proton and neutron were omitted, so the relevant channels are
$(\ell,L)$=(0,17), (2,15), (2,17) and  (2,19). For the CDCC-Bin calculations, the continuum was divided into 
$N_s=N_d=35$ bins up to a maximum excitation 
energy of 70 MeV. For the CDCC-PS calculation, a transformed harmonic oscillator (THO) basis of $N=50$ states was necessary to obtain 
full convergence at high excitation energies, although $N=30$ gives already rather good results. After diagonalization  of the internal Hamiltonian, only the eigenstates below 70~MeV were retained, reducing the actual size of the basis to only 24 for each partial wave  (along with the ground state).
 In Fig.~\ref{sbu_sd_j17}, the histogram represents  the 
CDCC-Bin calculation, the filled circles correspond to the CDCC-PS
calculation,  in which each discrete $S-$matrix 
has been divided by the square root of the pseudostate width, and the line is the
CDCC-PS calculation folded with the  
continuum wavefunctions, Eq.~(\ref{scont}). It is clearly seen that both
discretization methods  are in almost perfect agreement. Further details about these calculations can be found in Ref.~\cite{Mor09b}.

The connection between the discrete and continuous  breakup cross sections can be also done at the level of the transition amplitudes themselves. In both CDCC and XCDCC, the solution of the coupled-channel equations provides some discrete breakup  transition amplitudes, denoted $T_{M_0,M'}^{i,J_0,J'}(\theta_i,K_i)$, connecting an initial state $|J_0  M_0 \rangle$ with a three-body final state comprised of the target (assumed to be structureless), the valence particle and the core, at some discrete value of the final projectile-target c.m.\ momentum  $\vec{K}_i=\{\theta_i,K_i\}$. The first step of the formalism is to relate these discrete amplitudes with the  actual breakup scattering amplitudes, without continuum discretization, that we denote as   $T_{\mu \sigma; M_0}^{I s; J_0}(\vec{k},\vec{K})$, where  $\vec{k}$ is the projectile internal  relative momentum. 
Formally, these breakup transition amplitudes can be written in integral form as:
\be
T_{\mu \sigma; M_0}^{I s; J_0}(\vec{k},\vec{K})=
\langle \phi_{\vec{k}; I \mu; s \sigma}^{(-)} 
e^{i \vec{K} \cdot \vec{R}} | U | \Psi_{J_0,M_0}(\vec{K}_0)\rangle , 
\label{T-matrix1}
\ee
where $U=U_{bA}(\vec{r},\vec{R},\xi)+U_{xA}(\vec{r},\vec{R})$ and $\phi_{\vec{k}; I \mu; s \sigma}^{(-)}$ are two-body  {\it exact} scattering wave functions of the $b+x$ system for a relative final momentum $\vec{k}$ and  given core and valence spins. 

In order to relate the discrete and continuous amplitudes, one can 
approximate the exact wavefunction $\Psi_{J_0,M_0}$ in the equation above by its (X)CDCC counterpart and introduce the approximate completeness relation in the truncated discrete basis $\{ \phi^{(N)}_{n,J,M}; i=1,\ldots,N \}$: 
\begin{align}
T_{\mu \sigma; M_0}^{I s; J_0}(\vec{k},\vec{K}) & \simeq  \sum_{n,J',M'} \langle \phi_{\vec{k}; I \mu; s \sigma}^{(-)}| \phi^{(N)}_{n,J',M'}\rangle  \langle \phi^{(N)}_{n,J',M'}
e^{i \vec{K} \cdot \vec{R}} | U | \Psi_{J_0,M_0}^\mathrm{CDCC}(\vec{K}_0)\rangle  \nonumber \\
&= \sum_{n,J',M'} \langle \phi_{\vec{k}; I \mu; s \sigma}^{(-)}| \phi^{(N)}_{n,J',M'}\rangle T_{M_0,M'}^{n,J_0,J'}(\vec{K}) ,
\label{T-matrix2}
\end{align}
where the transition matrix elements $T_{M_0,M'}^{n,J_0,J'}(\vec{K})$ 
are to be interpolated from the discrete ones $T_{M_0,M'}^{n,J_0,J'}(\theta_i,K_i)$.   Expressions for the overlaps between the final scattering states and the discrete states, $\langle \phi_{\vec{k}; I \mu; s \sigma}^{(-)}| \phi^{(N)}_{n,J',M'}\rangle $, are given explicitly in Refs.~\cite{Tos01} and \cite{Die17} for bin and PS functions, respectively.

The transition amplitudes of Eq.~(\ref{T-matrix2}),
$T_{\mu \sigma; M_0}^{I s; J_0}(\vec{k},\vec{K})$,
contain the dynamics of the process in the coordinates
describing the relative and center of mass motion 
of the core and the valence particle. From these amplitudes one can
derive two-body observables for a fixed spin of the core, $I$,
the solid angles describing the orientations of $\vec{k}$ ($\Omega_k$) and $\vec{K}$ ($\Omega_K$),
as well as the relative energy between the valence
and the core, $E_\mathrm{rel}$.
These observables 
factorize into the transition matrix elements and a kinematical factor:
\begin{align}
\frac{d^3\sigma^{(I)}}{d\Omega_k d\Omega_K dE_\mathrm{rel}} = & 
     \frac{\mu_{bx} k_I}{(2\pi)^5 \hbar^6} \frac{K}{K_0} \frac{\mu_{aA}^2}{2J_0+1}  \sum_{\mu, \sigma, M_0} |T_{\mu \sigma; M_0}^{I s; J_0}(\vec{k},\vec{K})|^2 ,
\label{three-obsv-cm}
\end{align}
where $\mu_{bx}$ and $\mu_{aA}$ are the valence-core and projectile-target
reduced masses. 

\subsubsection{Three-body observables}
\label{sec:3b-obs}
Within the CDCC and XCDCC reaction formalisms, breakup is treated as an
excitation of the projectile to the continuum, so the theoretical cross sections are described in terms of
the c.m.\ scattering angle of the projectile and the relative energy of the constituents,
using two-body kinematics. For comparison with experimental data, it is
useful to have also the cross sections in terms of the angle and energy of the projectile fragments, since these quantities are more directly connected with the actual measurements.

In the case of the standard CDCC framework,  fivefold fully exclusive cross sections have been derived and presented by several authors \cite{Ise86,Tos01}. The method was  generalized in Ref.~\cite{Die17} to the case of XCDCC. We briefly review the main formulas of the latter, noting that the case without core excitation is recovered when a single core state is considered. For simplicity, we ignore the target spin. 

Using the two-body transition amplitudes discussed in the previous subsection, the three-body observables, assuming the energy of the core is measured, are given by \cite{Tos01}:
\begin{align}
\frac{d^3\sigma^{(I)} }{d\Omega_b d\Omega_x dE_b} &=
     \frac{2 \pi \mu_{aA}}{\hbar^2 K_0} \frac{1}{2J_0+1} \nonumber \\ 
 &\times \sum_{\mu, \sigma, M_0} |T_{\mu \sigma; M_0}^{I s; J_0}(\vec{k},\vec{K})|^2               
      \rho(\Omega_b, \Omega_x, E_b) ,
\label{three-obsv}
\end{align}
where the phase space term $\rho(\Omega_b, \Omega_x, E_b)$, i.e., the number of states per
 unit core energy interval at solid angles $\Omega_b$ and $\Omega_x$,
 takes the form \cite{Fuc82}:
\begin{align}
\rho(\Omega_b, \Omega_x, E_b) & =
     \frac{m_b m_x \hbar k_b \hbar k_x}{(2\pi\hbar)^6} \nonumber \\
     & \times  \left[\frac{m_A}{m_x+m_A+m_x(\vec{k}_b-\vec{K}_{tot})\cdot\vec{k}_x/k_x^2}\right] .
\label{phfac}
\end{align}
Here, the particle masses are given by $m_b$ (core), $m_x$ (valence), and $m_A$ (target)
while $\hbar \vec{k}_b$ and $\hbar \vec{k}_x$ are the core and valence particle momenta
 in the final state. The total momentum of the system corresponds to $\hbar \vec{K}_{tot}$ and
the connection with the momenta in Eq.~(\ref{T-matrix2}) is made through:
\be
\vec{K}=\vec{k}_b+\vec{k}_x-\frac{m_a}{M_{tot}}\vec{K}_{tot} ; \quad
\vec{k}=\frac{m_b}{m_a}\vec{k}_x-\frac{m_x}{m_a}\vec{k}_b
\ee
with $m_a=m_b+m_x$ and $M_{tot}=m_b+m_x+m_A$ the total masses of the projectile and the
three-body system, respectively.

\begin{figure}[!ht]
\begin{center}
\begin{minipage}[t]{.85\columnwidth}
\begin{center}\includegraphics[width=0.8\columnwidth]{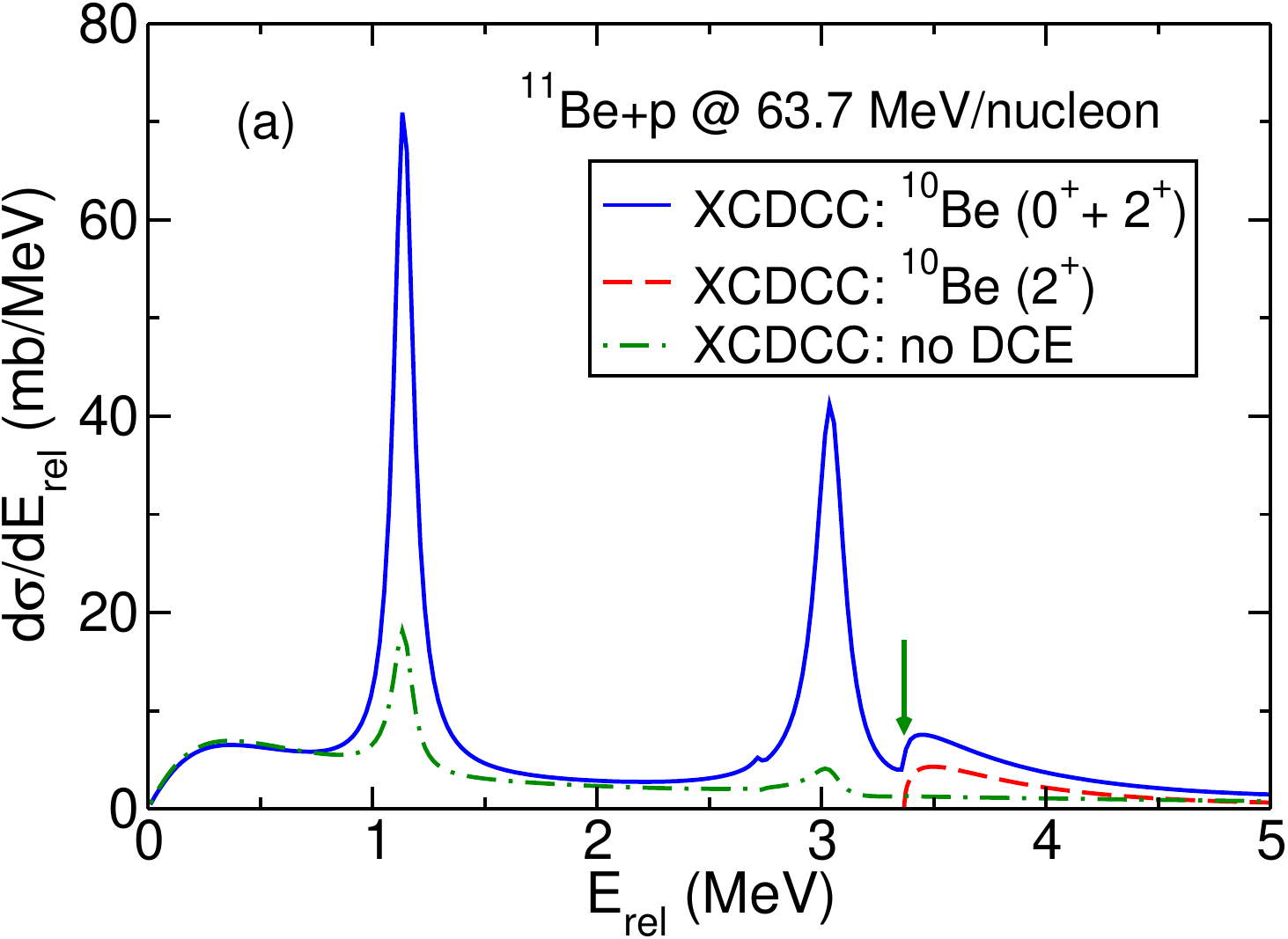} \end{center}
\end{minipage}
\begin{minipage}[t]{.85\columnwidth}
\begin{center}\includegraphics[width=0.8\columnwidth]{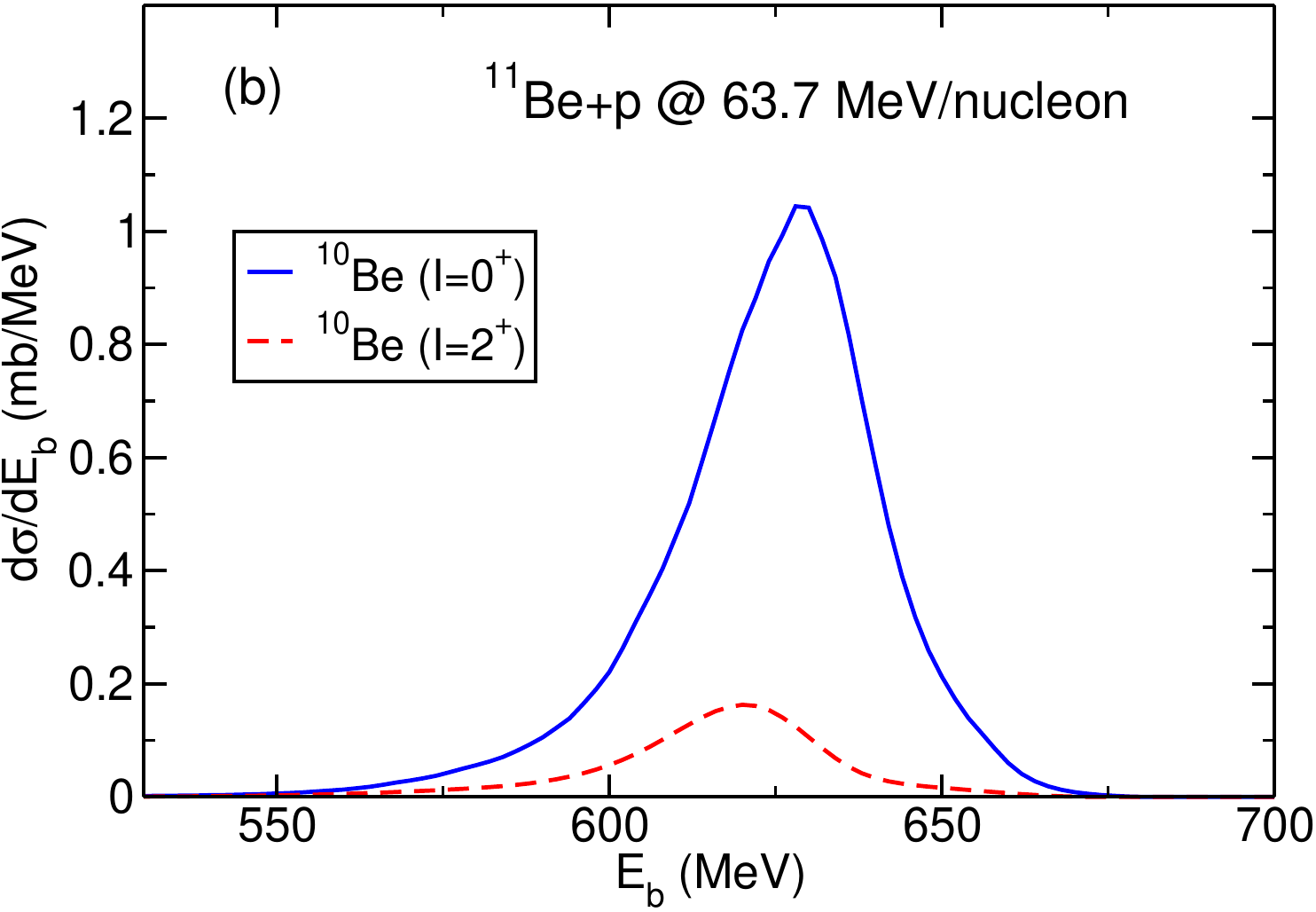}\end{center}
\end{minipage}
\caption{\label{fig:be11p_3b} Differential breakup cross  sections, with respect to the $n$-$^{10}$Be relative energy (a) and with respect to the final $^{10}$Be energy (b),  for the breakup of $^{11}$Be on protons at 63.7~MeV/nucleon. The contributions of the $^{10}$Be ground ($0^+$) and first excited ($2^+$) states are shown in each panel. In the top panel, the result of the calculation neglecting the $^{10}$Be excitation mechanism (labelled ``no DCE'') is also shown. The vertical arrow denotes the threshold for $^{10}$Be($2^+_1$)+n. Figure adapted from Ref.~\cite{Die17}. }
\end{center}
\end{figure}

An application of this formalism is presented in Fig.~\ref{fig:be11p_3b}, corresponding to the breakup of $^{11}$Be on a proton target at $E_p=63.7$~MeV/u. The top panel corresponds to the differential cross section with respect to the final $n$+$^{10}$Be relative energy and the bottom panel to the breakup with respect to the final $^{10}$Be energy (integrated in the $^{10}$Be and neutron angles). These observables were computed by  transforming the  XCDCC breakup transition amplitudes by means of Eqs.~(\ref{T-matrix2})-(\ref{three-obsv}), and integrating over the unmeasured variables. The XCDCC calculations were performed with a  $^{11}$Be model including the $^{10}$Be ground ($0^+$) and first excited ($2^+$) states. The formalism allows to separate and quantify the contribution of these two states of $^{10}$Be. Note that, due to energy conservation, in the relative energy distribution the contribution coming from the $^{10}$Be($2^+$) state contributes only above the excitation energy of this state ($E_x=3.367$~MeV). The importance of the $^{10}$Be excitation during the reaction due to its interaction with the target nucleus is illustrated also in the top panel by means of a calculation in which the deformed part of the $^{10}$Be+p interaction is omitted (dot-dashed line).

\subsubsection{Pseudostates versus bins}\label{sec:PSvsbin}
As discussed in the preceding sections,  discretization procedures employed in the CDCC method are either based on the pseudostate (PS) method or the binning method. Whereas reactions observables should be independent of the adopted discretization method (provided the calculations are fully converged), in practical applications one of the two methods might result more convenient than the other. 

\begin{figure}
\begin{center}
\includegraphics[width=0.5\textwidth]{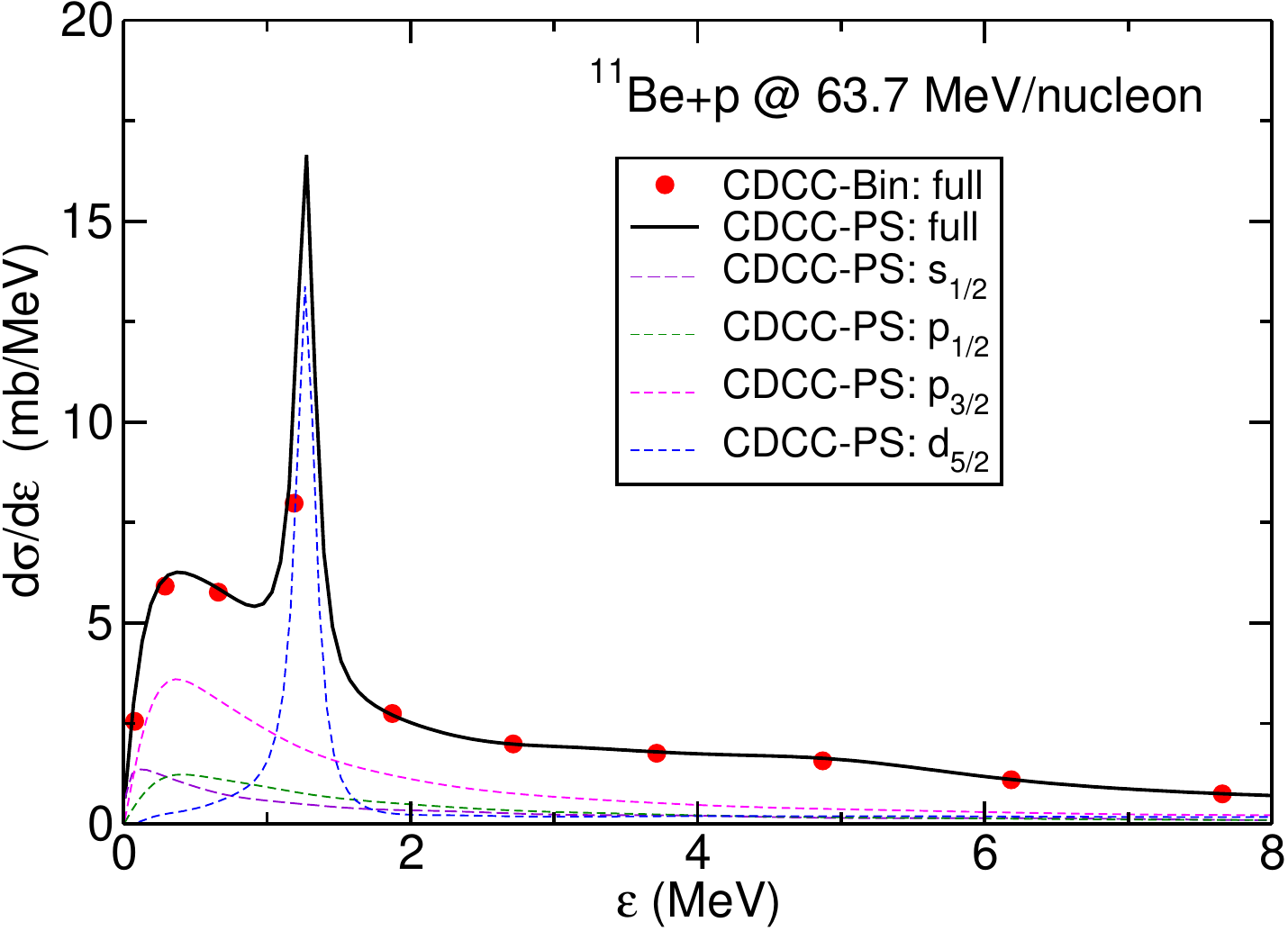}
\end{center}
\caption{\label{fig:be11p_dsde_tho_vs_bins} 
Comparison of pseudostate (PS) and binning methods for the $^{11}$Be+p $\to$ $^{10}$Be+n+p reaction at 63.7~MeV/nucleon. Both methods use the same structure model \cite{Cap04} and include continuum states with configuration $s_{1/2}$, $p_{1/2}$, $p_{3/2}$, $d_{3/2}$, $d_{5/2}$.  For the CDCC-PS calculation, the dominant individual contributions are shown separately.}
\end{figure}

The PS discretization method becomes particularly suitable when dealing with narrow resonances. In the case of bins, one needs fine discretization in order to obtain a detailed description of the resonance region in the relative energy spectrum. By contrast, using PS's one can obtain a detailed description of the resonance profile with a relatively small basis, with the aid of the convolution procedure discussed in Sec.~\ref{sec:3b-obs}. 
This is illustrated in Fig.~\ref{fig:be11p_dsde_tho_vs_bins}, where we show the $^{10}$Be+n relative energy differential cross section corresponding to the $^{11}$Be+p $\to$ $^{10}$Be+n+p reaction at an incident energy of 63.7~MeV/nucleon. The $^{11}$Be bound and continuum states were generated with the structure model of Ref.~\cite{Cap04}, which does not account for $^{10}$Be excitations. Continuum waves $s_{1/2}$, $p_{1/2}$, $p_{3/2}$, $d_{3/2}$, $d_{5/2}$ were included. For the CDCC-Bin calculation, the continuum was truncated at $E_\mathrm{rel}=12$~ MeV and divided into 12 bins for each partial wave, evenly spaced in momentum. The relative-energy differential cross section was obtained by dividing the cross section of each bin by the width of that bin [see Eq.~(\ref{eq:2body_bu})]. The CDCC-PS calculations employ a THO basis with $N=30$ ($N=25$) oscillator functions for $\ell=1,2$ ($\ell=0$). After diagonalization of the $^{11}$Be Hamiltonian on this basis, only the eigenvalues below 12 MeV were retained.  This results in 13 states for each partial wave (15 for $\ell=0$).  The calculated breakup scattering amplitudes were then convoluted using Eq.~(\ref{T-matrix2}), from which the differential cross sections, Eq.~(\ref{three-obsv-cm}), were then evaluated. Finally, the latter were integrated over the solid angles $\Omega_k$ and $\Omega_K$.

As seen in Fig.~\ref{fig:be11p_dsde_tho_vs_bins}, the binning and PS methods yield, as expected, almost identical results but the PS method allows for a finer description of the resonance region with a comparable, or even smaller, number of basis functions.

However, there are other situations where the binning procedure turns out to be more efficient than the PS method. One such situation is the case of reactions of highly polarizable weakly-bound nuclei (such as neutron-halo nuclei) on high-Z targets. These reactions are characterized by the presence of strong long-range Coulomb couplings, which tend to emphasize low-lying excitation states and probe large separations. In this case, the binning method has been shown to provide more stable results with respect to the basis size \cite{Rod11}.

\section{Transfer reactions with weakly bound nuclei \label{sec:transfer}}
Transfer reactions are key spectroscopic tools for both stable and unstable nuclei. Angular distributions of outgoing fragments provide information on the angular momentum content of the populated states and the magnitude is closely related to the single-particle content of these states, quantified in terms of spectroscopic factors. The standard tool for analysing transfer reactions, the distorted-wave Born approximation (DWBA) method, assumes that the transfer occurs in a single-step. Possible excitations of the colliding or outgoing nuclei are ignored or, at most, taken into account effectively through the choice of the effective interactions. These interactions are typically chosen so as to reproduce the corresponding elastic scattering cross sections in the incident and exit channels. 

\begin{figure}
\begin{center}
\includegraphics[width=0.75\columnwidth]{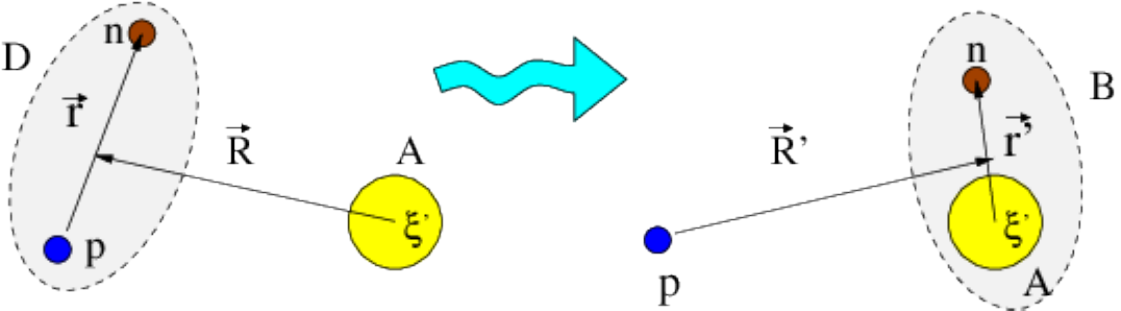}
\end{center}
\caption{\label{fig:transfer-dp} 
Relevant coordinates for a transfer reaction of the form $A(d,p)B$.}
\end{figure}

To see how this method works in practice, let us consider as a particular case a process of the form $A(d,p)B$, schematically depicted in Fig.~\ref{fig:transfer-dp}.  Using the post-form representation, the transition matrix for this process can be written as \cite{Sat83}
\begin{equation}
\label{eq:Tdp}
{\cal T}_{dp}= \langle \chi^{(-)}_p \Phi_B | V_{pn} + U_{pA}-U_{pB} | \Psi^{(+)}_d \rangle  ,
\end{equation}
where $V_{pn}$, $U_{pA}$ are the proton-neutron and proton-target interactions, $U_{pB}$ is an auxiliary (and, in principle, arbitrary) potential for the $p$-$B$ system, $\Phi_B$  is the internal wave function of the residual nucleus $B$ and $\Psi^{(+)}_d$ is the total wave function corresponding  to an incident deuteron beam of kinetic energy $E_d$ and binding energy $\varepsilon_d$. If target excitation is not considered explicitly, this total wavefunction can be  approximated as
\be
 \Psi^{(+)}_d(\vecr,\vecR,\xi_A) \approx \Psi^{3b}(\vecr,\vecR)_d \Phi^{(0)}_A(\xi_A)
\ee
where $\Phi^{(0)}_A(\xi_A)$ denotes the target ground-state wavefunction.

Ignoring antisymmetrization for simplicity, the wave function $\Phi_B$ for a total angular momentum $J$ and projection $M$ can be expanded in $A$ states using the usual parentage decomposition
\begin{equation}
\label{eq:AB}
\Phi^{JM}_B(\vecr_{nA},\xi)= \sum_{I,\ell,j} [ \Phi^I_A(\xi) \otimes \phi_{\ell j }(\vecr_{nA}) ]_{JM} ,
\end{equation}
where $I$ is the spin of $A$, $\ell$ and $j$ the orbital and total ($\vec{j}=\vec{\ell}+s$) angular momentum of the valence particle and  $\phi_{\ell j }(\vecr_{nA})$   is  a function describing the neutron-core relative motion. The normalization
\begin{equation}
\label{eq:SF}
{\cal S}_{\ell,j} = \int |\phi_{\ell j }(\vecr_{nA})|^2 d\vecr_{nA}
\end{equation}
can be regarded as a spectroscopic factor for the configuration $\{\ell,j\}$.\footnote{Strictly, the functions $\phi_{\ell j}$ and the spectroscopic factors ${\cal S}_{\ell j}$ depend in general on the core state $I$ but, in the present case, this quantum number can be readily inferred from the $\ell,j$ values, so it is omitted for brevity.}  

If the interactions $U_{pA}$ and $U_{pB}$ are assumed to be independent of the target degrees of freedom ($\xi_A$), the integral in $\xi_A$ can be readily performed, transforming Eq.~(\ref{eq:Tdp}) into
\begin{equation}
\label{eq:T3b}
{\cal T}^{3b}_{dp}= \sqrt{S_{\ell,j}} \langle \chi^{(-)}_p \phi_{\ell j} | V_{pn} + U_{pA}-U_{pB} | \Psi^{3b,(+)}_d \rangle  ,
\end{equation}

In  DWBA, the three-body wavefunction $\Psi^{3b,(+)}_d$ is further approximated by $\Psi^\mathrm{(+)}_d \approx \chi^{(+)}_{d}(\vecR) \varphi_{d}(\vecr)$. 
When the transfer reaction involves weakly-bound nuclei, including the deuteron discussed here, this choice is not well justified. Due to the presence of the short-range $V_{pn}$ interaction, the evaluation of DWBA matrix element is mostly sensitive to small $p$-$n$ separations. These configurations do not only contain contributions coming from the deuteron ground state, but also from $p$-$n$ unbound states. By contrast, the deuteron optical potential describing deuteron-target elastic scattering is not restricted to small $p$-$n$ distances. Therefore, the use of the deuteron optical potential in the $(d,p)$ or $(p,d)$ transfer amplitude is likely to lack important deuteron breakup components.  This problem was recognized long ago and several solutions have been proposed. One of the first and  most popular ones is the adiabatic method of Johnson and Soper \cite{Ron70}. 

\subsection*{The Johnson-Soper approximation}
In the Johnson-Soper (JS) approximation, this effect is approximately taken into account by means of the choice
\be
\Psi_d^{3b,(+)}(\vec R, \vec r) \simeq \chi_{JS}^{(+)}(\vecR) \phi_{d}(\vec r) ,
\ee
where $\chi_{JS}^{(+)}(\vec R)$ is the solution of a two-body scattering problem, on the coordinate $\vec R$, in which the interaction is given by a $p-n$ zero-range approximation:
\be
U_{JS}(R) = U_{pA}(R)+U_{nA}(R)  .
\label{eq:JS}
\ee
We see that, in this limit, the adiabatic theory of the transfer amplitude adopts a form akin to that found in  DWBA, but this analogy is only formal because the function $\chi_{JS}^{(+)}(\vecR)$ includes contributions from breakup and the potential  $U_{JS}(R)$ may have little to do with the optical potential describing the deuteron elastic scattering, so $\chi_{JS}^{(+)}$ does not provide a good description of the elastic channel.  Due to the adiabatic approximation, the JS theory is not expected to be accurate at low incident energies. 

\begin{figure}
\begin{center}
\includegraphics[width=0.9\columnwidth]{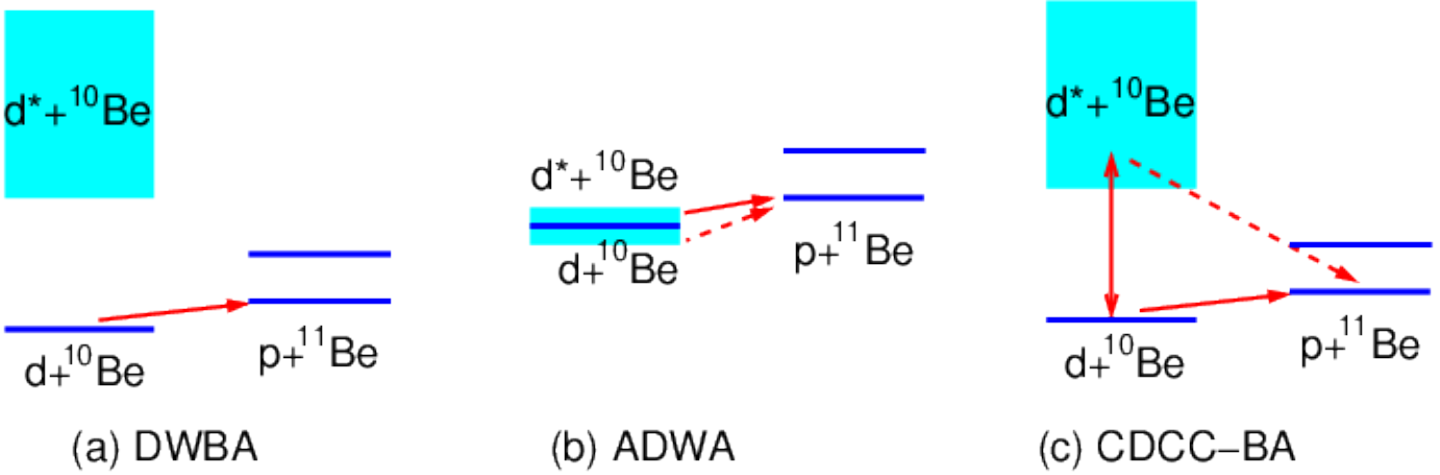}
\end{center}
\caption{\label{be10dp_schemes} Schematic representation of DWBA, ADWA and CDCC-BA approaches for a $(d,p)$ transfer reaction. }
 \end{figure}

The adiabatic approximation is equivalent to neglecting the excitation energy of the projectile states, \cite{Ron70}, which amounts at setting $H_{bx} \to \varepsilon_0$ in the three-body Hamiltonian (\ref{eq:Heff}). The adiabatic wave function takes into account the excitation to breakup channels, assuming that these states are degenerate in energy with the projectile ground state, as illustrated in Fig.~\ref{be10dp_schemes}(b). Therefore,  the ADWA approach 
takes into account, approximately, the effect of deuteron break-up on the transfer cross section, within the adiabatic approximation. So, it should be well suited to describe deuteron
scattering at high energies, around 100 MeV per nucleon. Systematic studies \cite{Har71,Sat71,Wal76} have shown that ADWA is superior to standard DWBA for $(d,p)$ scattering at these relatively high energies.

\subsection*{The Johnson-Tandy approximation}
\label{sec:JT}
Although the zero-range adiabatic model of JS provides a systematic improvement over the conventional DWBA, there are situations in which the former fails to reproduce the experimental data \cite{Ste86,Ron89}. Models which go beyond the zero-range and adiabatic approximations are therefore needed. One of such models is the Weinberg expansion method of Johnson and Tandy (JT) \cite{JT74}. The idea is to expand $\Psi_d^\mathrm{3b,(+)}$ in terms of a set of functions which are complete within the range of $V_{pn}$. A convenient choice is the set of Weinberg states (also called Sturmians), given by
\begin{equation}
\label{eq:Psi-W}
 \Psi_d^{3b,(+)}(\bR,\br)=\sum_{i=0}^{N} \phi^W_{i}(\br)\chi^W_{i}(\bR) \, ,
\end{equation}
where $\phi^W_i(\xi)$ are the Weinberg states, which are solutions of the eigenvalue equation
\be
[ T_\vecr  + \alpha_i V_{pn} ] \phi^W_i(\vecr) = - \varepsilon_d \phi^W_i(\vecr) ,
\ee
where $\varepsilon_d=2.225$~MeV is the deuteron binding energy and where $\alpha_i$ are the eigenvalues, to be determined along with the eigenfunctions. Beyond the range of the potential all the Weinberg states decay exponentially, like the deuteron ground-state wave function. For  $i=0$, $\alpha_0=1$ and so $\phi^W_0(\vecr)$ is just proportional to the deuteron ground state. As $i$ and $\alpha_i$ increase, they oscillate more and more rapidly at short distances. The Weinberg states form a complete set of functions of $\vecr$ within the range of the potential $V_{pn}$. They are well suited to expand $ \Psi_d^{3b,(+)}$ in this region, as is required by the amplitude in Eq.~(\ref{eq:Tdp}). They do not satisfy the usual orthonormality relation but the less conventional one
\be
\langle \phi^W_i | V_{pn} | \phi^W_j \rangle = - \delta_{ij} .
\ee
If we retain in  Eq.~(\ref{eq:Psi-W})  only the leading term, $\Psi_d^{3b,(+)}(\bR,\br) \approx \phi^W_0 \chi^W_{0}(\bR)$, one finds \cite{JT74} that $\chi^W_{0}$ verifies the single-channel equation
\be
[T_\vecR  + U_{JT}(\vecR) - E_d ]  \chi^W_{0}(\vecR) =0 ,
\ee
with $E_d= E - \varepsilon_d$ and where the potential $U_{JT}$ is given by:
\be
\label{eq:UJT}
U_{JT}(R)=  \frac{\langle \phi^W_{0}(\br) | V_{pn} (U_{nA}+U_{pA}) | \phi^W_{0}(\br) \rangle}{\langle \phi^W_{0}(\br) | V_{pn} | \phi^W_{0}(\br) \rangle} .
\ee
The bra and ket in this equation mean integration over $\vecr$, with fixed $\vecR$. Interestingly, in the zero-range limit, $U^{JT}(R)$ reduces to the JS potential, Eq.~(\ref{eq:JS}). Therefore, the zero-order result given by Eq.~(\ref{eq:UJT}) can be regarded as a finite-range version of the adiabatic (JS) potential. These two models are globally referred to as Adiabatic Distorted Wave Approximation (ADWA). However, it is worth noting that the full Weinberg expansion makes no reference to the incident energy and, as such, does not involve the adiabatic approximation. This suggests that a stripping theory based on this Weinberg expansion can be used at low energies, where the adiabatic condition is not well satisfied.  The inclusion of higher-order terms ($i>0$) in the Weinberg expansion has been investigated in Refs.~\cite{Lai93,Pan13}. 

\begin{figure}
\begin{center}
\includegraphics[width=0.5\textwidth]{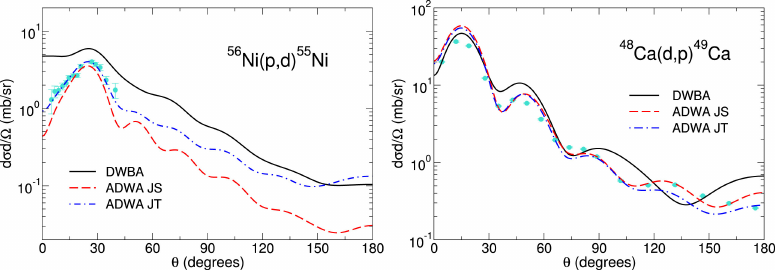}
\end{center}
\caption{\label{fig:dwba_vs_adwa} 
Comparison of the ADWA  and DWBA methods for the $^{56}$Ni($p$,$d$)$^{57}$Ni (left) and $^{48}$Ca($d$,$p$)$^{49}$Ca (right) reactions at incident energies of 37 MeV and 10 MeV, respectively. The nucleon optical potentials are taken from the global CH89 \cite{CH89} systematics 
while the Daehnick \cite{DCV80} global potential is used for the deuteron optical potential in the DWBA calculations. For the ADWA calculations, the prescriptions of Johnson-Soper (JS) and Johnson-Tandy (JT) are shown. Quoted from Ref.~\cite{Tim20}.}
\end{figure}

In Fig.~\ref{fig:dwba_vs_adwa} we present a comparison between the DWBA and ADWA methods for the $^{56}$Ni(p,d)$^{57}$Ni (left) and $^{48}$Ca(d,p)$^{49}$Ca (right) reactions. For the ADWA calculations, the prescriptions of JS and JT are shown.  It is clearly seen that the ADWA, in both its zero-range and finite-range forms, provides an improved description of the data as compared to the DWBA method.

An appealing feature of the ADWA method is that its ingredients are completely determined by experiments. These ingredients are the proton-target and neutron-target optical potentials, evaluated at half the deuteron incident energy,  as well as the well-known proton-neutron interaction. 
 On the negative side, the ADWA approach does not consistently describe elastic scattering and nucleon transfer. Although, physically, 
elastic scattering, transfer and break-up should be closely related by flux conservation, this connection is not present in ADWA. Furthermore, the arguments leading to ADWA are strongly dependent on the assumption that the transfer process is governed by
a short-range operator. Thus, it is not obvious that the approximations remain valid for other weakly bound systems, like \nuc{11}{Be}. Even in the case
of $(d,p)$ scattering, the transfer matrix element is determined, in addition to the  $n$-$p$ interaction, by the proton-target and proton-composite interactions which define the {\it remnant} term. The role of these terms, which would have contributions of three-body configurations in which proton and neutron are not so close together, is not clear {\it a priori}.  The latter problem can be avoided by using an alternative expression for the scattering amplitude. Following Goldberg and Watson \cite{Gol64}, one can choose the auxiliary interaction $U_{pB}$ that appears in the remnant term of Eq.~(\ref{eq:T3b}) to cancel exactly the core-target interaction, $U_{pA}$. This results in an alternative, but still exact, scattering amplitude (the spectroscopic factor is omitted for simplicity), namely,
\begin{equation}
\label{eq:T3b_GW}
{\cal T}^\mathrm{3b_{GW}}_{dp}= \langle \tilde{\Phi}^{(-)}_{pB} | V_{pn}  | \Psi^{3b,(+)}_d \rangle  ,
\end{equation}
where $\tilde{\Phi}^{(-)}$ is a solution of the three-body equation:
\begin{equation}
[T_{\vecr'} + T_{\vecR'} + U_{pB} + U - E ] \tilde{\Phi}^{(-)}_{pB} =0 .
\end{equation}

 In \cite{Tim99} Timofeyuk and Johnson use an adiabatic approximation for $\tilde{\Phi}^{(-)}_{pB}$ to produce a tractable expression that still includes recoil excitation and breakup effects, while keeping the matrix element constrained to the range of the $V_{pn}$ interaction:

\begin{equation}
    \tilde{\Phi}^{(-)}_{pB}\simeq \chi^{(-)}_{pA}(\boldsymbol{r}_{pA},\boldsymbol{k}_\alpha)\phi_{nA}(\boldsymbol{r}_{nA})e^{-i\alpha \boldsymbol{k}_\alpha \boldsymbol{r}_{nA}}.
\end{equation}

In this expression, $\boldsymbol{k}_\alpha$ is the relative momentum between $B$ and $p$ and $\alpha=m_n/m_B$, and it was found to produce a good agreement between this calculation and experimental data for the $^{16}$O$(d,p)^{17}$O and $^{10}$Be$(d,p)^{11}$Be reactions. Nevertheless, given that this expression is better suited for the transfer of weakly bound and halo nuclei, the ADWA model has enjoyed more widespread use.

\subsection*{The CDCC-BA approximation}
Another way  of accounting for the breakup channels in transfer reactions is by insertion of the CDCC wavefunction, Eq.~(\ref{PsiCDCC}), in the transition amplitude of Eq.~(\ref{eq:T3b}).  The resultant amplitude is formally analogous to that found in the CCBA method \cite{Sat83}, so we refer to it as CDCC-BA approximation.  This is expected to be a good approximation since the CDCC expansion is accurate for small $p$-$n$ separations, for which the matrix transition amplitude presents the largest magnitude. The resultant amplitude, however, will be clearly much more computationally demanding than that obtained with the DWBA or ADWA methods. 

\begin{figure}
\begin{center}
\includegraphics[width=0.75\columnwidth]{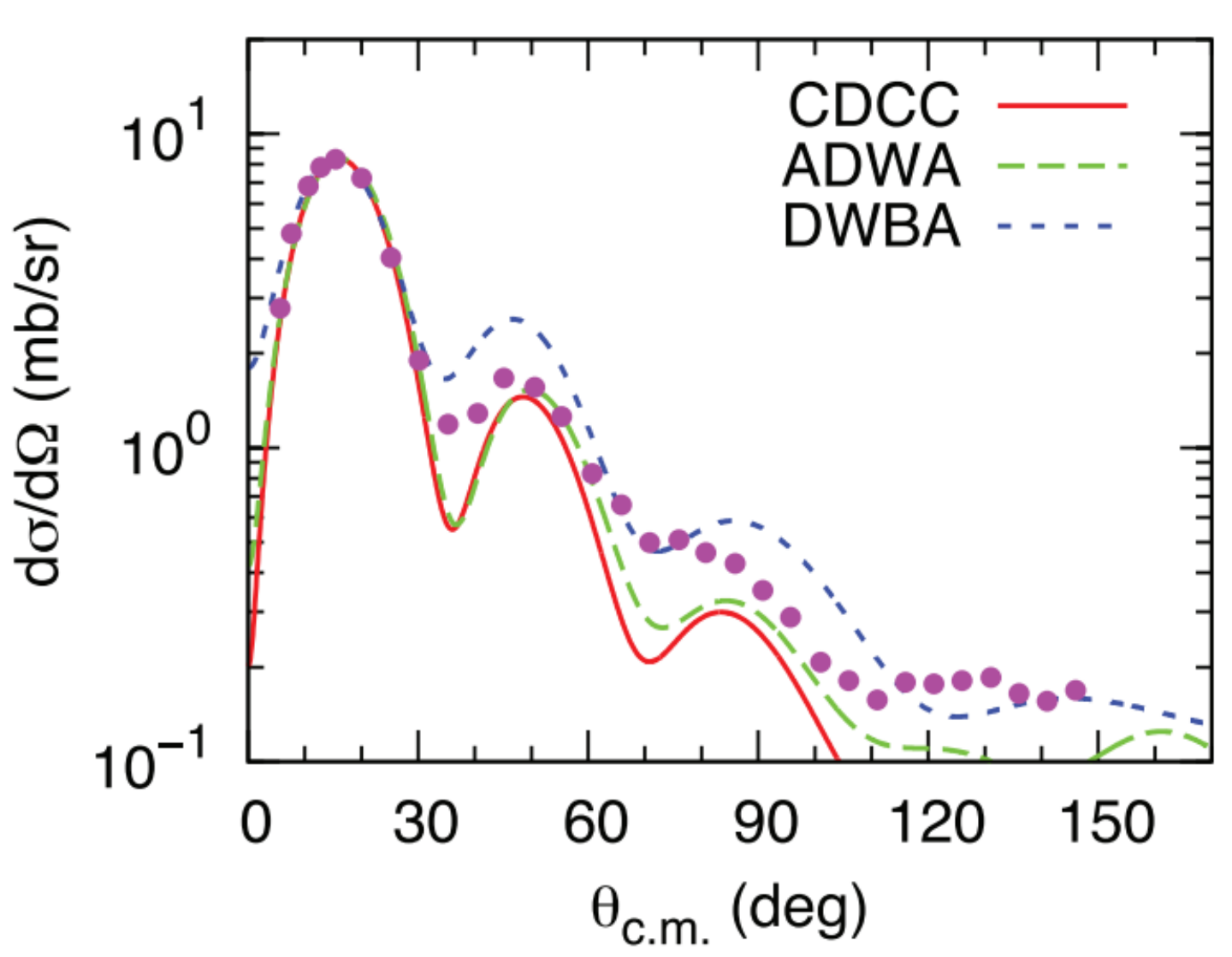}
\includegraphics[width=0.75\columnwidth]{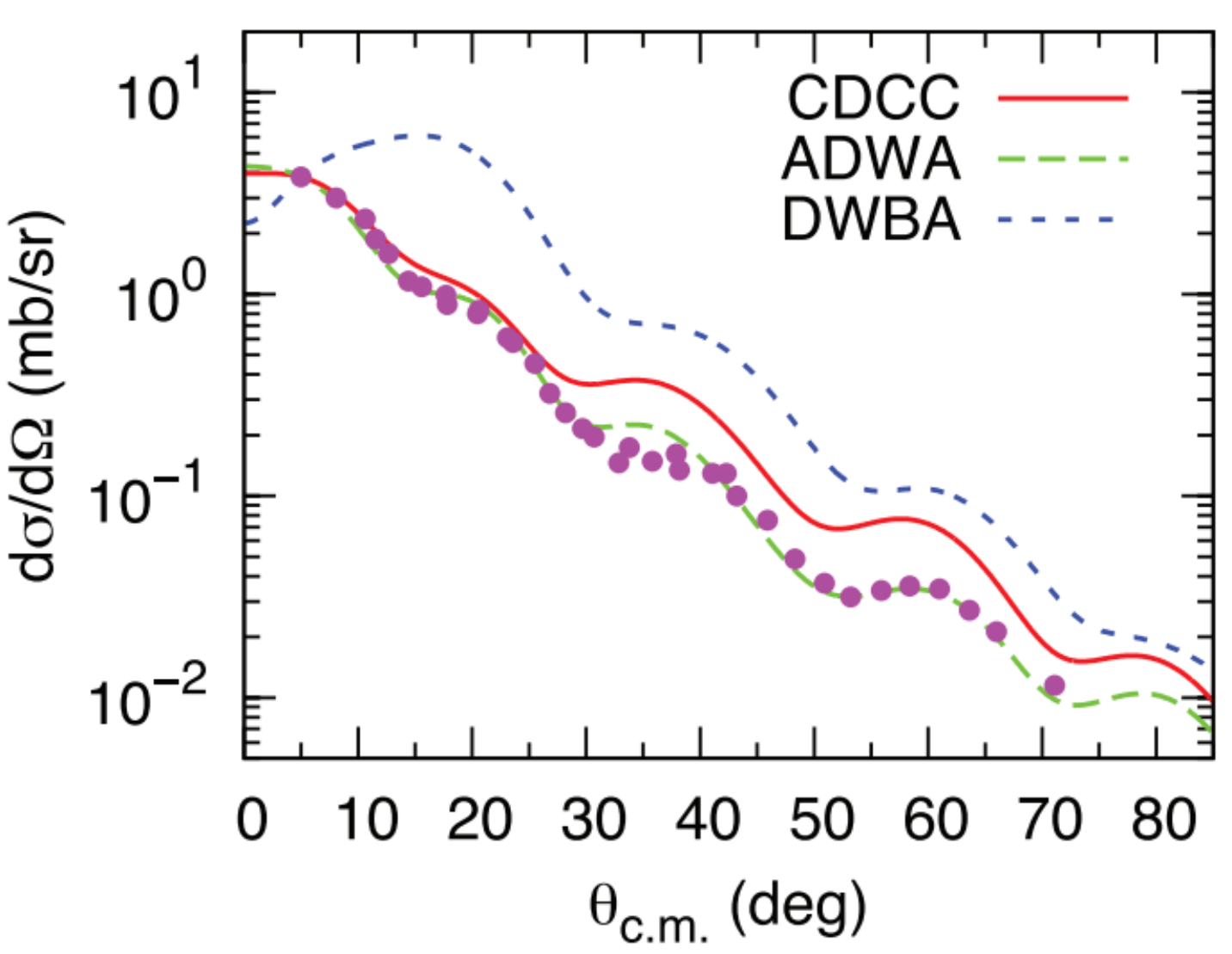}
\end{center}
\caption{\label{fig:adwa_vs_cdcc} 
Comparison of the DWBA, ADWA and CDCC-BA methods for $^{58}$Ni(d,p)$^{59}$Ni reaction at $E_d=10$~MeV (upper panel) and $E_d=56$~MeV (lower panel). Adapted from Ref.~\cite{Pan14}.}
\end{figure}

In Fig.~\ref{fig:adwa_vs_cdcc} we compare the performance of the DWBA, ADWA and CDCC-BA methods for the $^{58}$Ni($d$,$p$)$^{59}$Ni reaction at $E_d=10$~MeV (upper panel) and $E_d=56$~MeV (lower panel), taken from Ref.~\cite{Pan14}. It becomes clear that the ADWA and CDCC methods provide a more reliable description of the data compared to the traditional DWBA approach. In \cite{Cha17}, a systematic comparison between the CDCC-BA method and an adiabatic approximation (CDCC-AD) was performed for $(d,p)$ reactions on various targets, showing good agreement between both, except in some cases, such as when the deuteron beam energy is small and similar to the binding energy of the nucleon in the target or when the binding energy of the nucleon is very small ($S_n~\simeq 0.1$ MeV). It should be noted that in that work the adiabatic approximation is performed by setting the nominal energies of the continuum bins to that of the bound state, which is formally equivalent to the ADWA approach, although the calculation is rather different, so extensions of these results to standard ADWA calculations should be made with caution.

The CDCC-BA approximation is reasonable as long as the coupling to the rearrangement (transfer) channels is weak, typically, when the transfer cross section is much smaller than the reaction cross section. Otherwise, other methods that treat transfer on equal footing to breakup, such as the Coupled-Reaction-Channel method (CRC) \cite{Sat83}, are required.

\subsection{Transfer reactions populating unbound systems \label{sec:TransCont}}

So far, we have considered transfer reactions as a tool for investigating the bound states of a given nucleus. However, in a rearrangement process, the transferred particle can also populate unbound states of the final nucleus. This opens the possibility of studying and characterizing structures in the continuum, such as resonances or virtual states.   In fact, due to the matching conditions for this type of processes \cite{Bri72},  reactions induced by weakly-bound nuclei  favour the population of highly excited states of the residual nucleus, including those above the breakup threshold.

As in the case of transfer to bound states, the simplest formalism to analyze these processes is the DWBA method. In this case, the bound wavefunction $\phi_{\ell,j}(\vecr')$   appearing in the final state in Eq.~(\ref{eq:T3b}) should be replaced by a positive-energy wavefunction describing the state of the transferred particle (neutron in this case) with respect to the  core target. In principle, for this purpose, one could use the suitable scattering state of the $v$+$b$ system at the appropriate relative energy. However, this procedure tends to give numerical difficulties in evaluating the transfer amplitude due to the oscillatory behaviour of both the final distorted wave and the wavefunction $\phi_{\ell,j}(\br')$.  To avoid this problem, several alternative methods have been used. We enumerate here some of them:

\begin{enumerate}[(i)]
\item The bound state approximation \cite{Cok74}. In the case of transfer to a resonant state, this method replaces the scattering state $\varphi_{\ell,j}(\vecr')$ by a weakly bound wavefunction with the same quantum numbers $\ell$ and $j$. In practice, this can be achieved by starting with the potential that generates a resonance at the desired energy and increase progressively the depth of the central potential until the state becomes bound. 

\item Huby and Mines \cite{Hub65} used a scattering state for $\phi_{\ell,j}(\vecr')$ modulated by a  convergence factor $e^{-\alpha r'}$ (with $\alpha$ a positive real number), which is intended to  eliminate its contribution to the integral coming from large $r'$ values, and then extrapolate numerically to the limit $\alpha \rightarrow 0$.  

\item Vincent and Fortune \cite{VF70} put into question the validity of the bound state approximation arguing that, in general, the bound state and resonant form factors can be very different and, even in those cases in which the fictitious form factor gives the correct shape, they can lead to very different absolute cross sections. They suggest using the actual scattering state, but choosing an integration contour along the complex plane  in such a way that the oscillatory integrand is transformed into an exponential decay, thus palliating the slow convergence problem of the post-form transfer amplitude. 

\item In a real transfer experiment leading to positive-energy states, one does not have access to a definite final energy, but to a certain region of the continuum. That is to say, the extracted observables, such as energy differential cross sections, are integrated over some energy range which, at least, is of the order of the energy resolution of the experiment. This suggests a method of dealing with the unbound states consisting of 
discretizing the continuum states in energy bins, as in the CDCC approximation. 

\end{enumerate}

\begin{figure}
\begin{center}\includegraphics[width=0.75\columnwidth]{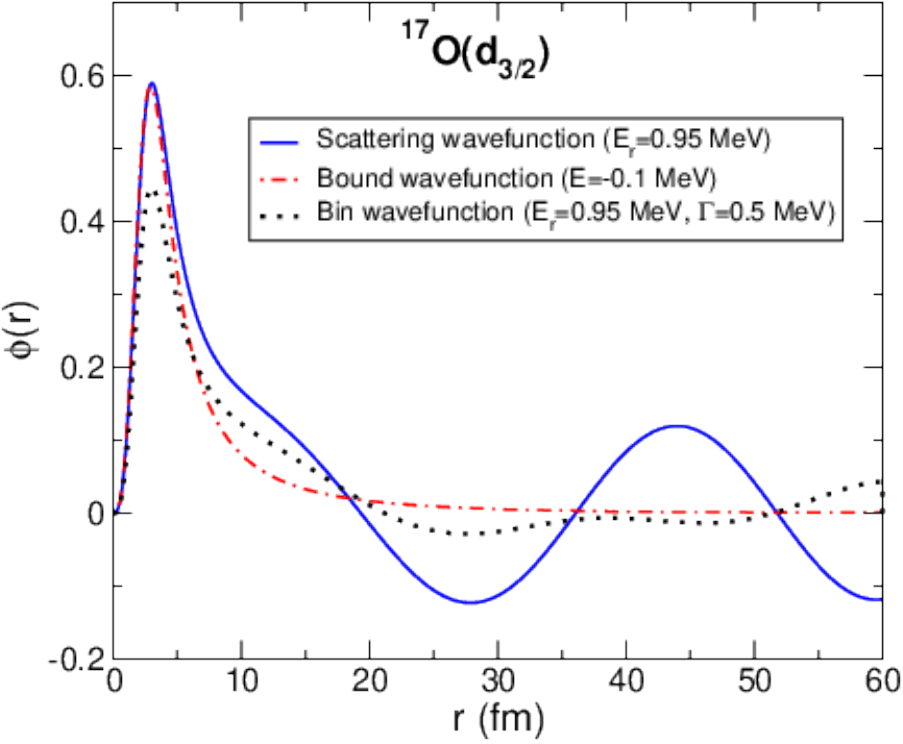}\end{center}
\caption{\label{wfres} Radial part of the  $d_{3/2}$ single-particle resonance wavefunction in $^{17}$O at $E_r=0.95$~MeV compared with a slightly bound wavefunction ($E=-0.1$~MeV) and a bin wavefunction, centered at the nominal energy of the resonance and with a width of 0.5 MeV.}
\end{figure}

In Fig.~\ref{wfres}, we show as an example the radial part of a $3/2^{+}$  resonance in $^{17}$O, described in terms of a $d_{3/2}$ neutron coupled to a zero-spin $^{16}$O core. The solid line is a scattering wavefunction evaluated at the nominal resonance energy ($E_{rel}=0.95$ MeV). Note the oscillatory behaviour at large distances. The dotted line is a bin wavefunction, constructed by a superposition of scattering states, within the range of 0.5 MeV around the resonance energy. It is seen that, asymptotically, the oscillations are damped with respect to the original scattering states. Finally, the dot-dashed line is a bound state wavefunction, with a $1d_{3/2}$ single-particle configuration, and a separation energy of 0.1 MeV. This wavefunction is very similar to the scattering state at short distances but decays exponentially at large distances.

An advantage of the method (iv) is that it can be applied to situations in which one is interested in the description of the population of a range of continuum energies possibly covering both resonant and non-resonant states. For that, it is convenient to resort to the prior-form expression of the transition amplitude.  Considering again the $A(d,p)B$ reaction for simplicity, this amplitude reads
\begin{equation}
\mathcal{T}^\mathrm{prior}_{if}= \langle \Psi_{f}^{(-)}(\vecRp,\vecrp)|V_{nA}+U_{pA}-U_{dA} | \phi_d(\vec{r}) \chi_{dA}^{(+)}(\vec{R}) \rangle .
\label{eq:tmatrix}
\end{equation}
The function $\chi^{(+)}_{dA}$ is the distorted wave generated by the optical potential $U_{dA}$ and $\Psi^{(-)}_f(\vecRp,\vecrp)$ is the exact three-body wave function for the final $p$+$n$+A state, with $\vecrp$ and $\vecRp$ denoting the $n$-$A$ and $p$-$B$ relative coordinates, respectively.  The wave function $\Psi_{f}^{(-)}(\vecRp,\vecrp)$ is the time-reversed of $\Psi_{f}^{(+)}(\vecRp,\vecrp)$, which satisfies the three-body equation:
\begin{equation}
\label{eq:sch}
[T_{\vecRp} + T_{\vecrp} + V_{pn} + U_{pA} + V_{nA} -E]    \Psi_{f}^{(+)}(\vecRp,\vecrp)=0 ,
 \end{equation}
with $E$ the total energy of the system. To solve this equation, the wavefunction $\Psi^{(-)}_f(\vecRp,\vecrp)$ can be expanded 
in  $n+A$ states with well-defined energy and angular momentum, as in the continuum-discretized coupled-channels (CDCC) method.

\begin{figure}
\begin{center}\includegraphics[width=0.75\columnwidth]{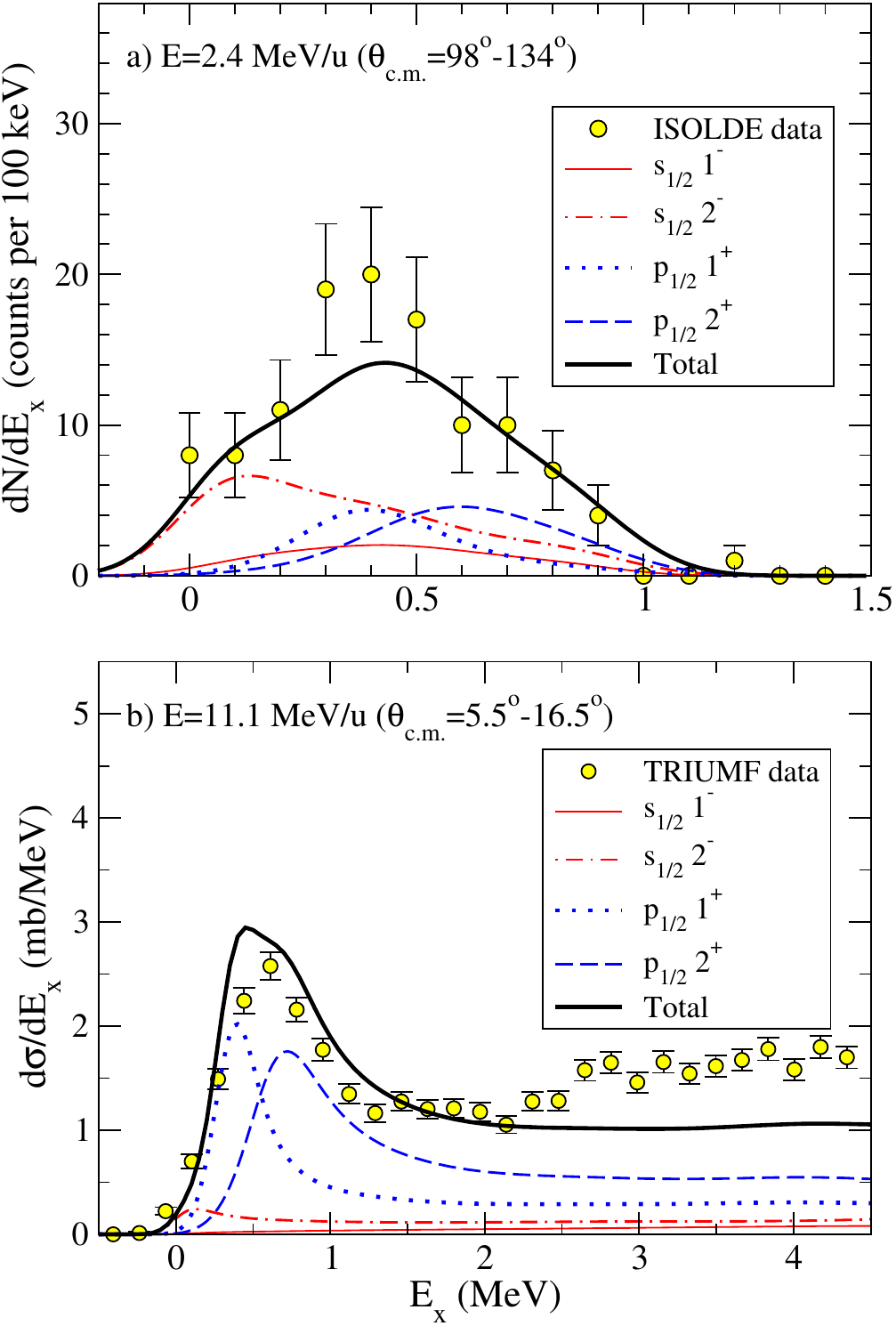}\end{center}
\caption{\label{li9dp} Illustration of the transfer-to-the-continuum method, using a {\it binning} discretization, for the reaction \nuc{2}{H}(\nuc{9}{Li},p)\nuc{10}{Li}$^{*}$. Calculations are compared with the experimental data from Refs.~\cite{Jep06} and \cite{Cav17}. Figure taken from Ref.~\cite{Mor19}.}
\end{figure}

An example is shown in Fig.~\ref{li9dp}, which corresponds to the  differential cross section, as a function of the $n$-$^{9}$Li relative energy, for the reaction \nuc{2}{H}(\nuc{9}{Li},$p$)\nuc{10}{Li}$^{*}$ at 2.36 MeV/u (upper) and 11.1 MeV/u (lower panel), corresponding to the measurements of ISOLDE~\cite{Jep06} and TRIUMF \cite{Cav17}, respectively. The lines are the results of transfer-to-the-continuum calculations populating  $^{10}$Li$^{*}$ continuum states using the same $^{10}$Li structure in both cases. The figure shows the separate contribution of the s-wave ($1^-$, $2^-$) and p-wave ($1^+$, $2^+$) continuum states. The strength of the measured cross section close to zero energy is due to the presence of a virtual state in the  n+\nuc{9}{Li} $s$-wave, whereas the peak around 0.4~MeV is due to a $p_{1/2}$ resonance. This is an example of how the use of transfer reactions can provide information on the continuum structure of weakly-bound or even unbound systems. For more details on the calculations, see Ref.~\cite{Mor19}.

\subsection{Transfer reactions involving three-body projectiles\label{sec:3b_transfer}}

\begin{figure}[t]
\centering
\includegraphics[width=0.7\columnwidth]{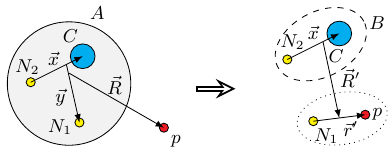}
\caption{Diagram for a $(p,d)$ or $(p,pN)$ reaction induced by a three-body projectile in inverse kinematics. Taken from Ref.~\cite{Gom17c}.}
\label{fig:scheme}
\end{figure}

The formalism presented in the previous sections can be extended and applied to more complex systems. We discuss here the interesting case of the stripping reaction induced by a three-body Borromean system, in which one of the projectile fragments is transferred to the target,  leaving a two-body residual unbound system. To be more specific, we consider the case of a two-neutron halo projectile, such as $^{11}$Li, $^{6}$He impinging on a proton target. In these cases, given the unstable nature of the projectile, the reaction experiment must be performed in inverse kinematics. 
This process can be schematically represented as (see Fig.~\ref{fig:scheme}).
\begin{equation}
\underbrace{(C+N_1+N_2)}_A + p \rightarrow \underbrace{(C+N_2)}_{B} + d,
\label{eq:scheme}
\end{equation}
The prior-form transition amplitude for this process can be formally reduced to an effective few-body problem, leading to
\begin{equation}
\mathcal{T}_{if}=\sqrt{2}~\langle \Psi_{f}^{(-)}(\vec{x},\vec{R}',\vec{r}')|V_{pN_1}+U_{pB}-U_{pA} |\Phi_A(\vec{x},\vec{y}) \chi_{pA}^{(+)}(\vec{R}) \rangle,
\label{eq:tmatrix3body}
\end{equation}
where $\Phi_A$ represents the ground-state wave function of the initial three-body composite, 
 $\chi^{(+)}_{pA}$ is the distorted wave generated by the auxiliary potential $U_{pA}$, and $\Psi^{(-)}_f$ is the exact four-body wave function for the outgoing $d$-$B$ system.  Notice the explicit factor $\sqrt{2}$ coming from the two identical neutrons in the initial wave function. 

In writing the transition amplitude in the form of Eq.~(\ref{eq:tmatrix3body}), one implicitly uses a {\it participant/spectator} approximation,  assuming that the reaction occurs due to the interaction of the incident proton with a single neutron $(N_1)$ of $A$ (the {\it participant}), leaving the
residual $B=N_2+C$ system unperturbed.

To reduce Eq.~(\ref{eq:tmatrix3body}) to a tractable form, one can approximate the exact wave function $\Psi^{(-)}_f$ by the factorized form,
\begin{equation}
\label{eq:dwba}
\Psi_{f}^{(-)}(\vec{x},\vec{R}',\vec{r}') \approx \varphi^{(-)}_{\vec{q},\sigma_2,\zeta}(\vec{x}) \chi_{dB}^{(-)}(\vec{R}')\phi_d(\vec{r}'),
\end{equation}
where $\phi_d$ is the deuteron wave function, $\varphi^{(-)}_{\vec{q},\sigma_2,\zeta}$ is a two-body continuum wave function with wave number $\vec{q}$ and spin projections ${\sigma_2,\zeta}$ of the binary subsystem $B$, and 
$\chi_{dB}^{(-)}$ is a distorted wave describing the $d$-$B$ relative motion in the exit channel. The function 
 $\varphi^{(-)}_{\vec{q},\sigma_2,\zeta}$ is the time-reversed of $\varphi^{(+)}_{\vec{q},\sigma_2,\zeta}$, and presents incoming scattering boundary conditions.

The ground state wavefunction of the initial three-body composite $A$ ($\Phi_A^{j\mu}(\vec{x},\vec{y})$) is conveniently described using a three-body model  \cite{Des03,FaCE,Rod05}, as introduced for the four-body CDCC calculations in Sec.~\ref{sec:4bcdcc}. 
Typically, the three-body wave function is most naturally obtained in the so-called Jacobi-T set, but for the purpose of computing the required overlap functions, it is then transformed into the Jacobi-Y set, where the $x$ coordinate relates the core and one neutron (see Fig.~\ref{fig:scheme}).

Consistently with the spectator approximation for the composite $B$, it is assumed that the interaction does not change the state of this system, and the overlap between the two- and three-body wave functions ($\psi(\vec{q},\vec{y})$)  
contains all the relevant structure information: 
\begin{equation}
\psi(\vec{q},\vec{y}) = \int \varphi^{(-)}_{\vec{q},\sigma_2,\zeta}(\vec{x})\Phi_A^{j\mu}(\vec{x},\vec{y})d\vec{x},
\label{eq:overlap1}
\end{equation}
from which the following transition amplitude can be defined:
\begin{equation}
\mathcal{T}_{if} \equiv \langle \chi_{dB}^{(-)}\phi_d|V_{pN_1}+U_{pB}-U_{pA} |\psi~\chi_{pA}^{(+)} \rangle.\label{eq:tJ}
\end{equation}
From the transition amplitude 
the differential cross section of the $C$-$N_2$ relative momentum $\vec{q}$ and the scattering angle of $B$ with respect to the incident direction can be written as
\begin{align}
\frac{d\sigma^2}{d\Omega_B d\vec{q}} & =  \frac{2}{(2j+1)}\frac{\mu_i\mu_f}{(2\pi\hbar^2)^2}\frac{k_f}{k_i} 
\sum_{\nu}\left|\mathcal{T}_{if}\right|^2,
\label{eq:doubleTj}
\end{align}
where 
$\mu_{i,f}$ the projectile-target reduced mass in the initial and final partitions, and $\nu\equiv\{M_T\sigma_d\}$ represents the spin projections of the final products. 

\begin{figure}[t]
\centering
\includegraphics[width=0.7\columnwidth]{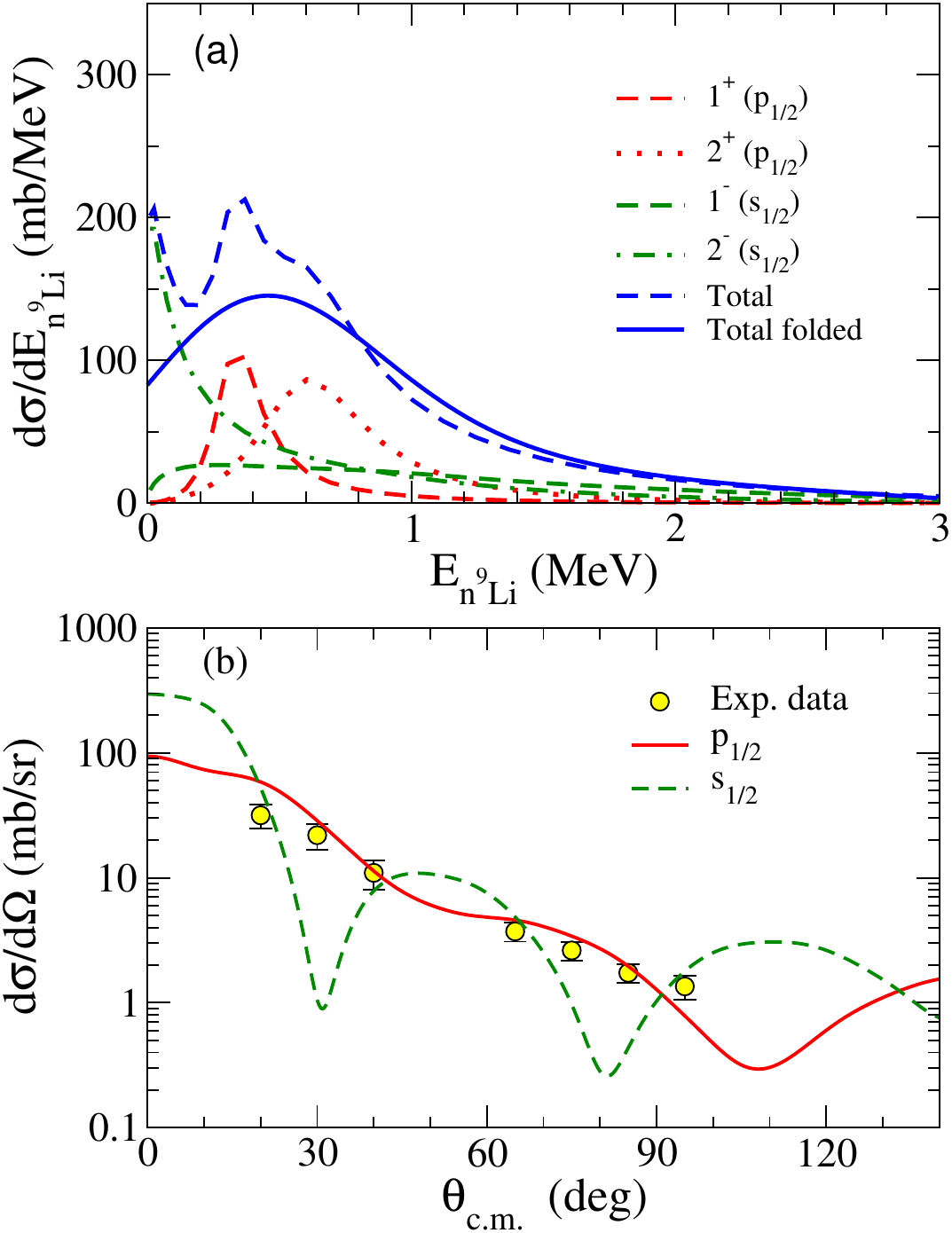}
\caption{(a) Relative $n$-$^{9}$Li energy spectrum, integrated in angle, for $s_{1/2}$ (green) and $p_{1/2}$ contributions (red) for the $^{11}$Li(p,d )$^{10}$Li reaction at 5.7~MeV/u. evaluated with the prior-form DWBA calculations described in the text. The summed contributions, before (blue dashed) and after (blue solid) convoluting with the experimental resolution are also shown.  The $1^-$ and $2^-$ contributions are shown separately as well as $1^+$  and $2^+$.
(b) Energy-integrated angular differential cross section populating the $s_{1/2}$ and $p_{1/2}$ orbitals in $^{10}$Li. The data of \cite{San16}, corresponding to the resonant contribution ($p$-wave) is also shown. Figure adapted from Ref.~\cite{Cas17}.}
\label{fig:li11pd_p1I}
\end{figure}

After integration over the angle of $\vec{q}$ and transforming from momentum to energy, expression (\ref{eq:doubleTj}) provides a double differential cross section with respect to the scattering angle and the excitation energy of the residual $N_2 +C$ system. This observable can in principle be reconstructed experimentally from the measured energy and angle of the outgoing deuteron. To illustrate this formalism, we show in Fig.~\ref{fig:li11pd_p1I} results for the reaction $^{11}$Li(p,d)$^{10}$Li \cite{Cas17}, recently measured at TRIUMF by Sanetullaev \emph{et al.}~\cite{San16}.

For the calculations presented, optical potentials adjusted to describe $p,d+^{11}$Li elastic scattering were used, following \cite{San16}. Besides, it was assumed that the two halo neutrons of $^{11}$Li have the same angular momentum $\vec{J}= \vec{L} + \vec{s}_2$. This condition is strictly fulfilled when the spin of the core is neglected and is also a good approximation when using $I=3/2^-$. Only the configurations $(p_{1/2})^2$ and $(s_{1/2})^2$ were considered, as other contributions, while present in the three-body calculations, give negligible components in the ground state of $^{11}$Li. The considered $^{9}$Li+n potential contains terms that couple the spins of the remaining neutron and $^9$Li in $^{10}$Li leading to splitting of the $p$-wave resonance into $1^+$ and $2^+$ states and of the virtual state in the $s$-wave in $1^-$ and $2^-$ components. These components are shown in the top panel of Fig.~\ref{fig:li11pd_p1I}, which depicts the relative energy spectrum between $n$ and $^9$Li. In red, the $p-$wave components are presented, where the splitting of  the $p$-wave resonance into $1^+$ and $2^+$ resonances at 0.37 and 0.61 MeV is evident, while in green the $s$-wave conributions are shown, presenting a virtual state (with scattering length $a=-37.9$ fm) only in the $2^-$ component. The total is shown in dashed blue, while its folding with the experimental resolution given in Ref.~\cite{San16} is presented in the solid blue line. It is rather remarkable how the low-energy peaks are smoothed out by the experimental resolution leading to a relatively featureless distribution, which could explain why the fine details of $^{10}$Li are not observed in this experiment. In the bottom panel, the energy-integrated angular distribution of the outgoing deuteron is presented. It has been assumed that the subtraction of the non-resonant background performed in the analysis of the data completely excludes the $s$-wave component, so experimental data are compared only with the cross section corresponding to neutrons transferred from the $p_{1/2}$ orbitals, finding an excellent agreement both in shape and magnitude. Note that the experimental data have been measured at angles for which the $s_{1/2}$ contribution is minimal. This may explain why in the calculated energy distribution the contribution from $s$ waves is quite important at low energies, while experimentally there is no direct sign of this effect.

\begin{figure}
\begin{center}\includegraphics[width=0.75\columnwidth]{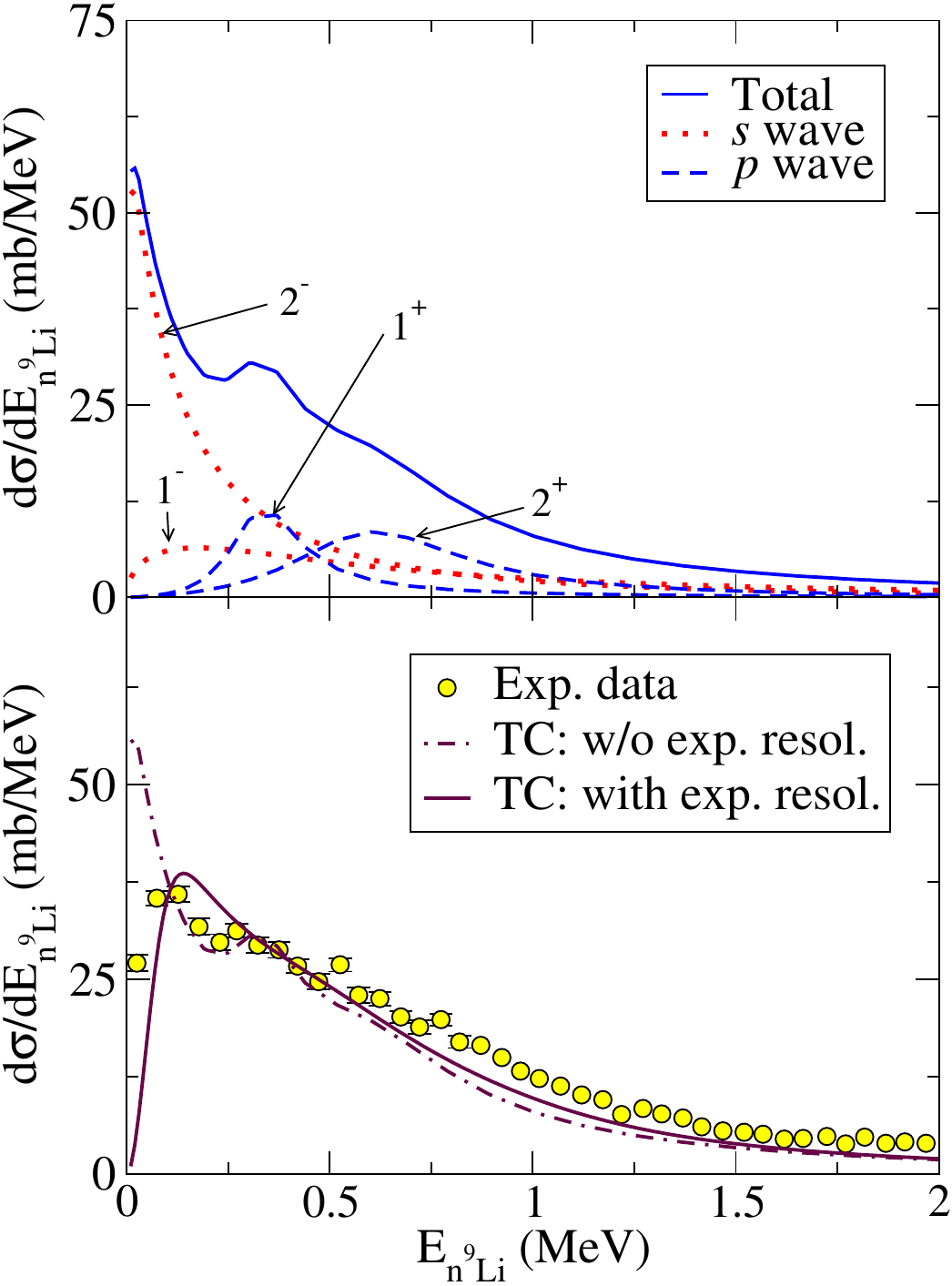}\end{center}
\caption{\label{fig:li11ppn} $n$-$^9\mathrm{Li}$ relative energy distribution for $^{11}$Li$(p,pn)^{10}$Li at 280~MeV/nucleon. In the top panel, the different contributions for the $s$ ($1^-,2^-$) and $p$ ($1^+,2^+$) waves within a three-body $^{11}$Li model including the $^{9}$Li spin \cite{Cas17}. The total cross section is given by the solid blue line.  In the bottom panel, the results are compared, with and without convoluting with the experimental resolution, with the data from  Ref.~\cite{Aks08}. Figure adapted from Ref.~\cite{Gom17c}, with authorization from APS. }
\end{figure}

Using the methods described in Sec.~\ref{sec:TransCont} to consider continuum states of the $p-N$ subsystem, a similar formalism can be used to study $(p,pN)$ reactions on three-body nuclei, simply considering the auxiliary amplitude  in Eq.~(\ref{eq:tJ}) for continuum $p-N$ states, which are typically chosen to have a definite $p-N$ energy, angular momentum and parity:
\begin{equation}
\mathcal{T}_{if}^{J^\pi_{pn}}(E_{pN}) \equiv \langle \chi_{pNB}^{(-)}\phi^{J^\pi_{pn}}_{pN}|V_{pN_1}+U_{pB}-U_{pA} |\psi~\chi_{pA}^{(+)} \rangle.
\label{eq:tJpn}
\end{equation}
and where the outgoing $\chi_{pNB}^{(-)}$ functions can be computed using a CDCC solution for the final 3-body system \cite{Mor15}. Figure \ref{fig:li11ppn} shows  results for the $n$-$^9\mathrm{Li}$ relative energy distribution in the $^{11}$Li$(p,pn)^{10}$Li reaction at 280~MeV/nucleon, using the same model for $^{11}$Li and $^{10}$Li as for the $(p,d)$ reaction in Fig.~\ref{fig:li11pd_p1I}. The contributions of the different components are shown on the top panel, where again a splitting of the $p-$wave resonance is visible, as well as the virtual state only in the $2^-$ component, the main difference to the $(d,p)$ case being the increased importance of the $s$-wave components. In the bottom panel, the theoretical distribution is presented convoluted with the experimental resolution and compared with the experimental data, showing an excellent agreement with it, although the experimental resolution again significantly smooths the distribution.

It is a success of the model model that it is able to reproduce experimental data at  $\sim$6 MeV per nucleon (the $(p,d)$ reaction) and at $\sim$280 MeV per nucleon (the $(p,pN)$ reaction), although perhaps not an unexpected one, given the similarities of both reactions when viewed as transfer to bound and continuum states of the $p-n$ system.

\subsection{Simultaneous inclusion of projectile breakup and target excitation in transfer reactions \label{sec:simuldtexc}}
The models discussed in the previous subsections are based on a strict three- or four-body description of the transfer reaction in which possible excitations of the target ($A$) are only accounted for approximately by the choice of the effective interactions. From the analysis of reactions with ordinary nuclei, it is known that an explicit account of collective excitations of the colliding nuclei is required for an accurate description of the reaction. 
For $(d,p)$ reactions on deformed targets, one may anticipate that both projectile breakup and target excitation can play a role and may require their explicit inclusion. This is not possible in the standard formulations of the aforementioned CCBA and ADWA methods, which tend to emphasize one of the two mechanisms. As a way of including the two mechanisms simultaneously, one may resort to the extended CDCC wavefunction (\ref{eq:CDCC_tarx}) which was devised to include in the model space the projectile breakup channel as well as some excited states of the target nucleus.

For the particular case of a $A(d,p)B$ reaction, and considering for simplicity only one excited state of the target, this extended CDCC wavefunction can be schematically expressed as 
\begin{align}
\label{eq:Phi}
\Psi^{(+)}_d(\vecr,\vecR,\xi) & = \phi_d(\vecr) \Phi^{0}_{A} (\xi) \chi^{(+)}_{d,0}(\vecR)    
              + \phi_d(\vecr) \Phi^{1}_{A} (\xi) \chi^{(+)}_{d,1}(\vecR) \nonumber \\
             & + \sum_{n} \phi^{n}_{pn}(\vecr) \Phi^\text{0}_{A} (\xi) \chi^{(+)}_{n,0}(\vecR) 
              + \sum_{n} \phi^{n}_{pn}(\vecr) \Phi^\text{1}_{A} (\xi) \chi^{(+)}_{n,1}(\vecR)
\end{align}
where $ \{\phi_d, \phi^{n}_{pn} \}$ denote the deuteron ground state and (discretized) continuum states,  and  $\{\chi^{(+)}_{n,0}(\vecR), \chi^{(+)}_{n,1}(\vecR) \}$ the functions describing the projectile-target relative motion with the target in either its ground state $\Phi^{0}_{A} (\xi)$ or in the excited state $\Phi^\text{1}_{A} (\xi)$, respectively. 
 Therefore, the first two terms of Eq.~(\ref{eq:Phi}) describe, respectively, elastic and inelastic scattering with the deuteron remaining in its ground state. The third  and fourth terms describe deuteron breakup with respect to the target in its ground state or first excited state, respectively.  When inserted into Eq.~(\ref{eq:tmatrix}) this gives rise also to four terms, 
\begin{equation}
\label{eq:Tdecomp}
{\cal T}_{dp}= {\cal T}^\mathrm{el}_{dp} + {\cal T}^\mathrm{inel}_{dp} +{\cal T}^\mathrm{elbu}_{dp} + {\cal T}^\mathrm{inbu}_{dp} ,
\end{equation}
which may be interpreted as (I) {\it elastic transfer}, i.e.,  direct transfer from the deuteron ground state leaving the target in its ground state, (II) {\it inelastic transfer}, i.e., target excitation followed by one-neutron transfer, (III) {\it elastic breakup transfer}, i.e.,  deuteron breakup followed by transfer, leaving the target in the ground state and (IV)  {\it inelastic breakup transfer}, i.e., deuteron breakup, accompanied by target excitation, and followed by neutron transfer.

 Note that these four terms are to be added coherently, giving rise to interference effects, as discussed below.  Note also that, in DWBA, only  the first term ({\it elastic transfer}) is explicitly  taken into account. One of our goals here is to study the contribution of the other terms.

As an example, we consider the reaction $^{10}$Be(d,p)$^{11}$Be(g.s.), including both the $^{10}$Be(g.s.) and its first excited state ($2^+$; $E_x=3.367$~MeV).  

Calculations for this reaction have been reported in Ref.~\cite{Gom17b}. The structure of the $^{11}$Be nucleus is treated within the particle-rotor model of Ref.~\cite{Nun96}, which assumes a deformation length of $\delta_2=1.664$~fm for  $^{10}$Be which allow for the coupling of the ground  ($0^+$) and first excited state ($2^+$) of this nucleus.

\begin{figure}
\centering
\includegraphics[width=0.75\columnwidth]{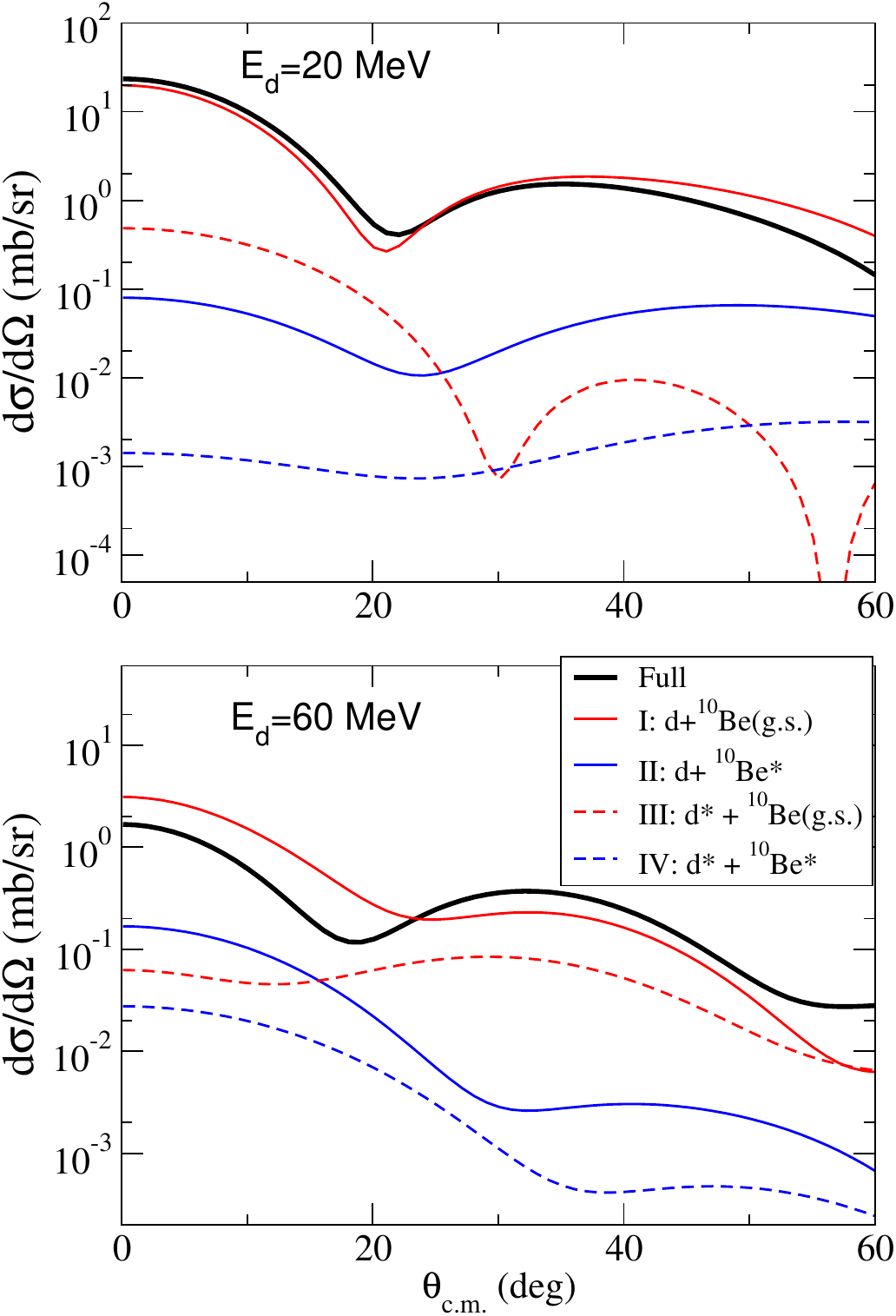}
 \caption{\label{fig:tr_desc}  Differential transfer cross section for $E_d=$20~MeV (top) and 60~MeV (bottom). All results correspond to full CDCC calculations for the elastic channel, while the transfer channel has been calculated through Born approximation from all paths, and paths I, II, III and IV (see Fig.~\ref{fig:transfer_paths})  for the black solid, red solid, blue solid, red dashed and blue dashed lines respectively. Figure taken from \cite{Gom17b}, with authorization from APS.} 
\end{figure}

\begin{figure}[tb]
\begin{center}
 {\centering \resizebox*{0.85\columnwidth}{!}{\includegraphics{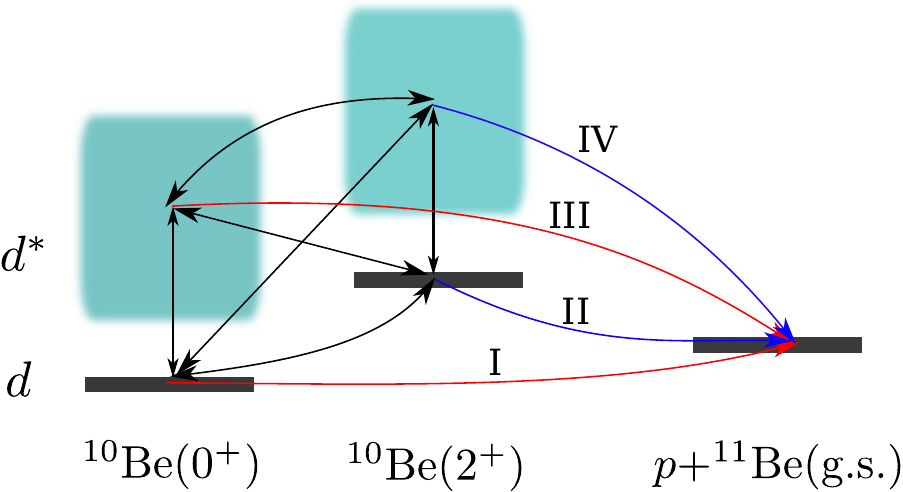}} \par}
\caption{\label{fig:transfer_paths}(Color online) Pictorial representation of the $^{10}$Be(d,p)$^{11}$Be reaction, highlighting the different contributions considered in this work. The initial $d+$\nuc{10}{Be} system includes breakup of the deuteron and excitation of \nuc{10}{Be} coupled to all orders (black lines). The transfer channel is then coupled in first order to these channels (red and blue lines), and four terms are distinguished: I (elastic transfer), II (inelastic transfer), III (breakup transfer) and IV (inelastic breakup transfer). Figure adapted from \cite{Gom17b}, with authorization from APS.}
\end{center}
\end{figure}

In Fig.~\ref{fig:tr_desc} we show the calculated angular distributions for the deuteron incident energies of $E_d=20$~MeV (top) and $E_d=60$~MeV (bottom). In each panel, we show the full calculation (thick solid black line) along with the calculations keeping only one of the terms in Eq.~(\ref{eq:Tdecomp}). The red solid line corresponds to ${\cal T}^{el}_{dp}$ (path I in Fig.~\ref{fig:tr_desc}), the blue solid to ${\cal T}^{inel}_{dp}$ (II), the red dashed to ${\cal T}^{elbu}_{dp}$ (III) and the blue dashed to ${\cal T}^{inbu}_{dp}$ (IV). As before, for the $d$+\nuc{10}{Be} partition all channels are coupled to all orders, while the transfer channel is coupled in first order. This is shown in a pictorial representation in Fig.~\ref{fig:transfer_paths}, where the black lines with two arrows correspond to couplings to all orders, while the red and blue lines with one arrow denote couplings to first order. The labels related to each of the terms in Eq.~\ref{eq:Tdecomp} are also included.
  It can be seen that for both energies the main contributor is the elastic transfer ${\cal T}^{el}_{dp}$, while, at small angles,  for $E_d=20$ MeV  the second main contributor is the breakup of the deuteron, ${\cal T}^{elbu}_{dp}$, whereas at 60~MeV it is the excitation of \nuc{10}{Be}, ${\cal T}^{inel}_{dp}$ .

\begin{figure}
\centering
\includegraphics[width=0.75\columnwidth]{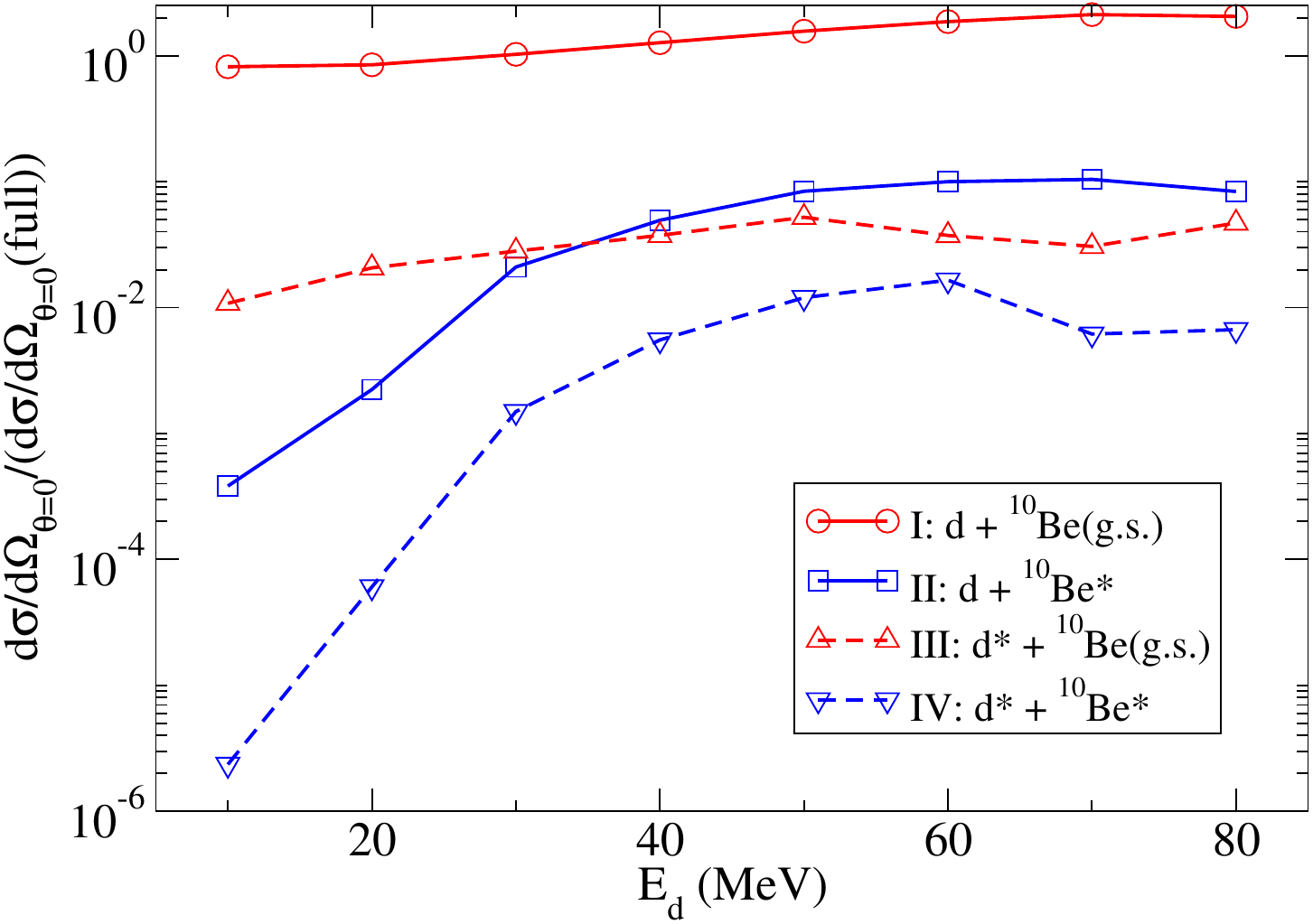}
 \caption{\label{fig:th0_vs_E} Ratio between cross sections at the peak  as a function of the deuteron incident energy. For each energy the ratio is between the calculation where one of the transfer terms is selected (as in Fig.~\ref{fig:tr_desc}) and the full calculation. As can be seen, term II, corresponding to the inelastic excitation of \nuc{10}{Be}, gains importance with increasing energy so that at the higher energies, it is of the same magnitude as term III, corresponding to deuteron breakup. Figure taken from \cite{Gom17b}, with authorization from APS.} 
\end{figure}

In Fig.~\ref{fig:th0_vs_E} we plot, as a function of incident energy, the ratio of the cross section at the peak for the calculation where one of the transfer paths was selected divided by the cross section at the peak for the full calculation, containing all transfer paths. As can be seen in the figure, path II, corresponding to ${\cal T}^{inel}_{dp}$ gains relevance as the energy increases, although it must be noted that this behaviour is found for this particular reaction and should not be generalized to other target nuclei.

\section{Inclusive breakup reactions}
\label{sec:inc}
\begin{figure}
\begin{center}\includegraphics[width=0.9\columnwidth]{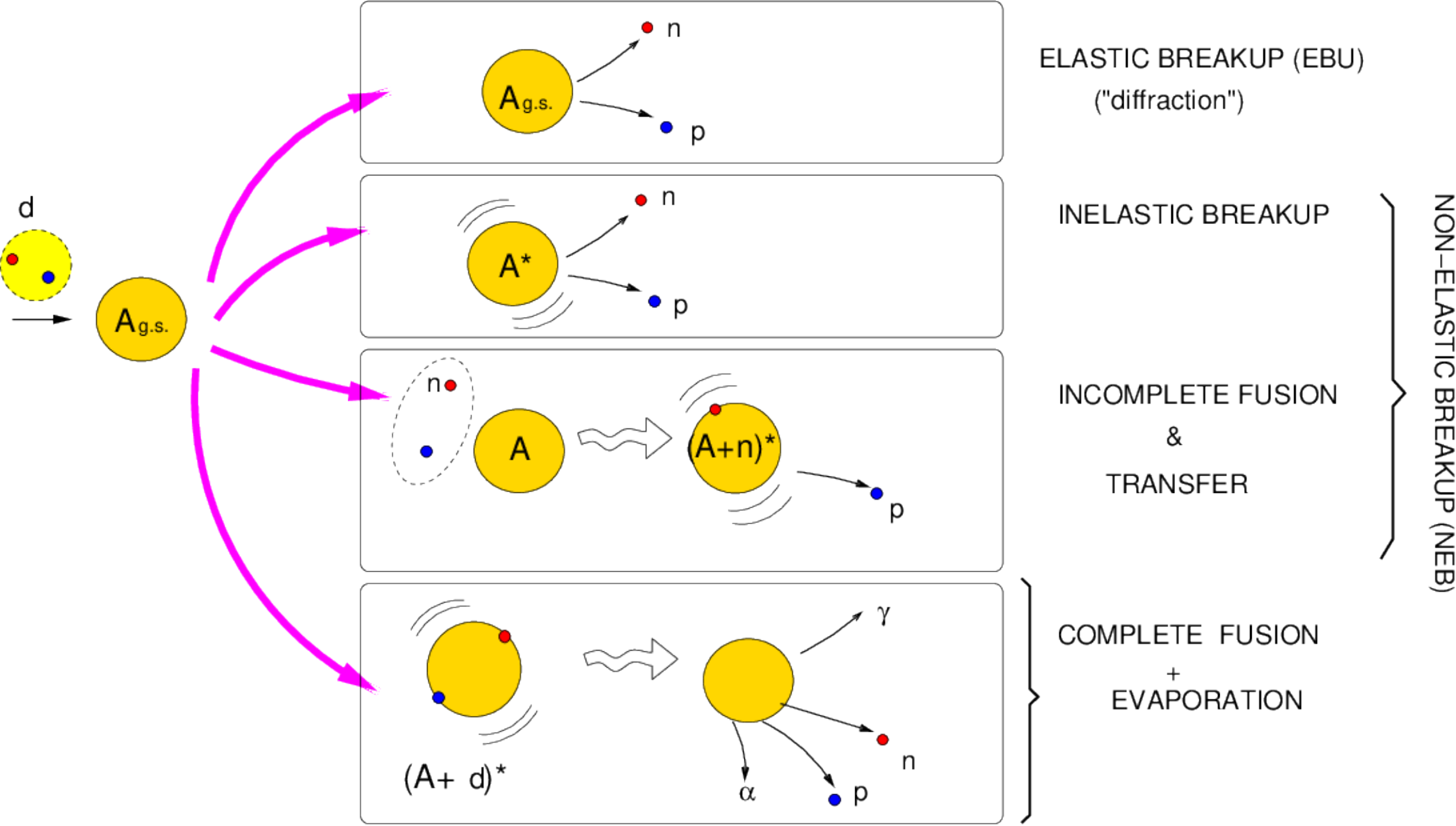} \end{center}
\caption{\label{fig:dA_neb}Pictorial representation of elastic and non-elastic breakup channels for a $d+A$ reaction.}  
\end{figure}

In previous sections, we have considered only exclusive breakup processes in which the three fragments ($b$, $x$ and $A$) survive after the collision and are observed in a definite internal state. In particular, when all fragments end up in their ground state, the process is called {\it elastic breakup}  (EBU).  In many experiments, however,  one or more fragments are not detected in the final state. This is the case, for example, of reactions of the form $A(a,b)X$, in which only one of the projectile constituents ($b$ in this case) is observed. The angular and energy distributions of the $b$ fragments will contain contributions from all possible final states of the $x+A$ system.  
This includes the EBU channel, in which $x$ and $A$ remain in their ground state, but also $x$ transfer, breakup accompanied by excitations of $A$, and  $x$+$A$ fusion [named {\it incomplete fusion} (ICF)].  These processes are schematically depicted in Fig.~\ref{fig:dA_neb} for a deuteron-induced reaction.   These non-elastic breakup components (NEB) must be added to the EBU component to give the total inclusive breakup.   Whereas the EBU contribution can be accurately calculated within the CDCC method and other approaches, the evaluation of NEB is more involved due to the large number of accessible states.

\subsection{The Ichimura-Austern-Vincent (IAV) model \label{sec:iav}}
\begin{figure}[tb]
\begin{center}
 {\centering \resizebox*{0.7\columnwidth}{!}{\includegraphics{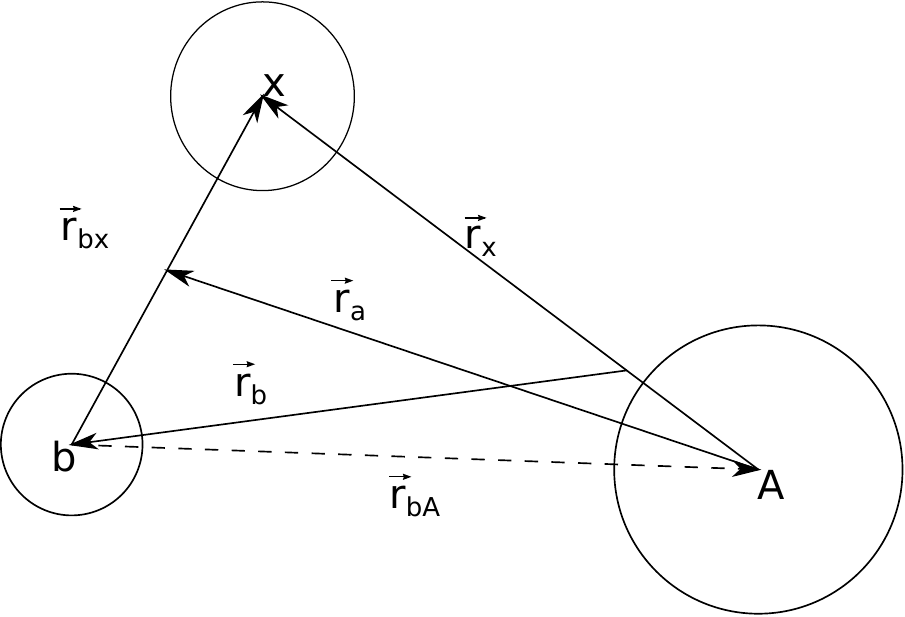}} \par}
\caption{\label{fig:zrcoor} Relevant coordinates for a $A(a,b)X$ reaction.}
\end{center}
\end{figure}

 The problem of inclusive breakup  was addressed in the 1980s by several groups, which developed formal techniques to reduce the sum over final states to a closed form \cite{HM85,Ich85}. These models are based on a spectator-participant picture. The $b$ fragment, which  acts as spectator, is assumed to scatter elastically by the target nucleus and hence its relative motion with the target is represented by some optical potential $U_{bA}$. The fragment $x$, the participant, is allowed to interact in any possible way with the target. When $x$ scatters elastically by the target, we simply have the elastic breakup defined above. Conversely, NEB encompasses those processes in which $x$ scatters nonelastically by the target.  Interestingly, all these early theories arrived at a similar expression for the NEB differential cross section with respect to the angle and energy of $b$ fragments, given by: 
\begin{equation}
\label{eq:neb}
\left . \frac{d^2\sigma}{dE_b d\Omega_b} \right |_\mathrm{NEB} = -\frac{2}{\hbar v_{a}} \rho_b(E_b)
 \langle  \varphi_x  | W_x |  \varphi_x  \rangle   ,
\end{equation}
where  $\rho_b(E_b)=k_b \mu_{b} /[(2\pi)^3\hbar^2]$ is the density of states (with $\mu_b$ the reduced mass of $b+B$ and $k_b$ their relative wave number),  $W_x$ is the imaginary part of the optical potential $U_x$, which describes $x+A$ elastic scattering. The function $\varphi_x $ ($x$-channel wavefunction) describes the final state of the $x-A$ system when the spectator $b$ is scattered with energy $E_b$ within the solid angle $\Omega_b$.  A sketch with the relevant coordinates is shown in Fig.~\ref{fig:zrcoor}.

The  proposed models differ in the prescription for the $x$-channel wavefunction. Austern \emph{et al.}~\cite{Aus87} proposed a formula based on the CDCC wavefunction for the $d+A$ system,
\be
\label{eq:chiPsi}
\psixaus (\vecr_x) = \langle \vecr_x \chi_b^{(-)}  |  \PsiTB \rangle .
\ee
This expression has the appealing feature of connecting the NEB cross section with the CDCC three-body wavefunction. As noted in Sec.~\ref{sec:interp}, the CDCC method does not provide detailed NEB breakup cross sections by itself. However, with the aid of Eqs.~(\ref{eq:chiPsi}) and (\ref{eq:neb}) one can effectively extract the NEB content of the CDCC wavefunction. 

Expression (\ref{eq:chiPsi}), albeit formally simple, possesses the  difficulty of requiring an accurate representation of the three-body wavefunction $\PsiTB$ in the full configuration space, since there is no natural cutoff in the integration variable $\vecr_b$. This difficulty can be sorted out recasting $\psixaus (\vecr_x)$ as a solution of the inhomogeneous equation
\begin{equation}
\label{eq:inh_3b}
(E^+_x - K_x - U_{xA})  \psixaus = \langle \vecr_x \chi_b^{(-)}| V_\mathrm{post}| \Psi^{\mathrm{3b}^{(+)}} \rangle.
\end{equation}
Although this equation is a priori more difficult to implement than Eq.~(\ref{eq:chiPsi}), it has the advantage that the $\PsiTB$ function appears multiplied by $V_\mathrm{post}$, that will tend to emphasize small $b-x$ separations and hence one requires only an approximate three-body wavefunction accurate within that range.   This can be achieved, for instance, expanding $\PsiTB$  in terms of $b-x$ eigenstates, as done in the CDCC method, or in terms of Weinberg states \cite{Pan13}, already discussed in Sec.~\ref{sec:JT}. The implementation of the method with CDCC wavefunctions is numerically challenging and the first calculation of this kind was reported only recently \cite{Jin19b}. 

Ichimura, Austern and Vincent \cite{Ich85} proposed a simpler DWBA version of the 3B formula above. In DWBA, the exact wavefunction $\Psi^{(+)}$  is approximated by the factorized form:
\begin{equation}
|\Psi^{(+)}\rangle \approx |\chi_a^{(+)}  \phi_a   \phi^{0}_A \rangle .
\end{equation}
where $\phi^0_A$ is the target ground state wavefunction. With this approximation, the NEB component of the $b$ singles cross section becomes
 \begin{equation}
\label{eq:iav}
\left . \frac{d^2\sigma}{dE_b d\Omega_b} \right |^\mathrm{IAV,post}_\mathrm{NEB} = 
-\frac{2}{\hbar v_{a}} \rho_b(E_b)   \langle \psixpost | W_x | \psixpost \rangle   ,
\end{equation}
with
 \begin{equation}
\label{eq:phix_post}
(E^+_x - K_x - {U}_{xA})  \psixpost(\vecr_x) =  \langle \vecr_x \, \chi_b^{(-)}| V_\mathrm{post}| \chi_a^{(+)} \phi_a  \rangle \, .
\end{equation}
which can be also expressed in integral form as
 \begin{equation}
\label{eq:phix_post_green}
 \psixpost(\vecr_x) =  G^{(+)}_x \langle \vecr_x \, \chi_b^{(-)}| V_\mathrm{post}| \chi_a^{(+)} \phi_a  \rangle .
\end{equation}
This latter form is more convenient for the application of Green's function techniques \cite{Pot15}.

For completeness we mention here an alternative formula proposed by Udagawa and Tamura \cite{Uda81}. Their final result can be again expressed in the form (\ref{eq:neb}),   
\begin{equation}
\label{eq:UT1}
\left . \frac{d^2\sigma}{dE_b d\Omega_b} \right |^\mathrm{UT}_\mathrm{NEB} = 
-\frac{2}{\hbar v_{i}} \rho_b(E_b) 
 \langle \psixprior | W_x | \psixprior \rangle ,
\end{equation}
where $ \psixprior$ is the solution of a $x-A$ inhomogeneous equation similar to Eq.~(\ref{eq:phix_post}) but replacing in the source term the post-form transition operator, $V_\mathrm{post}$, by its prior form counterpart, $V_\mathrm{prior} \equiv U_{xA} + U_{bA}-U_{a}$, i.e.:
\begin{equation}
\label{phix_prior}
(E^+_x - K_x - {U}_x)  \psixprior(\vecr_x) =  \langle \vecr_x \, \chi_b^{(-)}| V_\mathrm{prior}| \chi^{(+)}_{a} \phi_a \rangle.
\end{equation}
Although the final expressions for the IAV and UT models have the same formal structure, they lead to different predictions for
the NEB cross sections. This is in contrast to the DWBA
formula for transfer between bound states, where it is well
known that the post and prior formulas are fully equivalent.
This discrepancy led to a long-standing controversy between
these two groups. At the heart of the discussion was the fact that the transformation of the post-form DWBA expression of IAV to its prior form gave rise to additional terms, not present in the UT prior formula.
These additional terms guaranteed the post-prior equivalence
for NEB, but they were regarded as unphysical by UT. Currently, there is consensus that that  the IAV formula is the correct one, as corroborated by numerical comparisons to experimental data \cite{Jin15b}.

The IAV model has recently been revisited by several groups \cite{Car16,Jin15,Pot15} and its accuracy has been assessed against experimental data with considerable success \cite{Jin17,Pot17}.  

\begin{figure}
\begin{center}\includegraphics[width=0.85\columnwidth]{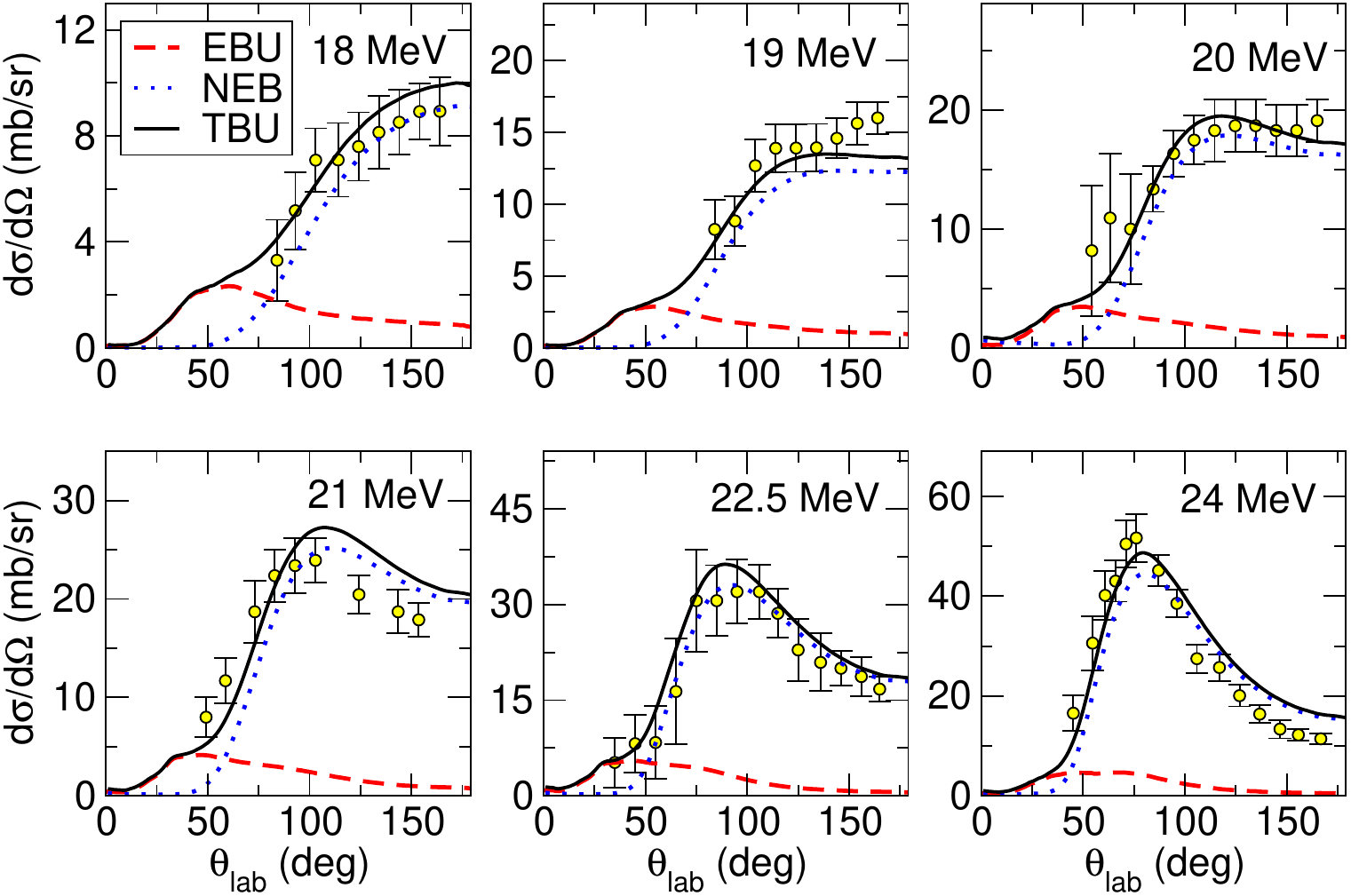} \end{center}
\caption{Energy distribution of $\alpha$ particles produced in the reaction $^{118}$Sn($^{6}$Li,$\alpha$)$X$ at the incident energies indicated by the labels. The dashed and dotted lines are the EBU and NEB contributions, and the solid line is their incoherent sum (TBU). Quoted from \cite{Jin17}, with authorization from APS.}  
\label{fig:neb} 
\end{figure}

 As an example, in Fig.~\ref{fig:neb} we show the experimental and calculated energy distribution of $\alpha$ particles emitted in the reaction induced by $^6$Li on $^{118}$Sn at the incident energies indicated by the labels (adapted from Ref.~\cite{Jin17}). The EBU contribution (dashed line) was evaluated with the CDCC method, whereas the NEB part (dotted line) was obtained with the IAV method, in its DWBA form.  Interestingly, one can see that the inclusive $\alpha$ yield is largely dominated by the NEB mechanism. The EBU is only important at small scattering angles (distant collisions, in a classical picture).  

\subsection*{Application to three-body projectiles}
The model has also been applied to projectiles with a genuine three-body structure, such as Borromean nuclei, in which one of the fragments is removed from the projectile, leaving a residual two-body unbound system. An example is the reaction A($^{9}$Be,$^8$Be)X which produces two alpha particles in the final state. 

Using the simplifying assumption that the transition operator $V_\mathrm{post}$ does not alter the state of the unbound two-body core, one can use the IAV formalism with the new $x$-channel wavefunction [c.f.\ Eq.~(\ref{eq:phix_post_green})]:
\begin{align}
\varphi_x(\vec{k}_b, \vecr_x) = G^{\mathrm{opt}(+)}_{xA}
 \langle \vec{r}_x \, \chi^{(-)}_{b} (\vec{k}_b)  |  V_\mathrm{post} | \chi_a f_{ab}(\vec{r}_{bx})  \rangle  
\end{align}
where we have introduced the overlap function between the three-body projectile and the two-body residual nucleus after the particle transfer, i.e.,:
\begin{equation}
    f_{ab}(\vec{r}_{bx})\equiv \int d\xi_b \, \phi^*_b(\xi_b)\phi_a(\vec{r}_{bx},\xi_b).
\end{equation}

\begin{figure}[ht]
\centering
\includegraphics[width=0.75\columnwidth]{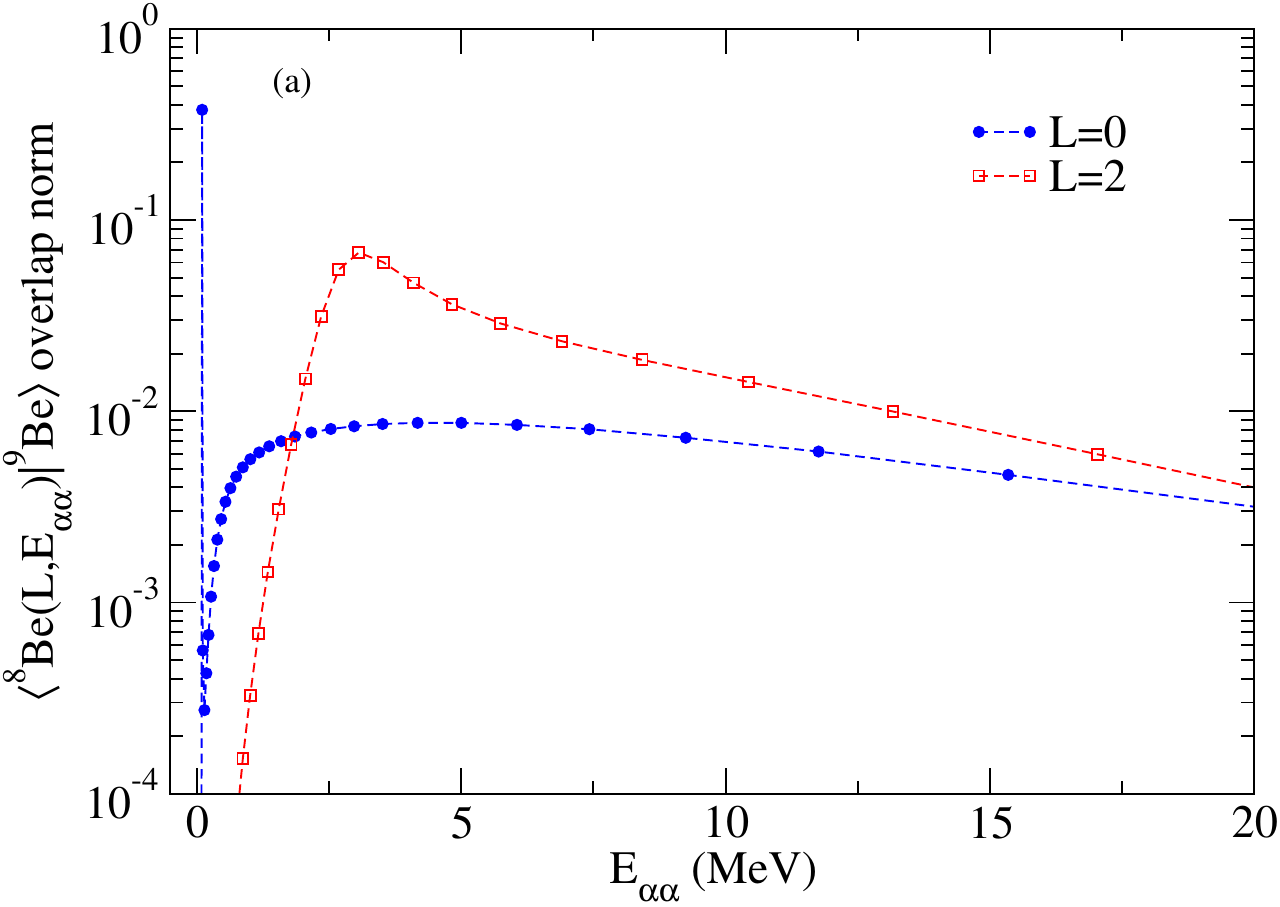}
\includegraphics[width=0.75\columnwidth]{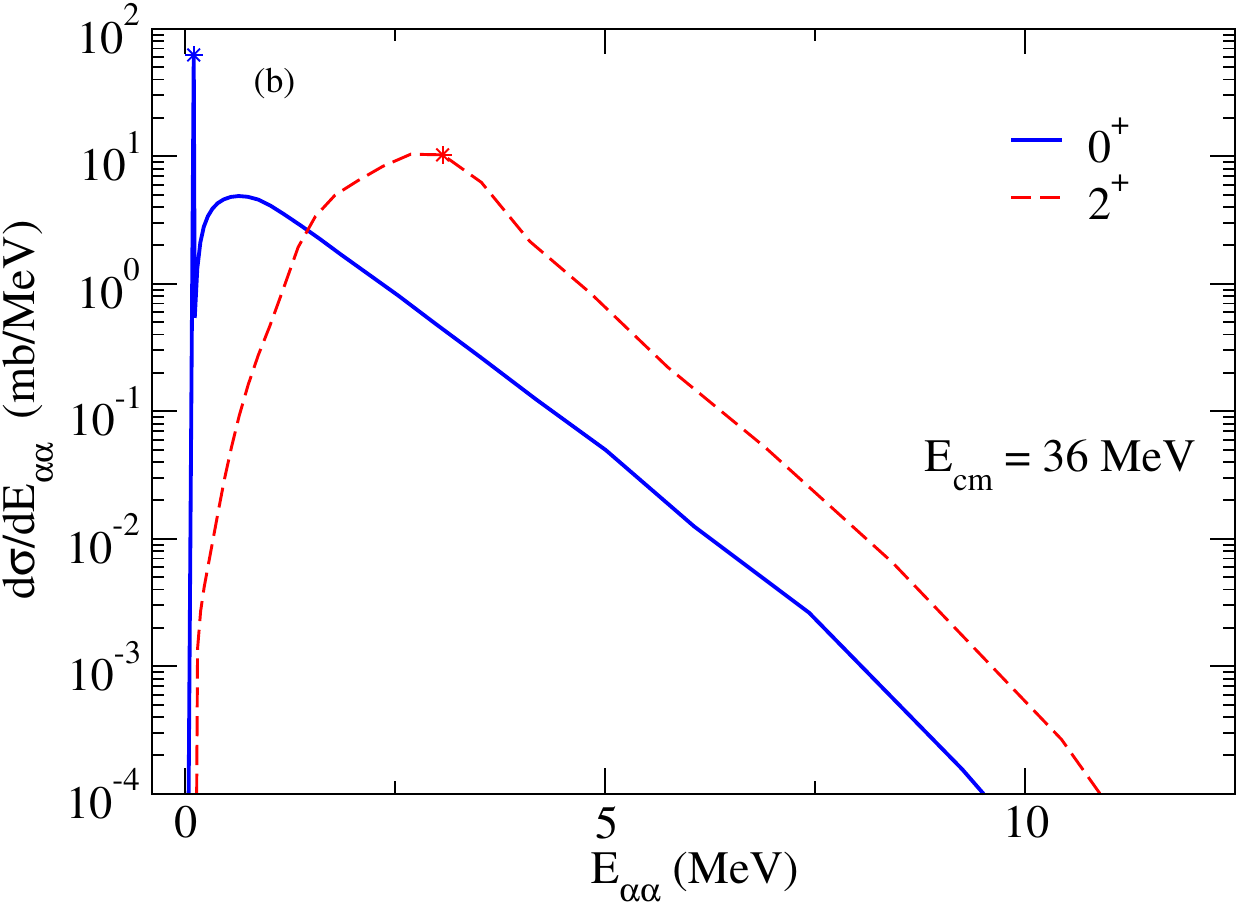}
\includegraphics[width=0.75\columnwidth]{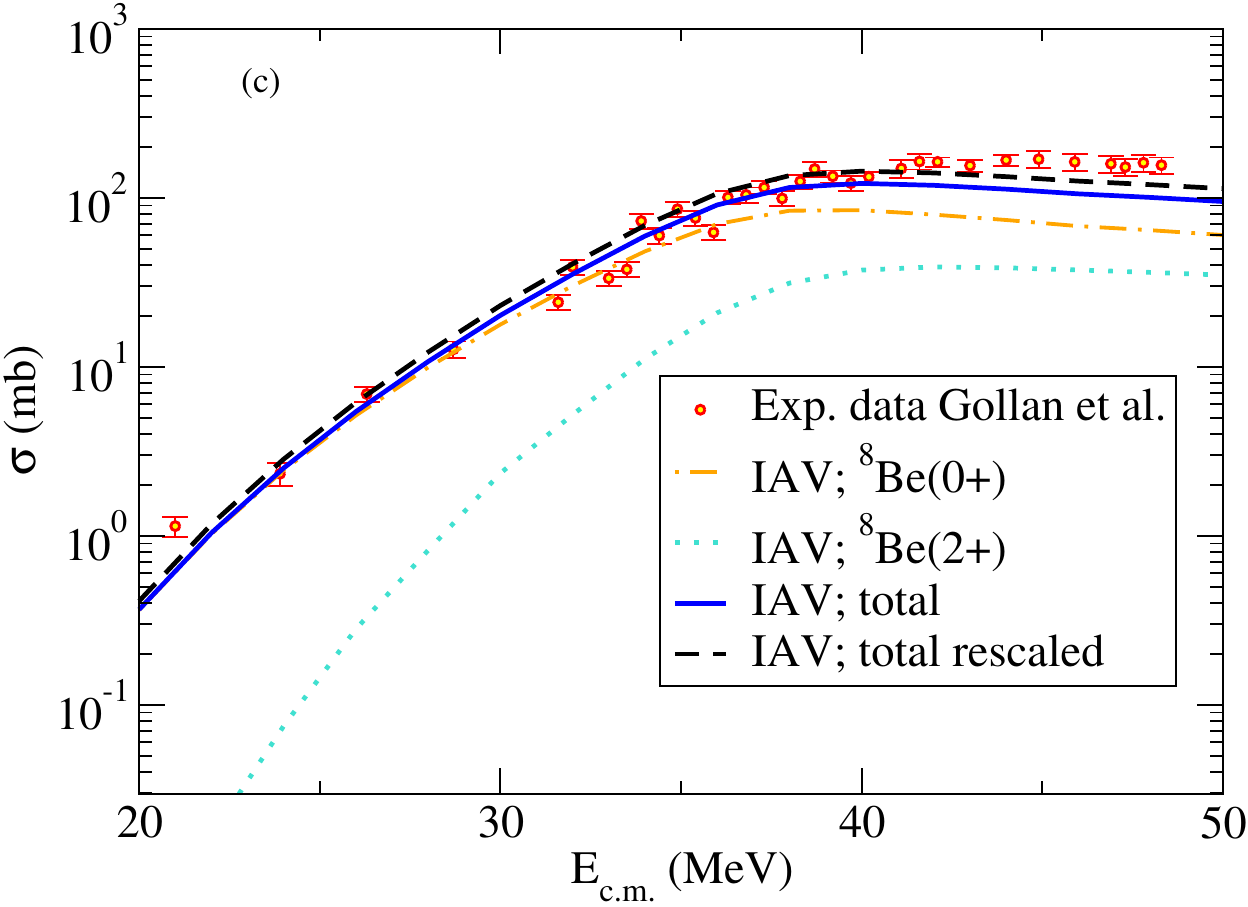}
\caption{\label{fig:be9au} 
IAV calculations for the reaction $^{197}$Au($^{9}$Be,$^8$Be*)$^{198}$Au populating bound states of the residual nucleus. (a) Norm of the overlaps between the $^9$Be g.s.\ and the $^8$Be continuum pseudostates for different relative angular momenta $L$, as a function of the $\alpha$-$\alpha$ relative energy up to 20 MeV. Dashed lines are included as a guide.
(b) Differential cross section, at $E_\mathrm{cm}=36$ MeV, as a function of the $\alpha$-$\alpha$ relative energy in the final $^8$Be$^*$ system, for 0$^+$ states (solid line) and $2^+$ states (dashed line). The symbols indicate the nominal energies of the g.s.\ ($0^+$) and the first $2^+_1$ resonance of $^{8}$Be. 
(c)  Angle-integrated total transfer cross section as a function of $E_\mathrm{cm}$. The contribution from the $0^+$ (dot-dashed line) and $2^+$ (dotted) states of $^8$Be, together with their sum (blue solid), are compared to the experimental data of Ref.~\cite{gollan21}. The IAV result rescaled by ab-initio Variational Monte Carlo (VMC) spectroscopic factors (dashed line) is also shown (see text). Adapted from Ref.~\cite{Vil24}. }
\end{figure}

The three-body wavefunction can be evaluated with the techniques discussed in previous sections, such as the hyperspherical expansion method employed in four-body CDCC calculations. An application is shown in Fig.~\ref{fig:be9au}, taken from \cite{Vil24}. The top panel (a) shows the norm of the $\langle{^9}\mathrm{Be}| ^8\mathrm{Be}\rangle$ overlaps as a function of the $\alpha-\alpha$ relative energy. The middle panel is the cross section, as a function of $\alpha-\alpha$ relative energy, leading to bound states of the $^{198}$Au system for an incident energy of $E_\mathrm{c.m.}$. Finally, the bottom panel is the stripping cross section, integrated in the relative energy of the two $\alpha$'s, as a function of the incident energy. The separate contributions due to the $^{8}$Be $0^+$ and $2^+$ states are shown, together with their sum. The dashed line is the result of rescaling the IAV calculations by the spectroscopic factors obtained in the ab-initio Variational Monte Carlo (VMC). See \cite{Vil24} for further details.

\subsection{The Eikonal Hussein-McVoy formula (EHM)}
\label{sec:EHM}

Hussein and McVoy \cite{HM85} obtained an alternative formula for the inclusive breakup cross section. Although this was not the rationale followed by Hussein and McVoy, this formula can be readily derived  making the same factorized approximation used by IAV, but using such approximation in Eq.~(\ref{eq:chiPsi}):
\begin{equation}
\label{eq:psixno}
\psixno(\vec{r}_x) = \langle \vecr_x | \psixno \rangle =   \langle \vecr_x \chi^{(-)}_b |  \chi_a^{(+)} \phi_a  \rangle .
\end{equation}

This result is referred to as Hussein-McVoy (HM) formula for inclusive breakup (or, more precisely, for NEB). At sufficiently high energies, the HM formula can be simplified using the following three approximations:
\begin{enumerate}[(i)]
\item the distorted waves $\chi^{(+)}_{a}(\vec{r}_a)$ and  $\chi^{(-)}_{b}(\vec{r}_b)$ are treated in the Glauber (also known as  {\it eikonal}) approximation. The Glauber approximation to an elastic-scattering distorted wave is:
\be
\label{eq:Glauber}
\chi_{\vec  k}^{(+)}({\vec r})=\mathrm{e}^{i \vec{k} \cdot \vec{r}} \exp \left[+i \int_{-\infty}^{z} \Delta k\left(z^{\prime}, b\right) \mathrm{d} z^{\prime}\right]
\ee
where the incident momentum $\vec{k}$ points along the positive $z$-axis, and $\vec{b}$ is the component of $\vec{r}$ perpendicular to $z$. The exponent in the second factor is 
\begin{equation}
\Delta k\left(z^{\prime}, b\right) \equiv-\frac{k}{2 E} U(z, b) ,
\end{equation}
which, when integrated along the entire trajectory, gives the optical phase shift
\begin{equation}
2 \delta(b)=\int_{-\infty}^{\infty} \Delta k\left(z^{\prime}, b\right) \mathrm{d} z^{\prime}=2 \int_{0}^{\infty} \Delta k\left(z^{\prime}, b\right) \mathrm{d} z^{\prime} ,
\end{equation}
that is related to the partial-wave optical S-matrix, i.e., $S(b)=\exp[{2 i \delta(b)}]$.

\item The $U_{a}$ potential, distorting the incident wave, is taken as the sum of the corresponding fragment-target potentials:
\begin{equation}
\label{eq:UaA}
U_{a}=U_{bA} + U_{xA} 
\end{equation}
This particular choice has the virtue of taking into account breakup effects in the entrance channel. However, since each potential in Eq.~(\ref{eq:UaA})  is evaluated in the corresponding fragment-target coordinate, the associated initial state  wavefunction of the system would be a solution of a complicated three-body equation. 

\item The above complication vanishes  thanks to the use of the eikonal approximation: the $x$ and $b$ fragments move with the same average velocity as the projectile and hence their momenta are given by 
\begin{equation}
\label{eq:kxb}
\vec{k}_{x}=\left(m_{x} / m_{\mathrm{a}}\right) \vec{k}_{\mathrm{a}}, \quad \vec{k}_{b}=\left(m_{b} / m_{a}\right) \vec{k}_{a} .
\end{equation}
\end{enumerate}

With the particular choice of Eq.~(\ref{eq:UaA}) and the assumption in Eq.~(\ref{eq:kxb}) one obtains the following result for the $x$-channel wavefunction:
\begin{equation}
\begin{aligned}
\varphi^\mathrm{EHM}_x(\vecr_x) = & 
\int \mathrm{d}^{3} \vecr_{b}
\chi_{b}^{(-)*}(\vecr_b)   \chi_{a}^{(+)}(\vecr_a) \phi_{a}(\vecr_{bx})\\
=& \mathrm{e}^{i \vec{k}_{x} \cdot \vecr_{x}} \exp \left[i \int_{-\infty}^{z_{x}} \Delta k_{x}
\left(z^{\prime}, b_{x}\right) \mathrm{d} z^{\prime}\right]  \\
& \times \int \mathrm{d}^{3} \vecr_{b} \mathrm{e}^{i \vec{q} \cdot \vecr_{b}} S_{bA}\left(b_{b}\right) \phi_{\mathrm{a}}\left(\vecr_{bx}\right)
\end{aligned}
\end{equation}
with $\vec{q}=\vec{k}_b - \vec{k}'_b$, the average momentum transferred in $b-A$ elastic scattering.  

Inserting this expression into the general expression (\ref{eq:iav}) (see details in \cite{HM85,Mar02}) one obtains for the double differential cross section: 
\begin{align}
\label{eq:EHM}
\left . \frac{d^2\sigma}{dE_b d\Omega_b} \right |^\mathrm{EHM}_\mathrm{NEB}  & = \frac{2}{\hbar v_{a}} \rho_b(E_b)
\frac{E_{x}}{k_{x}}  \nonumber \\ 
& \times \int \mathrm{d}^{2} \vec{b}_{x} \left| \tilde{\phi}_{a,b}(\vec{q},\vec{b}_{x}) \right|^{2}
\left[ 1-\left|S_{xA} (b_{x} ) \right|^{2} \right] .
\end{align} 
It should be noticed that the NEB depends only on the asymptotic properties, this is, the $S$ matrices, of the interaction of $b$ and $x$ with the target. There is no  sensitivity on the wavefunctions in the interaction region. This is a result of the eikonal approximation, plus the particular choice of the distorted interaction, which included the imaginary potential $W_{xA}$ which ultimately generates the NEB.

In many applications, one is interested in the total yield of fragment $b$, which is obtained upon integration of the previous formula over the angular and energy variables, resulting:
\begin{align}
\label{eq:EHM_total}
\sigma_\mathrm{NEB}^\mathrm{EHM}
= & \frac{2}{v_{\mathrm{a}}}(2 \pi)^{3} \frac{E_{x}}{\hbar k_{x}} 
 \int \mathrm{d}^{3} \vec{r}_{b} \mathrm{d}^{3} \vec{r}_{x}     \left|\phi_{a}\left(\vec{r}_{bx}\right)\right|^{2} \nonumber \\
\times & \left|S_{bA}(b_{b}) \right|^{2} \left[1-\left|S_{xA}\left(b_{x}\right)\right|^{2}\right] .
\end{align}

This equation has an appealing and intuitive form: the integrand contains the product of the probabilities for the core being elastically scattered by the target, $|S_{bA} (b_b) | ^2$, times the probability of the valence particle being  absorbed, $(1- |S_{xA}(b_x)|)^2$. These probabilities are weighted by the projectile wave function squared and integrated over all possible impact parameters. Due to the Glauber approximation, Eq.~(\ref{eq:EHM})  is expected to be accurate at high energies (above $\sim$100~MeV per nucleon). In fact, this formula has been extensively employed in the analysis of intermediate-energy knockout reactions (see e.g.~\cite{Han03,Tos01,Tos14} and references therein) mostly aimed at obtaining spectroscopic information of nucleon hole states.




\subsection{Interpretation of inclusive breakup data}

In the previous sections, we have considered elastic breakup and  nonelastic breakup (with the latter possibly including transfer and incomplete fusion) taking place in a breakup reaction. In actual inclusive breakup experiments these contributions will appear entangled in the data, although they produce some distinctive features. For example, the energy distribution of the observed fragment at a given scattering angle will exhibit a characteristic shape, as shown in Fig.~\ref{fig:dsdep_general} for a hypothetical $A(d,pX)$ reaction. To understand this spectrum, it is important to recall that a given proton energy and angle will univocally determine the excitation energy of the residual $n+A$ system. According to the proton energy (or, equivalently, the residual system excitation energy), we may distinguish the following regions:

\begin{figure}
\begin{center}\includegraphics[width=0.9\columnwidth]{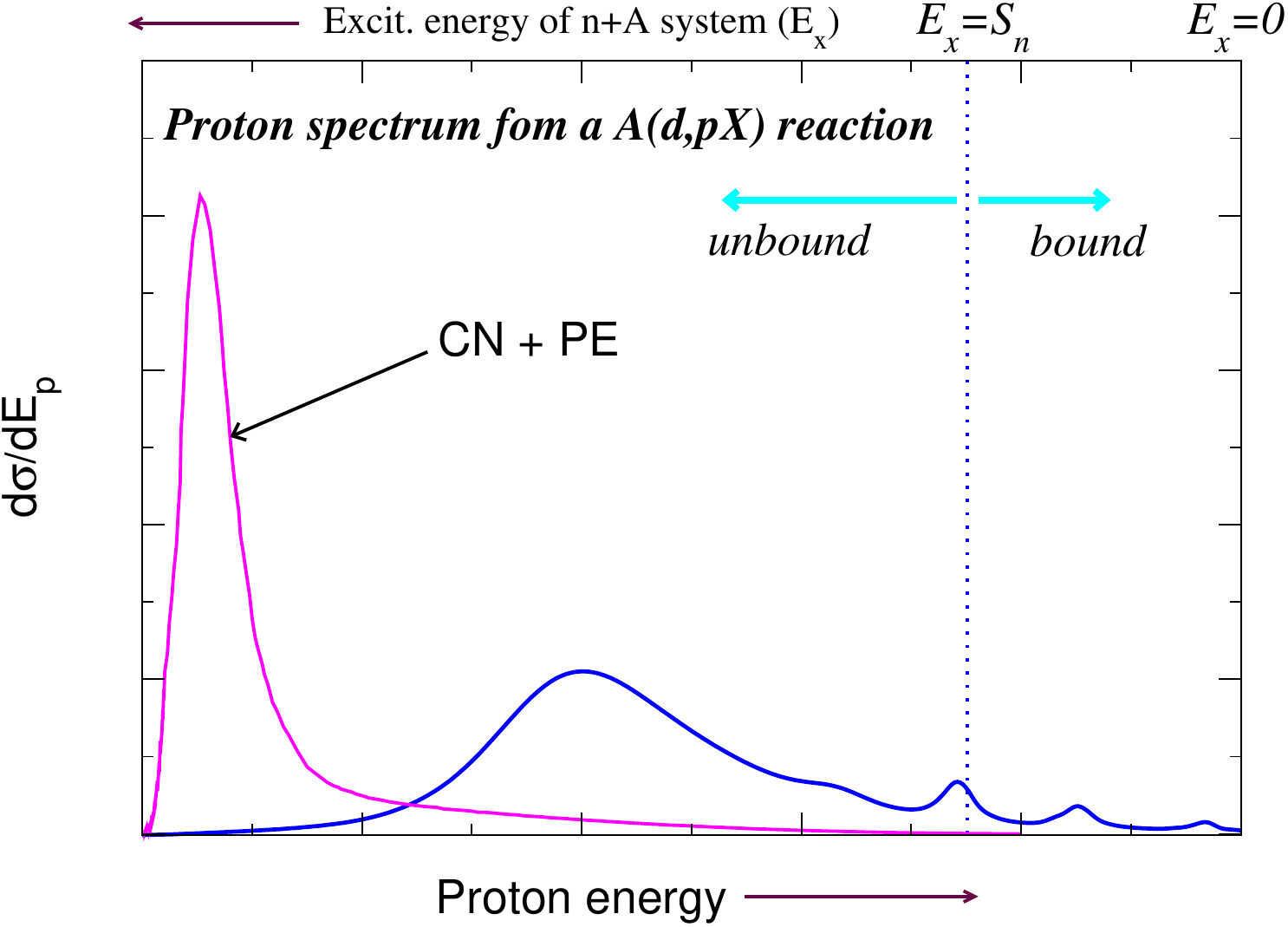} \end{center}
\caption{\label{fig:dsdep_general}  Proton energy spectrum from a $A(d,pX)$ inclusive breakup reaction (blue line). The vertical dotted line marks the neutron separation threshold in the $A+n$ system. The low-energy peak arises from compound nucleus (CN) and pre-equilibrium (PE)  processes followed by proton evaporation (magenta line).} 
\end{figure}

\begin{enumerate}[(i)]

\item The highest proton energies will be characterized by some narrow peaks corresponding to bound states of the $n+A$ system. Depending on the experimental resolution, these peaks will appear separated or, instead, will merge with neighbouring peaks.  Theoretically, the cross section for these isolated bound states can be evaluated with the standard formalisms for transfer reactions, such as DWBA, CCBA, ADWA or CRC.

\item As the excitation energy of the residual nucleus increases, so does the density of states and the low-lying bound states will merge to form a quasi-continuum. Here, the treatment with the aforementioned methods becomes more troublesome because of the impossibility of disentangling unambiguously the contribution of each state from the data. When one is interested in energy averaged cross sections, the combination of the IAV model of Sec.~\ref{sec:iav} with dispersive models provides an appealing alternative to the traditional methods.

\item At a certain excitation energy, the residual system will reach the neutron separation threshold ($E_x=S_n$). Just above this excitation energy, the proton spectrum will exhibit narrow peaks, corresponding to low-lying resonances, superimposed to a nonresonant continuum background. It is important to realize that the properties of the system just above the threshold ($E_x > S_n$) and just below it ($E_x < S_n$) states are qualitatively similar. In particular, the discrete energy levels extend above $S_n$ even if those states are not strictly stationary. Therefore, the wavefunctions of the narrow resonances resemble very much those for the bound states below threshold and correspond to situations in which the system remains bound for a long time before decaying by barrier penetration. This is important from the reaction point of view because it indicates that the {\it transfer} cross section should evolve smoothly from negative to positive energies, as it actually happens in the distribution shown in Fig.~\ref{fig:dsdep_general}

\item As the excitation energy of the residual nucleus increases above the threshold, the  narrow resonances disappear, the spectrum becomes continuous and structureless,  giving rise to a bell-shaped bump. The most probable energy of the emerging proton can be obtained by assuming that the proton gets half the kinetic energy of the deuteron at the point of breakup, plus the Coulomb energy of the deuteron at that place, and half the internal energy of the deuteron ($E_B=-2.22$~MeV) (see Ref.~\cite{Bla91}, p.509), i.e.,
\begin{equation}
\langle E_p \rangle \simeq \frac{1}{2}\left ( E_d -\frac{Z e^2}{R_\text{bu}}    \right) + \frac{Ze^2}{R_\text{bu}} - \frac{1}{2}E_B 
\end{equation}
where $E_d$ is the kinetic energy of the relative motion of the deuteron and the target nucleus A in the c.m.\ frame and $R_{bu}$ the deuteron-target separation at the time of the deuteron breakup. 
The bump will contain contributions coming from both EBU and NEB components discussed in previous sections. The EBU can be conveniently evaluated with the DWBA or CDCC methods whereas, for the NEB part, the IAV model provides a very convenient framework.

\item In the former contributions, we have implicitly assumed that the observed particle (proton in this case) scatters elastically by the target nucleus. There will be situations in which this not the case; for example, when the projectile fuses completely with the target nucleus forming a compound nucleus that will  eventually thermalize by emitting particles and gamma rays. Among these particles, there will be protons that will add up incoherently to the proton spectrum. These protons are typically emitted with low energy in the c.m.\ frame and will therefore contribute to the low energy part of the proton spectrum (see magenta line in Fig.~\ref{fig:dsdep_general}). Some phenomena related to the fusion of weakly-bound nuclei will be discussed in the next section. 
\end{enumerate}

\section{Fusion involving weakly bound nuclei \label{sec:fusion}}
The previous sections have been devoted to the modeling of direct nuclear reactions. Fusion reactions involving weakly-bound nuclei display also distinctive features which require adequate reaction formalisms and considerations with respect to the case of well-bound nuclei. The topic of fusion with both stable and unstable nuclei has motivated many works and excellent review papers have been published in recent years, so we refer the reader to these works for a detailed account of the present status of the description of fusion (e.g.~\cite{Can06,Can15,Can21}). We shall discuss two phenomena which are the focus of many theoretical and experimental efforts by several groups. One is the phenomenon of complete fusion suppression.  Complete fusion is conveniently defined as the process in which the whole {\it charge} of the projectile and target nuclei merge, giving rise to an excited compound nucleus that will subsequently decay by particle and/or gamma emission.  The other phenomenon is the large observed yields compatible with the partial fusion of the projectile (incomplete fusion, ICF). Experimental results and theoretical calculations indicate that these two phenomena are actually related since they appear simultaneously and so a plausible explanation of the CF suppression might be in fact the leak of flux going to the ICF channels. 

A variety of models have been proposed to evaluate the CF cross section,  from the simple single-barrier penetration model to more sophisticated coupled-channels methods, in which collective excitations of the projectile and/or target nucleus are taken into account explicitly. These models are very successful at predicting the cross section for well bound nuclei, but tend to overestimate them for weakly bound projectiles   at energies above the barrier.  For example, for the light weakly bound nuclei $^{6,7,8}$Li, $^{9}$Be the experimental CF cross sections are found to be suppressed by $\sim$20-30\% compared to the case of tightly bound nuclei \cite{Das99,Tri02,Das02,Das04,Muk06,Rat09,LFC15}.

Early analyses of these experiments tried to explain the phenomenon using coupled-channels calculations, including the coupling to low-lying excited states of the projectile and target \cite{Das99,Das02,Kum12,Zha14,Fan15}. Yet, these calculations  systematically failed to reproduce the experimental suppression.  This failure has been attributed to the omission in these calculations of the breakup of the projectile; a scenario was suggested in which the weakly bound projectile breaks up prior to reaching the fusion barrier, with the subsequent reduction of the complete fusion probability. This interpretation is supported by the presence of large $\alpha$ yields (in \nuc{6,7}{Li}-induced reactions) as well as target-like residues  which are consistent with the capture of one of the fragment constituents  of the projectile, that is, ICF.   
To account for these observations, some authors have proposed a two-step scenario \cite{Das02,Dia07}  consisting on the elastic dissociation of the projectile followed by the capture of one of the fragments by the target.  However,  calculations based on a three-dimensional classical dynamical model \cite{Dia07}, which incorporates this two-step breakup-fusion mechanism, can only explain a small fraction of the observed CF suppression for $^{9}$Be \cite{Coo16} and $^{8}$Li \cite{Coo18} reactions. More encouraging results have been obtained with different methods based on the CDCC formalism, as described in the following subsections.

\subsection{Computation of CF and ICF with CDCC \label{sec:cficf}}
Although the CDCC method was originally envisaged as a practical tool to evaluate the elastic and breakup observables, some works have been done to use this method to obtain complete and incomplete fusion cross sections. 


Some authors \cite{Hag00,Dia02} have proposed to identify the total fusion with the amount of flux that leaves the coupled
channels set due to a short-range imaginary potential $i W_F(R)$, while CF is identified with the absorption due to such potential, but restricted to bound states of the projectile only. Thus, total fusion is computed as 
\begin{equation}
\sigma_{TF}=\frac{\pi }{\hbar^2 K_0}\sum_{J_T}(2J_T+1)P_{J_T},  
\label{eq:CF_WF}
\end{equation}
where $K_0$ is the wavenumber of the incident channel and $P_{J_T}$ is the complete fusion  probability for total angular momentum $J_T$  
\begin{equation}
P_{J_T}=
-\frac{8 \mu}{\hbar^2K_0}
\sum_{\beta} \int \limits_{0}^{\infty } \mid \chi^{J_T}_{\beta,\beta_i}(R) \mid
^{2} ~W_{F}(R)  dR.  
\label{eq:PJ}
\end{equation}
where $\chi^{J_T}_{\beta,\beta_i}$ are the radial solutions obtained from the coupled equations (\ref{eq:cc_radial}).
In the case of CF, the expression for the cross section is identical but the sum in $\beta$ is restricted to those channels associated with bound states. 

The CF obtained in this way represents a
lower limit of the physical CF cross section, since one has assumed no capture of all projectile fragments  from breakup channels.
In reality, these events should contribute to the CF, but 
cannot be distinguished in this model from the capture of only one projectile fragment.
In this model, the incomplete fusion $\sigma _{ICF}$  is therefore 
defined as the absorption from breakup channels.

\begin{figure}[tb]
\begin{center}
 {\centering \resizebox*{0.94\columnwidth}{!}{\includegraphics{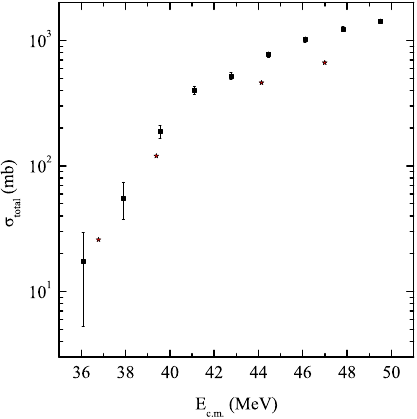}} \par}
\caption{\label{fig:be11pb_totfus} Calculated total fusion cross sections 
for $^{11}$Be + $^{208}$Pb (full stars) compared with 
the experimental data from \cite{Sig98} 
for $^{11}$Be + $^{209}$Bi (full squares) using the CDCC wavefunction. 
Adapted from Ref.~\cite{Dia02}, with permission from APS.}
\end{center}
\end{figure}

An application of this method is shown in Fig.~\ref{fig:be11pb_totfus}, where the calculations for the total fusion of $^{11}$Be + $^{208}$Pb (full stars) are compared with the total fusion data for the nearby reaction $^{11}$Be + $^{209}$Bi from Ref.~\cite{Sig98}. The agreement is reasonable around the Coulomb barrier, but the calculation underestimates the data by $\sim$ 41\% for energies well above the Coulomb
barrier.

\begin{figure}[!ht]
\centering
\includegraphics[width=0.7\columnwidth]{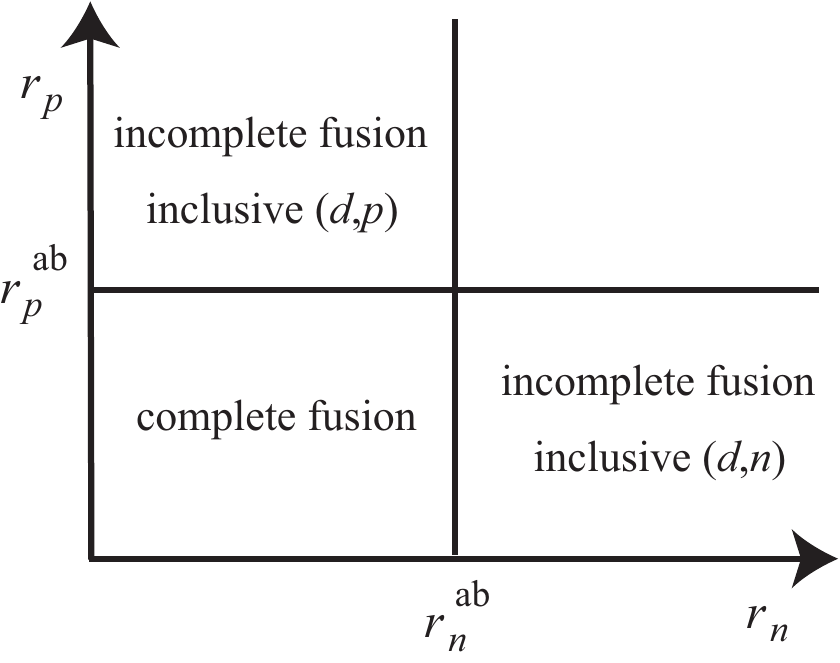}
\caption{\label{fig:hashimoto_regions} Schematic illustration of the four integration regions employed in the method of Hashimoto {\it et al.}~\cite{Hash09} for evaluating complete and incomplete fusion from CDCC wavefunction Quoted with authorization from Oxford University Press.}
\end{figure}

A limitation of this method is that it can only be applied to projectiles composed of a heavy charged fragment and a light uncharged one (such as $^{11}$Be), since it relies on the assumption that the center of mass of the projectile is close to that
of the heavy fragment and far from the light one.   Thus, it cannot
be used for projectiles like $^{6,7}$Li that break up into two fragments of comparable masses. 
To overcome this difficulty, Hashimoto {\it et al.}~\cite{Hash09}  proposed an alternative approach based also on the CDCC method and applied it to the case of deuteron scattering.  Their idea is to transform the CDCC wavefunction from its natural coordinates 
\{$\vecr$, $\vecR$\} to the coordinates \{$\vecr_p$, $\vecr_n$\}, 
\begin{equation}
|\Psi(\vecr,\vecR) |^2 \, d\vecr d\vecR =  |\widetilde{\Psi}(\vecr_p,\vecr_n) |^2  d\vecr_p  d\vecr_n    \, ,
\end{equation}
and then associate the CF and ICF cross sections with the absorption taking place in different regions of the $\{ r_p , r_n \}$ space, as  illustrated in Fig.~\ref{fig:hashimoto_regions}. The distances $r^\mathrm{ab}_p$ and $r^\mathrm{ab}_n$ denote the absorption radii for the proton and neutron, such that for $r_p > r^\mathrm{ab}_p$ ($r_n > r^\mathrm{ab}_n$) the proton (neutron) absorption becomes negligible. The  CF is identified with the absorption taking place when both the proton and neutron are inside their respective absorption radii, i.e.,
\begin{align}
\sigma_{\rm CF} & =
\frac{2 \mu}{\hbar^2 K_0}
  \int_{r_p<r_p^{\rm ab}}  d\vecr_p
  \int_{r_n<r_n^{\rm ab}}  d\vecr_n
  |\widetilde{\Psi}(\vecr_p,\vecr_n) |^2 \nonumber \\
  &\times \{ W_p(\vecr_p)+W_n(\vecr_n) \},
\label{eq-CFX}
\end{align}
where $K_0$ is the incident wave number, and $E$ is the energy.   
Similarly, the ICF cross section is obtained from the absorption occurring in the region where one of the two fragments is inside its absorption radius while the other is outside it, i.e.,
\begin{eqnarray}
\sigma_{\rm ICF}^{(p)}=
  \frac{2 \mu}{\hbar^2 K_0}
  \int_{r_p<r_p^{\rm ab}}  d\vecr_p
  \int_{r_n>r_n^{\rm ab}}  d\vecr_n
  |\widetilde{\Psi}(\vecr_p,\vecr_n) |^2
  W_p(\vecr_p),
\label{eq-IFXp}
\end{eqnarray}
\begin{eqnarray}
\sigma_{\rm ICF}^{(n)}=
\frac{2 \mu}{\hbar^2 K_0}
  \int_{r_p>r_p^{\rm ab}}  d\vecr_p
  \int_{r_n<r_n^{\rm ab}}  d\vecr_n
  |\widetilde{\Psi}(\vecr_p,\vecr_n) |^2
  W_n(\vecr_n) .
\label{eq-IFXn}
\end{eqnarray}

\begin{figure}[!ht]
\centering
\includegraphics[width=0.7\columnwidth]{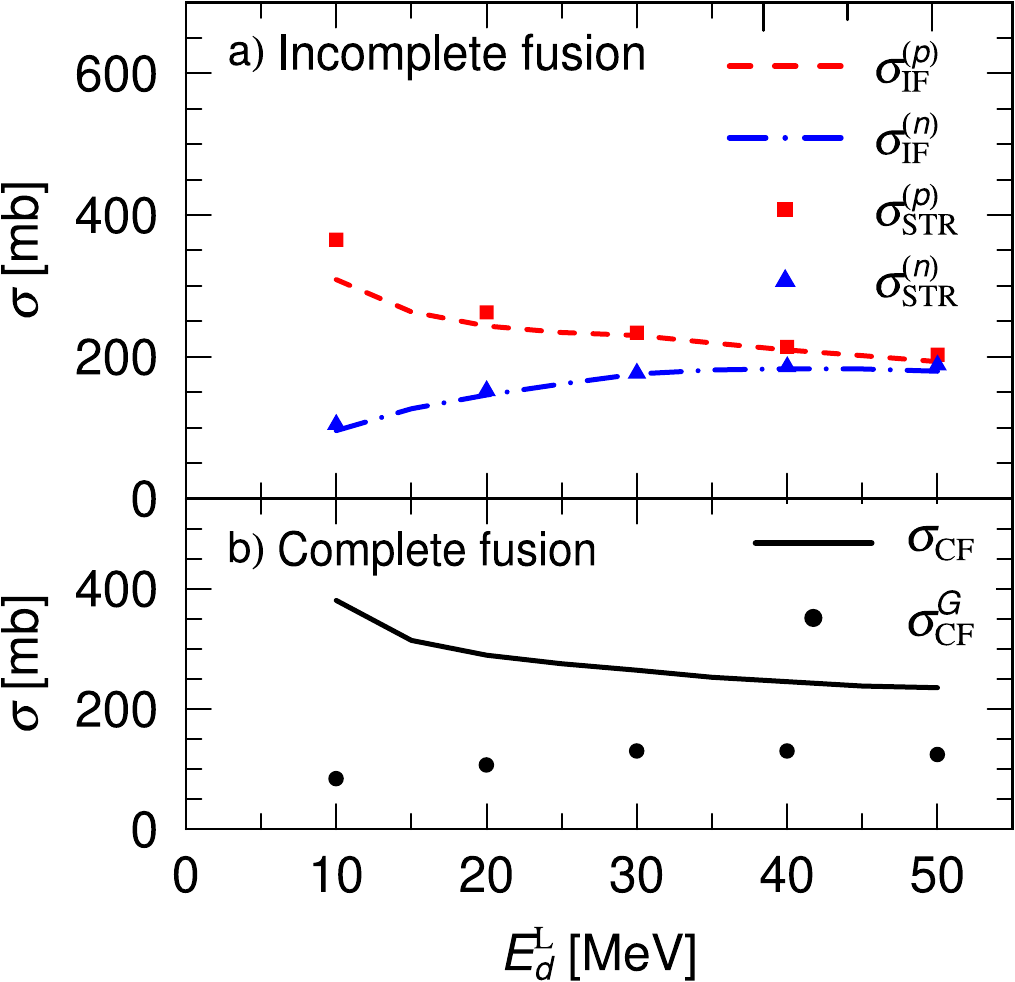}
\caption{\label{fig:hashimoto}  
a) Comparison of $\sigma_{\rm ICF}^{(p)}$ (short-dashed line)
and $\sigma_{\rm ICF}^{(n)}$ (dash-dotted line) obtained in \cite{Hash09}  with $\sigma_{\rm STR}^{(p)}$ (squares)
and $\sigma_{\rm STR}^{(n)}$ (triangles)
given by the Glauber model~\cite{Ye09} \
b) Complete fusion cross sections calculated with
the Glauber model (dots) and with the method of \cite{Hash09}
(solid line). Quoted with authorization from Oxford University Press.
} 
\end{figure}

The method was applied to the reaction $d$+$^{7}$Li \cite{Hash09}. Although no comparison with experimental data was attempted, the authors compared their results with those computed with the Glauber calculations of Ref.~\cite{Ye09}. 
This comparison is shown in  Fig.~\ref{fig:hashimoto} with the upper and lower panels corresponding respectively to the ICF and CF cross sections as a function of the deuteron incident energy. 
The agreement for the ICF is found to be very satisfactory for energies as low as $E_d$=10~MeV. Since the Glauber model is a high-energy approximation, the agreement at these relatively low energies is somewhat unexpected. By contrast, for the CF part (bottom panel), the Glauber method is found to give significantly smaller cross sections. A discussion of these results is provided in~\cite{Ye09}.

Another procedure to extract the CF and ICF cross section from the CDCC method was proposed by Parkar and co-workers \cite{Par16}. The central idea of their method is to perform a series of CDCC calculations with different choices of the fragment-target potentials. In particular, to obtain the ICF cross section for the capture of a given fragment, they perform a CDCC calculation using a short-range imaginary potential for the interaction of that fragment and the target, and a real potential for the other fragment. Using this method, these authors have been able to obtain a reasonable account of the CF, ICF and TF cross sections of reactions induced by $^{6,7}$Li projectiles \cite{Par16,Par18}, although the short-range  fusion potential needs to be adjusted for each reaction.

More recently, the authors of Refs.~\cite{Ran20,Cor20} have proposed an alternative method which requires a single CDCC calculation to compute the TF, CF and ICF cross sections. In this CDCC calculation, the fragment-target interactions are also modeled with optical potentials with a short-ranged imaginary part. For a two-body projectile, the total fusion cross section is computed as:
\begin{equation} 
\sigma_\mathrm{TF} = 
\frac{2 \mu}{\hbar^2 K_0}
\left\langle\ \Psi^{(+)}\, \left| \, W^{(1)} + W^{(2)}\,  \right| \Psi^{(+)}\  \right\rangle \, ,
\label{sigTF-1}
\end{equation}
where $W^{(1,2)}$ represent the imaginary parts of the fragment-target interactions. This CDCC wavefunction can be split into bound ($\Psi^{B}$) and continuum ($\Psi^{C}$) components:
\begin{equation} 
{\Psi}^{(+)}({\vecR},\vecr) = {\rm \Psi}^{B}({\vecR},{\vecr})  + {\Psi}^{C}({\vecR},{\vecr}) ,
\end{equation}
where ${\Psi}^{B}$ and ${\Psi}^{C}$ are given by the channel expansions
\begin{eqnarray} 
{\Psi}^{B}({\vecR},{\vecr}) &=& \sum_{\beta\, \in\, B} \  \chi_{\beta}({\vecR})  \phi_\beta ({\vecr}) \label{PsiB-1}\\
{\Psi}^{C}({\vecR},{\bf r}) &=& \sum_{\gamma\, \in\,  C} \  \chi_{\gamma}({\vecR})  \phi_\gamma ({\vecr}) , 
\label{PsiC-1}
\end{eqnarray}
where $\phi_\beta$ and $\phi_\gamma$ are, respectively, the bound and unbound states of the projectile, and $\chi_\beta$ and $\chi_\gamma$ are
the corresponding wave function describing the projectile-target relative motion.

Assuming that matrix elements of the imaginary potentials connecting bound channels to bins are negligible,  Eq.~(\ref{sigTF-1}) can  be put in the form
\begin{equation} 
\sigma_\mathrm{TF} = \sigma_\mathrm{TF}^{B} \ +\  \sigma_\mathrm{TF}^{C},
\end{equation}
with
\begin{eqnarray} 
 \sigma_\mathrm{TF}^{B} &=& 
 \frac{2 \mu}{\hbar^2 K_0}
 \sum_{\beta, \beta^\prime\, \in\, {B}} \
 \left\langle \chi_\beta \left|  W_{\beta \beta^\prime}^{(1)}+ W_{\beta \beta^\prime}^{(2)}\,
 \right|   \chi_{\beta^\prime} \right\rangle \label{TF-B}\\
 \sigma_{TF}^{C} &=&  
 \frac{2 \mu}{\hbar^2 K_0}
\sum_{\gamma,\gamma^\prime \in\,{C}} \
 \left\langle \chi_\gamma \left|  W_{\gamma \gamma^\prime}^{(1)} + W_{\gamma \gamma^\prime}^{(2)}\,
 \right|   \chi_{\gamma^\prime} \right\rangle .
 \label{TF-C}
\end{eqnarray}
where $W_{\alpha \alpha^\prime}^{(i)} =\left(\phi_{\alpha} \left| {W}^{(i)} \right| \phi_{\alpha^\prime} \right),$
with $\alpha, \alpha^\prime$ standing for either $\beta, \beta^\prime$ or $\gamma, \gamma^\prime$, are the matrix elements of the imaginary potentials.\\

Then, by performing an angular momentum expansion of the wave functions and the imaginary potentials,  Eqs.~(\ref{TF-B}) and
(\ref{TF-C}) become
\begin{eqnarray}
\sigma_{TF}^{B} &=& \frac{\pi}{K_0^2}\,\sum_{J_T} (2J_T+1)\ \mathcal{P}_{B}^{TF}(J_T)  \label{TF-B1} \\
\sigma_{TF}^{C} &=& \frac{\pi}{K_0^2}\,\sum_{J_T} (2J_T+1)\ \mathcal{P}_{C}^{TF}(J_T) \label{TF-C1},
\end{eqnarray}
with
\begin{eqnarray}
\mathcal{P}_{B}^{TF}(J_T)  &=& \mathcal{P}_{B}^{(1)}(J_T) +  \mathcal{P}_{B}^{(2)}(J_T) \label{P_TF-B} \\
\mathcal{P}_{C}^{TF}(J_T)  &=& \mathcal{P}_{C}^{(1)}(J_T) +  \mathcal{P}_{C}^{(2)}(J_T) \label{P_TF-C} .
\end{eqnarray}
where $\mathcal{P}_{B}^{(i)}(J_T)$ and $\mathcal{P}_{C}^{(i)}(J_T)$ are the probabilities of absorption of fragment $c_i$ in
bound channels and in the continuum, respectively, resulting from the contributions of ${W}^{(i)}$ to the TF cross section.

In terms of these probabilities,  the authors of Ref.~\cite{Ran20,Cor20} introduce  the ICF probabilities 
\begin{eqnarray}
 {\mathcal P}^{ICF1}(J_T)  &=& {\mathcal P}^{(1)}_{C} (J_T) \times \left[\ 1 - {\mathcal P}^{(2)}_{C} (J_T)\ \right]   \label{PICF1}\\
{\mathcal P}^{ICF2}(J_T)  &=&  {\mathcal P}^{(2)}_{C} (J_T) \times \left[\ 1 - {\mathcal P}^{(1)}_{C} (J_T)\ \right]   \label{PICF2},
\end{eqnarray}
and the  {\it sequential complete fusion} probability 
\begin{equation}
{\mathcal P}^\mathrm{SCF}(J_T)  =    {\mathcal P}^{(1)}_{C} (J_T) \times  {\mathcal P}^{(2)}_{C} (J_T).
 \label{TSCF}
\end{equation}

In terms of the introduced probabilities, the following fusion cross sections are defined:
\begin{itemize}
    \item Direct complete fusion (CF):
 \begin{equation}
 \sigma_{DCF} = \sigma_{TF}^{B} ,
 \label{DCF}
 \end{equation}
  which describes the simultaneous capture of the two fragments. 
  
  \item Sequential complete fusion (SCF):
   \begin{equation}
\sigma_{SCF} = \frac{\pi}{K_0^2}\,\sum_{J_T} (2J_T+1)\ \mathcal{P}^{SCF}(J_T) .
\label{SCF}
\end{equation}

\item  ICF of fragment $c_{i}$ (ICFi) 
\begin{equation}
\sigma_{ICFi} = \frac{\pi}{K_0^2}\,\sum_{J_T} (2J_T+1)\ \mathcal{P}^{ICFi}(J_T) .
\label{sigICFi}
\end{equation}

\end{itemize}

In this formalism, the  CF, ICF and TF cross sections are then given by
\begin{eqnarray}
\sigma_\mathrm{CF} &=& \sigma_{DCF}\,+\,\sigma_{SCF},  \label{sigCF} \\
\sigma_\mathrm{ICF} &=&\sigma_{ICF1} \,+\, \sigma_{ICF2}, \label{sigICF}\\
\sigma_{TF} &=& \sigma_{CF} + \sigma_{ICF} . \label{sigmaTF-sum} 
\end{eqnarray}

\begin{figure}
\begin{center}
\begin{minipage}{0.45\columnwidth}
\includegraphics[width=0.95\columnwidth]{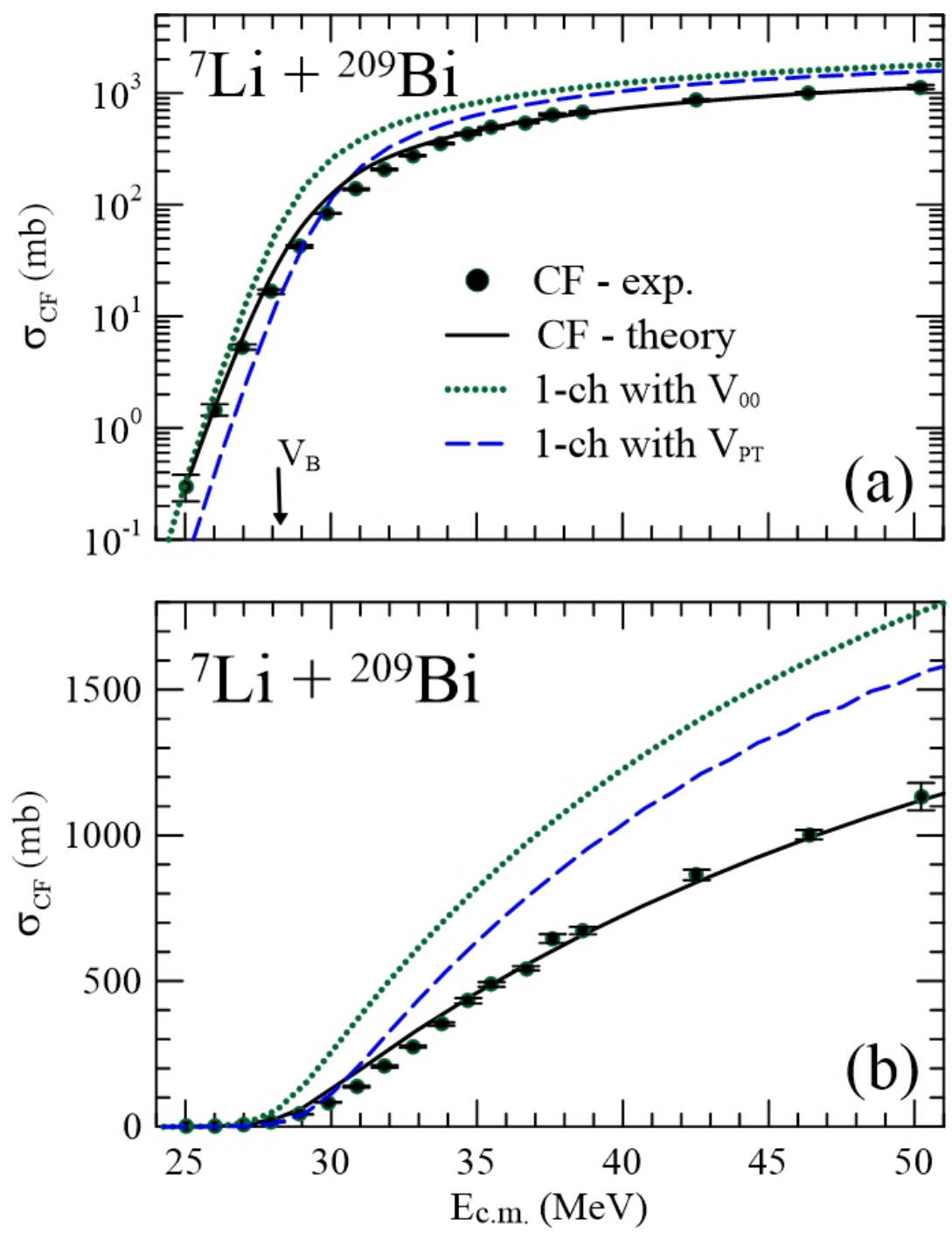}
\end{minipage}
\begin{minipage}{0.45\columnwidth}
\includegraphics[width=0.95\columnwidth]{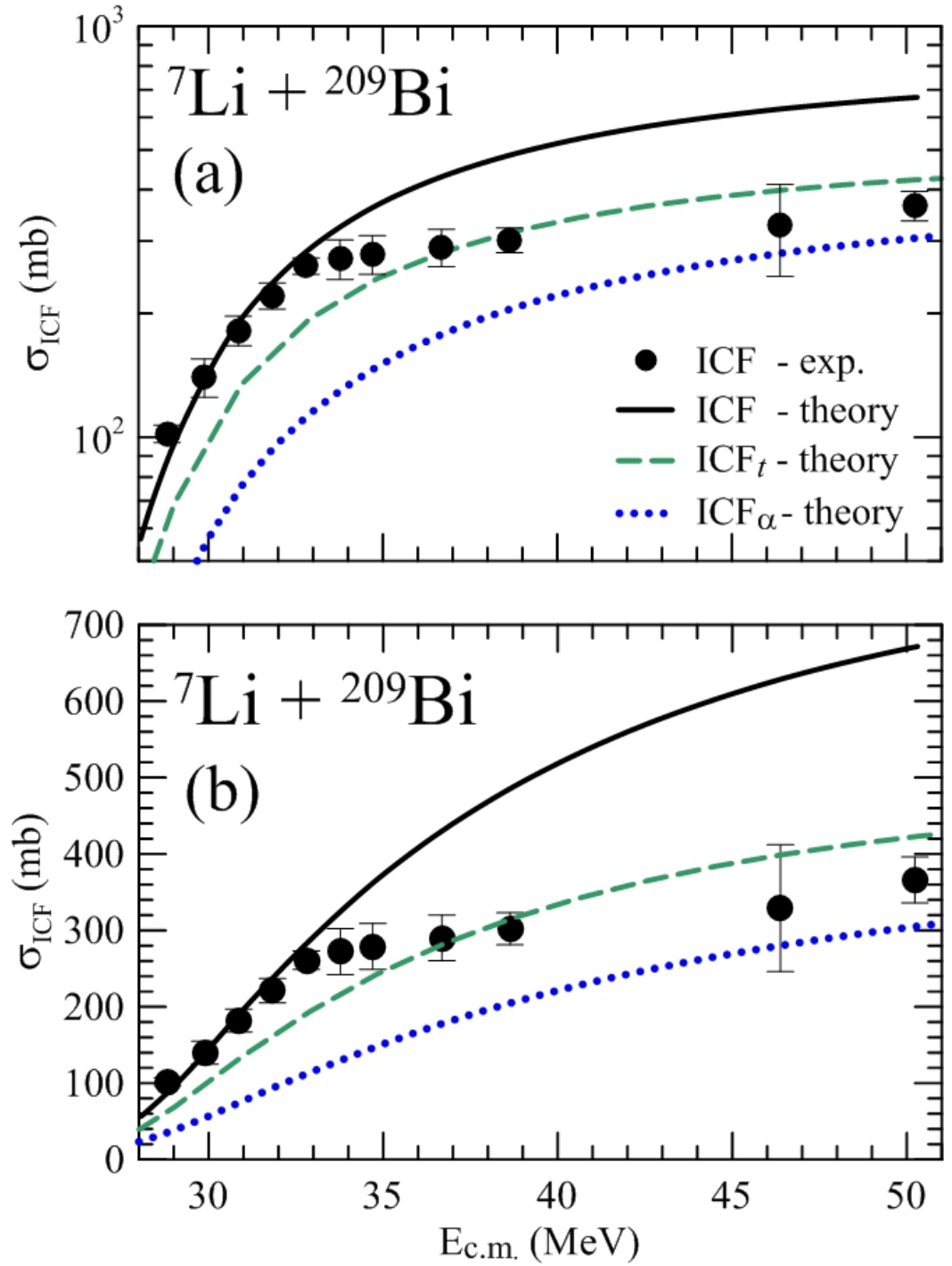}
\end{minipage}
\end{center}
\caption{\label{fig:li7bi_canto} CF (left) and ICF (right) cross sections for $^{7}$Li+$^{209}$Bi. Experimental data from Refs.~\cite{Das02,Das04} are compared with the calculations of Ref.~\cite{Cor20}. With permission of APS.} 
\end{figure}

An application of this model is shown in Fig.~\ref{fig:li7bi_canto}, where CF (left panel) and ICF (right panel) data for the reaction $^{7}$Li+$^{209}$Bi \cite{Das02,Das04} are compared with the predictions of the model. The upper and lower panels display the same results in logarithmic and linear scale, respectively. For CF, the agreement with the data is very satisfactory, both above and below the barrier (indicated by the arrow). For ICF, the separate contributions for triton capture and $\alpha$ capture are shown, with the former  giving a much larger cross section. The calculations are found to reproduce well the data up to $E_\mathrm{c.m.} \approx 34$~MeV, but overestimate them at higher energies. A detailed discussion of these results can be found in Ref.~\cite{Cor20}.

\subsection{Evaluation of CF and ICF cross sections with the IAV model}

As discussed in Sec.~\ref{sec:inc}, the IAV model  provides the total inclusive cross section corresponding to the detection of the $b$ fragment in  reactions of the form $A(a,b)X$. This results from the fact that the imaginary part that appears in the expectation value of Eq.~(\ref{eq:iav}) accounts in principle for {\it all} processes in which the participant fragment $x$ interacts nonelastically with the target nucleus. This will include the ICF cross section, but also other NEB processes not associated with the formation of a compound nucleus of the $x+A$ system, such as target excitation.  The isolation of the ICF cross section from the total NEB cross section is indeed not a trivial problem. An intuitive approach consists in identifying the ICF with the absorption due to a short-ranged imaginary potential. 

\begin{figure}
\begin{center}
\includegraphics[width=0.7\columnwidth]{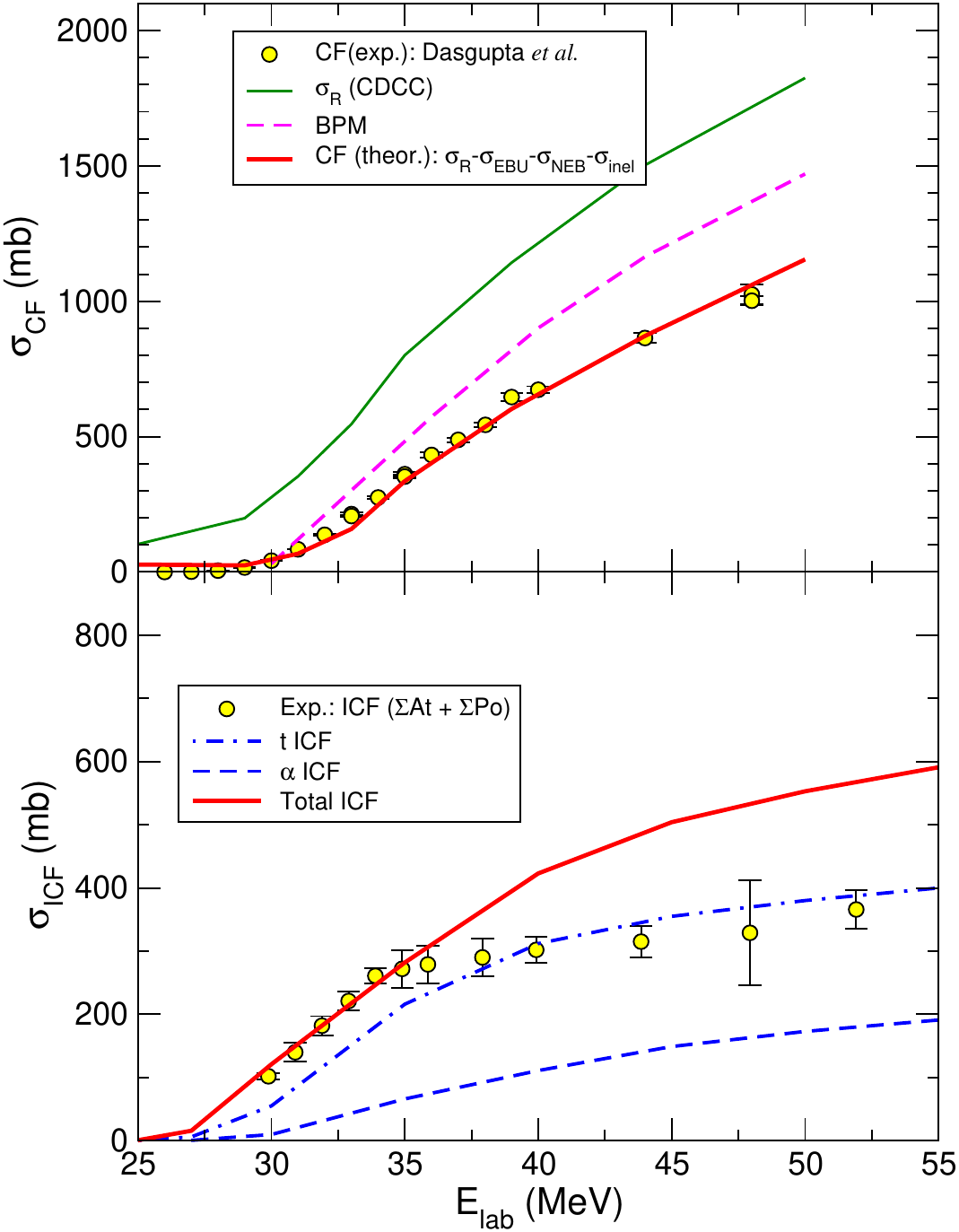}
\end{center}
\caption{\label{fig:li7bi_iav} CF (top) and ICF (bottom) cross sections for $^{7}$Li+$^{209}$Bi. Experimental data from Dasgupta {\it et al.}~\cite{Das02,Das04} are compared with the calculations  based on the IAV model. In the CF plot, we include also the reaction cross section (from a CDCC calculation) and the fusion computed with the barrier penetration model (BPM). Top panel adapted from \cite{Jin19b}. Bottom panel quoted from \cite{Mor22}. 
}
\end{figure}

A tentative application of this idea is shown in the bottom panel of Fig.~\ref{fig:li7bi_iav}, quoted from \cite{Mor22}, corresponding to the reaction  $^{7}$Li+$^{209}$Bi at energies below and above the barrier. The symbols correspond to the ICF data of Dasgupta {\it et al.}~\cite{Das02,Das04}. For the calculations we show the individual contributions to the ICF cross section, namely,  $\alpha$-ICF (i.e., $\alpha$ absorbed) and $t$-ICF ($t$-absorbed) as well as their sum.  To compute the $\alpha$-ICF ($t$-ICF), the imaginary part of the $\alpha$+$^{209}$Bi ($t$+$^{209}$Bi) system was replaced by a short-range imaginary potential of Woods-Saxon form and parameters $W_0=-50$~MeV, $r_i=1.0$~fm, $a=0.2$~fm.  The results are very similar to those reported  in Ref.~\cite{Cor20} and shown in Fig.~\ref{fig:li7bi_canto}, in which the authors made use of the absorption and survival probabilities extracted from the CDCC calculations. Further calculations are needed to elucidate the usefulness and applicability of the IAV model to evaluate ICF cross sections. 

The IAV model has also been used to infer CF cross sections of weakly bound nuclei \cite{Jin19a,Jin19b}. The idea of this method is to decompose the reaction cross section as follows    
 \begin{equation}
 \label{eq:decomp}
 \sigma_R \approx    \sigma_\mathrm{CF} + \sigma_\mathrm{inel} + \sigma_\mathrm{EBU} +
 \sigma^{(b)}_\mathrm{NEB} + \sigma^{(x)}_\mathrm{NEB} .
 \end{equation}
In this expression, $\sigma_\mathrm{inel}$ corresponds to the excitation of the projectile and/or target without dissociation (i.e., inelastic scattering). The terms $\sigma_\mathrm{EBU}$ and $\sigma^{(b,x)}_\mathrm{NEB}$ correspond to the elastic breakup (EBU) and nonelastic breakup (NEB) contributions already discussed in Sec.~\ref{sec:inc}. In the latter, one distinguishes the cases in which either the fragment $x$ or $b$ interacts nonelastically with the target whereas the other scatters elastically.  
A successful determination of the CF cross section from the decomposition (\ref{eq:decomp}) requires that all other quantities involved in this formula can be evaluated accurately. The pure inelastic scattering cross sections ($\sigma_\mathrm{inel}$) are standardly computed by means of coupled-channels calculations including low-lying collective excitations of the projectile and target.  The EBU part  can be accurately calculated with the continuum-discretized coupled-channels (CDCC) method. Finally, the NEB contributions can be evaluated with the IAV model, at least for projectiles with a developed two-boy structure, such as  $^{6,7}$Li.

An application of the method to the $^{7}$Li+$^{209}$Bi reaction is shown in the top panel of Fig.~\ref{fig:li7bi_iav}, adapted from Ref.~\cite{Jin19b}. The circles are the CF data from Ref.~\cite{Das04}, the solid green line is the reaction cross section obtained from a CDCC calculation and the solid red line is the calculated CF cross section inferred from Eq.~(\ref{eq:decomp}) assuming a two-body model ($\alpha+t$) for $^{7}$Li. For comparison, a single-channel barrier penetration model (BPM) calculation (dashed lined) is shown. As can be seen, the data are largely suppressed with respect to the BPM. In contrast, the CF extracted from Eq.~(\ref{eq:decomp}) explains very well the data.  As discussed in~\cite{Jin19a}, the reduction with respect to the BPM is found to be mainly due to the competition due to the $^{209}$Bi($^{7}$Li, $\alpha$)X channel, which includes, among others, the $t$-ICF channel (see the bottom panel).

\subsection{Application to surrogate reactions}
\label{sec:surrogate}
The gradual improvement in models oriented to the computation of ICF cross sections has driven their applicability to the so-called {\it surrogate method} (SRM). This method provides an indirect way of evaluating compound-nucleus cross sections in reactions for which the direct measurement is difficult or even unpossible. An example is the extraction of neutron-induced cross sections of the form $(n,\chi)$, where $\chi$ is a given decay channel product ($\gamma$, fission fragment, etc). Following the Bohr hypothesis, it is customarily assumed that in these reactions the formation and decay of a compound nucleus take place independently of each other. To obtain information on the decay of the compound nucleus ($B^*$) that occurs in the reaction of interest ($n + A \to B^* \to \chi + C$), one uses the alternative (surrogate) reaction $d + D \to b+ B^* \to b + \chi + C $ that involves a projectile-target combination ($d+D$) that is experimentally more accessible. For example, one can use the stripping reaction $d+ A \to p + B^* \to p + \chi + C$, in which $p$ and $\chi$ are measured in coincidence.  

Compound nuclear reactions are adequately described in the Hauser-Feshbach formalism, which considers the conservation of angular momentum $J$ and parity $\pi$.  The cross section for the ``desired'' reaction $A(n,\chi)C$ is given by
\begin{eqnarray}
\sigma_{(n,\chi)}(E_{n}) &=& \sum_{J_T,\pi}  \sigma^{CN}(E_{ex},J_T,\pi) \;\; G_{\chi}^{CN}(E_{ex},J_T,\pi) \; ,
\label{eq:DesReact}
\end {eqnarray}
\noindent
where $ \sigma_{J_T,\pi}^{CN}(E_{ex},J_T,\pi)$ is the cross section for the CN formation and  $G_{\chi}^{CN}(E_{ex},J_T,\pi)$ the branching ratio for the decay to channel $\chi$.
The objective of the surrogate method is to experimentally determine  the decay probabilities $G_{\chi}^{CN}(E_{ex},J_T,\pi)$, which are often difficult to calculate accurately.

In the surrogate reaction, $d + D \to b+ B^* \to b + \chi + C $,  the same CN nucleus $B^*$ is formed and the decay product of interest ($\chi$) is measured in coincidence with the outgoing particle $b$. The probability for this process can be written as:
\begin{equation}
P_{S,\chi}(E_{ex}) = \sum_{J_T,\pi} F_{S}^{CN}(E_{ex},J_T,\pi) \; G_{\chi}^{CN}(E_{ex},J_T,\pi) \; ,
\label{Eq:SurReact}
\end{equation}
where the subscript $S$ denotes the specific surrogate reaction, $F_{S}^{CN}(E_{ex},J_T,\pi)$ is the probability of forming $B^*$ in this surrogate reaction (with specific values of $E_{ex}$, $J$ and $\pi$) and where $G_{\chi}^{CN}(E_{ex},J_T,\pi)$ are the same branching ratios appearing in   equation (\ref{eq:DesReact}). The probability $P_{S, \chi}(E_{ex})$ can be obtained {\it experimentally} as the ratio between  the number of coincidences between the $b$ particle and the decay particle $\chi$, $N_{S,\chi}$, and the total number of surrogate events, $N_{S}$, i.e.:
\begin{equation}
P^{exp}_{S,\chi}(E_{ex}) = \frac{N_{S,\chi}}{N_{S} \, \epsilon_\chi} \; ,
\label{Eq:CoincProb}
\end{equation}
where $\epsilon_\chi$ is the efficiency of detecting the exit-channel $\chi$ for the reactions in which $b$ is detected. 

Ideally, if a reliable prediction of $F_{S}^{CN}(E_{ex},J_T,\pi)$ is possible, with an accurate determination of $P^{exp}_{S,\chi}(E_{ex})$ for a range of energies and angles of $b$, it might be possible to extract the branching ratios $G_{\chi}^{CN}(E_{ex},J_T,\pi)$ which can then be used to calculate the desired cross section using Eq.~(\ref{eq:DesReact}). In practice, this approach is not always feasible due to the lack of some of this required information and the approach has relied on additional approximations. In particular,  early applications  made use of the so-called ``Weisskopf-Ewing  approximation''  , which assumes that the branching ratios $ G_{\chi}^{CN}(E_{ex},J_T,\pi)$ are independent of the angular momentum and spin, giving rise to the simplified cross section:
\begin{eqnarray}
\sigma_{(n,\chi)}(E_{n}) &=& \sigma^{CN}(E_{ex}) \; G_{\chi}^{CN}(E_{ex}) \; ,
\label{eq:WE}
\end {eqnarray}
where $\sigma^{CN}(E_{ex})$ is the CN cross section summed over all possible $J_T,\pi$ values. Applying the same approximation to the surrogate reaction, and using $\sum_{J_T,\pi} F_{S}^{CN}(E_{ex},J_T,\pi)=1$ we have
\begin{equation}
P_{S,\chi}(E_{ex}) = G_{\chi}^{CN}(E_{ex}) \; ,
\label{Eq:SurReact_WE}
\end{equation}
allowing the determination of the desired cross section as
\begin{eqnarray}
\sigma_{(n,\chi)}(E_{n}) &=& \sigma^{CN}(E_{ex}) \; P_{S,\chi}(E_{ex}) \; ,
\label{eq:WE_2}
\end {eqnarray}
which avoids the need of the probabilities $F_{S}^{CN}(E_{ex},J_T,\pi)$.

So, under the validity of theWeisskopf-Ewing approximation, the neutron $(n,\chi)$ cross section can be readily inferred from the measured probabilities for the surrogate reaction. Note also that the CN cross section must be estimated in some way, using, for example, an optical model calculation. 

\begin{figure}[!ht]
\begin{center}
\begin{minipage}{0.5\textwidth}
\includegraphics[width=0.85\columnwidth]{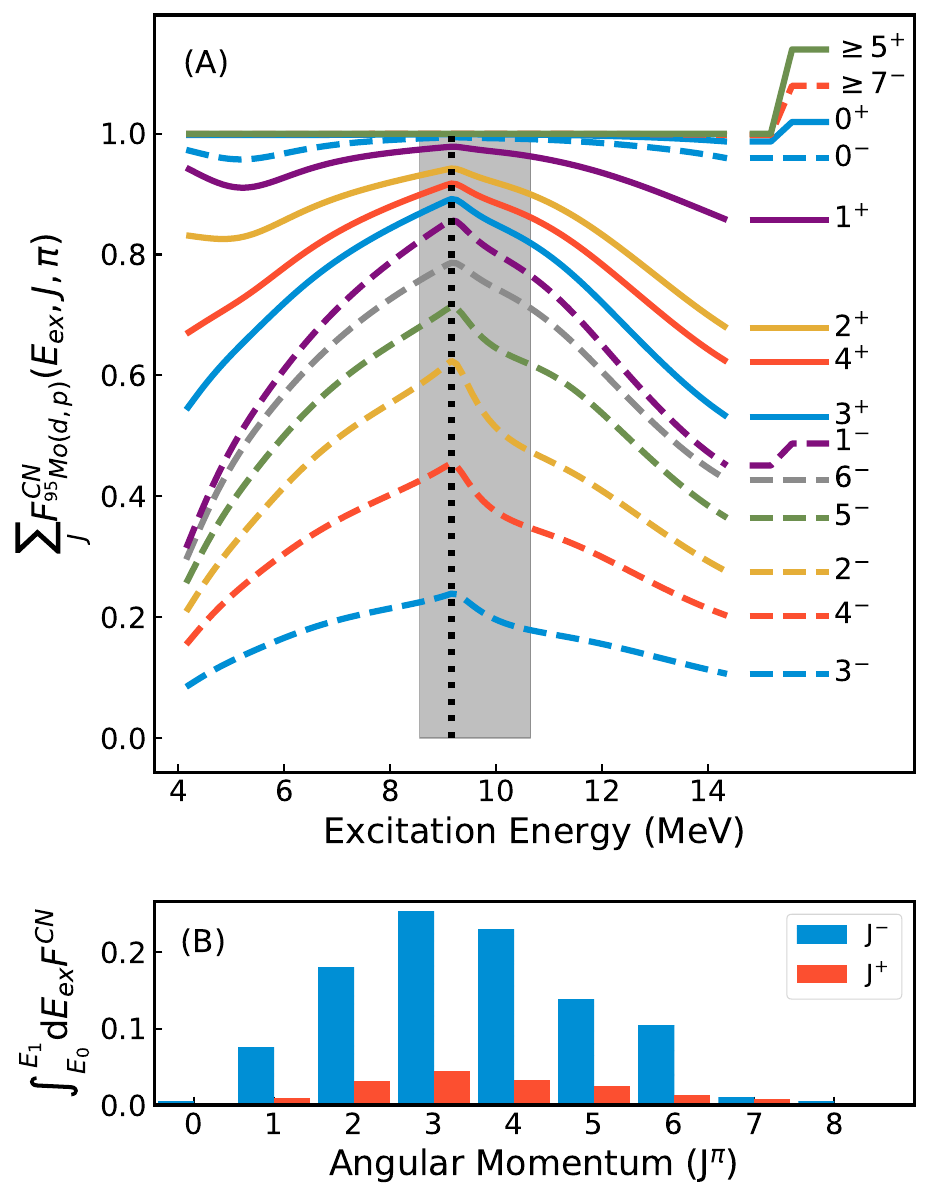}
\end{minipage}
\begin{minipage}{0.5\textwidth}
\includegraphics[width=0.85\columnwidth]{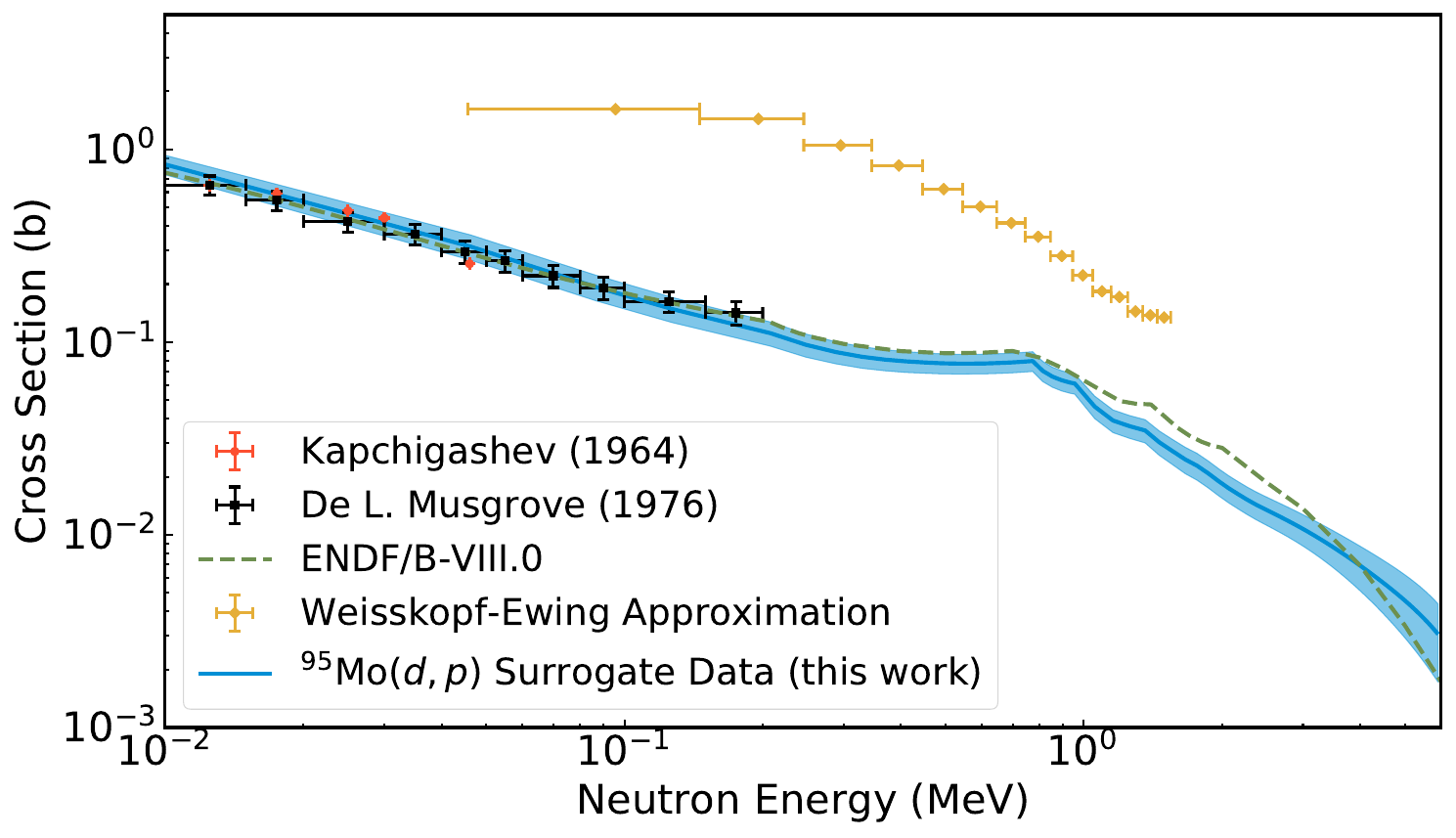}
\end{minipage}
\end{center}
\caption{ \label{fig:surrogate_mo95} Application of the surrogate method  to  the $^{95}$Mo$(n,\gamma)$ reaction. Top (panel (A)):
  Calculations of the cumulative probability of forming the CN through $^{95}$Mo$(d,p)$ ($\sum F^{CN}_{{}^{95}Mo(d,p)}$). The shaded region, which spans excitation energies in the range from $E_{x} = 8.55$ MeV to $E_{x} = 10.65$ MeV, indicates the excitation energies over which the surrogate data are fit.  The vertical dotted line represents the neutron evaporation threshold. Middle (panel (B)): histogram of the total contribution to CN formation over the shaded range in (A) as a function of angular momentum, decomposed into positive and negative parities, and normalized to one over the integration region. 
Bottom: The solid blue region is the $(n,\gamma)$ cross section obtained from the SRM, the red circles and black squares are the direct measurements \cite{Abr08,Mus76}. The uncertainty resulting from experimental data and fitting error is indicated by the shaded band. The result obtained using the Weisskopf-Ewing   approximation is also shown (gold diamonds). Quoted from Ref.~\cite{Rat19}, with permission from APS.}
\end{figure}

In practice,   the Weisskopf-Ewing approximation is rarely justified in most cases so one needs to resort to the more general expressions (\ref{eq:DesReact}) and (\ref{Eq:SurReact}). The probabilities $F_{S}^{CN}(E_{ex},J_T,\pi)$ that appear in the latter can be estimated with the IAV model discussed in Sec.~\ref{sec:iav}. This idea has been successfully applied to the  $^{95}$Mo$(n,\gamma)$ reaction \cite{Rat19}. The direct $(n,\gamma$) cross section for this reaction are compared in Fig.~\ref{fig:surrogate_mo95} with the cross section extracted from the surrogate reaction $^{95}$Mo(d,p) using the aforementioned formalism. As can be seen, they are in excellent agreement. As also shown in this figure, the result using the Weisskopf-Ewing approximation departs significantly from the direct measurement.

\section{Semiclassical description of breakup and transfer reactions}
\label{sec:AW}
When the de Broglie wavelength of the projectile is small compared to some characteristic distance of the collision process  one may describe its motion in terms of classical trajectories. This provides a more intuitive and usually mathematically simpler description of the reaction. This approximation cannot be applied to the internal motion of the nucleons inside the nucleus because their typical wavelength is of the same order as the size of the nucleus and therefore quantum effects are important (for example, for an  energy of 30~MeV, $v\sim c/5$ and so $\lambdabar=\hbar/p \approx$ 1~fm). The methods in which the internal excitations are treated quantum-mechanically, while the projectile--target relative motion is treated classically, are called semiclassical methods. A large variety of such models have been proposed in the literature \cite{Typ94,Esb96,Kid94,Typ01,Cap04,Gar06}.  As examples, we discuss here the one developed  by Alder and Winther and the semiclassical transfer-to-the-continuum model of Bonaccorso and Brink. 

\subsection{The semiclassical formalism of Alder and Winther}

In its simplest form, the theory of Alder and Winther \cite{Ald75} assumes that the projectile moves along a classical trajectory, which is weakly affected by the internal excitations of the colliding nuclei. This means that: 
$$
\frac{\Delta \ell}{\ell} \ll 1
~ ~
\textrm{and}
~~
 \frac{\Delta \varepsilon_n}{E} \ll 1 .
 $$
The projectile--target interaction is split as: $V(\bR,\xi)= V_0(\bR) + V_\mathrm{coup}(\bR,\xi)$, where $V_0(\bR)$ is independent of the internal coordinates and determines the classical trajectory $\vec{R}(t)$.  The time evolution of the total wave function of the system verifies  the 
 Schr\"odinger equation 
\be
i \hbar {d \Psi(\xi,\theta,t) \over dt} = \left[ V(\vecR (\theta,t), \xi) + H_p(\xi) \right] \Psi(\xi,\theta, t) 
\label{eq:TDSE}
\ee
subject to the initial condition $|\Psi (- \infty)\rangle= |0 \rangle$. 

In the spirit of the coupled-channels method, the total wave function is expanded in a basis of internal states of the projectile Hamiltonian $[H_p(\xi) - \varepsilon_n]\phi_n(\xi) $=0   :
\be
\Psi(\xi,\theta,t) = \sum_{n=0} c_n(\theta, t) e^{-i \varepsilon_n t/\hbar} \phi_n(\xi)
\ee
which, when inserted in Eq.~(\ref{eq:TDSE}), gives rise to a set of coupled equations for the expansion coefficients: 
\be
\label{eq:cc_clas}
i \hbar {d c_n(\theta, t) \over dt} = \sum_m e^{-i (\varepsilon_m-\varepsilon_n) t/\hbar} 
V_{nm}(\theta,t) c_m(\theta, t)
\ee
with the initial condition $c_n(\theta,- \infty)= \delta_{n0}$. 
The time-dependent coupling potentials $V_{nm}(\theta,t)$ are given by:
\be
V_{nm}(\theta,t) = \int d \xi \phi_n^*(\xi) V(\bR(\theta, t), \xi)
\phi_m(\xi) .
\ee

Once the coefficients are obtained, the excitation probability for a $0\rightarrow n$ transition is given by:
$$
  P_n(\theta)=|c_n(\theta, \infty)|^2 ,
$$
and the differential cross section by: 
 $$
 \left({d \sigma \over d \Omega}\right)_{0\to n} = 
 \left({d \sigma \over d \Omega}\right)_\mathrm{clas} P_n(\theta) .
 $$
where $(d \sigma / d \Omega)_\mathrm{clas}$ is the classical differential elastic cross section which, for a pure Coulomb case, coincides with the Rutherford cross section.

Due to the conservation of the total probability (flux), one has
$$
\sum_n P_n(t) = \sum_{n} |c_n(t)|^2 =1 .
$$

When the couplings are weak, one may solve Eq.~(\ref{eq:cc_clas}) perturbatively, assuming that $c_0 \approx 1$ and $c_n \ll 1$ for $n>0$. This gives the first-order solution
\be
c_n(\theta) \equiv c_n(\theta, \infty) \simeq 
 {1 \over i \hbar}\int_{-\infty}^{\infty}
 e^{-i (\varepsilon_0-\varepsilon_n) t/\hbar}
 V_{n0}(\theta,t) dt .
\ee

In the important case of pure Coulomb scattering, which was the case studied in detail by Alder and Winther \cite{Ald75}, one finds analytical expressions for the excitation probabilities. In particular, the first-order excitation probability for a $0\rightarrow n $ transition, due to the electric Coulomb operator E$\lambda$ , results
%
%
\be
\left( {d \sigma \over d \Omega}\right)_{0\rightarrow n}= \left({ Z_t  e^2 \over   \hbar v}\right)^2
{B(E \lambda, 0 \to n) \over  e^2 a_0^{2 \lambda-2}} f_\lambda(\theta,\xi)
\ee
which is valid only for angles smaller than the grazing%
\footnote{The grazing angle refers to the angle at which the projectile interacts with the surface of the target in such a way that the projectile barely ``grazes'' the surface of the target, rather than fully penetrating or colliding head-on.}
one ($\theta <\theta_\text{gr}$) and 
where 
$a_0$ is half the distance of closest approach in a classical head-on collision, $\xi_{0 \rightarrow n} =  \frac{(\varepsilon_n-\varepsilon_0)}{\hbar} \frac{a_0}{v}$ is the adiabaticity parameter and $f_\lambda(\theta,\xi)$ is an analytic function, depending on the kinematical conditions, but independent of the structure of the projectile. 

For weakly bound nuclei, the excitation will typically populate unbound (continuum) states. The previous formula can be generalized to:
\be
\label{eq:dsdwde}
\frac{d \sigma(E\lambda)}{d \Omega d \varepsilon} = \left({ Z_t  e^2 \over   \hbar v}\right)^2
{1 \over  e^2 a_0^{2 \lambda-2}}
\frac{dB(E \lambda)}{d \varepsilon} 
{df_\lambda(\theta,\xi) \over d\Omega}
\quad \quad \ 
(\theta < \theta_\text{gr})
\ee
where  $df_\lambda(\theta,\xi) / d\Omega$ is also a well-defined analytic function and   $dB(E \lambda)/d\varepsilon$ is the  electric reduced probability to the continuum states.

It is common to express (\ref{eq:dsdwde}) in terms of the so-called number of equivalent photons,
\be
\label{eq:dsdwde_ne}
\frac{d \sigma(E\lambda)}{d \Omega d \varepsilon} =\frac{16 \pi^3}{9 \hbar c}
\frac{dN_{E\lambda}(E_x\theta)}{d\Omega}
\frac{dB(E \lambda)}{d \varepsilon} 
\ee
where ${dN_{E\lambda}(E_x\theta)}/{d\Omega}$ is the number of equivalent photons per solid angle. This is typically referred to as Equivalent Photon Method (EPM).

\begin{figure}
\begin{center}\includegraphics[width=0.75\columnwidth]{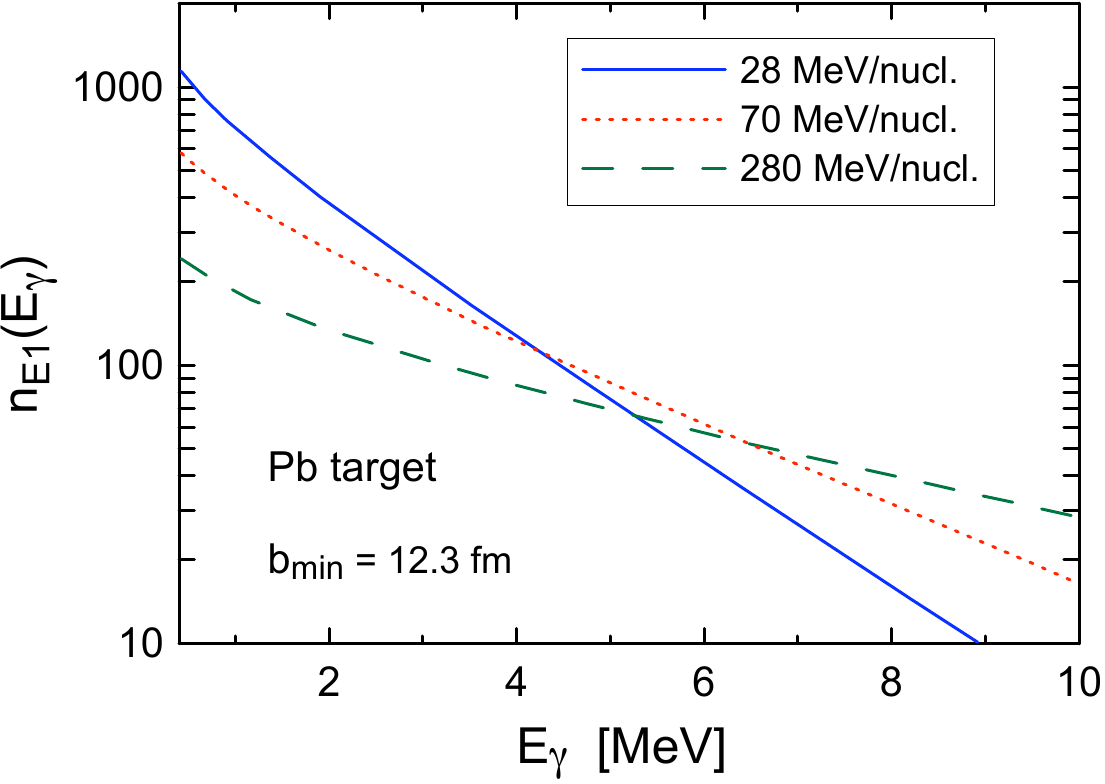} \end{center}
\caption{\label{fig:virtual_photons} Total number of virtual photons for the E1 multipolarity, corresponding to the scattering of a projectile by a $^{208}$Pb target at different incident energies and integrated for impact parameters $b > 12.3$~fm. Quoted from Ref.~\cite{Ber09}.}  
\end{figure}

Figure \ref{fig:virtual_photons},  quoted from \cite{Ber09}, shows the number of equivalent virtual photons, integrated for impact parameters beyond $b=12.3$~fm (note that $b=a_0\cot{(\theta/2)}$), for the collision of a projectile with a $Z=82$ target, for three different incident energies. It can be seen that increasing the incident energy decreases the population of low-lying states and enhances the population of high-energy ones.

\begin{figure}
\begin{center}
\includegraphics[width=0.75\columnwidth]{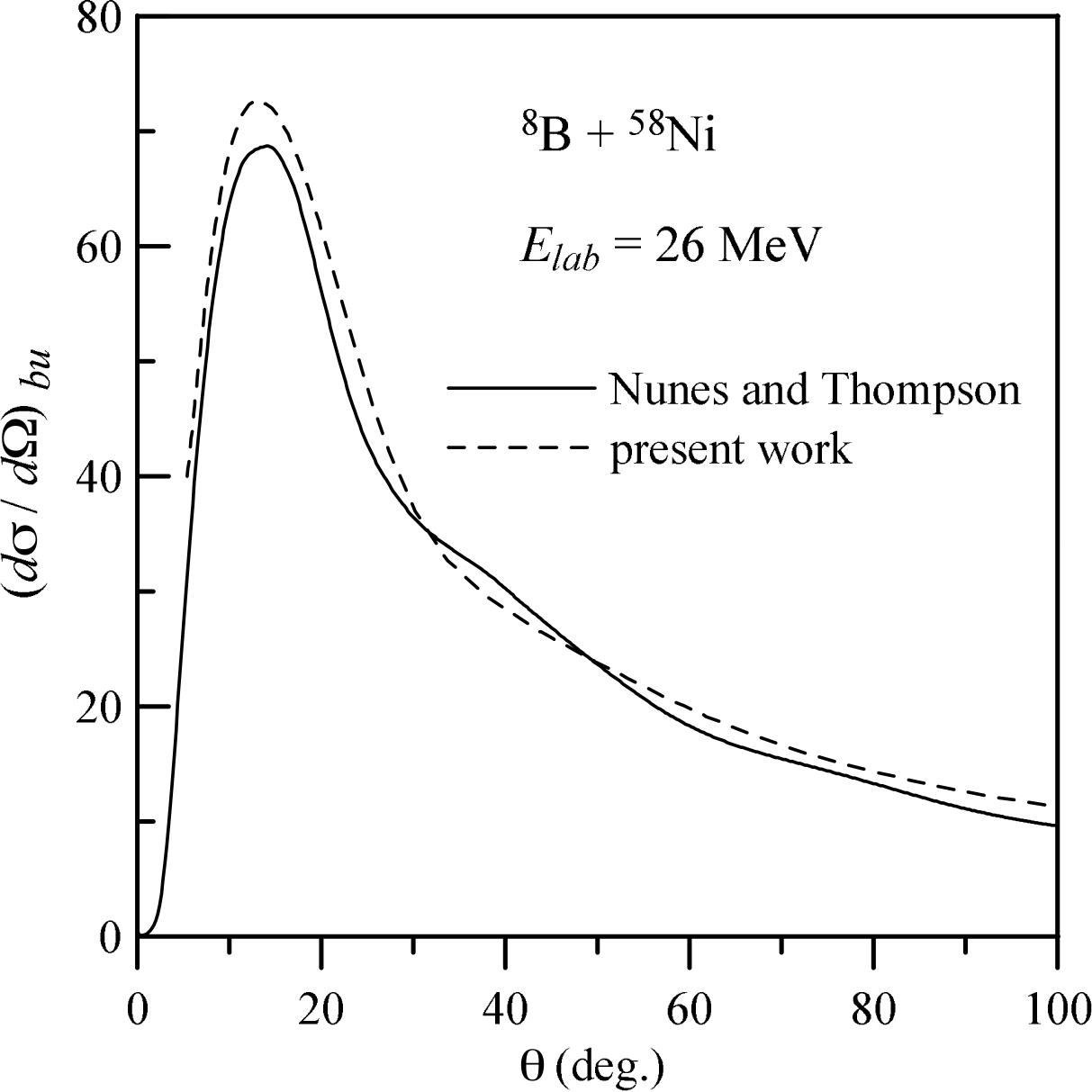} \end{center}
\caption{\label{fig:b8ni_dsdw}Breakup of $^{8}$B$\rightarrow$ $^{7}$Be+p on a $^{58}$Ni target at 26~MeV. Coupled-channel semiclassical calculations (labeled ``present work''), using Coulomb+nuclear trajectories, are compared with CDCC calculations (labeled ``Nunes and Thompson''), which include also nuclear and Coulomb couplings. Quoted from \cite{Mar02} with permission from APS.}  
\end{figure}

The assumption of pure Coulomb trajectories can be relaxed, at the expense of losing some of the simplicity of the method. A compelling application is shown in Fig.~\ref{fig:b8ni_dsdw} (taken from Ref.~\cite{Mar02}) where semiclassical coupled-channels calculations, using trajectories modified by the nuclear interaction, are compared with CDCC calculations for the breakup angular distribution of the $^{8}$B+$^{58}$Ni reaction at the near-barrier energy of 26 MeV, finding a nice agreement between both.

\subsection{Dynamic Coulomb polarization potential  from the AW theory}
\label{sec:DPP}
As discussed in Sec.~\ref{sec:om}, the Feshbach theory of nuclear reactions provides a formal expression for the nucleus-nucleus  optical potential. This can be expressed as a sum of a bare potential (that is, the static part of the interaction due to the ground-state densities for the projectile and target) and a polarization potential. Under suitable approximations, it is possible to derive simple,  analytical expressions for specific parts of the polarization potential. One example is the adiabatic polarization potential describing the effect of the dipole Coulomb interaction discussed in Sec.~\ref{sec:vad}.



This  adiabatic expression assumes that the excitation energies are large and, therefore,  is not applicable to the Coulomb breakup of very weakly bound nuclei, for which the average excitation energies are typically small (of the order of 1~MeV or less). 

\begin{figure}
\begin{center}\includegraphics[%
  width=0.8\columnwidth]{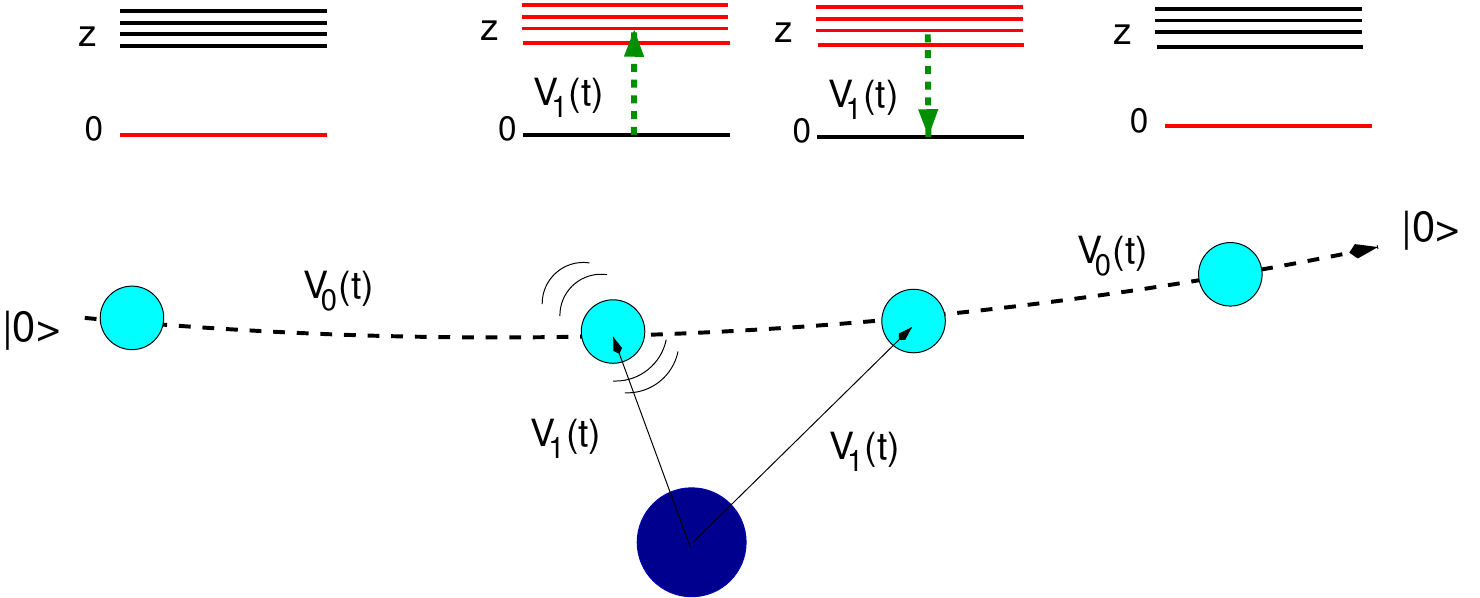}\end{center}
\caption{\label{fig:a1_traj} Second-order process describing the excitation and de-excitation of a nucleus moving along a classical trajectory determined by potential $V_0(t)$, due to the action of potential $V_1(t)$.}
\end{figure}


A non-adiabatic (also called {\it dynamic}) polarization potential can be derived from the Alder and Winther theory \cite{And94,And95}.
These works consider the second-order process in which the projectile is excited to a continuum state due to the dipole Coulomb interaction ($V_1(t)$) and then de-excites, returning to the ground state. The situation is schematically depicted in Fig.~\ref{fig:a1_traj}.
The amplitude probability for this process is given by:
\begin{align}
c^{(2)}_n = & \sum_{z} \left({-i \over \hbar}\right)^2 
  \int_{-\infty}^{+\infty} dt  \langle  n | V_1(t) | z \rangle
  \exp \left\{ {i \over \hbar}(\varepsilon_n - \varepsilon_z) t \right\} 
\nonumber \\  & \times 
  \int_{-\infty}^{t} dt'  \langle  z | V_1 (t') | 0 \rangle
  \exp \left\{ {i \over \hbar}(\varepsilon_z - \varepsilon_0) t' \right\} ,
\end{align}
where $|z\rangle$ denotes the intermediate states populated during the reaction. 
Then, one requires that this second-order amplitude coincides with the first-order amplitude associated with the polarization potential  for all classical trajectories corresponding to a given scattering energy. 
This gives rise to the following expression:
\begin{align}
\label{eq:Upol}
U_\mathrm{pol}(R)&=-\frac{4\pi }{9}\frac{Z_t^{2}e^{2}}{\hbar v}\frac{1}{(R-a_{o})^{2}R} 
\nonumber \\ 
&\times 
\int ^{\infty }_{0}d\varepsilon \frac{dB(E1,\varepsilon )}{d\varepsilon }
\left[ g\left(\frac{R}{a_{o}}-1,\xi \right)+if\left(\frac{R}{a_{o}}-1,\xi \right) \right] , 
\end{align}
where \emph{g} and \emph{f} are analytic functions defined as
\begin{eqnarray}
\label{eq:ucdp}
f(z,\xi ) &=& 4\xi ^{2}z^{2}\exp{(-\pi \xi )}K_{2i\xi }''\left(2\xi z\right), \\
g(z,\xi ) &=& \frac{P}{\pi }\int _{-\infty }^{\infty }\frac{f(z,\xi ')}{\xi -\xi '}d\xi ',
\end{eqnarray}
where $K_{2i\xi }(2\xi z)$ is a modified Bessel functions
of imaginary order and $\xi$ is the Coulomb adiabaticity 
parameter corresponding to the excitation to the continuum energy $\varepsilon$ of the nucleus. An important feature of this potential is that when the breakup energy \( \varepsilon _{b} \) is large enough, the purely real adiabatic dipole potential given by Eq.~(\ref{eq:vad}) is recovered. In the opposite limit, for small breakup energies, \( f\left(\frac{R}{a_{o}}-1,\xi \right)\rightarrow 1 \) and \( g\left(\frac{R}{a_{o}}-1,\xi \right)\rightarrow 0 \), and the polarization potential becomes purely imaginary, depending on $R$ as \( {1}/[(R-a_{o})^{2}R]. \)

\begin{figure}
\begin{center}\includegraphics[width=0.8\columnwidth]{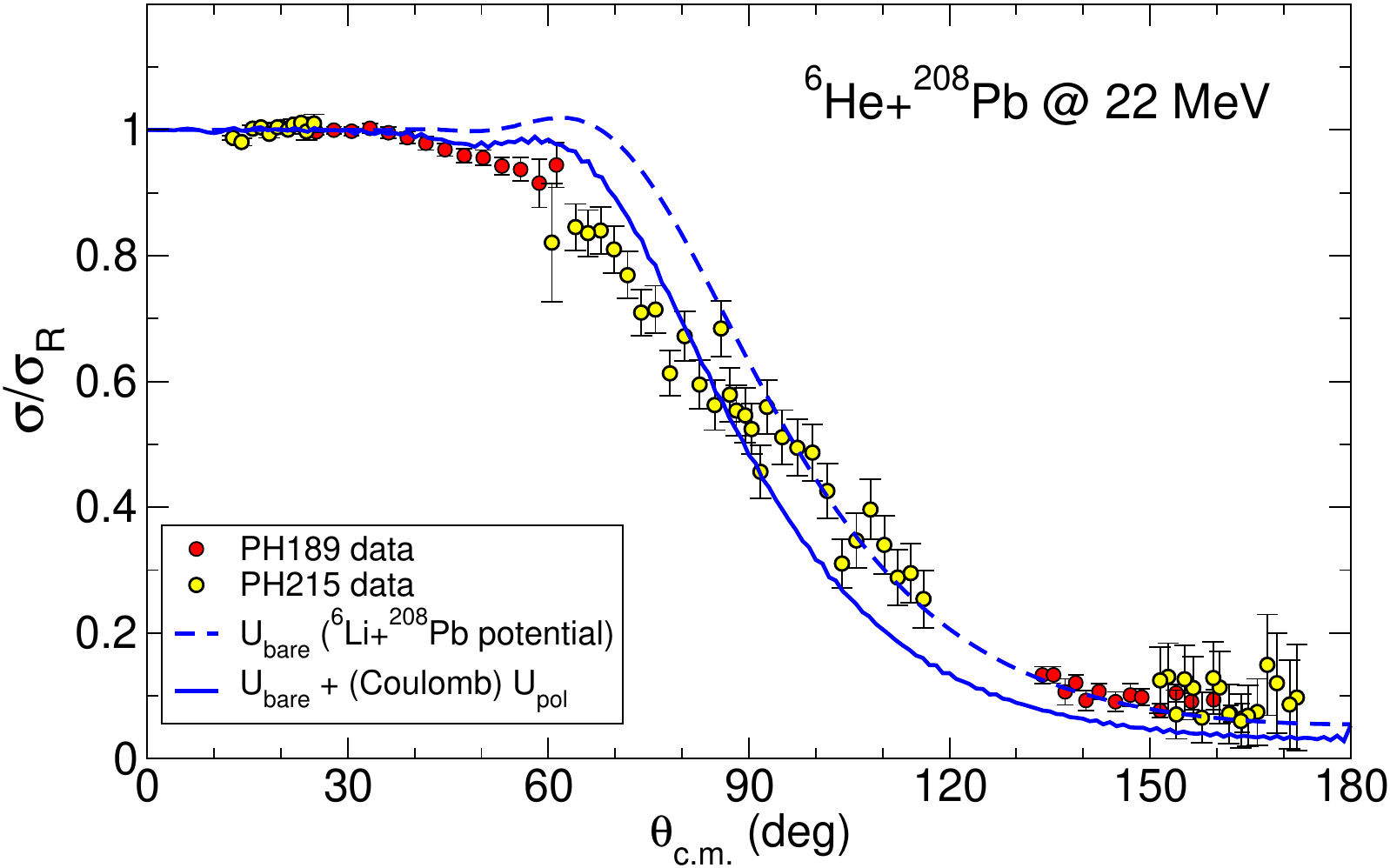}\end{center}
\caption{\label{fig:he6pb_e22_om} Elastic scattering differential cross section for  $^{6}$He+$^{208}$Pb at $E_\mathrm{lab}=22$~MeV. The dashed line is a single-channel calculation performed with a {\it bare} potential, given by a $^{6}$Li+$^{208}$Pb optical potential. The solid line is the optical model calculation obtained with the bare plus dynamic Coulomb polarization potential of Eq.~(\ref{eq:Upol}). }
\end{figure}

\begin{figure}
\begin{center}\includegraphics[%
  width=0.8\columnwidth]{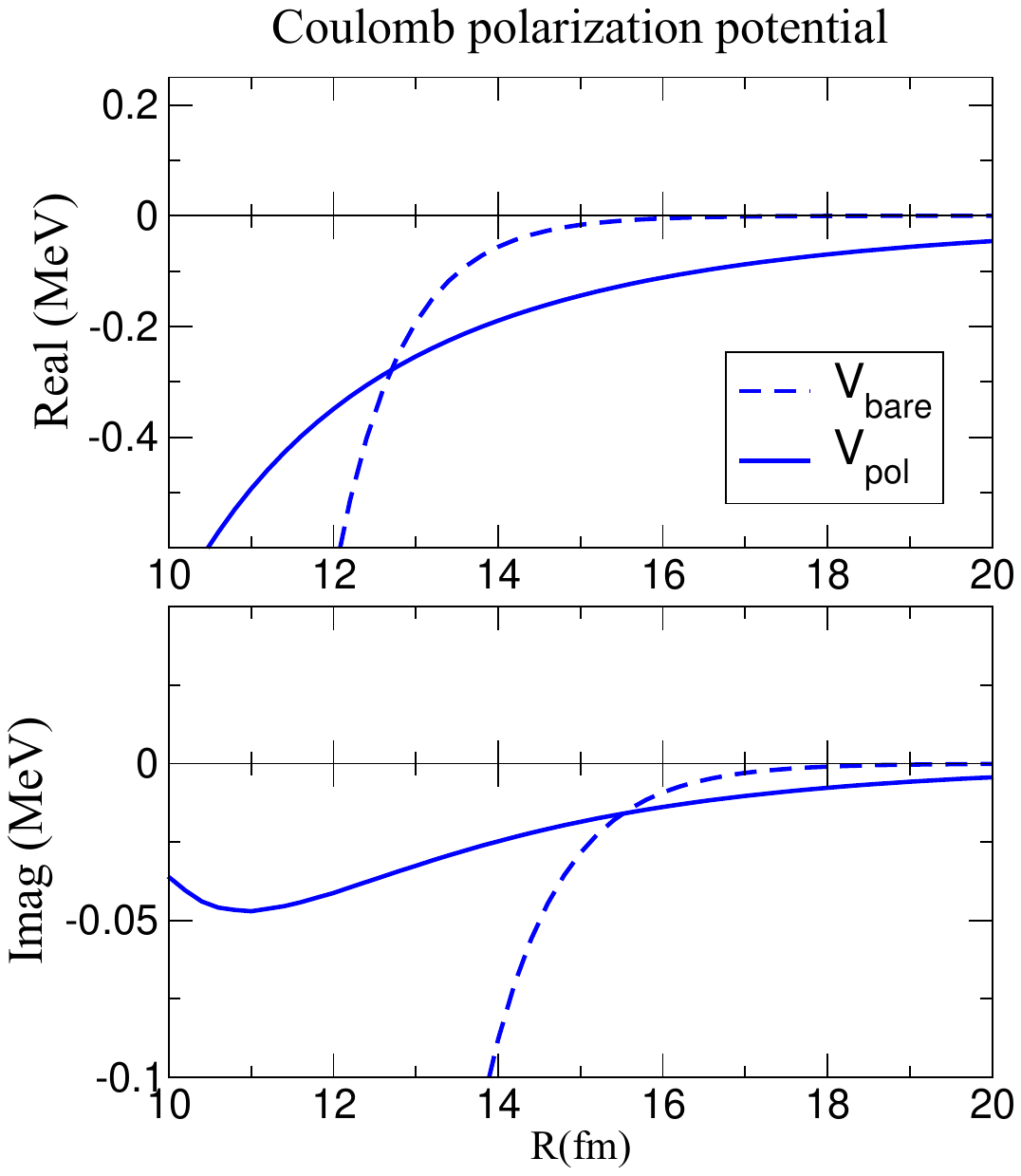}\end{center}
\caption{\label{fig:he6pb_e22_om_cdp} Real and imaginary parts of the bare and CDP potentials for  $^{6}$He+$^{208}$Pb at $E_\mathrm{lab}=22$~MeV computed with the dynamic Coulomb polarization potential of Eq.~(\ref{eq:Upol}).}
\end{figure}

As an example, we show in Fig.~\ref{fig:he6pb_e22_om} the elastic scattering data for $^{6}$He+$^{208}$Pb at 22~MeV compared with the optical model calculations obtained with a bare potential alone (dashed) and with the bare plus the Coulomb dipole polarization (CDP) potential (solid line). For the bare interaction, we have used an optical potential for the nearby projectile $^{6}$Li, which has a structure similar to that of $^{6}$He but which does not exhibit the strong dipole excitation mechanism. The elastic cross section predicted by the bare potential presents a Fresnel-like diffraction pattern, not observed in the data. The inclusion of the CDP significantly reduces the elastic cross section and suppresses the Fresnel peak. This plot shows that the Coulomb polarization effect (encoded here in the form of a polarization potential), is able to account for part of the effects of the breakup channels on the elastic channel. The remaining disagreement with the data can be explained as due to other couplings not accounted for by this Coulomb polarization potential, namely, nuclear couplings and possible interferences between nuclear and Coulomb. These effects are naturally accommodated within the CDCC formalism discussed in \ref{sec:cdcc}. 

In Fig.~\ref{fig:he6pb_e22_om_cdp}, the real and imaginary parts of the bare and CDP potentials are shown. Their most remarkable feature is their long range, which was anticipated from the optical model analysis of Sec.~\ref{sec:om},  and is associated with the long range of the dipole Coulomb couplings, as we have already discussed in the context of the CDCC method.

\subsection{Semiclassical transfer-to-the-continuum model}

The Semiclassical Transfer-to-the-Continuum (STC) model \cite{Bon88} developed by Bonaccorso and Brink  is a generalization of the model of  Brink and collaborators \cite{Has78,Has79,Mon85,Bon87} to the case of final unbound states. These methods provide a more transparent and often analytical interpretation of the quantum-mechanical counterparts.    
The method is valid for peripheral reactions based on the core spectator model as in the IAV model discussed in previous sections. To obtain an analytic expression for the transfer amplitude the STC replaces the internal state wavefunctions by their asymptotic forms, given analytically in terms of Hankel functions.  


Considering the case of neutron transfer, the starting point of the STC is the semiclassical transfer to the continuum scattering amplitude, given by \cite{Bon88}
\begin{equation}
A_{fi}=\frac{1}{ i\hbar}
\int_{-\infty}^{\infty}dt\langle\phi_{f} (\boldsymbol{r}_n)|U_{nT}(\boldsymbol{r}_n)|\phi_{i}(\boldsymbol{r}_n-\boldsymbol{R}(t))\rangle e^{-i\bar\omega}\label{1}
\end{equation}
with $\bar\omega=(\omega t-mvz/\hbar)$. 
The time dependent nucleon initial and final wave functions in their respective reference frames are  \begin{equation}\psi_{i,f}(\boldsymbol{r}_n,t)=\phi_{i,f} (\boldsymbol{r}_n)e^{-i\varepsilon_{i,f} t/ \hbar}\label{wf1}\end{equation}
The initial and final radial wave functions are approximated by Hankel functions according to \cite{Bon88}. The initial state is therefore approximated as:
\begin{eqnarray}
\phi_{l_i}(\boldsymbol{r}_{Cn}) \simeq -C_i i^l\gamma_i h^{(1)}_{l_i}(i\gamma_i r_{Cn})Y_{l_i,m_i}(\Omega_i).  
\label{wf}\end{eqnarray}
with $\gamma_i^2=-2 m \varepsilon_i/\hbar^2$,   $C_i$ the asymptotic normalization coefficient of the initial bound state wavefunction. The final (continuum) state is approximated by: 
\begin{eqnarray}
\phi_{l_f}(\boldsymbol{r}_n) \simeq C_f k_f\frac{i}{2}(h^{(+)}_{l_f}(k_fr_n)- S_{l_f} h^{(-)}_{l_f}(k_fr_n))Y_{l_f,m_f}(\Omega_f),
\end{eqnarray}
where $\ S_{l_f} (\varepsilon_f)$ is the  neutron-target elastic scattering S-matrix at energy $\varepsilon_f$ and $C_f$ a normalization constant.

With these approximate initial and final wavefunctions, the one-neutron removal probability results:
\begin{align} 
\frac{dP_{-n} }{d\varepsilon_f} & \approx \frac{1}{2}\sum_{j_f}(2j_f+1)|1- S_{j_f} |^2
(1+R_{if})
\left [\frac{\hbar}{mv}\right ]^2 
\nonumber \\
& \times \frac{m}{\hbar^2 k_f}
|C_i|^2 
\frac{e^{-2\eta b_c}}{2\eta b_c}
M_{l_fl_i}, \label{dpde}
\end{align} 
The factor $M_{l_fl_i}$ is due to the overlap of the angular parts, and $R_{if}$ are spin-coupling coefficients. Further definitions and discussion can be found in Refs.~\cite{Bon88}.

To calculate the elastic $S$-matrix a choice must be made for the neutron-target potential. In general, such potential will be complex and energy dependent. At low neutron-target relative energies, narrow resonances are expected that will progressively become broad and overlapping at higher energies. In the spirit of the optical model, an energy averaging procedure is assumed, resulting:
\begin{equation}
\frac{dP_{-n} }{d\varepsilon_f}(j_f,j_i)=\left ( |1 - \bar S_{j_f} |^2+ (1- |\bar S_{j_f} |)^2 \right ) B_{if}\label{dpde1}
\end{equation}
The first and second terms correspond to the elastic and nonelastic breakup contributions discussed in the context of the IAV model.  
$\bar S_{j_f}$ are nucleon-target S-matrices calculated for each nucleon final energy according to the optical model with an energy-dependent optical potential. $B_{if}$ 
 are kinematical factors depending
on the energies  ($\varepsilon_i$ and $\varepsilon_f$), momenta ($\gamma_i$
and $k_f$),  and orbital angular momenta ($l_i$ and $l_f$) of the initial and final
states of a single neutron particle, respectively, and on the incident energy per
particle, $mv^2/2$ at the distance of closest approach $d$, which coincides with the projectile impact parameter in a straight line trajectory. Their explicit form can be found in the original works \cite{Bon88}.  

To obtain the relative energy distribution of the final states, one integrates the probabilities over the core impact parameters 
\begin{align}
\frac{d\sigma}{d\epsilon_f} =\int d {\vec b}_c \,  |S_c(b_c)|^2 \frac{dP_{-n} (b_c) }{d\varepsilon_f}
\end{align}
where $S_c(b_c)$ is the core-target S-matrix and $|S_c(b_c)|^2$ is therefore a core-survival probability.

For sufficiently high incident energies, the $S$-matrix can be evaluated in the eikonal approximation and one can also use the connection between angular momentum and impact factor to transform the sum in Eq.~(\ref{dpde}) to an integral over the impact parameter, i.e,  $\Sigma_{j_f}\to \int d^2\boldsymbol{b}_n$.

\begin{figure}[h!t]
\center
 {\centering \resizebox*{0.85\columnwidth}{!}{\includegraphics{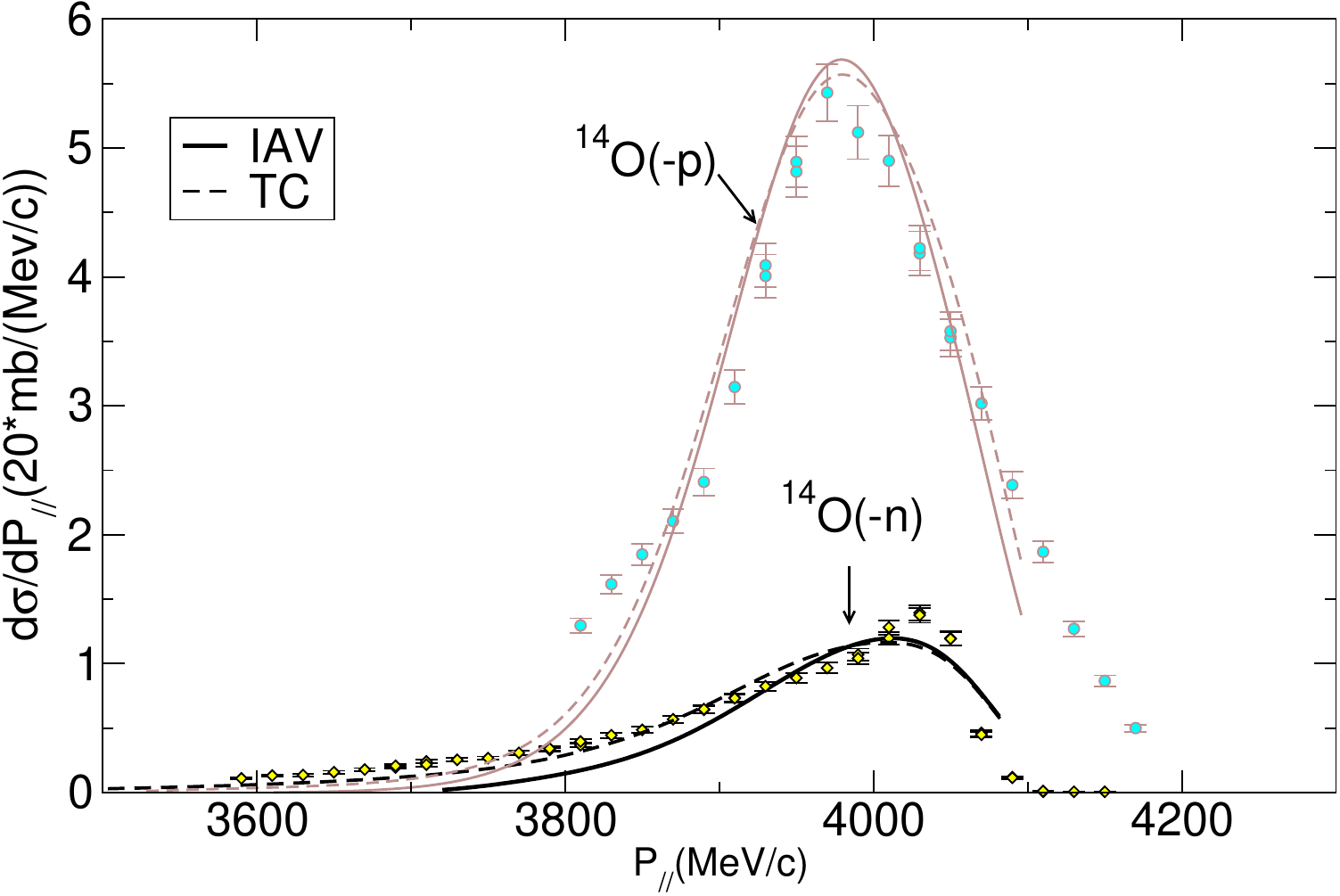}} \par}
\caption{Experimental and calculated cross section momentum distribution for the proton and neutron knockout reactions  of $^{14}$O on $^{9}$Be at 55~MeV/a.  The solid lines are the results obtained by the IAV model and the dashed lines are calculated by the semiclassical transfer to the continuum (TC) method.  Adapted from  Ref.~\cite{Jin21}. Experimental data are from Ref.~\cite{Fla12}.}
\label{fig:o14_tc}
\end{figure}

An example of the application of the method is given in Fig.~\ref{fig:o14_tc}. It corresponds to the knockout reactions $^{9}$Be($^{14}$O,$^{13}$O)X $^{9}$Be($^{14}$O,$^{13}$N)X at 55~MeV/u.  The semiclassical transfer to the continuum  calculations are compared to those obtained with the IAV model discussed in Sec.~\ref{sec:iav} and with the experimental data from \cite{Fla12}. A nice agreement between the purely quantum mechanical and semiclassical methods is observed.

\subsection{Dynamical Eikonal approximation}


The dynamical eikonal approximation (DEA) \cite{Bay05,Gol06} unifies the semiclassical time-dependent and eikonal methods. It
allows for the calculation of differential cross sections for elastic scattering and breakup in a quantal way by taking
into account interference effects. 

As in the CDCC method, the idea of the method is to provide an approximate solution of the three-body Schrodinger equation with the effective Hamiltonian in Eq.~(\ref{eq:Heff}). The three-body scattering solution for this equation is expressed in the factorized form:
\begin{align}
\label{eq:Psi_DEA}
\Psi^{\rm DEA}(\vec{R},\vec{r}) &=
e^{i K_0 Z}\widehat\Psi(\vec{R},\vec{r}).
\end{align}
where $K_0$ stands for the incident momentum and $Z$ is the third component of the $\vec{R}$ vector. 
At sufficiently high energies, the deviation from the initial plane wave $e^{i K_0 Z}$ of the projetile-target relative motion 
is expected to be small and hence the dependence on $\vec{R}$ of $\widehat\Psi$ is expected to be smooth. This enables to neglect its second-order
derivative in $\bR$ with respect to its first-order derivative
\begin{align}
\nabla^2_{\bR}\widehat\Psi(\vec{R},\vec{r})\ll K_0 \frac{\partial \widehat\Psi(\vec{R},\vec{r})}{\partial Z}.
\label{eq:DEA}
\end{align}
Therefore, introducing the factorization (\ref{eq:Psi_DEA})  into the
three-body Schrodinger equation leads to the so-called DEA equation 
\begin{align}
\lefteqn{i\frac{\hbar^2K_0}{\mu_{aA}}\frac{\partial}{\partial Z}
\widehat\Psi(Z,\vec{b},\vec{r})=}\nonumber \\
& & \left[(H_0-\varepsilon_0)+U_{bA}(\br_{bA})+U_{xA}(\br_{xA})\right]
\widehat{\Psi}(Z,\vec{b},\br),
\end{align}
where the three-body wavefunction has been conveniently expressed in terms of the  $Z$ and transverse $\vec{b}$ parts of the projectile-target coordinate $\vec{R}$.

The DEA equation (\ref{eq:DEA}) is mathematically equivalent to a time-dependent
Schrodinger equation for a straight-line trajectory.
It can therefore be solved using similar numerical techniques as in the time-dependent model (see e.g. \cite{Kid94,Cap03}). 
The DEA equation must be solved  for each transverse component $\vec{b}$
of the $a$-$A$ coordinate with the initial boundary condition
\begin{align}
\widehat{\Psi}(Z\rightarrow-\infty,\vec{b},\vec{r})=\phi_0(\vec{r}).
\end{align}
Breakup amplitudes can then be extracted from the wave function as \cite{Bay05}:
\begin{align}
T_{fi} &= i 2 \pi \hbar \nu \int_{0}^{\infty} b db J_0(qb) S(\vec{k}, \vec{b})
\label{eq:Tbu_DEA}
\end{align}
where $\vec{q}= \vec{K} - K \hat{Z}$ is the transferred momentum and 
\begin{align}
S(\vec{k}, \vec{b}) = \langle \phi_{\vec{k}}^{(-)}(\vec{r}) | \hat{\Psi}(\vec{R}, \vec{r}) \rangle_{Z = + \infty}
\end{align}
where $\phi_{\vec{k}}^{(-)}(\vec{r}) $ is a $b+x$ scattering state with incoming boundary conditions evaluated at the desired final energy. 

Despite its formal analogy, it is important to note that there is no semiclassical approximation involved in the derivation of the method and so  the coordinates $Z$ and $\vec{b}$ are quantal variables. This enables to take into account interferences between
trajectories. 

\begin{figure}
\begin{center}
\includegraphics[width=0.7\columnwidth]{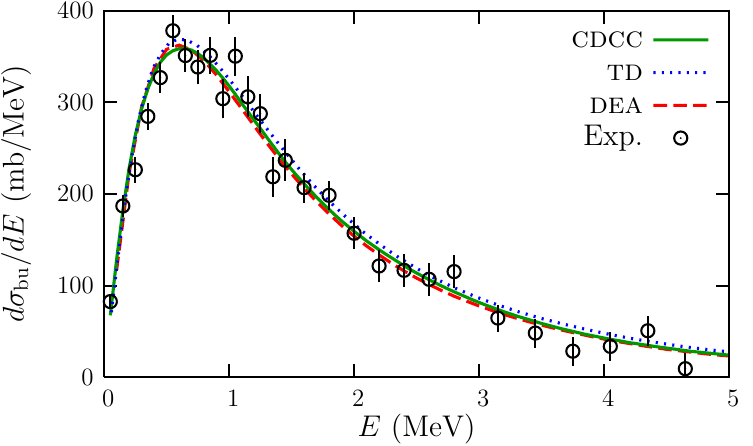} 
\vspace{5pt}
\includegraphics[width=0.7\columnwidth]{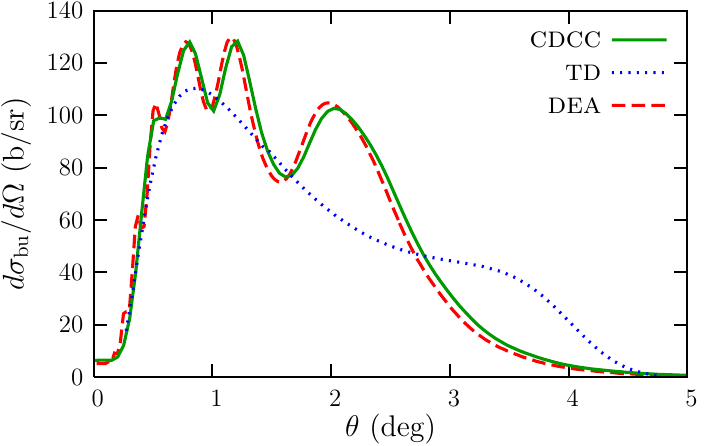}
\caption{\label{fig:DEA} Top: Differential breakup cross  sections, with respect to the $n$-$^{14}$C relative energy,  for the breakup of $^{15}$C on Pb at 68 MeV/nucleon. The CDCC, DEA and a semiclassical time-dependent method (TD) are compared. Bottom: Breakup angular distributions from the same calculations. Quoted from \cite{Cap13}.
}
\end{center}
\end{figure}

The DEA method has been tested against the CDCC and time-dependent semiclassical methods for the breakup of halo nuclei. These studies indicate that the DEA predictions agree very well with those of the  CDCC method at energies of around 100~MeV/u and even below. Moreover, it is able to reproduce the angular pattern of breakup, that is not given by the semiclassical approximations. An example of this benchmark comparison is given in Fig.~\ref{fig:DEA}.  

\section{Extraction the electric transition probabilities from Coulomb dissociation data}
\begin{figure}
\begin{center}
\includegraphics[width=0.75\columnwidth]{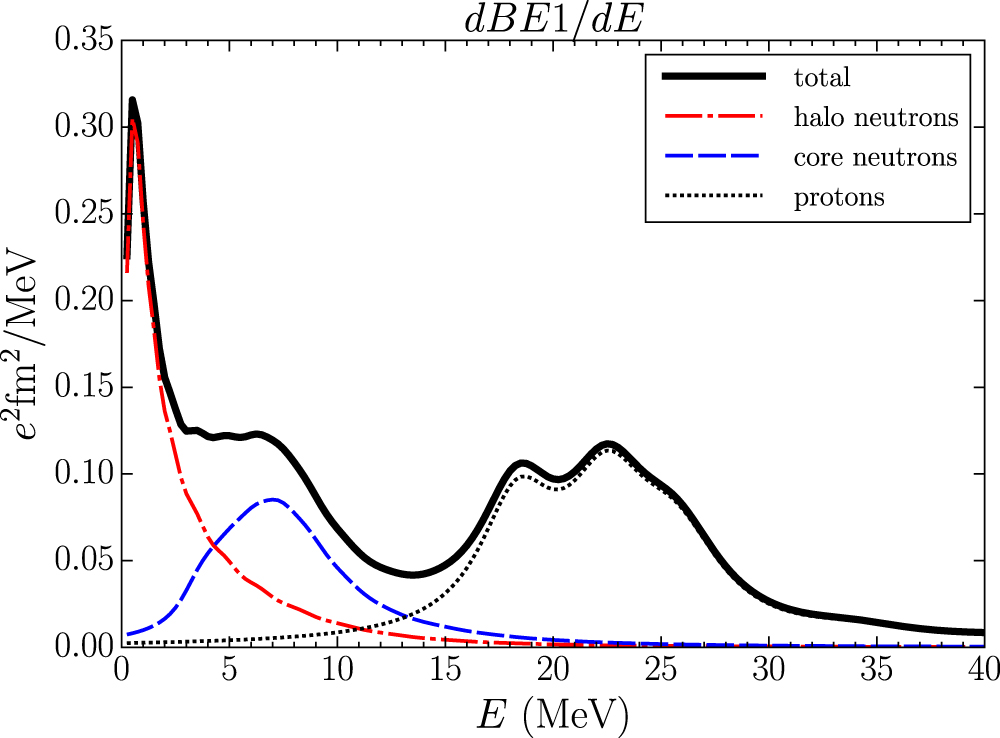} \end{center}
\caption{\label{fig:softdipole} B(E1) distribution for nucleus $^{11}$Li. At low energies, the peak associated to the weakly-bound neutrons forms the so-called ``pygmy'' or ``soft dipole'' mode. Figure taken from Ref.~\cite{Bro19}.}  
\end{figure}

Coulomb excitation has been used for many years as a powerful spectroscopic tool to extract the electromagnetic response of nuclei to a Coulomb field, provided by a heavy nucleus. For stable nuclei and low bombarding energies (tens of MeV per nucleon), Coulomb excitation is dominated by transitions to low-lying excited states connected by quadrupole and octupole multipolarities to the ground state, whereas at sufficiently high energies, Coulomb excitation is dominated by the so-called giant dipole resonance, characterized by the collective motion of protons and neutrons moving in an out-of-phase motion. In the case of weakly bound nuclei and, more specifically, for neutron halo nuclei, the differential force of the Coulomb field acting on the valence neutron(s) with respect to the core produces a strong dipole force that may eventually break the system. This mode has been customarily termed as ``soft dipole E1'' excitation, as opposed to the giant dipole excitation, see Fig.~\ref{fig:softdipole}.  

\begin{figure}
\begin{center}
\includegraphics[width=0.75\columnwidth]{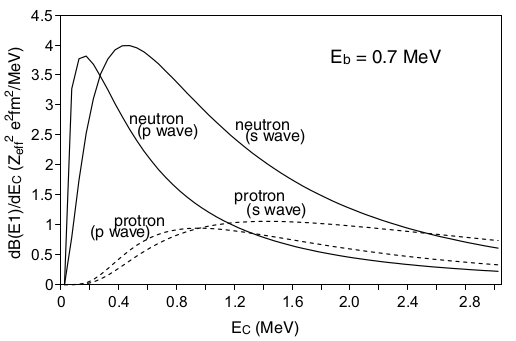} \end{center}
\caption{\label{fig:be1_nag} B(E1) strength distribution for different situations, all
associated with single-particle excitation from a bound state to
the continuum. In all cases the initial particle is bound with
the same energy ($E_b$ = 0.7 MeV). The labels ``neutron''
and  ``proton'' denote, respectively, the neutron and proton
excitations. For both cases, the plot shows the situations
of initial $s$-state (with dipole transition to $p$-state in the continuum) and of initial $p$-state (with final $s$-state in the continuum). Bound and continuum single-particle states were
generated from a Woods-Saxon potential, with parameters
adjusted to reproduce the correct binding energy of
the initial bound state. Taken from \cite{Nag05}.}  
\end{figure}

The concentration of dipole strength at low-excitation energies is a general property of weakly-bound nuclei, particularly those with a neutron-halo structure. This can be verified considering a simple situation of a two-body system described with a single-particle model in which the $B(E1)$ can be readily evaluated numerically. Considering such a model, 
Fig.~\ref{fig:be1_nag} displays the $B(E1)$ transition for a hypothetical (and unrealistic) system
with mass $A = 8$ and a nucleon binding energy $E_b$ = 0.7~MeV for different assumptions of the initial state orbital angular momentum. In the valence-proton case, the strength distribution is shifted to higher energies, together with a damping of the absolute value. The concentration of strength becomes very large near the breakup threshold for an initial state with $\ell=0$.  It is worth noting that the presence of this near-threshold peak does not necessarily imply the existence of a resonant state at this energy.

In the limit of vanishing binding energy, one can derive analytical expressions for the $B(E1)$ distribution. The basic approximation is that most of the contribution of the $B(E1)$ comes from the external part of the initial and final wavefunctions so that they can be replaced by their asymptotic forms. For example, in the case of a $s$ to $p$ transition this leads to the simple expression
\begin{equation}
    \frac{dB(E1)}{d\varepsilon}(s \rightarrow p) = \frac{3 \hbar^2}{\pi^2 \mu} \left( Z_{\text{eff}} e \right)^2 \frac{\sqrt{E_b} \varepsilon^{3/2}}{\left( \varepsilon + E_b \right)^4}
    \label{eq:be1_anal}
\end{equation}
where $Z_{\text{eff}}=-Z/A$ is the effective charge for the case of a neutron halo. The expression above predicts a maximum at $\varepsilon = 3/5E_b$ and hence, the smaller the binding energy, the closer the position of the maximum to the breakup threshold.

The aim of a Coulomb dissociation experiment with dripline nuclei is to extract the electric response to the continuum, which, in the case of halo nuclei, is expected to be largely dominated by its $E1$ distribution.

\subsection{Semiclassical analysis of Coulomb dissociation data}
An attractive feature of the semiclassical formulas discussed in Sec.~\ref{sec:AW} is that they provide a neat separation between the structure of the projectile (through $B(E\lambda)$) and the reaction dynamics. In practical situations, nuclear couplings will be also present and there will not be such a simple connection between the measure breakup cross sections and the underlying ${dB(E \lambda)}/{d\varepsilon}$ distribution. 

To overcome this difficulty, two alternative strategies have been adopted. The first consists in measuring the breakup data at sufficiently small c.m.\ scattering angles, under the assumption that, for angles well below the grazing one ($\theta_\mathrm{gr}$), nuclear effects are negligible (the so-called ``safe-Coulomb'' region). This method has been adopted, for example, in several exclusive breakup experiments performed at RIKEN, at typical energies of around 100~MeV per nucleon (e.g.~\cite{Nak99,Nak06,Fuk04}). A second, alternative, strategy is to collect the breakup data in a wider angular range. This has the advantage of increasing the statistics, but the disadvantage of including nuclear contributions. The latter must be estimated and subtracted from the breakup data for a meaningful extraction of the electric contribution. 

The ``safe-Coulomb method'' has the disadvantage of considering only a fraction of the breakup data. In fact, nuclear breakup can extend down to very small angles. For example, in the analysis of the $^{11}$Li+$^{208}$Pb Coulomb dissociation experiment performed at RIKEN at an incident energy of 70 MeV per nucleon \cite{Nak06}, the authors considered data below $1.46^\circ$. The alternative method of considering a wider angular range has the obvious challenge of requiring a realistic method to estimate the nuclear contribution. In the remainder of this section we discuss  further these two methods.

For the practical application of the ``safe-Coulomb'' method, one integrates the double differential cross section given by Eq.~(\ref{eq:dsdwde}) up to a maximum angle ($\theta_\mathrm{max}$) such that for $\theta < \theta_\mathrm{max}$ breakup can be assumed to be purely Coulomb and, therefore, (\ref{eq:dsdwde}) is valid ($ \theta_\mathrm{max} <\theta_\text{gr}$, with $\theta_\text{gr}$ the grazing angle). This gives:
\be
\frac{d\sigma}{d\varepsilon} (\theta < \theta_\mathrm{max}) = 
\int_{0}^{\theta_\mathrm{max}}  \frac{d \sigma(E\lambda)}{d \Omega d \varepsilon} d\Omega
\propto \frac{dB(E \lambda)}{d \varepsilon}  \, .
\label{eq:dsde_epm}
\ee

\begin{figure}
\begin{center}\includegraphics[width=0.8\columnwidth]{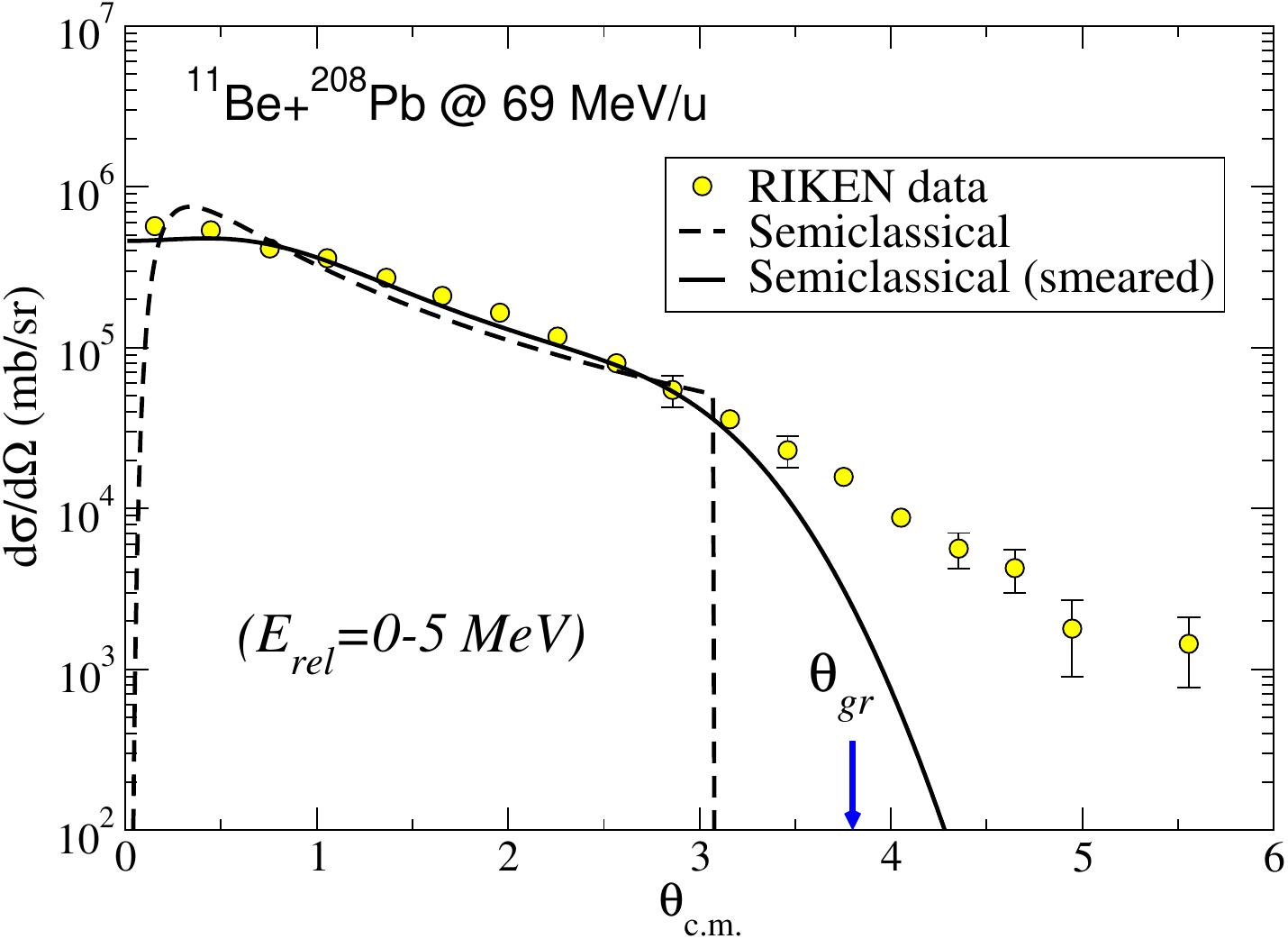} \end{center}
\begin{center}\includegraphics[width=0.8\columnwidth]{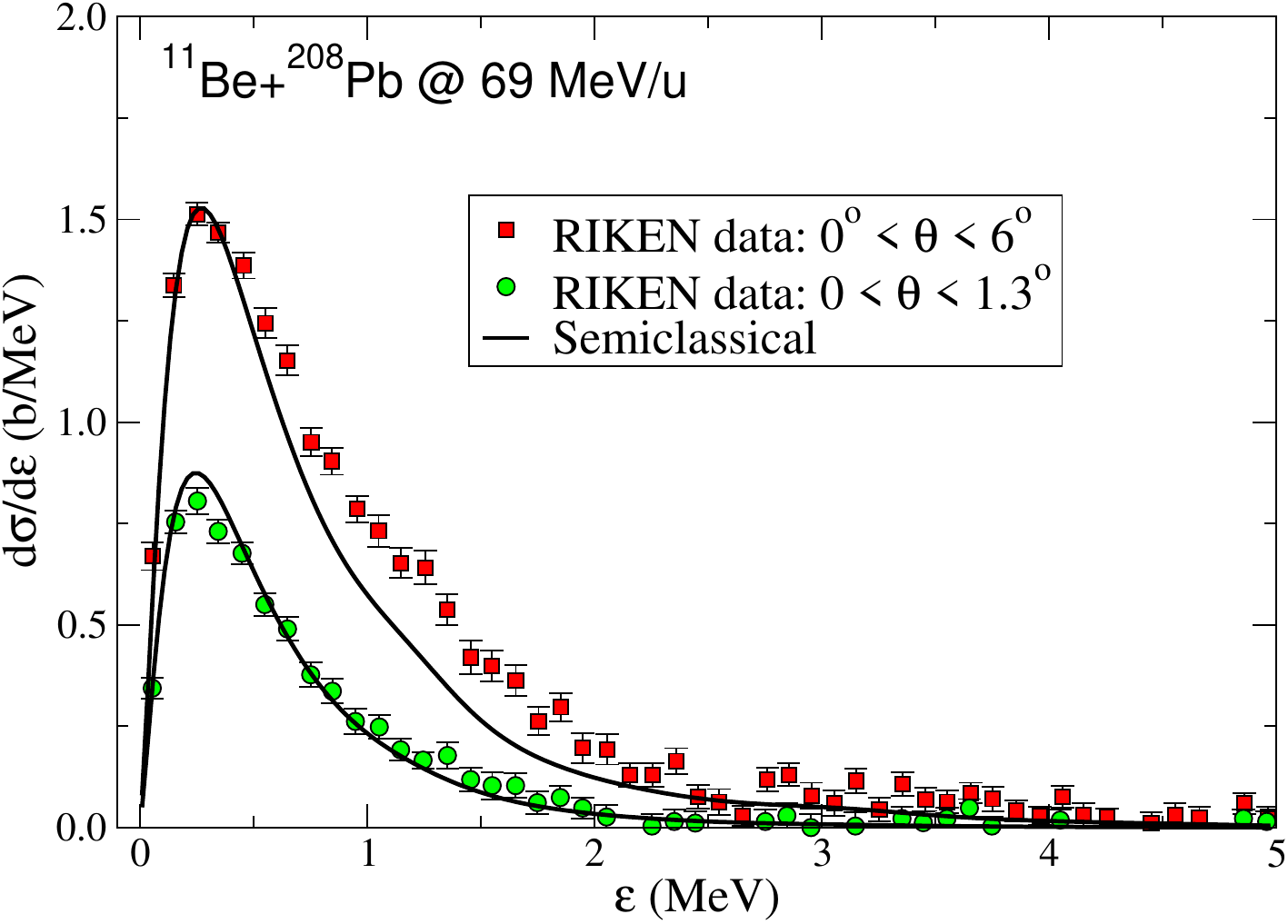} \end{center}
\caption{Top: angular distribution of the c.m.\ of the outgoing $n$+$^{10}$Be system, following the breakup of  $^{11}$Be on $^{208}$Pb at 69~MeV/u. Bottom: relative energy distribution between the $n$+$^{10}$Be  fragments, integrated up to two different maximum angles, as indicated by the labels. The curves are semiclassical calculations based on the theory of Alder and Winther with a model $B(E1)$ distribution obtained from a simple two-body potential model and smeared with the experimental resolution. Experimental data are from \cite{Fuk04}.}
\label{fig:be11pb_epm} 
\end{figure}

\begin{figure}
\begin{center}\includegraphics[width=0.8\columnwidth]{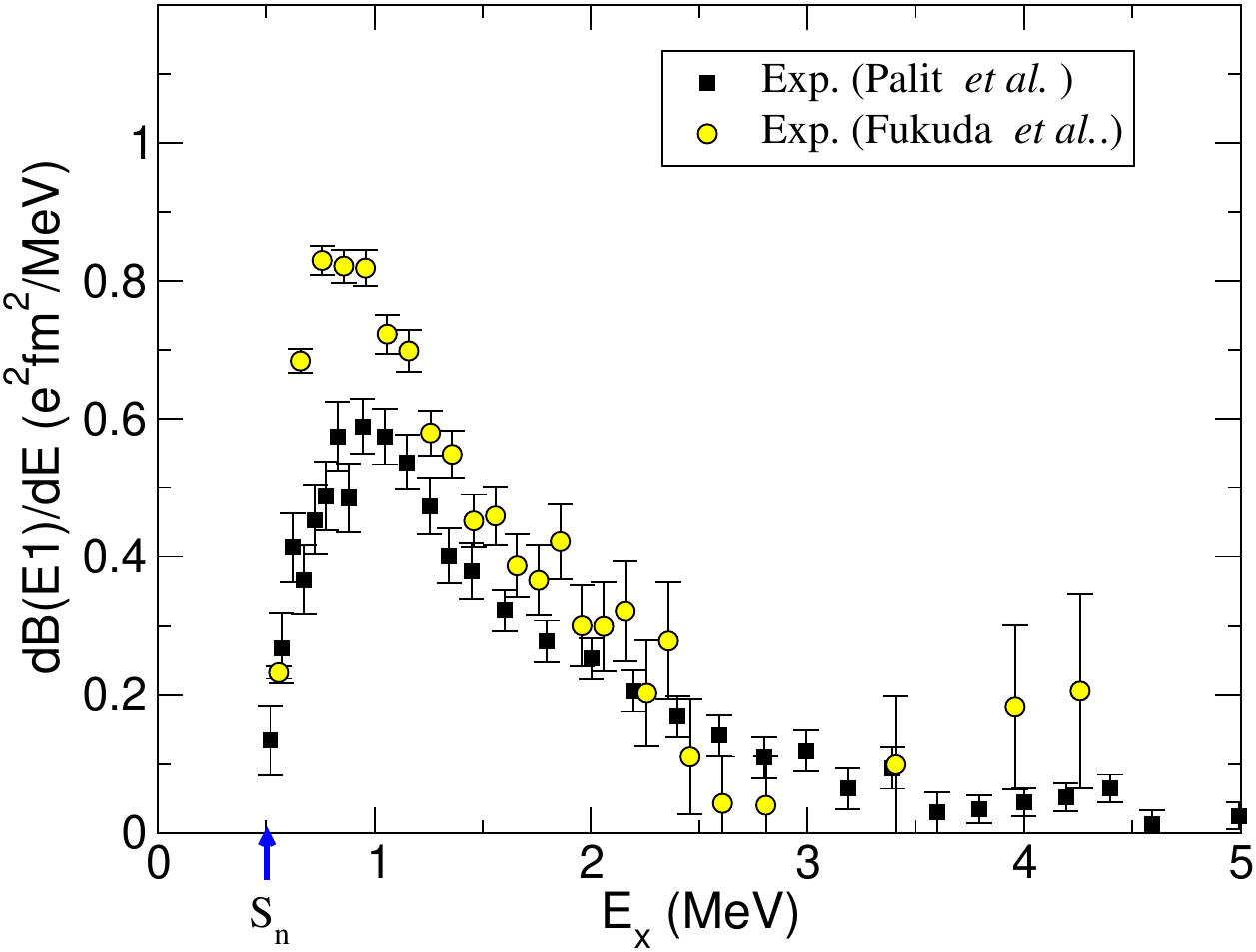} \end{center}
\caption{\label{fig:dbde_be11_exp} Electric transition probability for $^{11}$Be as a function of the excitation energy extracted from Coulomb dissociation experiments of $^{11}$Be on $^{208}$Pb at 69~MeV/u \cite{Fuk04} and 520~MeV/u \cite{Pal03}, measured at RIKEN and GSI, respectively. }
\end{figure}

An application of this method to the  $^{11}$Be+$^{208}$Pb  reaction discussed above is shown in Fig.~\ref{fig:be11pb_epm}. The data correspond to the measurements of Fukuda {\it et al.} \cite{Fuk04} performed at RIKEN at 69~MeV/u.  The curves are the result of semiclassical calculations using the EPM with a two-body structure model for $^{11}$Be and convoluted with the experimental energy and angular resolutions. In the top panel, it is apparent that the semiclassical calculation reproduces well the data up to $\theta_\text{c.m.} \sim 3^\circ$. For larger angles, the assumption of pure Coulomb scattering is no longer valid (the grazing angle is estimated to be $\theta_\text{gr}=3.8^\circ$). In the bottom panel, the corresponding relative energy distributions are compared with the data, for two angular cuts. It is seen that, for the wider angular range ($\theta_\mathrm{c.m.} < 6^\circ$), there is some underestimation of the data, which may be attributed to the omission of nuclear effects and higher Coulomb multipoles. 

Restricting the analysis to the lower cut ($\theta_\mathrm{c.m.} < 1.3^\circ$), and assuming that for these angles the dissociation probability is dominated by the Coulomb dipole couplings, one can use the approximation in Eq.~(\ref{eq:dsde_epm}) and extract the $B(E1$) distribution. The result of such an analysis is shown in Fig.~\ref{fig:dbde_be11_exp}, which displays the $B(E1)$ extracted in \cite{Fuk04} (circles).

The alternative strategy, in which angle-integrated cross sections are considered for the analysis, has the advantage of profiting from the entire breakup data, but deals with the uncertainties associated with the estimation of the nuclear contribution. For that, measurements with a high-$Z$  target nucleus are combined with those for a light nucleus, like $^{12}$C, for which the nuclear contribution is expected to be dominant. Thus, the nuclear contribution in the reaction with the heavy target is accounted for by subtracting a properly scaled cross section measured with the light
target. The scaling factor can be estimated from the respective radii of the heavy and light targets. Alternatively, one may resort to eikonal formulas for stripping and diffraction to estimate the cross section ratio between heavy and light targets, as done in \cite {Pal03}. 

In addition to the additional source of uncertainty  caused by the need to estimate the nuclear contribution, another caveat of this method is the neglect of possible interference effects between nuclear and Coulomb contributions. The importance of these effects can be assessed by means of quantum-mechanical calculations, such as those given by the CDCC method.  

If the approximations implied by the EPM are well fulfilled, the extracted $B(E\lambda$) distributions should be independent of the incident energy and the target as well as of the adopted analysis method.  A consistency test can be found in the  $B(E1)$ distributions extracted for the $^{11}$Be nucleus obtained from the analysis of two $^{11}$Be+$^{208}$Pb  experiments performed at GSI by Palit {\it et al.}~\cite{Pal03} at 70~MeV/u and  520~MeV/u, shown in Fig.~\ref{fig:dbde_be11_exp}. (The origin of this discrepancy is further discussed below). The two distributions are compatible but with clear differences at the peak.

\subsection{Quantum-mechanical effects: CDCC analysis of Coulomb dissociation data}
\label{sec:beyond_epm}
As an alternative to the semiclassical methods discussed in the previous section, one can use quantum-mechanical methods, such as CDCC, discussed in Sec.~\ref{sec:cdcc}.  In CDCC, nuclear and Coulomb breakup are included on an equal footing to all orders, and their interference is properly taken into account.  

\begin{figure}
\begin{center}\includegraphics[width=0.75\columnwidth]{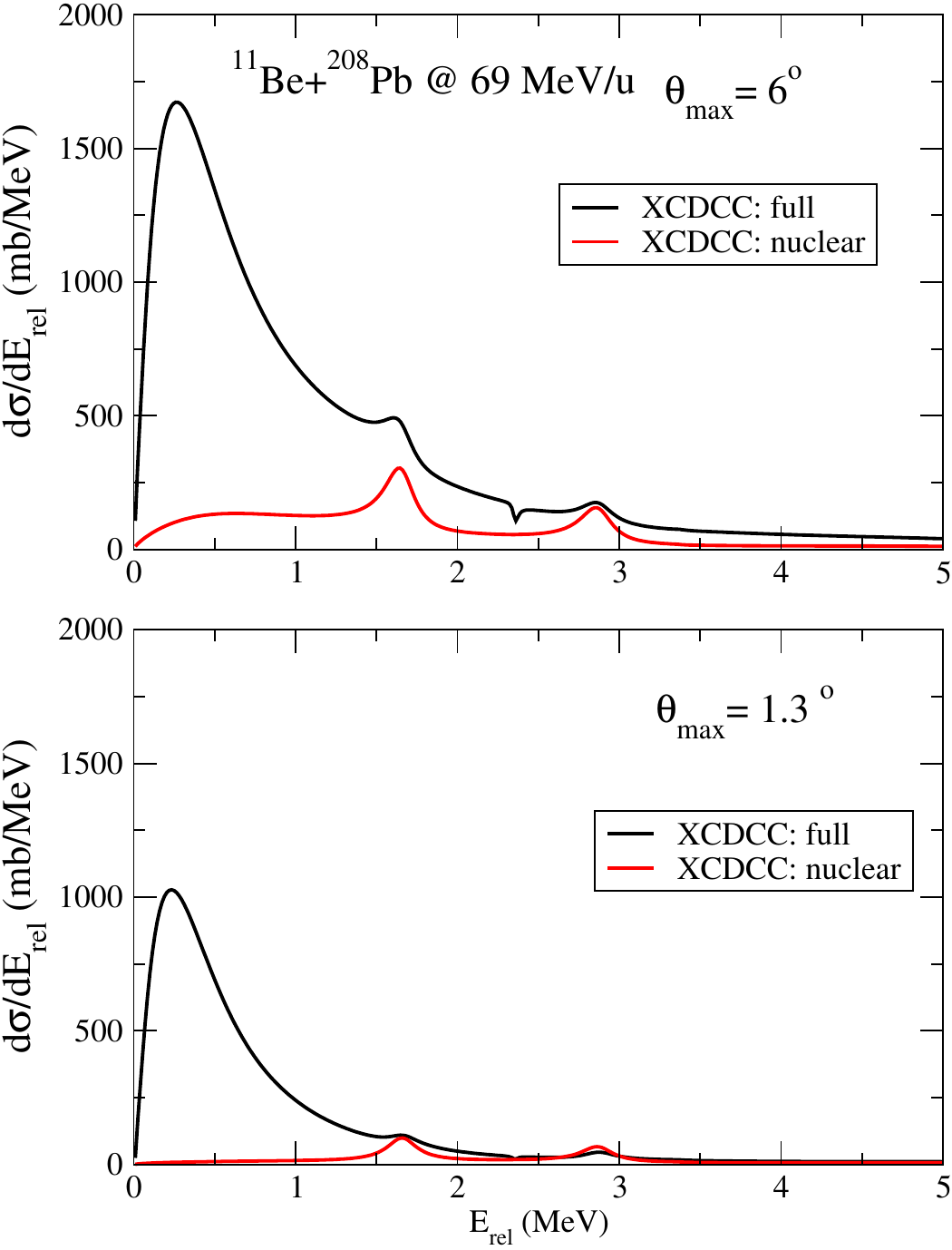} \end{center}
\caption{\label{fig:be11pb_e69A_cn} Relative energy distribution of the fragments following the breakup of $^{11}$Be on a lead target at 69 MeV/u, integrated in the scattering angle up to a maximum value of $\theta_\mathrm{max}=6^\circ$ (top) and $1.3^\circ$ (bottom). Full CDCC calculations, including both Coulomb and nuclear couplings are compared with CDCC calculations including only nuclear couplings.
}  
\end{figure}

In addition, the CDCC method can be used to assess the validity of the approximations involved in the semiclassical approaches. 
For example, the ``safe-Coulomb'' assumption is tested in Fig.~\ref{fig:be11pb_e69A_cn} for the  $^{11}$Be +$^{208}$Pb reaction, which shows the n-$^{10}$Be relative energy distribution for two upper cuts of the scattering angle, $\theta_\mathrm{max}=6^\circ$ (top) and $1.3^\circ$ (bottom).  The lines are XCDCC calculations based on a particle-plus-rotor model of $^{11}$Be, including both Coulomb and nuclear couplings or only nuclear couplings. It is evident from the figure that nuclear effects are still very relevant for $\theta_\mathrm{max}=6^\circ$ but are almost suppressed (but not negligible) for the lower cut. This gives an idea of the accuracy and validity of the EPM analysis performed in \cite{Fuk04} to extract the $B(E1)$ distribution for $^{11}$Be. 

\begin{figure}
\begin{center}\includegraphics[width=0.85\columnwidth]{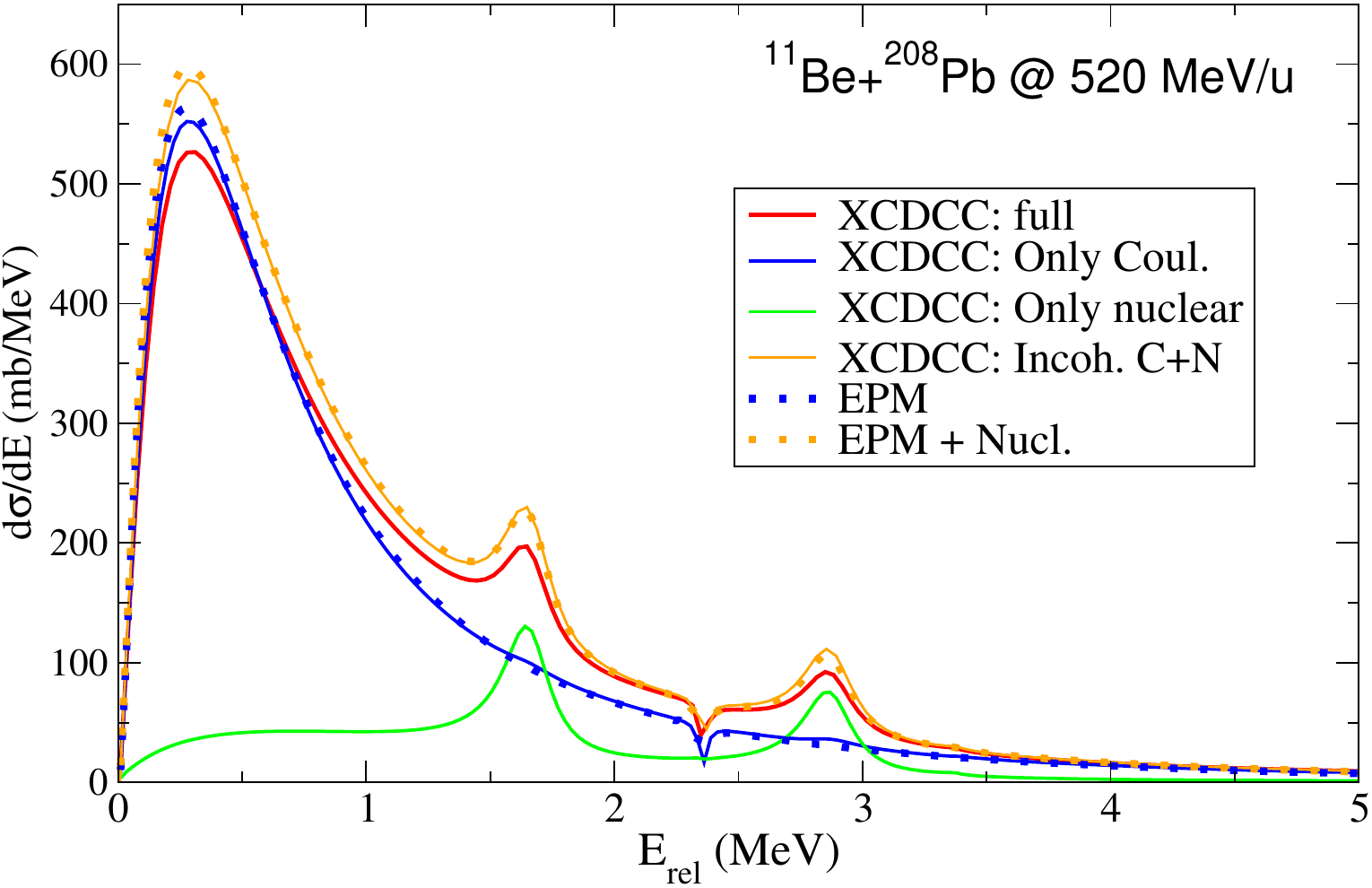} \end{center}
\caption{\label{fig:be11pb_e69A_cn_interf} Relative energy distribution of the fragments following the breakup of $^{11}$Be on a lead target at 520 MeV/u. EPM and XCDCC calculations, based on the same structure model, are compared. The effect of the nuclear/Coulomb interference is illustrated.}  
\end{figure}

The CDCC method can be also used to test the importance of Coulomb/nuclear interference effects. This is illustrated in  Fig.~\ref{fig:be11pb_e69A_cn_interf} for the $^{11}$Be+$^{208}$Pb reaction at 520~MeV/u.
The plot shows EPM and XCDCC calculation. In the latter, different choices for the nuclear and Coulomb couplings are shown. It is seen that (i) nuclear breakup is rather significant (consistently with the estimate of \cite{Pal03}) and, more importantly,  (ii) the incoherent sum of the nuclear and Coulomb contributions overestimates the full XCDCC calculations with interference effects. From these calculations, it becomes apparent that the procedure of summing incoherently the nuclear and Coulomb contributions introduces some error in the extraction of the electric transition probabilities from Coulomb dissociation experiments. The error will become larger at lower incident energies.

Ideally, one may also analyze directly the experimental cross sections with the CDCC method to extract the electric transition probabilities and other structure properties. The practical applicability of this idea is hindered by the fact that there is no simple connection between the reaction observables (cross sections) and the underlying $B(E\lambda)$ distributions. In particular, breakup cross sections are no longer proportional to the $B(E\lambda)$. 

\begin{figure}
\begin{center}\includegraphics[width=0.85\columnwidth]{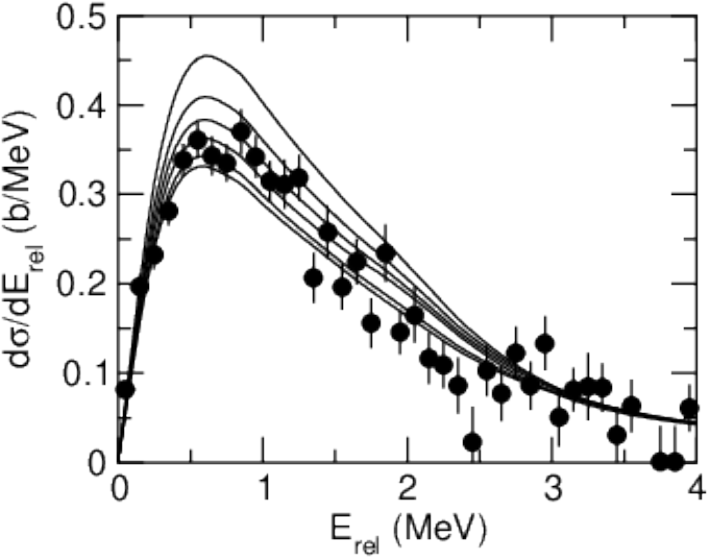} \end{center}
\caption{\label{fig:c15pb_dsde} Comparison of the Coulomb dissociation data of $^{15}$C on $^{208}$Pb at 68~MeV/u \cite{Nak03} with CDCC calculations using different $^{14}$C+n models for $^{15}$C.  Figure from \cite{Sum08}. }  
\end{figure}

One possibility is to compare the measured cross sections with a series of CDCC calculations assuming different structure models,  with different electric strength distributions. This is exemplified in Fig.~\ref{fig:c15pb_dsde} where the data correspond to the relative energy distribution of $^{15}$C on $^{208}$Pb at 68~MeV/u measured at RIKEN \cite{Nak03}  and the lines are  CDCC calculations for different potential models of the $^{15}$C nucleus. From the comparison one may select the {\it best} structure model among those considered. 

Recently, an alternative to this procedure to extract $B(E\lambda)$ distributions from the comparison of CDCC calculations with breakup cross section data has been proposed \cite{Mor20}. The strategy can be summarized as follows:
\begin{itemize}
\item Start with some trial structure model of the projectile. This will predict some distribution $B^0(E1;\varepsilon)$ depending on the continuum energy. 
\item Introduce some small changes ({\it perturbation}) in the model, by multiplying all Coulomb dipole matrix elements by arbitrary factors $(1+ 2\delta(\varepsilon_i))$, where $\varepsilon_i$ are the measured energies. This will modify the $B(E1)$ at each measured $\varepsilon_i$
\begin{equation}
B^\text{mod}(E1,\varepsilon_i) \simeq  B^0(E1,\varepsilon_i)  
\left(1+ 2 \delta(\varepsilon_i) \right),
\label{eq:Bmod}
\end{equation}

\item For small perturbations, the calculated breakup cross sections are modified simply as \cite{Mor20}
\begin{equation}
   \sigma^\text{mod}_i \simeq   \sigma^0_i + \delta(\varepsilon_i) \,{\sigma^\prime}_i .
   \label{neweqSdelta}
\end{equation}
where $\sigma^0_i$ and ${\sigma^\prime}_i$ are constants which can be determined by performing several calculations for different $\delta$ values. 
\item By comparing the model calculations with the measured breakup data, one may infer the optimal value of $\delta(\varepsilon_i)$ for each excitation energy and, from them, the values of $B(E1;\varepsilon^i)$ that best describe the data within the CDCC framework.
\end{itemize}

\begin{figure}
\begin{center}\includegraphics[width=0.85\columnwidth]{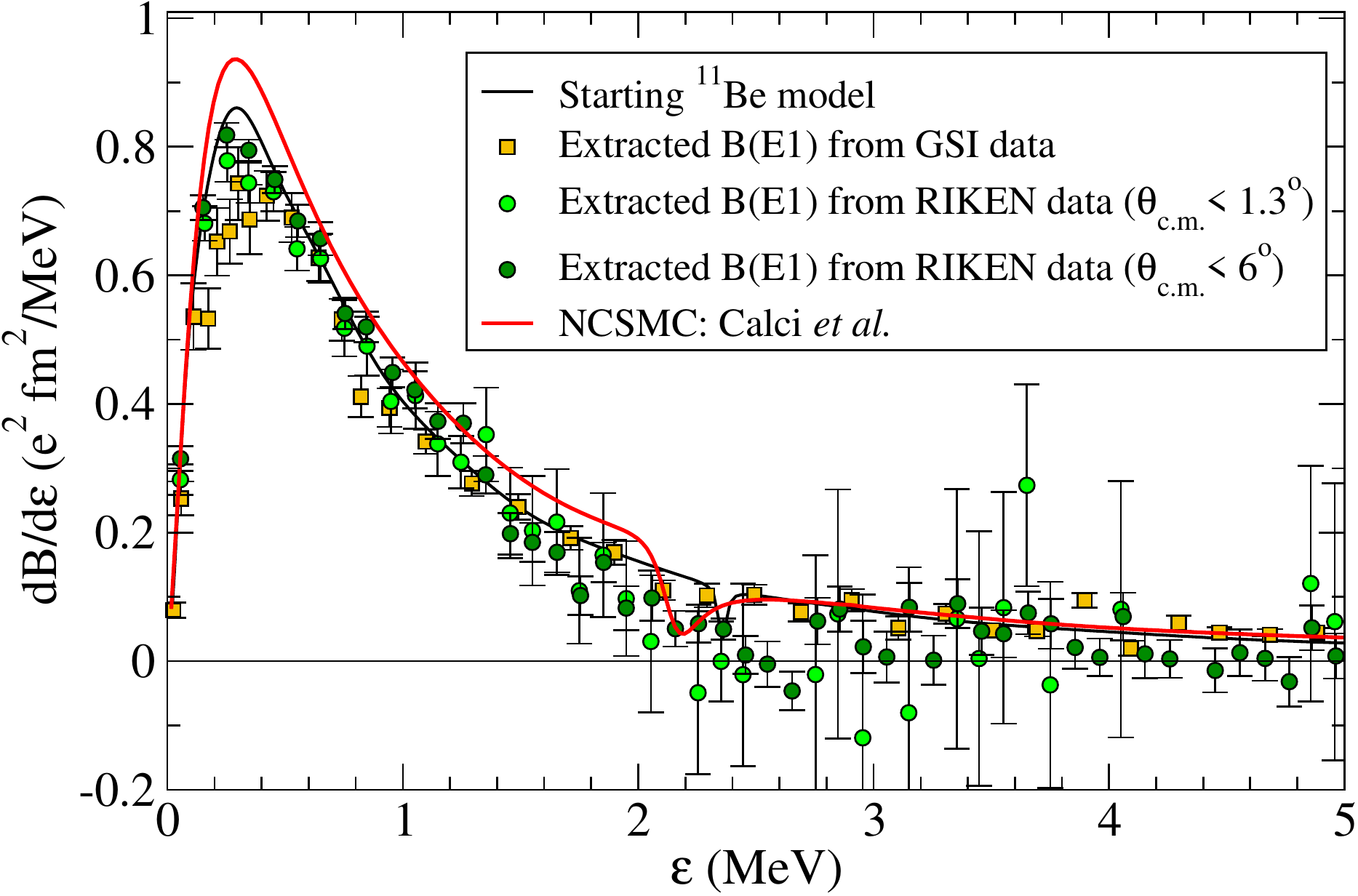} \end{center}
\caption{\label{fig:dbde_be11_xcdcc}Unfolded $B(E1)$ distributions extracted from the experimental breakup cross sections measured in \cite{Pal03} (squares) and \cite{Fuk04} (circles) using the procedure based on the XCDCC method proposed in \cite{Mor20}. For comparison, the starting  $B(E1)$ distribution is shown by the black line and a NCSMC {\it ab-initio} calculation of Ref.~\cite{Cal16} (red solid line) is also shown.  Figure adapted from \cite{Mor20}. }  
\end{figure}

An application of this method is shown in Fig.~\ref{fig:dbde_be11_xcdcc}, adapted from~\cite{Mor20}. The solid black line is the $B(E1)$ distribution provided by the starting (trial) $B(E1)$ model used in a XCDCC calculation. Following the outlined procedure, this $B(E1)$ is corrected for each excitation energy by comparing the calculated and measured cross sections. The procedure is applied separately to the data of Ref.~\cite{Fuk04}, for the two angular cuts discussed above, as well as to the data of \cite{Pal03}. The corrected $B(E1)$ distributions are displayed in this figure with the symbols. It is seen that the three of them agree reasonably well within the estimated errors.  This is in contrast to the $B(E1)$ extracted in the original works and shown in Fig.~\ref{fig:dbde_be11_exp}. It is important to note that the latter are affected by the experimental resolution, while those shown in Fig.~\ref{fig:dbde_be11_xcdcc} do not include the effect of the experimental energy resolution. Differences between these experimental resolutions in the two experiments are partially responsible for the differences observed in the $B(E1)$ distributions reported in the original works.  

\section{Study of nucleon-nucleon correlations from nucleon removal reactions \label{sec:corr}}
One particular point of interest for nucleon-Borromean systems of the form  N+N+C  is the study of the correlation between their two valence nucleons, since the nucleon-nucleon interaction is essential to hold the Borromean system bound. As many Borromean systems also present nucleon halos, the study of the correlation between their valence nucleons permits the exploration of the nucleon-nucleon interaction in low-density, isospin-asymmetric environments. A quantity related to the nucleon-nucleon correlation is the average opening angle between nucleons with respect to the core $\left\langle\theta_{NN}\right\rangle$. If both nucleons follow independent uncorrelated orbits, this angle will on average be $90^\circ$ \cite{Esb92,Ber07}. A smaller angle corresponds to a small average distance between nucleons compared to the distance of their center of mass to the core, in which is known as a ``dineutron'' (or diproton) configuration \cite{mig73}, which is associated with an attractive correlation between nucleons. Meanwhile, an angle larger than 90$^\circ$ indicates a larger internucleon distance compared to their distance to the core, which corresponds to both nucleons being on opposite sides of the core, in the so-called ``cigar'' configuration, which corresponds to a repulsive nucleon-nucleon correlation. Note that three-body calculations show that both ``dineutron'' and ``cigar'' components appear in the wave function~\cite{Zhu93}, and it is their relative strength which leads to a more dominant configuration.

An experimental probe for $\left\langle\theta_{NN}\right\rangle$ in the case of neutron-Borromean systems is the dipole electric strength $B(E1)$: assuming point-like particles, its expression reduces to:
\begin{equation}
   \begin{split}
    B(E1)&=\dfrac{3}{\pi}\left(\frac{Z_c}{A}\right)^2e^2\left\langle r_n^2+r_{n'}^2+2r_nr_{n'}\cos\theta_{NN}\right\rangle\\&\simeq \dfrac{6}{\pi}\left(\frac{Z_c}{A}\right)^2e^2\left\langle r_n^2\right\rangle(1+\cos\left\langle\theta_{NN}\right\rangle),
    \end{split}
\end{equation}
where $\left\langle r_n^2\right\rangle$ is the mean square distance between neutron and core, $Z_c$ is the charge of the core and $A$ the mass of the nucleus \cite{Bert91}, so an enhanced $B(E1)$ indicates a smaller angle and dineutron correlations while a reduced $B(E1)$ indicates the existence of a cigar-like configuration. $B(E1)$ strengths have been measured for various Borromean nuclei, from which values for $\left\langle\theta_{NN}\right\rangle$ can be extracted. The values extracted in \cite{Ber07} are presented in Table~\ref{tab:corr}, where the angles can be seen to suffer from significant uncertainties, due both to experimental uncertainties and the model dependence due to $\left\langle r_n^2\right\rangle$.

\begin{table}[ht]
    \centering
    \begin{tabular}{c|c|c|c|c}
    \hline
         & \nuc{6}{He} & \nuc{11}{Li} & \nuc{14}{Be} & \nuc{17}{Ne}  \\
         \hline \hline B(E1) (e$^2$fm$^2$)  & 1.20(20)  & 1.42(18)  &  1.69  & 1.56 \\
         &\cite{Aum99}&\cite{Nak06}&\cite{Zhu95}&\cite{Gri06}\\
       \hline
       $\left\langle \theta_{NN}\right\rangle$ (deg.)
 \cite{Ber07} & 83$^{+20}_{-10}$&66$^{+22}_{-18}$&64$^{+9}_{-10}$&110\\
 \hline
    \end{tabular}
    \caption{B(E1) strength function and deduced average opening angle $\left\langle\theta_{NN}\right\rangle$ for various Borromean nuclei. Table adapted from \cite{Ber07}.} 
    \label{tab:corr}
\end{table}

\begin{figure}
\begin{center}
\includegraphics[width=0.3\columnwidth]{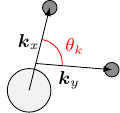} \end{center}
\caption{\label{fig:schemecorr} Opening angle and momenta measured in nucleon-removal reactions on Borromean systems. Figure adapted from \cite{Cor23}.}
\end{figure}

Another method of extracting the opening angle is the use of nucleon knockout reactions, where one of the valence nucleons is removed via a sudden reaction, such as nucleon-knockout with $^{9}$Be or $^{12}$C targets or a $(p,pN)$ reaction with proton targets, assuming the validity of the spectator approximation for the remaining $(C+N)$ system. Within this approximation, in the projectile rest frame the Borromean nucleus starts at rest, so that when the nucleon is removed, the final momentum of the $(C+N)$ system will be equal and opposite to that of the removed nucleon within the Borromean system ($\boldsymbol{k}_y$). Since the $(C+N)$ system is unbound, the relative momentum between $C$ and the remaining nucleon can also be measured $(\boldsymbol{k}_x)$, and the angle between them $\theta_k$ determined. A scheme of the quantities of interest is shown in Fig.~\ref{fig:schemecorr}.

It should be noted that, as $\theta_k$ is an angle between the momenta associated with the nucleons instead of their positions, it can be viewed as a Fourier transform of the opening angle $\theta_{NN}$. As such, a value of $\theta_k$ larger than $90^\circ$ is associated with a value of $\theta_{NN}$ lower than $90^\circ$, and thus to a dineutron configuration, while the cigar configuration corresponds to $\theta_k<90^\circ$ and $\theta_{NN}>90^\circ$. Various measurements of $\theta_k$ have been performed for Borromean nuclei \cite{Chu97,Sim99,Mar01,Sim07}, finding for its average value $\left\langle\theta_k\right\rangle$  the following results (in degrees): $90$~(\nuc{6}{He}), $90$~(\nuc{8}{He}), $103.6^{+0.7}_{-0.9}$~(\nuc{11}{Li}), $96.2^{+1.8}_{-1.6}$~(\nuc{14}{Be}) \cite{Sim07}. It can be seen that, indeed, values for $\left\langle\theta_k\right\rangle>90^\circ$ correspond to $\left\langle\theta_{NN}\right\rangle<90^\circ$, as shown in Table~\ref{tab:corr}. It should also be noted that an asymmetric distribution in either $\theta_k$ or $\theta_{NN}$     (resulting in $\left\langle\theta\right\rangle \neq 90^\circ$) requires the interference of waves with opposite parities \cite{Cat84}, which neatly explains the symmetry for $^{6,8}$He, since the valence neutrons find themselves in the negative-parity $p_{3/2}$ and $p_{1/2}$ waves, so their angular distribution must be symmetric. This also sets the average opening angle as a measure of the admixture of opposite parity components \cite{Kub20}.

\begin{figure}
\begin{center}
\includegraphics[width=0.9\columnwidth]{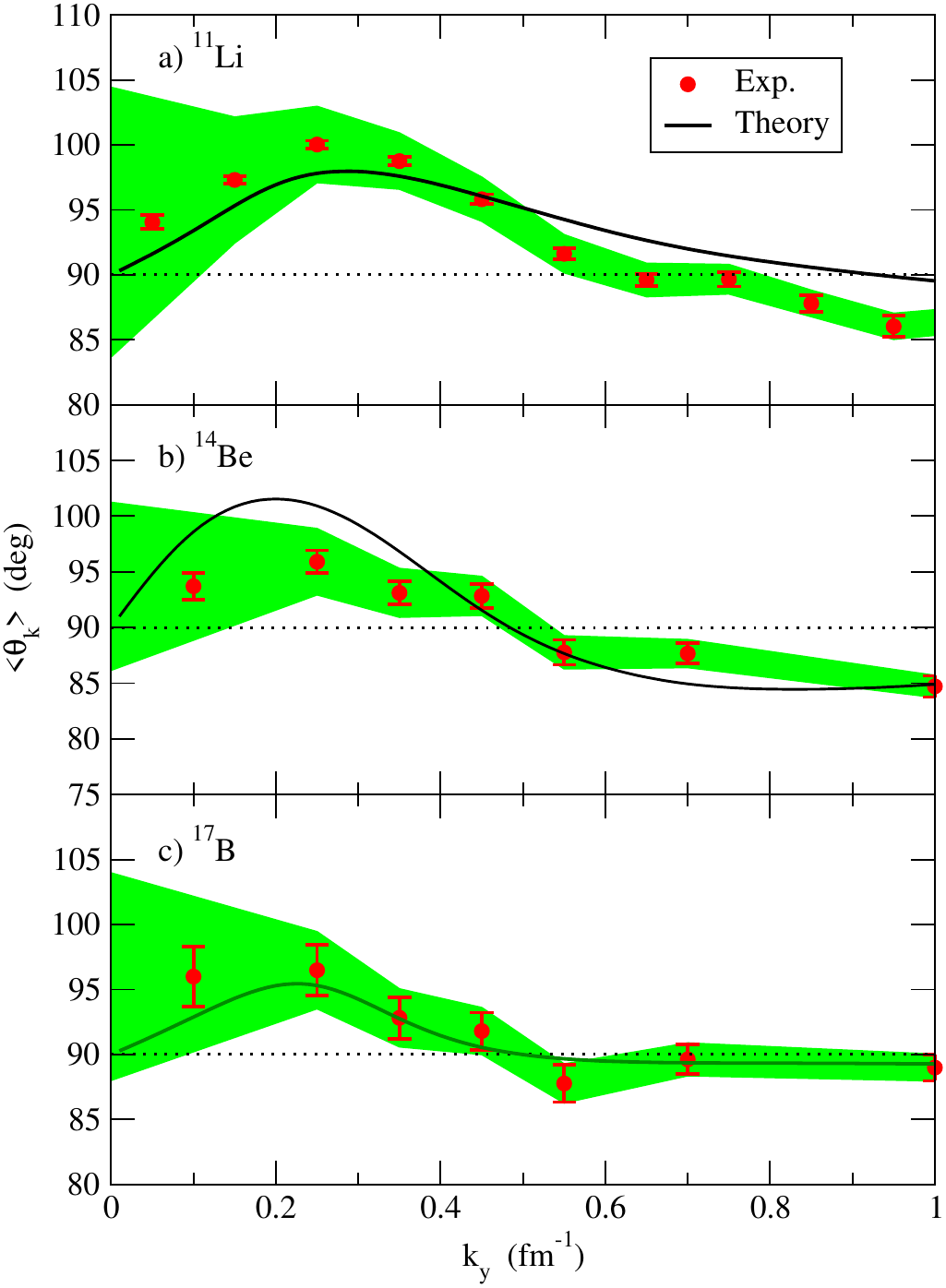} \end{center}
\caption{\label{fig:corr_corsi} Average opening angle in momentum space $\left\langle\theta_k\right\rangle$ as a function of $k_y$. Green bands correspond to the systematic errors in the experimental measurement. In all nuclei the maximum average momentum is found at similar values of $k_y$ for all considered nuclei. Figure adapted from Ref.~\cite{Cor23}.}  
\end{figure}

As seen above, for neutron-Borromean systems, the dineutron correlation seems to dominate over the cigar distribution. Recent experiments have tried to explore the spatial location of this dineutron correlation, as some nuclear matter calculations point to the low-density surface of Borromean systems as a region where these correlations should be favoured \cite{Mats06}. Recent $(p,pn)$ experiments on Borromean nuclei \cite{Kub20,Cor23} have used the modulus of the momentum of the removed neutron ($k_y$) as a proxy for the nuclear location (smaller momenta correspond to the nuclear halo and surface, while larger momenta correspond to the nuclear interior) and the average opening angle as a proxy for dineutron correlations. Through the use of an eikonal zero-range approximation for the $(p,pn)$ nucleon-nucleon collision process, the cross section can be expressed as \cite{Kik16,Cas21}: 
\begin{equation}
    \sigma \propto\left|\left\langle\phi_{Cn}(\boldsymbol{k}_x,\boldsymbol{x}) \otimes e^{i\boldsymbol{k}_y\cdot\boldsymbol{y}}|S(y)\phi_{gs}(\boldsymbol{x},\boldsymbol{y})\right\rangle\right|^2,
\end{equation}
where $\boldsymbol{x}$ and $\boldsymbol{y}$ are defined as in Sec.~\ref{sec:3b_transfer}, $\phi_{gs}$ describes the bound state of the Borromean nucleus, $\phi_{Cn}$ the final-state wave-function between the core and the remaining neutron, and $S(y)$ is a $S-$matrix that includes the absorptive potential between target proton and core $C$. As such, the cross section can be interpreted as the square of the Fourier transform of the bound state, distorted by the absorption of the target proton and the final-state interaction between neutron and core. In \cite{Kub20}, the maximum $\left\langle\theta_k\right\rangle$ for $^{11}$Li was found for small values of $k_y\sim0.3$~fm$^{-1}$, which was interpreted as the dineutron being located at the nuclear surface where the nuclear density is low, as predicted by previous models \cite{Mats06,Hag07}. As the low nuclear density is a feature of the surface of all neutron-Borromean nuclei, it is expected that the dineutron also appears in the surface of other such systems. Indeed, in \cite{Cor23}, the maximum of $\left\langle\theta_k\right\rangle$ was found for roughly the same value of $k_y$ also for \nuc{14}{Be} and \nuc{17}{B}, as shown in Fig.~\ref{fig:corr_corsi}, making a tantalizing case for the universality of dineutron correlations in the surface of neutron-Borromean nuclei, as predicted in \cite{Mats06}, although experiments on more nuclei, such as $^{19}$B or $^{22}$C, are required in order to confirm this feature. In the figure the green bands correspond to systematic errors in experimental data, while statistical errors are indicated by the error bars. The theoretical calculations are able to match experimental results for \nuc{11}{Li} and \nuc{17}{B} but overestimate the maximum average angle for \nuc{14}{Be}. This has been associated with the contribution of excited states of the \nuc{12}{Be} beyond the $2^+$ state included in the theoretical model \cite{Cor23}.

\section{Uncertainty evaluation \label{sec:uncertainty}}
Given the amount of ingredients required for the computation of reaction observables (bound-state wavefuncions, optical potentials, spectroscopic factors,...) there has been a push for the quantification of the uncertainties in the calculated observables due to the ambiguities of these ingredients (the Igo ambiguity \cite{Igo58} of the optical potentials is a classical example of such ambiguities). While the classical response to this demand was the comparison of calculations using different ingredients (for example, optical potentials from different global parametrizations), this evaluation of the uncertainties is at most qualitative and highly dependent on the choice of inputs. Recently,  Bayesian methods have been proposed to quantify the uncertainties that optical potentials suffer when they are obtained from the fitting of elastic scattering data (as is usual for optical potentials) and to study the propagation of these uncertainties to observables such as $(d,p)$ transfer angular or charge exchange $(p,n)$ differential cross sections \cite{Lov17b,Lov18,Kin18,Kin19,Cat19,Cat20,Phi21,Whi22} or nucleon-knockout momentum distributions \cite{Heb23}. The determination of Bayesian posterior distributions and confidence intervals requires the evaluation of the reaction observables for a range of values of every considered parameter, which rapidly results in a humongous number of calculations, which are feasible in facilities with high computing power for light-computation models such as DWBA and ADWA, but become prohibitive for models which require heavier computations, such as CDCC or CRC. As such, methods to accelerate these calculations using emulators \cite{Fur20,Dri21} to predict and interpolate the result of these heavier calculations for the required parameter values are being explored. Through these methods, uncertainty analyses have been performed using CDCC \cite{Sur22}, although admittedly only the uncertainties due to the valence-core interaction were studied, further proof of the computational magnitude of the task. In this regard, new global potential parametrizations with quantified uncertainties have recently been developed, with uncertainties originating either from the chiral interactions used to generate the potentials \cite{Whi21} or from fitting to extensive elastic scattering data \cite{Pru23}, although it should be noted that in \cite{Pru23} the fit was not performed within a Bayesian framework. The latter potential has been used to study the uncertainty in the reduction factors for transfer and knockout reactions \cite{Heb23b}, which is still an open problem \cite{Aum21}.

\section{Inclusion of non-local potentials}
\label{sec:nonlocal}

It has long been known that nucleon-nucleus and nucleus-nucleus optical potentials are non-local \cite{Fes58a}, both due to antisymmetrization and the effect of non-elastic channels (encoded in the second term of the Feshbach potential in Eq.~(\ref{eq:u_fesh})). The use of non-local potentials transforms the Schr\"odinger equation into an integro-differential equation, which is far more cumbersome to solve, so local potentials have traditionally been favoured in reaction calculations.  In fact, for a long time only a few non-local parametrizations for optical potentials have been published \cite{Per62,Gia76,Gia80,Tia15} and non-locality is not explicitly considered in the majority of potential parametrizations \cite{CH89,KD03}. In recent years there has been a development of dispersive optical potentials based on the nuclear self-energy, which is inherently non-local \cite{Cha07,Dic10,Mah16,Mor24}, thus reigniting the interest in fully non-local potentials. 

For the study of elastic scattering, the neglect of non-locality is not problematic, as one can always find a local equivalent potential which reproduces the phase-shifts and observables of the non-local one \cite{Fie66,Fie67}. Analogously, reactions that are only sensitive to the asymptotics of the wave functions are usually well described by local potentials. However, non-locality reduces wavefunctions in the nuclear interior, in what is called the Perey effect \cite{Aus65,Fie66}, so non-locality should be taken into consideration in reactions that are sensitive to the nuclear interior, such as transfer reactions. This has usually been approximated through the multiplication of wavefunctions obtained from a local potential by the so-called Perey factor \cite{Fie66}, which reduces the wavefunctions in the interior without altering their asymptotics. Although the Perey factor can be obtained for any non-locality shape, the prescriptions typically used assume a Gaussian dependence on non-locality \cite{Fra57} (usually referred to as Perey-Buck geometry). These approximations (local equivalent potential and Perey factor) have been compared to the computation of the wavefunction using the full non-local potential in \cite{Tit14} for $(d,p)$ transfer reactions, finding a difference in spectroscopic factors of $\sim 10\%$, although these results did not consider non-locality in the outgoing deuteron wavefunction. 

The study of non-locality in $(d,p)$ transfer reactions has received particular attention, due to the larger effect non-locality has on the observables, the fact that there exist more reliable non-local nucleon-nucleus potentials than nucleus-nucleus ones and the relative simplicity of their description, which allows for more manageable computations. As such, non-locality has been explicitly included in Faddeev calculations for $(d,p)$ transfer reactions \cite{Del09c} as well as in the widespread ADWA calculations \cite{Tim13,Tim13b,Wal16}, with a special focus on the local-equivalent adiabatic deuteron potential, whose main feature was the energy shift in nucleon-nucleus potentials from the nominal value of half the deuteron energy due to the internal kinetic energy of the nucleons in the deuteron. Later implementations included non-locality in the deuteron channel fully \cite{Tit16,Bai16,Bai17}, with the intriguing finding \cite{Bai16,Bai17} that cross sections for the $^{26}\mathrm{Al}(d,p)^{27}\mathrm{Al}$ reaction were significantly dependent on the deuteron model, and in particular on its $d$-wave component, a dependence that was not observed in calculations with local potential. This dependence was later found to be an artifact due to the interplay of the ADWA formalism and the d-wave component of the deuteron, as the inclusion of non-locality in CDCC calculations (through the use of first-order local equivalent potentials \cite{Gom18}) and Faddeev calculations \cite{Del18} resulted in a reduced dependence on the deuteron model, showing the need for precise reaction theories to extract reliable information from experimental data. The CDCC formalism has been further extended to include non-locality through the consideration of the Perey effect \cite{Gom19} and the extension to $^{6}$Li$\rightarrow \alpha +d$ breakup reactions \cite{tim23}. The effect of non-locality in other reactions, such as $(d,p)X$ surrogate reactions \cite{Li18} and eikonal descriptions of knockout reactions  \cite{Heb21}, has also been recently studied, as corresponds to the mounting interest in non-locality arising from the new ab initio and dispersive potentials.

\section{Conclusions and perspectives}

The present review aimed to provide a glimpse into the unique physics behind reactions involving weakly-bound nuclei and the formalisms and methods commonly used in the modeling and analysis of these reactions. We have shown the importance of breakup channels for weakly-bound systems and the methods developed to properly consider them, such as the CDCC method (Sec.~\ref{sec:cdcc}), several semiclassical descriptions (Sec. \ref{sec:AW}),  or the ADWA and CDCC-BA formalisms for transfer reactions (Sec.~\ref{sec:transfer}), as well as their extension to consider collective excitations and/or 3-body projectiles, where the particular characteristics of Borromean systems have been exploited to explore nucleon-nucleon correlations (Sec.~\ref{sec:corr}). As well, inclusive breakup (Sec.~\ref{sec:inc})  and incomplete and complete fusion (Sec.~\ref{sec:fusion}) reactions with weakly-bound systems have been explored due to the unique phenomenon of complete fusion suppression found for weakly-bound species. Thanks to the increase in computational power, demanding methods such as CDCC and IAV become more accessible and useful in the analysis of experimental data. In addition, questions which require larger numerical efforts, such as the inclusion of non-locality in these sophisticated methods (Sec.~\ref{sec:nonlocal}) and the study of theoretical uncertainties (Sec.~\ref{sec:uncertainty}), are starting to be explored.

Admittedly, the selection of reactions and methods presented in this work has been heavily curated by the expertise and preferences of the authors and many other methods for the study of reactions with weakly-bound nuclei exist and/or are currently being developed:

 \begin{itemize}
 \item Some of the most promising advances are those related to few-body and ab-initio methods. The application of the Faddeev method to nuclear reactions, while potentially providing an exact,  rigorous solution of a given few-body Hamiltonian, has only been possible in recent years thanks to the increasing computational capabilities,  the inclusion of complex optical potentials and the development of techniques to deal with the long-range Coulomb couplings. The method has been very useful for benchmarking some of the methods discussed in this review, such as the CDCC and ADWA methods \cite{Del07,Upa12}. It has also proven very useful for analyzing nucleon knockout reactions \cite{Cre11,Cre14,Cre16}.

\item There has also been remarkable progress in the No-Core-Shell-Model (NCSM) and its extension, the NCSM with Continuum (NCSMC)  \cite{Bar13}. While originally developed to study nuclear structure,  the method has been applied with great success to the calculation of scattering observables \cite{Vor19,Baroni13,Cal16}. 
 
\item Ab-initio methods provide not only alternative ways for evaluating scattering observables by themselves, but they also furnish very useful structure inputs for traditional reaction frameworks. For example, the Variational  Monte Carlo (VMC) \cite{Car15} and its variants, such as the Green's function Monte Carlo and the Cluster Variational Monte Carlo \cite{Pie92}, provide  microscopic overlap functions which can be directly {\it plugged} into reaction formalisms  for transfer \cite{Fla13} and knockout \cite{Gri11} reactions. 
 


\item The expansion of experimental studies, reaching heavier exotic nuclei, has also shown the need to extend and upgrade the existing reaction models to incorporate the new features found in the structure and reactions of the newly discovered nuclei. For example, it has become clear that a proper description of newly discovered halo nuclei in the islands of inversion requires the simultaneous inclusion of deformation, pairing and Pauli blocking. Reaction formalisms, such as DWBA or CDCC, must therefore be extended to accommodate such effects.  Admittedly, the XCDCC method discussed in this review has used so far relatively simple particle-plus-rotor models, and hence it is mandatory to resort to more sophisticated models.  

\item The challenge is even larger for the case of three-body systems which, in addition to the properties cited above, require specific formalisms to account for their three-body nature. For example, knockout experiments with $^{14}$Be have clear evidenced the need of including the excitations of the $^{12}$Be ``core'' \cite{Cor19,Cor23}. Even beyond, models including four- and five-body systems may be necessary, as the recent discovery of a four-neutron correlated system in \nuc{8}{He} attests \cite{Due22}.
\end{itemize}

All of these avenues for advancement show that, despite its numerous and significant successes, the study of nuclear reactions with weakly-bound nuclei remains a developing and active field, pushed by theoretical and technical advances and by new experimental measurements in ever more exotic nuclei. While we hope that this work has transmitted to the reader at least a glimpse of the current methods in this fascinating field, we also hold the paradoxical hope that new developments promptly render this work somewhat obsolete thanks to new and perhaps surprising advancements.

\section*{Acknowledgements}
We are grateful to Angela Bonaccorso and Jin Lei for their feedback on the semiclassical transfer to the continuum calculations. This work has been partially funded  by the Ministerio de Ciencia e Innovaci\'on,  MCIN/AEI/10.13039/501100011033 under I+D+i project No.\ PID2020-114687GB-I00. M.G.-R. acknowledges financial support by MCIN/AEI/10.13039/501100011033 under grant IJC2020-043878-I (also funded by ``European Union NextGenerationEU/PRTR'').

%
\bibliographystyle{epj}
 \bibliography{refer}

\begin{thebibliography}{258}

\bibitem{Cle62b}
C.F. Clement, Phys. Rev. \textbf{128}, 2728 (1962)

\bibitem{Tan85a}
I.~Tanihata, H.~Hamagaki, O.~Hashimoto, S.~Nagamiya, Y.~Shida, N.~Yoshikawa,
  O.~Yamakawa, K.~Sugimoto, T.~Kobayashi, D.~Greiner et~al., Phys. Lett. B
  \textbf{160}, 380 (1985)

\bibitem{Tan85b}
I.~{Tanihata et al}, Phys. Rev. Lett. \textbf{55}, 2676 (1985)

\bibitem{Kob88}
T.~Kobayashi, O.~Yamakawa, K.~Omata, K.~Sugimoto, T.~Shimoda, N.~Takahashi,
  I.~Tanihata, Phys. Rev. Lett. \textbf{60}, 2599 (1988)

\bibitem{San08}
A.~S{\'a}nchez-Ben{\'\i}tez, D.~Escrig, M.~{\'A}lvarez, M.~Andr{\'e}s,
  C.~Angulo, M.~Borge, J.~Cabrera, S.~Cherubini, P.~Demaret, J.~Espino et~al.,
  Nucl. Phys. A \textbf{803}, 30 (2008)

\bibitem{Pie12}
A.~{Di Pietro} et~al., Phys. Rev. C \textbf{85}, 054607 (2012)

\bibitem{Cub12}
M.~Cubero et~al., Phys. Rev. Lett. \textbf{109}, 262701 (2012)

\bibitem{Pes17}
V.~Pesudo et~al., Phys. Rev. Lett. \textbf{118}, 152502 (2017)

\bibitem{Das99}
M.~Dasgupta, D.J. Hinde, R.D. Butt, R.M. Anjos, A.C. Berriman, N.~Carlin,
  P.R.S. Gomes, C.R. Morton, J.O. Newton, A.~Szanto~de Toledo et~al., Phys.
  Rev. Lett. \textbf{82}, 1395 (1999)

\bibitem{Tri02}
V.~Tripathi, A.~Navin, K.~Mahata, K.~Ramachandran, A.~Chatterjee, S.~Kailas,
  Phys. Rev. Lett. \textbf{88}, 172701 (2002)

\bibitem{Das02}
M.~{Dasgupta}, D.J. {Hinde}, K.~{Hagino}, S.B. {Moraes}, P.R.S. {Gomes}, R.M.
  {Anjos}, R.D. {Butt}, A.C. {Berriman}, N.~{Carlin}, C.R. {Morton} et~al.,
  Phys. Rev. C \textbf{66}, 041602 (2002)

\bibitem{Das04}
M.~Dasgupta et~al., Phys. Rev. C \textbf{70}, 024606 (2004)

\bibitem{Muk06}
A.~Mukherjee, S.~Roy, M.~Pradhan, M.S. Sarkar, P.~Basu, B.~Dasmahapatra,
  T.~Bhattacharya, S.~Bhattacharya, S.~Basu, A.~Chatterjee et~al., Phys. Lett.
  B \textbf{636}, 91 (2006)

\bibitem{Rat09}
P.K. Rath, S.~Santra, N.L. Singh, R.~Tripathi, V.V. Parkar, B.K. Nayak,
  K.~Mahata, R.~Palit, S.~Kumar, S.~Mukherjee et~al., Phys. Rev. C \textbf{79},
  051601 (2009)

\bibitem{LFC15}
L.~Canto, P.~Gomes, R.~Donangelo, J.~Lubian, M.~Hussein, Phys. Rep.
  \textbf{596}, 1 (2015)

\bibitem{Aus87}
N.~Austern, Y.~Iseri, M.~Kamimura, M.~Kawai, G.~Rawitscher, M.~Yahiro, Phys.
  Rep. \textbf{154}, 125 (1987)

\bibitem{Sat83}
G.~Satchler, \emph{Direct Nuclear Reactions} (Oxford University Press, New
  York, (1983))

\bibitem{Fesh58}
H.~Feshbach, Ann. of Phys. \textbf{5}, 357 (1958)

\bibitem{Ald75}
K.~Alder, A.~Winther, \emph{Electromagnetic excitation: Theory of Coulomb
  excitation with heavy ions} (North-Holland, Amsterdam, 1975)

\bibitem{Rod82}
N.L. Rodning, L.D. Knutson, W.G. Lynch, M.B. Tsang, Phys. Rev. Lett.
  \textbf{49}, 909 (1982)

\bibitem{Fri89}
S.~Fricke, P.~Hatchell, K.~McVoy, G.~Satchler, Nucl. Phys. A \textbf{500}, 399
  (1989)

\bibitem{Bar74}
A.~Barnett, J.~Lilley, Phys. Rev. C \textbf{9}, 2010 (1974)

\bibitem{Aco11}
L.~Acosta, A.~S{\'a}nchez-Ben{\'\i}tez, M.~G{\'o}mez, I.~Martel,
  F.~P{\'e}rez-Bernal, F.~Pizarro, J.~Rodr{\'\i}guez-Quintero, K.~Rusek,
  M.~Alvarez, M.~Andr{\'e}s et~al., Phys. Rev. C \textbf{84}, 044604 (2011)

\bibitem{Raw74}
G.H. Rawitscher, Phys. Rev. C \textbf{9}, 2210 (1974)

\bibitem{Yah86}
M.~Yahiro, Y.~Iseri, H.~Kameyama, M.~Kamimura, M.~Kawai, Prog. Theor. Phys.
  Supp. \textbf{89}, 32 (1986)

\bibitem{Kaw86a}
M.~Kawai, Prog. Part. Nucl. Phys. Suppl. \textbf{89}, 11 (1986)

\bibitem{Mor09b}
A.M. Moro, J.M. Arias, J.~G{\'o}mez-Camacho, F.~P{\'e}rez-Bernal, Phys. Rev. C
  \textbf{80}, 054605 (2009)

\bibitem{Joa75}
C.J. Joachain, \emph{Quantum collision theory} (North-Holland; Amsterdam, The
  Netherlands, 1975), ISBN 0720402948

\bibitem{fresco}
I.J. Thompson, Comp. Phys. Rep. \textbf{7}, 167 (1988)

\bibitem{Hag22}
K.~Hagino, K.~Ogata, A.M. Moro, Prog. Part. Nucl. Phys. \textbf{125}, 103951
  (2022)

\bibitem{Glen04}
N.K. Glendenning, \emph{Direct nuclear reactions} (World scientific, 2004)

\bibitem{Maj78}
Z.~Majka, H.J. Gils, H.~Rebel, Zeitschrift f{\"u}r Physik A Atoms and Nuclei
  \textbf{288}, 139 (1978)

\bibitem{Oga16}
K.~Ogata, K.~Yoshida, Phys. Rev. C \textbf{94}, 051603 (2016)

\bibitem{AGS}
E.~Alt, P.~Grassberger, W.~Sandhas, Nucl. Phys. B \textbf{2}, 167 (1967)

\bibitem{Upa12}
N.J. {Upadhyay}, A.~{Deltuva}, F.M. {Nunes}, Phys. Rev. C \textbf{85}, 054621
  (2012)

\bibitem{Cre11}
R.~{Crespo}, A.~{Deltuva}, A.M. {Moro}, Phys. Rev. C \textbf{83}, 044622 (2011)

\bibitem{Mor12}
A.M. Moro, R.~Crespo, Phys. Rev. C \textbf{85}, 054613 (2012)

\bibitem{Sum06}
N.C. {Summers} et~al., Phys. Rev. C \textbf{74}, 014606 (2006)

\bibitem{Die14}
R.~{de Diego}, J.M. {Arias}, J.A. {Lay}, A.M. {Moro}, Phys. Rev. C \textbf{89},
  064609 (2014)

\bibitem{Sum07}
N.C. {Summers}, F.M. {Nunes}, Phys. Rev. C \textbf{76}, 014611 (2007)

\bibitem{Sum14}
N.C. Summers, F.M. Nunes, I.J. Thompson, Phys. Rev. C \textbf{89}, 069901
  (2014)

\bibitem{Die17}
R.~de~Diego, R.~Crespo, A.M. Moro, Phys. Rev. C \textbf{95}, 044611 (2017)

\bibitem{Lay12}
J.A. Lay, A.M. Moro, J.M. Arias, J.~G\'omez-Camacho, Phys. Rev. C \textbf{85},
  054618 (2012)

\bibitem{Tar03}
T.~Tarutina, L.C. Chamon, M.S. Hussein, Phys. Rev. C \textbf{67}, 044605 (2003)

\bibitem{Mor12b}
A.M. Moro, J.A. Lay, Phys. Rev. Lett. \textbf{109}, 232502 (2012)

\bibitem{Chau11}
P.~{Chau Huu-Tai}, Jour. of Phys.: Conf. Series \textbf{312}, 082018 (2011)

\bibitem{Gom17a}
M.~G\'omez-Ramos, A.M. Moro, Phys. Rev. C \textbf{95}, 034609 (2017)

\bibitem{Kis76}
A.~Kiss, O.~Aspelund, G.~Hrehuss, K.~Knöpfle, M.~Rogge, U.~Schwinn, Z.~Seres,
  P.~Turek, C.~Mayer-Böricke, Nucl. Phys. A \textbf{262}, 1  (1976)

\bibitem{Del16}
A.~Deltuva, Nucl. Phys. A \textbf{947}, 173  (2016)

\bibitem{CH89}
R.~Varner, W.~Thompson, T.~McAbee, E.~Ludwig, T.~Clegg, Phys. Rep.
  \textbf{201}, 57  (1991)

\bibitem{Mat04a}
T.~{Matsumoto}, E.~{Hiyama}, M.~{Yahiro}, K.~{Ogata}, Y.~{Iseri},
  M.~{Kamimura}, Nucl. Phys. A \textbf{738}, 471 (2004)

\bibitem{Mat04b}
T.~{Matsumoto}, E.~{Hiyama}, K.~{Ogata}, Y.~{Iseri}, M.~{Kamimura}, S.~{Chiba},
  M.~{Yahiro}, Phys. Rev. C \textbf{70}, 061601(R) (2004)

\bibitem{manoli08}
M.~Rodr{\'i}guez-Gallardo, J.M. Arias, J.~G{\'o}mez-Camacho, R.C. Johnson, A.M.
  Moro, I.J. Thompson, J.A. Tostevin, Phys. Rev. C \textbf{77}, 064609 (2008)

\bibitem{manoli09}
M.~Rodr\'{\i}guez-Gallardo, J.M. Arias, J.~G\'omez-Camacho, A.M. Moro, I.J.
  Thompson, J.A. Tostevin, Phys. Rev. C \textbf{80}, 051601 (2009)

\bibitem{casal15}
J.~Casal, M.~Rodr\'{\i}guez-Gallardo, J.M. Arias, Phys. Rev. C \textbf{92},
  054611 (2015)

\bibitem{Mat06}
T.~Matsumoto, T.~Egami, K.~Ogata, Y.~Iseri, M.~Kamimura, M.~Yahiro, Phys. Rev.
  C \textbf{73}, 051602 (2006)

\bibitem{Matsumoto19}
T.~Matsumoto, J.~Tanaka, K.~Ogata, Prog. Theor. Exp. Phys. \textbf{2019},
  123D02 (2019)

\bibitem{Zhu93}
M.~{Zhukov et al}, Phys.\ Rep. \textbf{231}, 151 (1993)

\bibitem{Desc15}
P.~Descouvemont, T.~Druet, L.F. Canto, M.S. Hussein, Phys. Rev. C \textbf{91},
  024606 (2015)

\bibitem{Rod09}
M.~Rodr\'{\i}guez-Gallardo, J.M. Arias, J.~G\'omez-Camacho, A.M. Moro, I.J.
  Thompson, J.A. Tostevin, Phys. Rev. C \textbf{80}, 051601 (2009)

\bibitem{Rod08}
M.~Rodr\'{\i}guez-Gallardo, J.M. Arias, J.~G\'omez-Camacho, R.C. Johnson, A.M.
  Moro, I.J. Thompson, J.A. Tostevin, Phys. Rev. C \textbf{77}, 064609 (2008)

\bibitem{Descouvemont20}
P.~Descouvemont, Phys. Rev. C \textbf{101}, 064611 (2020)

\bibitem{Singh21}
J.~Singh, T.~Matsumoto, T.~Fukui, K.~Ogata, Phys. Rev. C \textbf{104}, 034612
  (2021)

\bibitem{tanaka17}
J.~Tanaka, R.~Kanungo, M.~Alcorta, N.~Aoi, H.~Bidaman, C.~Burbadge,
  G.~Christian, S.~Cruz, B.~Davids, A.~{Diaz Varela} et~al., Phys. Lett. B
  \textbf{774}, 268 (2017)

\bibitem{mrg11proc}
M.~Rodr\'{i}guez-Gallardo, A.M. Moro, Int. J. Mod. Phys. E \textbf{20}, 947
  (2011)

\bibitem{Wooll2004}
R.J. Woolliscroft, B.R. Fulton, R.L. Cowin, M.~Dasgupta, D.J. Hinde, C.R.
  Morton, A.C. Berriman, Phys. Rev. C \textbf{69}, 044612 (2004)

\bibitem{Yu2010}
N.~Yu, H.Q. Zhang, H.M. Jia, S.T. Zhang, M.~Ruan, F.~Yang, Z.D. Wu, X.X. Xu,
  C.L. Bai, J. Phys. G: Nucl. Part. Phys. \textbf{37}, 075108 (2010)

\bibitem{Des13}
P.~Descouvemont, M.S. Hussein, Phys. Rev. Lett. \textbf{111}, 082701 (2013)

\bibitem{Des18}
P.~Descouvemont, N.~Itagaki, Phys. Rev. C \textbf{97}, 014612 (2018)

\bibitem{Mar96}
I.~Martel, J.~Gómez-Camacho, K.~Rusek, G.~Tungate, Nucl. Phys. A \textbf{605},
  417  (1996)

\bibitem{Descouvemont2018}
P.~Descouvemont, E.C. Pinilla, Few-Body Syst. \textbf{60}, 11 (2018)

\bibitem{Coo18}
K.J. Cook, I.P. Carter, E.C. Simpson, M.~Dasgupta, D.J. Hinde, L.T. Bezzina,
  S.~Kalkal, C.~Sengupta, C.~Simenel, B.M.A. Swinton-Bland et~al., Phys. Rev. C
  \textbf{97}, 021601 (2018)

\bibitem{Barioni09}
A.~Barioni, V.~Guimar\~aes, A.~L\'epine-Szily, R.~Lichtenth\"aler, D.R. Mendes,
  E.~Crema, K.C.C. Pires, M.C. Morais, V.~Morcelle, P.N. de~Faria et~al., Phys.
  Rev. C \textbf{80}, 034617 (2009)

\bibitem{Des17}
P.~Descouvemont, L.F. Canto, M.S. Hussein, Phys. Rev. C \textbf{95}, 014604
  (2017)

\bibitem{Mor07}
A.M. Moro, K.~Rusek, J.M. Arias, J.~G{\'o}mez-Camacho,
  M.~Rodr{\'\i}guez-Gallardo, Phys. Rev. C \textbf{75}, 064607 (2007)

\bibitem{Bon04}
A.~Bonaccorso, D.M. Brink, C.A. Bertulani, Phys. Rev. C \textbf{69}, 024615
  (2004)

\bibitem{Kum12}
H.~Kumawat, V.~Jha, V.V. Parkar, B.J. Roy, S.K. Pandit, R.~Palit, P.K. Rath,
  C.S. Palshetkar, S.K. Sharma, S.~Thakur et~al., Phys. Rev. C \textbf{86},
  024607 (2012)

\bibitem{Yan16}
Y.Y. Yang, X.~Liu, D.Y. Pang, Phys. Rev. C \textbf{94}, 034614 (2016)

\bibitem{Mor02}
A.M. Moro, R.~Crespo, F.~Nunes, I.J. Thompson, Phys. Rev. C \textbf{66}, 024612
  (2002)

\bibitem{Mat03}
T.~{Matsumoto}, T.~{Kamizato}, K.~{Ogata}, Y.~{Iseri}, E.~{Hiyama},
  M.~{Kamimura}, M.~{Yahiro}, Phys. Rev. C \textbf{68}, 064607 (2003)

\bibitem{Tos01}
J.~Tostevin, Nucl. Phys. A \textbf{682}, 320c  (2001)

\bibitem{Ise86}
Y.~Iseri, M.~Yahiro, M.~Kamimura, Prog. Theor. Phys. Supp. \textbf{89}, 84
  (1986)

\bibitem{Fuc82}
H.~Fuchs, Nucl. Instrum. Methods Phys. Res. \textbf{200}, 361 (1982)

\bibitem{Cap04}
P.~Capel, G.~Goldstein, D.~Baye, Phys. Rev. C \textbf{70}, 064605 (2004)

\bibitem{Rod11}
M.~RODR{\'I}GUEZ-GALLARDO, A.~Moro, International Journal of Modern Physics E
  \textbf{20}, 947 (2011)

\bibitem{Ron70}
R.C. Johnson, P.J.R. Soper, Phys. Rev. C \textbf{1}, 976 (1970)

\bibitem{Har71}
J.D. Harvey, R.C. Johnson, Phys. Rev. C \textbf{3}, 636 (1971)

\bibitem{Sat71}
G.R. Satchler, Phys. Rev. C \textbf{4}, 1485 (1971)

\bibitem{Wal76}
G.L. Wales, R.C. Johnson, Nucl. Phys. A \textbf{274}, 168 (1976)

\bibitem{Ste86}
E.J. Stephenson, R.C. Johnson, J.A. Tostevin, V.R. Cupps, J.D. Brown, C.C.
  Foster, J.A. Gering, W.P. Jones, D.A. Low, D.W. Miller et~al., Phys. Lett. B
  \textbf{171}, 358  (1986)

\bibitem{Ron89}
R.C. Johnson, E.J. Stephenson, J.A. Tostevin, Nucl. Phys. A \textbf{505}, 26
  (1989)

\bibitem{JT74}
R.C. Johnson, P.C. Tandy, Nucl. Phys. A \textbf{235}, 56 (1974)

\bibitem{Lai93}
A.~Laid, J.A. Tostevin, R.C. Johnson, Phys. Rev. C \textbf{48}, 1307 (1993)

\bibitem{Pan13}
D.Y. Pang, N.K. Timofeyuk, R.C. Johnson, J.A. Tostevin, Phys. Rev. C
  \textbf{87}, 064613 (2013)

\bibitem{DCV80}
W.W. Daehnick, J.D. Childs, Z.~Vrcelj, Phys. Rev. {\bf C21}, 2253  ((1980))

\bibitem{Tim20}
N.K. Timofeyuk, R.C. Johnson, Prog. Part. Nucl. Phys. \textbf{111}, 103738
  (2020)

\bibitem{Gol64}
M.~Goldberger, K.~Watson, \emph{Collision Theory} (Wiley, New York, 1964)

\bibitem{Tim99}
N.K. {Timofeyuk}, R.C. {Johnson}, Phys. Rev. \textbf{C59}, 1545 (1999)

\bibitem{Pan14}
D.~Pang, A.~Mukhamedzhanov, Physical Review C \textbf{90}, 044611 (2014)

\bibitem{Cha17}
Y.~Chazono, K.~Yoshida, K.~Ogata, Phys. Rev. C \textbf{95}, 064608 (2017)

\bibitem{Bri72}
D.~Brink, Phys. Lett. B \textbf{40}, 37 (1972)

\bibitem{Cok74}
W.R. Coker, Phys. Rev. C \textbf{9}, 784 (1974)

\bibitem{Hub65}
R.~Huby, J.R. Mines, Rev. Mod. Phys. \textbf{37}, 406 (1965)

\bibitem{VF70}
C.M. {Vincent}, H.T. {Fortune}, Phys. Rev. C \textbf{2}, 782 (1970)

\bibitem{Jep06}
H.B. Jeppesen et~al., Phys. Lett. B \textbf{642}, 449 (2006)

\bibitem{Cav17}
M.~Cavallaro et~al., Phys. Rev. Lett. \textbf{118}, 012701 (2017)

\bibitem{Mor19}
A.M. Moro, J.~Casal, M.~G{\'o}mez-Ramos, Phys. Lett. B \textbf{793}, 13 (2019)

\bibitem{Gom17c}
M.~G{\'o}mez-Ramos, J.~Casal, A.M. Moro, Phys. Lett. B \textbf{772}, 115 (2017)

\bibitem{Des03}
P.~Descouvemont, C.~Daniel, D.~Baye, Phys. Rev. C \textbf{67}, 044309 (2003)

\bibitem{FaCE}
I.J. Thompson, F.M. Nunes, B.V. Danilin, Comput. Phys. Commun. \textbf{161}, 87
   (2004)

\bibitem{Rod05}
M.~Rodr\'{\i}guez-Gallardo, J.M. Arias, J.~G\'omez-Camacho, A.M. Moro, I.J.
  Thompson, J.A. Tostevin, Phys. Rev. C \textbf{72}, 024007 (2005)

\bibitem{Cas17}
J.~Casal, M.~G{\'o}mez-Ramos, A.M. Moro, Phys. Lett. B \textbf{767}, 307 (2017)

\bibitem{San16}
A.~Sanetullaev, R.~Kanungo, J.~Tanaka, M.~Alcorta, C.~Andreoiu, P.~Bender, A.A.
  Chen, G.~Christian, B.~Davids, J.~Fallis et~al., Phys. Lett. B \textbf{755},
  481  (2016)

\bibitem{Mor15}
A.M. Moro, Phys. Rev. C \textbf{92}, 044605 (2015)

\bibitem{Aks08}
Y.~Aksyutina, H.T. Johansson, P.~Adrich, F.~Aksouh, T.~Aumann, K.~Boretzky,
  M.J.G. Borge, A.~Chatillon, L.V. Chulkov, D.~Cortina-Gil et~al., Phys. Lett.
  B \textbf{666}, 430  (2008)

\bibitem{Gom17b}
M.~G{\'o}mez-Ramos, A.M. Moro, Phys. Rev. C \textbf{95}, 044612 (2017)

\bibitem{Nun96}
F.M. Nunes, J.A. Christley, I.J. Thompson, R.C. Johnson, V.~Efros, Nucl. Phys.
  A \textbf{609}, 43  (1996)

\bibitem{HM85}
M.S. Hussein, K.W. McVoy, Nucl. Phys. A \textbf{445}, 124  (1985)

\bibitem{Ich85}
M.~Ichimura, N.~Austern, C.M. Vincent, Phys. Rev. C \textbf{32}, 431 (1985)

\bibitem{Jin19b}
J.~Lei, A.M. Moro, Phys. Rev. Lett. \textbf{122}, 042503 (2019)

\bibitem{Pot15}
G.~Potel, F.M. Nunes, I.J. Thompson, Phys. Rev. C \textbf{92}, 034611 (2015)

\bibitem{Uda81}
T.~Udagawa, T.~Tamura, Phys. Rev. C \textbf{24}, 1348 (1981)

\bibitem{Jin15b}
J.~Lei, A.M. Moro, Phys. Rev. C \textbf{92}, 061602 (2015)

\bibitem{Car16}
B.V. Carlson, R.~Capote, M.~Sin, Few-Body Systems \textbf{57}, 307 (2016)

\bibitem{Jin15}
J.~Lei, A.M. {Moro}, Phys. Rev. C \textbf{92}, 044616 (2015)

\bibitem{Jin17}
J.~Lei, A.M. Moro, Phys. Rev. C \textbf{95}, 044605 (2017)

\bibitem{Pot17}
G.~Potel, G.~Perdikakis, B.V. Carlson, M.C. Atkinson, W.H. Dickhoff, J.E.
  Escher, M.S. Hussein, J.~Lei, W.~Li, A.O. Macchiavelli et~al., Eur. Phys. J.
  A \textbf{53}, 178 (2017)

\bibitem{Vil24}
G.~Villanueva, A.M. Moro, J.~Casal, J.~Lei, Phys. Lett. B \textbf{855} (2024)

\bibitem{gollan21}
F.~Gollan, D.~Abriola, A.~Arazi, M.A. Cardona, E.~de~Barbar\'a, J.~de~Jes\'us,
  D.~Hojman, R.M.I. Betan, J.~Lubian, A.J. Pacheco et~al., Phys. Rev. C
  \textbf{104}, 024609 (2021)

\bibitem{Mar02}
H.D. Marta, L.F. Canto, R.~Donangelo, P.~Lotti, Phys. Rev. C \textbf{66},
  024605 (2002)

\bibitem{Han03}
P.G. Hansen, J.A. Tostevin, Annu. Rev. Nucl. Sci. \textbf{53}, 219 (2003)

\bibitem{Tos14}
J.A. Tostevin, A.~Gade, Phys. Rev. C \textbf{90}, 057602 (2014)

\bibitem{Bla91}
B.~{Blank et al}, Z. Phys. {\bf A340}, 41  (1991)

\bibitem{Can06}
L.~Canto, P.~Gomes, R.~Donangelo, M.~Hussein, Phys. Rep. \textbf{424}, 1 (2006)

\bibitem{Can15}
L.F. Canto, P.R.S. Gomes, R.~Donangelo, J.~Lubian, M.S. Hussein, Phys. Rep.
  \textbf{596}, 1 (2015)

\bibitem{Can21}
{L.F. Canto}, {K. Hagino}, {M. Ueda}, Eur. Phys. J. A \textbf{57}, 11 (2021)

\bibitem{Zha14}
N.T. Zhang, Y.D. Fang, P.R.S. Gomes, J.~Lubian, M.L. Liu, X.H. Zhou, G.S. Li,
  J.G. Wang, S.~Guo, Y.H. Qiang et~al., Phys. Rev. C \textbf{90}, 024621 (2014)

\bibitem{Fan15}
Y.D. Fang, P.R.S. Gomes, J.~Lubian, M.L. Liu, X.H. Zhou, D.R. Mendes~Junior,
  N.T. Zhang, Y.H. Zhang, G.S. Li, J.G. Wang et~al., Phys. Rev. C \textbf{91},
  014608 (2015)

\bibitem{Dia07}
A.~Diaz-Torres, D.J. Hinde, J.A. Tostevin, M.~Dasgupta, L.R. Gasques, Phys.
  Rev. Lett. \textbf{98}, 152701 (2007)

\bibitem{Coo16}
K.J. Cook, E.C. Simpson, D.H. Luong, S.~Kalkal, M.~Dasgupta, D.J. Hinde, Phys.
  Rev. C \textbf{93}, 064604 (2016)

\bibitem{Hag00}
K.~Hagino, A.~Vitturi, C.H. Dasso, S.M. Lenzi, Phys. Rev. C \textbf{61}, 037602
  (2000)

\bibitem{Dia02}
A.~Diaz-Torres, I.J. Thompson, Phys. Rev. C \textbf{65}, 024606 (2002)

\bibitem{Sig98}
C.~Signorini, Z.~Liu, A.~Yoshida, T.~Fukuda, Z.~Li, K.~L{\"o}bner,
  L.~M{\"u}ller, Y.~Pu, K.~Rudolph, F.~Soramel et~al., Eur. Phys. J. A
  \textbf{2}, 227 (1998)

\bibitem{Hash09}
S.~Hashimoto, K.~Ogata, S.~Chiba, M.~Yahiro, Prog. Theor. Phys. \textbf{122},
  1291 (2009)

\bibitem{Ye09}
T.~Ye, Y.~Watanabe, K.~Ogata, Phys. Rev. C \textbf{80}, 014604 (2009)

\bibitem{Par16}
V.V. Parkar, V.~Jha, S.~Kailas, Phys. Rev. C \textbf{94}, 024609 (2016)

\bibitem{Par18}
V.V. Parkar, S.K. Pandit, A.~Shrivastava, R.~Palit, K.~Mahata, V.~Jha,
  K.~Ramachandran, S.~Gupta, S.~Santra, S.K. Sharma et~al., Phys. Rev. C
  \textbf{98}, 014601 (2018)

\bibitem{Ran20}
J.~Rangel, M.R. Cortes, J.~Lubian, L.F. Canto, Phys. Lett. B \textbf{803},
  135337 (2020)

\bibitem{Cor20}
M.R. Cortes, J.~Rangel, J.L. Ferreira, J.~Lubian, L.F. Canto, Phys. Rev. C
  \textbf{102}, 064628 (2020)

\bibitem{Mor22}
A.M. Moro, J.~Lei, E.C. Simpson, J. Phys.: Conf. Ser. \textbf{2340}, 012034
  (2022)

\bibitem{Jin19a}
J.~Lei, A.M. Moro, Phys. Rev. Lett. \textbf{123}, 232501 (2019)

\bibitem{Rat19}
A.~Ratkiewicz, J.A. Cizewski, J.E. Escher, G.~Potel, J.T. Burke, R.J.
  Casperson, M.~McCleskey, R.A.E. Austin, S.~Burcher, R.O. Hughes et~al., Phys.
  Rev. Lett. \textbf{122}, 052502 (2019)

\bibitem{Abr08}
D.~Abriola, A.A. Sonzogni, Nucl. Data Sheets \textbf{109}, 2501 (2008)

\bibitem{Mus76}
A.R.d.L. Musgrove, B.J. Allen, J.W. Boldeman, R.L. Macklin, Nucl. Phys. A
  \textbf{270}, 108 (1976)

\bibitem{Typ94}
S.~{Typel}, G.~{Baur}, Phys.\ Rev. C \textbf{50}, 2104 (1994)

\bibitem{Esb96}
H.~Esbensen, G.F. Bertsch, Nucl. Phys. A \textbf{600}, 37 (1996)

\bibitem{Kid94}
T.~Kido, K.~Yabana, Y.~Suzuki, Phys. Rev. {\bf C50}, R1276  ((1994))

\bibitem{Typ01}
S.~{Typel}, G.~{Baur}, Phys.\ Rev. C \textbf{64}, 024601 (2001)

\bibitem{Gar06}
A.~{Garc{\'i}a-Camacho}, A.~{Bonaccorso}, D.M. {Brink}, Nucl. Phys. A
  \textbf{776}, 118 (2006)

\bibitem{Ber09}
C.~Bertulani (2009), preprint, \texttt{arXiv:0908.4307}

\bibitem{And94}
M.V. Andr{\'e}s, J.~G{\'o}mez-Camacho, M.A. Nagarajan, Nucl. Phys. A
  \textbf{579}, 273 (1994)

\bibitem{And95}
M.V. Andr{\'e}s, J.~G{\'o}mez-Camacho, M.A. Nagarajan, Nucl. Phys. A
  \textbf{583}, 817 (1995)

\bibitem{Bon88}
A.~Bonaccorso, D.M. Brink, Phys. Rev. C \textbf{38}, 1776 (1988)

\bibitem{Has78}
H.~Hasan, D.M. Brink, J. Phys. G: Nucl. Phys. \textbf{4}, 1573 (1978)

\bibitem{Has79}
H.~Hasan, D.M. Brink, J. Phys. G: Nucl. Phys. \textbf{5}, 771 (1979)

\bibitem{Mon85}
L.L. Monaco, D.M. Brink, J. Phys. G: Nucl. Phys. \textbf{11}, 935 (1985)

\bibitem{Bon87}
A.~Bonaccorso, D.M. Brink, L.L. Monaco, J. Phys. G: Nucl. Phys. \textbf{13},
  1407 (1987)

\bibitem{Jin21}
J.~Lei, A.~Bonaccorso, Phys. Lett. B \textbf{813}, 136032 (2021)

\bibitem{Fla12}
F.~Flavigny, A.~Obertelli, A.~Bonaccorso, G.F. Grinyer, C.~Louchart, L.~Nalpas,
  A.~Signoracci, Phys. Rev. Lett. \textbf{108}, 252501 (2012)

\bibitem{Bay05}
D.~Baye, P.~Capel, G.~Goldstein, Phys. Rev. Lett. \textbf{95}, 082502 (2005)

\bibitem{Gol06}
G.~Goldstein, D.~Baye, P.~Capel, Phys. Rev. C \textbf{73}, 024602 (2006)

\bibitem{Cap03}
P.~Capel, D.~Baye, V.S. Melezhik, Phys. Rev. C \textbf{68}, 014612 (2003)

\bibitem{Cap13}
P.~Capel, F.M. Nunes, H.~Esbensen, R.C. Johnson, \emph{Mechanisms of direct
  reactions with halo nuclei}, in \emph{J. Phys.: Conf. Ser.} (IOP Publishing,
  2013), Vol. 436, p. 012040

\bibitem{Bro19}
R.A. Broglia, F.~Barranco, A.~Idini, G.~Potel, E.~Vigezzi, Phys. Scr.
  \textbf{94}, 114002 (2019)

\bibitem{Nag05}
M.A. Nagarajan, S.M. Lenzi, A.~Vitturi, Eur. Phys. J. A \textbf{24}, 63 (2005)

\bibitem{Nak99}
T.~Nakamura, N.~Fukuda, T.~Kobayashi, N.~Aoi, H.~Iwasaki, T.~Kubo, A.~Mengoni,
  M.~Notani, H.~Otsu, H.~Sakurai et~al., Phys. Rev. Lett. \textbf{83}, 1112
  (1999)

\bibitem{Nak06}
T.~Nakamura, A.M. Vinodkumar, T.~Sugimoto, N.~Aoi, H.~Baba, D.~Bazin,
  N.~Fukuda, T.~Gomi, H.~Hasegawa, N.~Imai et~al., Phys. Rev. Lett.
  \textbf{96}, 252502 (2006)

\bibitem{Fuk04}
N.~Fukuda et~al., Phys. Rev. C \textbf{70}, 054606 (2004)

\bibitem{Pal03}
R.~Palit et~al. (LAND/FRS Collaboration), Phys. Rev. C \textbf{68}, 034318
  (2003)

\bibitem{Nak03}
T.~Nakamura, N.~Fukuda, N.~Aoi, H.~Iwasaki, T.~Kobayashi, T.~Kubo, A.~Mengoni,
  M.~Notani, H.~Otsu, H.~Sakurai et~al., Nucl. Phys. A \textbf{722}, C301
  (2003)

\bibitem{Sum08}
N.C. Summers, F.M. Nunes, Phys. Rev. C \textbf{78}, 011601 (2008)

\bibitem{Mor20}
A.M. Moro, J.A. Lay, J.~G{\'o}mez-Camacho, Phys. Lett. B \textbf{811}, 135959
  (2020)

\bibitem{Cal16}
A.~Calci, P.~Navr\'atil, R.~Roth, J.~Dohet-Eraly, S.~Quaglioni, G.~Hupin, Phys.
  Rev. Lett. \textbf{117}, 242501 (2016)

\bibitem{Esb92}
H.~Esbensen, G.~Bertsch, Nucl. Phys. A \textbf{542}, 310 (1992)

\bibitem{Ber07}
C.A. Bertulani, M.~S.~Hussein, Phys. Rev. C \textbf{76}, 051602 (2007)

\bibitem{mig73}
A.B. Migdal, Yadern. Fiz. \textbf{16}, 427 (1973), English translation Sov. J.
  Nucl. Phys., \textbf{16} 238. (1973)

\bibitem{Bert91}
G.F. Bertsch, H.~Esbensen, Ann. Phys. (N. Y.) \textbf{209}, 327 (1991)

\bibitem{Aum99}
T.~Aumann, D.~Aleksandrov, L.~Axelsson, T.~Baumann, M.J.G. Borge, L.V. Chulkov,
  J.~Cub, W.~Dostal, B.~Eberlein, T.W. Elze et~al., Phys. Rev. C \textbf{59},
  1252 (1999)

\bibitem{Zhu95}
M.V. Zhukov, B.~Jonson, Nucl. Phys. A \textbf{589}, 1 (1995)

\bibitem{Gri06}
L.V. Grigorenko, K.~Langanke, N.B. Shul'gina, M.V. Zhukov, Phys. Lett. B
  \textbf{641}, 254 (2006)

\bibitem{Cor23}
A.~Corsi, Y.~Kubota, J.~Casal, M.~G{\'o}-Ramos, A.M. Moro, G.~Authelet,
  H.~Baba, C.~Caesar, D.~Calvet, A.~Delbart et~al., Phys. Lett. B \textbf{840},
  137875 (2023)

\bibitem{Chu97}
L.V. Chulkov, T.~Aumann, D.~Aleksandrov, L.~Axelsson, T.~Baumann, M.J.G. Borge,
  R.~Collatz, J.~Cub, W.~Dostal, B.~Eberlein et~al., Phys. Rev. Lett.
  \textbf{79}, 201 (1997)

\bibitem{Sim99}
H.~Simon, D.~Aleksandrov, T.~Aumann, L.~Axelsson, T.~Baumann, M.J.G. Borge,
  L.V. Chulkov, R.~Collatz, J.~Cub, W.~Dostal et~al., Phys. Rev. Lett.
  \textbf{83}, 496 (1999)

\bibitem{Mar01}
K.~Markenroth, M.~Meister, B.~Eberlein, D.~Aleksandrov, T.~Aumann, L.~Axelsson,
  T.~Baumann, M.J.G. Borge, L.V. Chulkov, W.~Dostal et~al., Nuc. Phys. A
  \textbf{679}, 462 (2001)

\bibitem{Sim07}
H.~Simon, M.~Meister, T.~Aumann, M.J.G. Borge, L.V. Chulkov, U.D. Pramanik,
  T.W. Elze, H.~Emling, C.~Forss{\'e}n, H.~Geissel et~al., Nucl. Phys. A
  \textbf{791}, 267 (2007)

\bibitem{Cat84}
F.~Catara, A.~Insolia, E.~Maglione, A.~Vitturi, Phys. Rev. C \textbf{29}, 1091
  (1984)

\bibitem{Kub20}
Y.~Kubota, A.~Corsi, G.~Authelet, H.~Baba, C.~Caesar, D.~Calvet, A.~Delbart,
  M.~Dozono, J.~Feng, F.~Flavigny et~al., Phys. Rev. Lett. \textbf{125}, 252501
  (2020)

\bibitem{Mats06}
M.~Matsuo, Phys. Rev. C \textbf{73}, 044309 (2006)

\bibitem{Kik16}
Y.~Kikuchi, K.~Ogata, Y.~Kubota, M.~Sasano, T.~Uesaka, Prog. Theor. Exp. Phys.
  \textbf{2016}, 103D03 (2016)

\bibitem{Cas21}
J.~Casal, M.~G\'omez-Ramos, Phys. Rev. C \textbf{104}, 024618 (2021)

\bibitem{Hag07}
K.~Hagino, H.~Sagawa, J.~Carbonell, P.~Schuck, Phys. Rev. Lett. \textbf{99},
  022506 (2007)

\bibitem{Igo58}
G.~Igo, Phys. Rev. Lett. \textbf{1}, 72 (1958)

\bibitem{Lov17b}
A.E. Lovell, F.M. Nunes, J.~Sarich, S.M. Wild, Phys. Rev. C \textbf{95}, 024611
  (2017)

\bibitem{Lov18}
A.E. Lovell, F.M. Nunes, Phys. Rev. C \textbf{97}, 064612 (2018)

\bibitem{Kin18}
G.B. King, A.E. Lovell, F.M. Nunes, Phys. Rev. C \textbf{98}, 044623 (2018)

\bibitem{Kin19}
G.B. King, A.E. Lovell, L.~Neufcourt, F.M. Nunes, Phys. Rev. Lett.
  \textbf{122}, 232502 (2019)

\bibitem{Cat19}
M.~Catacora-Rios, G.B. King, A.E. Lovell, F.M. Nunes, Phys. Rev. C
  \textbf{100}, 064615 (2019)

\bibitem{Cat20}
M.~Catacora-Rios, G.B. King, A.E. Lovell, F.M. Nunes, Phys. Rev. C
  \textbf{104}, 064611 (2021)

\bibitem{Phi21}
D.R. Phillips, R.J. Furnstahl, U.~Heinz, T.~Maiti, W.~Nazarewicz, F.M. Nunes,
  M.~Plumlee, M.T. Pratola, S.~Pratt, F.G. Viens et~al., J. Phys. G: Nucl.
  Part. Phys. \textbf{48}, 072001 (2021)

\bibitem{Whi22}
T.R. Whitehead, T.~Poxon-Pearson, F.M. Nunes, G.~Potel, Phys. Rev. C
  \textbf{105}, 054611 (2022)

\bibitem{Heb23}
C.~Hebborn, T.R. Whitehead, A.E. Lovell, F.M. Nunes, Phys. Rev. C \textbf{108},
  014601 (2023)

\bibitem{Fur20}
R.J. Furnstahl, A.J. Garcia, P.J. Millican, X.~Zhang, Phys. Lett. B
  \textbf{809}, 135719 (2020)

\bibitem{Dri21}
C.~Drischler, M.~Quinonez, P.~Giuliani, A.~Lovell, F.~Nunes, Phys. Lett. B
  \textbf{823}, 136777 (2021)

\bibitem{Sur22}
O.~S\"urer, F.M. Nunes, M.~Plumlee, S.M. Wild, Phys. Rev. C \textbf{106},
  024607 (2022)

\bibitem{Whi21}
T.R. Whitehead, Y.~Lim, J.W. Holt, Phys. Rev. Lett. \textbf{127}, 182502 (2021)

\bibitem{Pru23}
C.D. Pruitt, J.E. Escher, R.~Rahman, Phys. Rev. C \textbf{107}, 014602 (2023)

\bibitem{Heb23b}
C.~Hebborn, F.M. Nunes, A.E. Lovell, Phys. Rev. Lett. \textbf{131}, 212503
  (2023)

\bibitem{Aum21}
T.~Aumann, C.~Barbieri, D.~Bazin, C.~Bertulani, A.~Bonaccorso, W.H. Dickhoff,
  A.~Gade, M.~G{\'o}mez-Ramos, B.P. Kay, A.M. Moro et~al., Prog. Part. Nucl.
  Phys. \textbf{118}, 103847 (2021)

\bibitem{Fes58a}
H.~Feshbach, Annu. Rev. Nucl. Sci. \textbf{8}, 49 (1958)

\bibitem{Per62}
F.~Perey, B.~Buck, Nucl. Phys. \textbf{32}, 353  (1962)

\bibitem{Gia76}
M.~Giannini, G.~Ricco, Ann. Phys. (N. Y.) \textbf{102}, 458 (1976)

\bibitem{Gia80}
M.~Giannini, G.~Ricco, A.~Zucchiatti, Ann. Phys. (N. Y.) \textbf{124}, 208
  (1980)

\bibitem{Tia15}
Y.~Tian, D.Y. Pang, Z.Y. Ma, Int. J. Mod. Phys. E \textbf{24}, 1550006 (2015)

\bibitem{KD03}
A.J. Koning, J.P. Delaroche, Nucl. Phys. A \textbf{713}, 231  (2003)

\bibitem{Cha07}
R.J. Charity, J.M. Mueller, L.G. Sobotka, W.H. Dickhoff, Phys. Rev. C
  \textbf{76}, 044314 (2007)

\bibitem{Dic10}
W.H. Dickhoff, D.~Van~Neck, S.J. Waldecker, R.J. Charity, L.G. Sobotka, Phys.
  Rev. C \textbf{82}, 054306 (2010)

\bibitem{Mah16}
M.H. Mahzoon, R.J. Charity, W.H. Dickhoff, H.~Dussan, S.J. Waldecker, Phys.
  Rev. Lett. \textbf{112}, 162503 (2014)

\bibitem{Mor24}
B.~Morillon, G.~Blanchon, P.~Romain, H.F. Arellano (2024), preprint,
  \texttt{arXiv:2403.05843}

\bibitem{Fie66}
H.~Fiedeldey, Nucl. Phys. \textbf{77}, 149 (1966)

\bibitem{Fie67}
H.~Fiedeldey, Nucl. Phys. A \textbf{96}, 463 (1967)

\bibitem{Aus65}
N.~Austern, Phys. Rev. \textbf{137}, B752 (1965)

\bibitem{Fra57}
W.E. Frahn, R.H. Lemmer, Il Nuovo Cimento (1955-1965) \textbf{5}, 1564 (1957)

\bibitem{Tit14}
L.J. Titus, F.M. Nunes, Phys. Rev. C \textbf{89}, 034609 (2014)

\bibitem{Del09c}
A.~{Deltuva}, Phys. Rev. C \textbf{79}, 021602 (2009)

\bibitem{Tim13}
N.K. Timofeyuk, R.C. Johnson, Phys. Rev. Lett. \textbf{110}, 112501 (2013)

\bibitem{Tim13b}
N.K. Timofeyuk, R.C. Johnson, Phys. Rev. C \textbf{87}, 064610 (2013)

\bibitem{Wal16}
S.J. Waldecker, N.K. Timofeyuk, Phys. Rev. C \textbf{94}, 034609 (2016)

\bibitem{Tit16}
L.J. Titus, F.M. Nunes, G.~Potel, Phys. Rev. C \textbf{93}, 014604 (2016)

\bibitem{Bai16}
G.W. Bailey, N.K. Timofeyuk, J.A. Tostevin, Phys. Rev. Lett. \textbf{117},
  162502 (2016)

\bibitem{Bai17}
G.W. Bailey, N.K. Timofeyuk, J.A. Tostevin, Phys. Rev. C \textbf{95}, 024603
  (2017)

\bibitem{Gom18}
M.~G\'omez-Ramos, N.K. Timofeyuk, Phys. Rev. C \textbf{98}, 011601 (2018)

\bibitem{Del18}
A.~Deltuva (2018), preprint, \texttt{arXiv:1806.00298}

\bibitem{Gom19}
M.~G\'omez-Ramos, N.K. Timofeyuk, J. Phys. G: Nucl. Part. Phys. \textbf{46},
  085102 (2019)

\bibitem{tim23}
N.K. Timofeyuk, M.~G\'omez-Ramos, Front. Phys. \textbf{11} (2023)

\bibitem{Li18}
W.~Li, G.~Potel, F.~Nunes, Phys. Rev. C \textbf{98}, 044621 (2018)

\bibitem{Heb21}
C.~Hebborn, F.M. Nunes, Phys. Rev. C \textbf{104}, 034624 (2021)

\bibitem{Del07}
A.~{Deltuva}, A.M. {Moro}, E.~{Cravo}, F.M. {Nunes}, A.C. {Fonseca}, Phys. Rev.
  C \textbf{76}, 064602 (2007)

\bibitem{Cre14}
R.~Crespo, A.~Deltuva, E.~Cravo, Phys. Rev. C \textbf{90}, 044606 (2014)

\bibitem{Cre16}
E.~Cravo, R.~Crespo, A.~Deltuva, Phys. Rev. C \textbf{93}, 054612 (2016)

\bibitem{Bar13}
B.R. Barrett, P.~Navrátil, J.P. Vary, Prog. Part. Nucl. Phys. \textbf{69}, 131
  (2013)

\bibitem{Vor19}
M.~Vorabbi, P.~Navr\'atil, S.~Quaglioni, G.~Hupin, Phys. Rev. C \textbf{100},
  024304 (2019)

\bibitem{Baroni13}
S.~Baroni, P.~Navr\'atil, S.~Quaglioni, Phys. Rev. C \textbf{87}, 034326 (2013)

\bibitem{Car15}
J.~Carlson, S.~Gandolfi, F.~Pederiva, S.C. Pieper, R.~Schiavilla, K.E. Schmidt,
  R.B. Wiringa, Rev. Mod. Phys. \textbf{87}, 1067 (2015)

\bibitem{Pie92}
S.C. Pieper, R.B. Wiringa, V.R. Pandharipande, Phys. Rev. C \textbf{46}, 1741
  (1992)

\bibitem{Fla13}
F.~Flavigny, A.~Gillibert, L.~Nalpas, A.~Obertelli, N.~Keeley, C.~Barbieri,
  D.~Beaumel, S.~Boissinot, G.~Burgunder, A.~Cipollone et~al., Phys. Rev. Lett.
  \textbf{110}, 122503 (2013)

\bibitem{Gri11}
G.F. Grinyer, D.~Bazin, A.~Gade, J.A. Tostevin, P.~Adrich, M.D. Bowen, B.A.
  Brown, C.M. Campbell, J.M. Cook, T.~Glasmacher et~al., Phys. Rev. Lett.
  \textbf{106}, 162502 (2011)

\bibitem{Cor19}
A.~Corsi, Y.~Kubota, J.~Casal, M.~G{\'o}mez-Ramos, A.~Moro, G.~Authelet,
  H.~Baba, C.~Caesar, D.~Calvet, A.~Delbart et~al., Phys. Lett. B \textbf{797},
  134843 (2019)

\bibitem{Due22}
M.~Duer, T.~Aumann, R.~Gernh{\"a}user, V.~Panin, S.~Paschalis, D.M. Rossi, N.L.
  Achouri, D.~Ahn, H.~Baba, C.A. Bertulani et~al., Nature \textbf{606}, 678
  (2022)

\end{thebibliography}

\end{document}